\newcommand{\arxivversion}{}
\documentclass[11pt]{article}
\usepackage[a4paper,margin=1in]{geometry}
\usepackage[T1]{fontenc}
\usepackage[utf8]{inputenc}
\usepackage{lmodern,microtype}
\usepackage{amsmath,amssymb,mathtools}
\usepackage{booktabs,tabularx,array,longtable}
\usepackage{graphicx}
\usepackage{placeins}
\usepackage{caption}
\graphicspath{{../figures/}}
\usepackage[numbers,sort&compress]{natbib}
\ifdefined\articleonly\else\ifdefined\sionly
  \newcommand{\citeSI}{\cite}
  \newcommand{\bibliographystyleSI}{\bibliographystyle}
  \newcommand{\bibliographySI}[1]{\renewcommand{\refname}{SI References}\bibliography{#1}}
\else
  \usepackage[resetlabels]{multibib}
  \newcites{SI}{SI References}
\fi\fi
\usepackage{xr-hyper}
\usepackage[hidelinks]{hyperref}
\usepackage{lineno}
\ifdefined\arxivversion\else\linenumbers\fi 

\usepackage{xcolor}
\usepackage{calc,fvextra,newunicodechar}
\newcounter{sisection}\renewcommand{\thesisection}{S\arabic{sisection}}
\newcommand{\sisection}[2]{\refstepcounter{sisection}\section*{SI Appendix, section \thesisection: #1}\label{#2}}
\providecommand{\tightlist}{\setlength{\itemsep}{0pt}\setlength{\parskip}{0pt}}
\newunicodechar{λ}{\ensuremath{\lambda}}\newunicodechar{τ}{\ensuremath{\tau}}
\newunicodechar{Φ}{\ensuremath{\Phi}}\newunicodechar{ε}{\ensuremath{\varepsilon}}
\newunicodechar{β}{\ensuremath{\beta}}\newunicodechar{Δ}{\ensuremath{\Delta}}
\newunicodechar{Σ}{\ensuremath{\Sigma}}\newunicodechar{−}{\ensuremath{-}}
\newunicodechar{≥}{\ensuremath{\ge}}\newunicodechar{≤}{\ensuremath{\le}}
\newunicodechar{→}{\ensuremath{\rightarrow}}\newunicodechar{←}{\ensuremath{\leftarrow}}
\newunicodechar{∈}{\ensuremath{\in}}\newunicodechar{≈}{\ensuremath{\approx}}
\newunicodechar{≠}{\ensuremath{\ne}}\newunicodechar{∓}{\ensuremath{\mp}}
\newunicodechar{√}{\ensuremath{\surd}}\newunicodechar{⁻}{\ensuremath{^{-}}}
\newunicodechar{⁹}{\ensuremath{^{9}}}\newunicodechar{¹}{\ensuremath{^{1}}}
\newunicodechar{²}{\ensuremath{^{2}}}

\newcommand{\papertitle}{A nudge among nudges:\\when the field pushes back}
\newcommand{\paperauthor}{Giuseppe Alessandro Veltri}
\newcommand{\paperemail}{gaveltri@nus.edu.sg}
\newcommand{\paperorcid}{0000-0002-9472-2236}
\newcommand{\paperaffiliations}{Behavioural and Implementation Science Interventions (BISI), Yong Loo Lin School of Medicine,\\National University of Singapore, Singapore\\[0.3em]Department of Sociology and Social Research, University of Trento, Trento, Italy\\[0.3em]\paperemail}

\graphicspath{{figures/}{../figures/}} 
\let\origfigure\figure\let\endorigfigure\endfigure
\renewenvironment{figure}[1][p]{\origfigure[tbp]}{\endorigfigure}

\title{\papertitle}
\author{\paperauthor\\[0.5em]\small\paperaffiliations\ \textperiodcentered\ ORCID \paperorcid}
\date{Preprint, November 2026}

\begin{document}
\maketitle

\begin{abstract}
Behavioural interventions are designed and tested as if the setting they enter stays put. But other actors push the same people, watch the behaviour and respond when it moves. The paper asks what becomes of a nudge in such a field. In an agent-based model, people act when their inclination and the pushes they feel cross a threshold, nudges compete for limited attention, and other actors steer the behaviour's rate towards their own goals without seeing the nudge. Calibrated to published effect sizes, it reproduces four documented regularities, one of them by calibration. The same nudge meets five fates, set by who else acts: attenuated, transient, null, reversed, or a blind site, where a sound field experiment reports nearly the full effect while its site gains little. Where the field reacts, a pilot's own comparison does not forecast the rollout: it cannot see the response the pilot provokes, and a small pilot does not exhaust the competitor as the rollout does. The fate follows from a few quantities, with nothing fitted after the calibration; the derivation matched the simulation on eight of ten maps of settings. Across 49,152 random settings, each with a competitor, the expected effect was the exception; what shifted the fate most depended on how attention is shared: the nudge's strength when the strongest nudge is heard first, the number of competing nudges when it is split. Whether a real field reacts can be read from marks a trial could record, such as drift among the untreated.
\end{abstract}

\section*{Significance Statement}

Behavioural interventions such as nudges often work in pilots and disappoint at scale. The usual explanations are poor implementation and selective publication. There is another: the setting pushes back. Other actors with their own goals watch the behaviour and respond when it moves. In a model, a pilot's own comparison cannot see that response, so it does not forecast the rollout, and a well-run trial can report nearly the full effect while its site gains little. The same nudge can be weakened, erased or reversed, depending on who else is acting. Each outcome leaves a trace a trial could record, such as drift in the untreated group. This suggests evaluations should record the other actors, not only the intervention.

\section*{Introduction}

Behavioural interventions, nudges among them~\cite{thaler2008}, are used to change how people exercise~\cite{milkman2021}, vote~\cite{mann2024}, use energy~\cite{allcott2015} and act on climate change~\cite{vlasceanu2024}, and governments have set up units to design and test them~\cite{dellavigna2022}. Four findings sit uneasily with how they are designed. Effects found in pilots shrink when an intervention is scaled up~\cite{alubaydli2020}. Trials run by two government nudge units in the United States reported average effects about a sixth the size of those in published academic trials~\cite{dellavigna2022}. Meta-analyses find effects that vary widely between studies, and that shrink once publication bias is allowed for, in one analysis to no evidence of an effect at all~\cite{mertens2022,maier2022,szaszi2025}. And the same intervention gives effects that differ from site to site by more than chance allows~\cite{bryan2021}.

Each finding has a standard reading. The loss at scale is put down to implementation~\cite{alubaydli2020}; the gap between government and academic trials, and the small averages, to low power and selective publication~\cite{dellavigna2022,maier2022}; the differences between sites to unmeasured moderators~\cite{bryan2021}, or to early sites chosen because they responded well~\cite{allcott2015}. The readings share an assumption. The setting an intervention enters is a backdrop: it may differ between pilot and rollout, or between sites, but it does not act on what the intervention does. Exceptions are known. Firms undo defaults they oppose~\cite{willis2013}, counter-nudges answer nudges~\cite{sunstein2017}, market reactions can reverse a nudge's predicted effect~\cite{spiegler2015}, and prices shift as a programme grows~\cite{heckman1998,lise2004}. Such responses are mostly treated as a threat to scaling~\cite{alubaydli2020} and seldom traced through to the treated and untreated groups of a trial, with exceptions~\cite{crepon2013,munro2025}.

The idea that behaviour is held in place by the forces around it is older than the nudge. Lewin described behaviour as an equilibrium of driving and restraining forces, in which added pressure tends to raise counter-pressure~\cite{lewin1951}. Following Lewin, the forces acting on a group of people, and the actors behind them, are called the field here. His idea has returned in several forms. A campaign's measured lift in online advertising depends on what competitors test at the same time~\cite{waisman2025}. Policies introduced together bias each other's evaluations~\cite{matthay2021}. In these versions the forces are given: they do not respond to what the intervention does to behaviour.

A field that responds is hard to study with the data that exist. Few trials record the other forces acting on the people they enrol. Three screenings of open repositories, about 120 candidate datasets in all, found none that could test what is specific to a responding field (\textit{SI Appendix}, section~\ref{si:methods}).

This paper builds a model that places a behavioural intervention among actors who watch the behaviour it targets and respond, and asks one question: what becomes of a nudge when the field reacts? The other actors are a competitor, an ally and a bystander; the bystander pushes some other behaviour and competes only for attention. Each pursues a goal for the rate of a behaviour; they watch that rate and never see the intervention. The main result is that the fate of the nudge is set by the field it enters, and that the trial which tests it can miss that fate. Where the field reacts, a pilot provokes a response in proportion to its size, and its own comparison cannot see that response. A small pilot also does not push the other actors to their limit, as the rollout does, so it does not show the rollout's fate. A field experiment can report nearly the full expected effect while its site gains little, a fate called a blind site here. Depending on the field, the same nudge is also attenuated, made transient, nulled or reversed. The fate follows from a few quantities by a short derivation, with nothing fitted, and the derivation agrees with the simulation on eight of ten simulated maps, grids of settings in which two properties of the field vary. Across random settings, what shifts the fate most depends on how attention is shared: the nudge's own strength where the strongest nudge is heard first, the number of bystanders where attention is shared in proportion. Each fate leaves a mark a trial could record. The model is calibrated to published effect sizes and fitted to no data. It shows what a reacting field can do, not how often real fields react.

\section*{Results}

\subsection*{A model of a field that reacts}

The model places one behavioural intervention, called the new nudge here, among other actors who respond to what they see (Fig.~1). Each person has a standing inclination towards the behaviour, and acts when that inclination, plus the pushes of the nudges that reach them and some chance, crosses a threshold (Fig.~1a). Each person also has a limited attention budget (Fig.~1b). Nudges that need attention, the new one among them, compete for it. A person attends first to the nudge they feel most strongly; this is the default rule, called priority sharing. Under the alternative, proportional sharing, every nudge that reaches a person gets an equal share of the budget (Methods). Automatic nudges, such as a pre-set option that applies unless a person opts out, act without attention.
\ifdefined\arxivversion\begin{figure}[p]\centering\includegraphics[width=\textwidth]{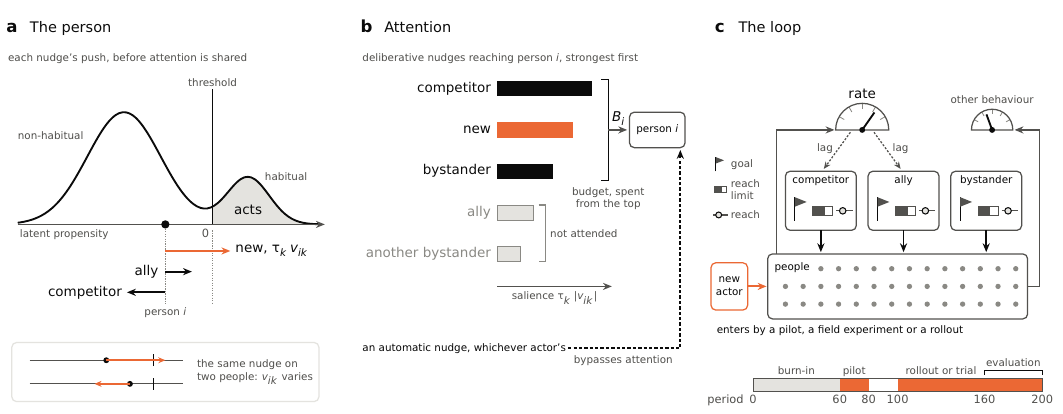}\caption{\textbf{A nudge is a force on a threshold; it shares a person's attention with other nudges; and the actors who own those nudges steer the rate of the behaviour with their own goals and reach limits, reacting to the rate, not to the new nudge.}
\textbf{a}, The person. The curve is the population's baseline propensity (non-habitual and habitual people); those beyond the threshold at 0 (shaded) act. Person $i$ is the dot. The arrows are three nudges on person $i$, the new nudge, an ally's and a competitor's, each its own push, $\tau_k v_{ik}$, before attention is shared; they are not summed. Inset: the same nudge on two people, longer for one and flipped for the other ($v_{ik}$ varies).
\textbf{b}, Attention. Person $i$'s deliberative nudges (those that need attention), ranked by salience $\tau_k|v_{ik}|$, spend the attention budget $B_i$ from the top (priority). Nudges beyond it (light) are not attended: here the bystander's nudge crowds out the ally's. An automatic nudge, whichever actor's (dashed), bypasses the budget.
\textbf{c}, The loop. The people of a site feed the rate (dial). A competitor, an ally and a bystander, each with a goal (flag), a limit on its reach (bar) and a reach (slider), observe a rate with a lag (dashed) and nudge the people (solid). The bystander promotes a different behaviour (second dial), keeps a fixed reach and competes for attention; nothing is fed back to it. No arrow runs from the new actor, the nudge's owner, to another actor. The timeline, in periods, marks burn-in, the pilot (60--79), the rollout or trial from 100 and the evaluation window (160--199); the new nudge is off over periods 80--99.
Response rules, not drawn: a thermostat steers the rate towards its goal; a maximiser pushes while the rate is above its reference. Lengths, the density and the dials are illustrative.}\end{figure}
\fi

A site has three kinds of other actors (Fig.~1c). A competitor pushes the same behaviour the other way, an ally pushes it the same way as the new nudge, and a bystander promotes a different behaviour and affects this behaviour only through attention. Each competitor or ally has a goal and chooses how many people its nudge reaches, up to a limit; a bystander keeps a fixed reach. A competitor or an ally watches the rate of its behaviour, the share of people who act in a period, with a delay; a period is read as a week. It never sees the new nudge. A competitor either steers, adjusting how many people its nudge reaches until the rate sits at its goal, or ratchets, reaching more people while the rate stays above the lowest level it has seen. Some competitors escalate: when the rate has stayed off target for ten periods in a row and they are at their limit, they switch on an automatic nudge that stays on (Methods gives the rule for each kind of competitor). The field is the other actors of a site and the nudges they send to its people.

A pilot treats a small random share of people. A field experiment treats a share $p$ of the site and compares them with the rest; the term is used for this design only. A rollout reaches everyone and is judged before and after.

With no other actors, the new nudge's effect is what the designer expected, written $E$. Any shortfall from $E$ splits in two, found by running the simulation three times on the same people with the same random numbers: with no other actors, with the actors held at what they were doing before the rollout, and with the actors reacting. The shortfall with the actors held still is the static part of the loss: what the field costs by standing there. The rest, which appears when the actors react, is the reaction part (Methods).

\subsection*{The model reproduces four known regularities}

The model was first checked against four regularities documented for nudges (\textit{SI Appendix}, Fig.~S3). Effects shrink at scale~\cite{alubaydli2020}. With a competitor or an ally present, the rollout keeps 0.62--0.74 of the pilot's effect, under either attention rule. Government trials report smaller effects than academic ones~\cite{dellavigna2022}. In the model the same trial reports less in a crowded field than in a quiet one. One crowded field, a stalemate in which a competitor and an ally are both at their limit, was calibrated on this gap. There the quiet field's reported effect is 2.25 times the crowded field's, on a scale that allows for how common the behaviour already is; the academic--government gap, measured the same way, is about 2. Effects differ between sites by more than chance~\cite{bryan2021}. Across six sites with different other actors, the spread is 11 times the sampling error within a site. And effects are largest where the rate without the nudge lies a little below one half~\cite{saccardo2025}. The model's peak is a plateau over a rate of 0.356--0.414, near the 0.423 at which a fixed push moves the most people across the threshold. These four are checks of consistency, not evidence for the model: the gap between government and academic trials is built in by calibration, the spread between sites is all but guaranteed by sites chosen to differ in their actors, and the peak near one half follows largely from the threshold.

\subsection*{The same nudge meets five fates}

The nudge is the same in every panel of Fig.~2. What differs is the other actors, how crowded attention is and how the nudge is tested. Each panel shows what the designer expected, what the trial reported and what the rollout delivered, averaged over 100 runs.
\ifdefined\arxivversion\begin{figure}[p]\centering\includegraphics[width=\textwidth]{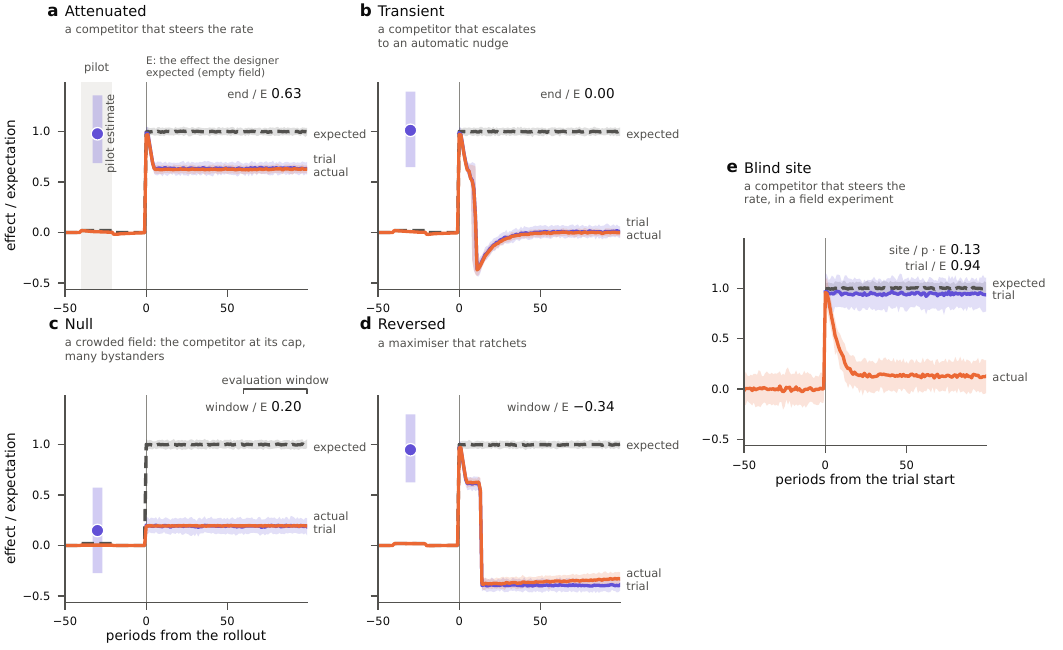}\caption{\textbf{The same nudge deployed into fields that differ in the other actors, in how crowded attention is and in how the nudge is tested: what the designer expected, what the trial reported and what the rollout delivered diverge, and the gap's sign and size follow from the other actors and the attention load.}
\textbf{a}--\textbf{d}, A pilot, then a rollout; periods are counted from the rollout at 0 (period 100 of the run). \textbf{a}, Attenuated: a thermostat competitor (one that steers the rate towards its goal), response gain 2. \textbf{b}, Transient: the same competitor, escalating to a persistent automatic nudge of strength $\tau$ (automatic) 0.30. \textbf{c}, Null: a crowded field, the competitor ($\tau$ 0.35) at its limit, 16 bystanders, attention budget 0.7. \textbf{d}, Reversed: a maximiser competitor (one that pushes while the rate is above its reference) with a persistent automatic nudge, $\tau$ (automatic) 0.30, and its reference the lowest rate it has ever seen. \textbf{e}, Blind site: a field experiment against the competitor of \textbf{a}, with 40\% of the site treated.
Lines show the expected (grey, dashed), trial-reported (violet) and actual (orange) effects over $E$, the effect the designer expected: the new nudge's effect on the site in the field with no other actors. In \textbf{a}--\textbf{c} the trial-reported line, a before-and-after estimate, lies under the actual line (within 0.02\,$E$ in the evaluation window); the two diverge in \textbf{d} and \textbf{e}. The violet dot with a bar is the pilot estimate: the pilot contrast as a share of $E$, averaged over the pilot window (periods $-40$ to $-21$, shaded in \textbf{a}); in \textbf{c} the pilot already runs in the crowded field. The evaluation window is periods 60--99 (bracketed over \textbf{c}). Each panel prints the quantity its outcome class reads: end\,/\,$E$, the mean of the last 10 periods (\textbf{a}, \textbf{b}); window\,/\,$E$, the evaluation-window mean (\textbf{c}, \textbf{d}); in \textbf{e}, site\,/\,$p\cdot E$, the site's effect over the site's expected effect ($p$, the share treated), and trial\,/\,$E$, the treated $-$ control contrast over $E$. Lines are replicate means; bands span the 10th--90th percentiles of $n = 100$ replicate runs per panel. Every fate drawn holds across runs and neighbouring settings (Fig.~3).}\end{figure}
\fi

\emph{Attenuated.} Against a competitor that steers the rate, the rollout keeps 0.63\,$E$ (Fig.~2a). The pilot had reported nearly the full expected effect: the response it provoked was small and fell on treated and untreated alike, so its comparison could not see it. The rollout moves that rate, and within a few periods the competitor reaches more people, up to its limit (\textit{SI Appendix}, Fig.~S2a).

\emph{Transient.} When the same competitor can escalate to an automatic nudge, the effect does not last (Fig.~2b). About ten periods after the rollout the automatic nudge comes on in every run (\textit{SI Appendix}, Fig.~S2b). The effect then dips below zero, to $-0.37\,E$, and climbs back as the competitor lowers its own reach. By the end it is erased: it stands at 0.00\,$E$.

\emph{Null.} An effect below a quarter of $E$ counts as null. Fig.~2c shows a heavily crowded field: one competitor already at its limit, and 16 bystanders competing for a tight attention budget. Even the pilot sees little (0.15\,$E$), because the crowd has taken the attention the new nudge needs before the rollout begins. The rollout delivers 0.20\,$E$. This crowded field is not the stalemate calibrated above: it has a competitor and bystanders but no ally (Methods), and the load is heavy enough to drive the competitor to its limit.

\emph{Reversed.} Against a competitor that ratchets, the behaviour ends below where it would have been without the new nudge, at $-0.34\,E$, although the pilot had again reported nearly the full effect (Fig.~2d). This competitor switches on its automatic nudge about 14 periods after the rollout. On the maps below (Fig.~3), a reversal that persists across runs and neighbouring settings needs such a competitor. Across the random settings described below, a competitor that steers the rate reverses the effect too, in a few per cent of them.

\emph{Blind site.} Only a field experiment shows the fifth fate (Fig.~2e). Facing the competitor of Fig.~2a, a trial that treats 40\% of the site reports 94\% of the effect expected for each treated person. The site as a whole gets 13\% of the gain expected from treating 40\% of it. The trial's comparison is intact, yet the site gains little. Across random settings this fate occurs in 4--16\% of the field experiments that received a fate (below).

In the rollouts of Fig.~2a--c a before-and-after comparison tracks the actual effect to within 0.02\,$E$; in Fig.~2d and 2e the reported and actual effects diverge. Two further fates were looked for, a delayed effect and an oscillating one. On the maps neither held across neighbouring settings. Across random settings both were rare: under 1\% of settings whatever the competitor's rule, the attention rule or the trial design, and under 2\% of the settings where the model also reproduces the known regularities (\textit{SI Appendix}, section~\ref{si:sweep}).

The loss splits into its static and reaction parts in proportions set by the field (\textit{SI Appendix}, Fig.~S1). In the null field of Fig.~2c the whole loss, 0.80\,$E$, is static: no actor there can respond. Wherever a competitor can respond, the field costs only about 0.03\,$E$ by standing there (the static part, plotted as $-0.03\,E$ in Fig.~S1), and the reaction carries almost all of the loss. An ally mixes the two. As the rate rises the ally withdraws its nudge, and the effect falls to 0.46\,$E$. The withdrawal accounts for 41\% of that loss; the rest is static.

\subsection*{What sets the fate}

Changing two properties of the field, with the new nudge held fixed, walks the nudge through all five fates (Fig.~3).
\ifdefined\arxivversion\begin{figure}[p]\centering\includegraphics[width=\textwidth]{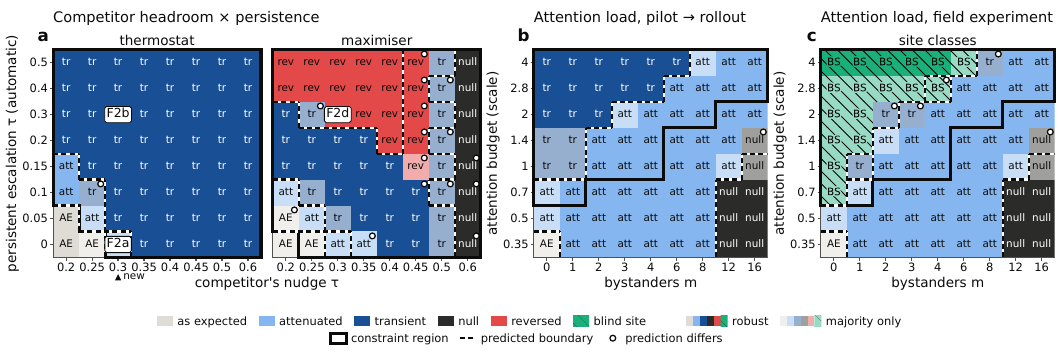}\caption{\textbf{Which fate a nudge meets is set by the other actors' response rule and headroom and by the attention load of the field, and, where the strongest nudge is heard first, by the nudge's own strength (\textit{SI Appendix}, Fig.~\ref{fig:sf1}): moving two field parameters walks the same nudge through attenuation, a transient effect, a null effect, reversal and a blind site, within the region where the model reproduces the known regularities.}
Each cell is the majority outcome over $n = 100$ replicate runs at that setting.
\textbf{a}, Competitor headroom and persistence, pilot then rollout: the strength $\tau$ of the competitor's nudge (the new nudge's is 0.30, the triangle labelled ``new'') against the strength of its persistent automatic nudge, $\tau$ (automatic), for a thermostat competitor (left) and a maximiser (right).
\textbf{b}, Attention load, pilot then rollout: the number of bystanders $m$ against the attention budget (scale). The scale is the mean of the people's attention budgets: each person's budget is the scale times a random multiplier that averages 1. The model's default scale is 1.
\textbf{c}, The same map in a field experiment, classed by what happens at the site.
A solid cell is robust: its outcome holds along $\tau$ (automatic) within each $\tau$ column in \textbf{a}, and along $m$ within each budget row in \textbf{b} and \textbf{c}; a faded cell is the majority outcome at that point only. The outline marks the cells where the rollout total stays below 0.8 of the pilot contrast (the first of the four checks, the loss at scale; \textit{SI Appendix}, Fig.~S3) and the new nudge's effect with no other actors lies within the model's magnitude bounds (Methods). Where the outline follows a map's edge, the region reaches that edge. The dashed line is the boundary between outcomes as the derivation predicts them from expected rates, without simulating runs (Methods); an open dot marks a cell where the prediction and the simulation differ in outcome or by more than 0.1 in size (as a share of $E$; in \textbf{c}, of the site's and the trial's expected effects), and the cell's colour is the simulation's. Cells that agree: \textbf{a}, 63 of 64 (thermostat) and 49 of 64 (maximiser, predicted in part); \textbf{b}, 71 of 72; \textbf{c}, 67 of 72. A map counts as predicted when at least 90\% of its cells agree. F2a, F2b and F2d mark the fields of Fig.~2a, b and d. The axes are categorical and unevenly spaced, drawn as equal cells with nothing interpolated. Cell labels name the outcomes for grayscale: AE, as expected; att, attenuated; tr, transient; null; rev, reversed; BS, blind site (hatched).}\end{figure}
\fi

Against a competitor that steers the rate, the fate depends on its strength and on whether it can escalate (Fig.~3a, left). With an automatic nudge to escalate to, the effect is transient over most of the map. Without one, a competitor as strong as the new nudge attenuates the effect. Under priority sharing a stronger competitor makes the effect fade; none remains once it is about one and a half times as strong as the new nudge. A competitor that is not yet at its limit has room left to push back, called headroom here. Its goal decides whether it has headroom before the rollout. In the calibrated settings the goal lies 0.03 below the rate without actors, a distance chosen so that the competitor has room before the rollout and runs out of it after. The random settings below vary that distance from 0 to 0.10. Against a competitor that ratchets, the effect reverses once the competitor's automatic nudge is strong enough (Fig.~3a, right).

On the map that varies the number of bystanders and the attention budget, the fate moves in a fixed order (Fig.~3b). As bystanders multiply and the attention budget tightens, the effect passes from transient to attenuated to null. Until the competitor reaches its limit, the effect barely moves with the number of bystanders (0.39--0.42\,$E$ at the default level of attention). Once there are enough bystanders to drive the competitor to its limit, the effect falls towards null. A likely reading is that crowding works in two stages: load first takes away the other actors' capacity to react, and only then the new nudge's own effect. In a field experiment on the same map, the blind site appears where attention is plentiful and bystanders are few, so that the competitor keeps room to respond (Fig.~3c).

The outline on each map marks the settings where the model also reproduces the loss at scale and published effect sizes. Inside the outline, no group of neighbouring settings on these maps delivers the expected effect. The definitions largely build this in: the outline requires the rollout to deliver less than 0.8 of what the pilot reported, and an effect counts as expected only from 0.75\,$E$ up, so a setting could meet both only if its pilot reported at least 0.94\,$E$. The null field of Fig.~2c lies outside the outline.

For a competitor that responds to any gap from its goal, the fate can be derived as a formula of four quantities, with no parameter fitted after the calibration (Methods). They are the new nudge's effect with no other actor present, $E$; how much the competitor's nudge lowers the rate among the people the new nudge reaches, called its pull on them; its pull on everyone else; and how far the competitor's goal lies below the rate with no other actors. A trial at the site could estimate the first three by switching the competitor's nudge on and off in a factorial design, which needs the competitor's cooperation. The fourth is not observed and would have to be inferred from where the competitor stops responding. When such a competitor has room to respond before the rollout but not after, the share of $E$ kept is
\begin{equation}
\frac{\text{effect kept}}{E} = 1 - \frac{\begin{array}{@{}c@{}}\text{pull on the people the new nudge reaches}\\ {}-\text{distance of the goal}\end{array}}{E}.
\end{equation}
So a rollout loses nothing at scale unless the competitor still had room to push before the rollout. The effect is erased once that pull reaches the new nudge's effect plus the distance of the competitor's goal. Where the competitor can escalate, the strength of its automatic nudge is a fifth quantity. A competitor that ratchets, or a steering one that moves only when the rate strays more than a small margin from its goal (a tolerance band), has no such formula: the nudge's fate against it is found by applying the competitor's own rules to the expected rates, again with nothing fitted.

The derivation was compared with the simulation by a rule fixed in advance. Each setting on a map is a cell. A cell agrees when the derivation and the simulation give the same fate and sizes within 0.1\,$E$, and a map agrees when at least nine of every ten cells do. By that rule the derivation agrees with the simulation on eight of the ten maps (Fig.~3 and \textit{SI Appendix}, Fig.~S4, dashed lines). It agrees in part on the other two: the ratcheting competitor under priority sharing (49 of 64 cells) and the tolerance-band competitor (50 of 64). In both, the suspected cause is chance: a small random change in the rate decides whether the competitor's trigger, the lowest rate it has seen or the edge of its band, is crossed. On these two maps, and on one that passes, the simulation gives a slightly larger effect than the derivation, by a median of about 0.05 to 0.07 of $E$, except against the strongest competitors (\textit{SI Appendix}, section~\ref{si:reduction}).

The fates survive a change of rules, but their boundaries move (\textit{SI Appendix}, Fig.~S4). When attention is shared in proportion instead of strongest first, or when the competitor acts only once the rate leaves a tolerance band, the same fates appear in different places. Of 114 comparisons between rules, each the fate at one setting, the sequence of fates along a line of a map or the point where a fate first appears, 32 agree in fate and size; in 76 of the other 82 the rules differ in where a fate holds, and in 6 only in its size. In one corner a fate changes (\textit{SI Appendix}, Fig.~S4c): a weak tolerance-band competitor, given the strongest automatic nudge tried, reverses the effect in just over half of the runs, short of the 90\% required for a fate to count as holding.

Last, the derivation was used to find the fate at 49,152 random settings: 4,096 for each of twelve combinations of three response rules for the competitor (steering to a goal, ratcheting, and steering only outside a tolerance band), two attention rules and two designs (Methods). Each setting draws nine or ten parameters at random. The ranges are only partly tied to the calibration, so the shares describe the model, not real fields. Three things stand out. First, the expected effect is the exception. A rollout against a steering competitor under priority sharing delivered it in 9.0\% of settings, and no combination in more than 11.7\%; the effect was attenuated in 32.1\%, null in 31.9\%, transient in 14.6\% and reversed in 2.9\%. Second, what shifts the fate most depends on how attention is shared. Where the strongest nudge is heard first, it is the new nudge's own strength, in all six combinations; the expectation fixed in advance was that strength would rank first in none of them. Where attention is shared in proportion, it is the number of bystanders, and the new nudge's strength matters much less. Third, a steering competitor reverses the effect in 3--6\% of the settings that received a fate; in rollouts its automatic nudge was on in almost all of these (322 of 323). Against this competitor, 96 random settings were also simulated. The simulation agreed with the derivation in fate and size at 90 of them. The derivation predicted a reversal at six, and the simulation reversed the effect at all six. Agreement was lower under the other two response rules (55 and 79 of 96). Five expectations had been fixed before the sample was drawn. One was met; \textit{SI Appendix}, section~\ref{si:sweep} reports all five.

\subsection*{Signatures a trial can record}

The fates a trial cannot see in its comparison of treated and untreated people leave marks in quantities it can record (Fig.~4).
\ifdefined\arxivversion\begin{figure}[p]\centering\includegraphics[width=\textwidth]{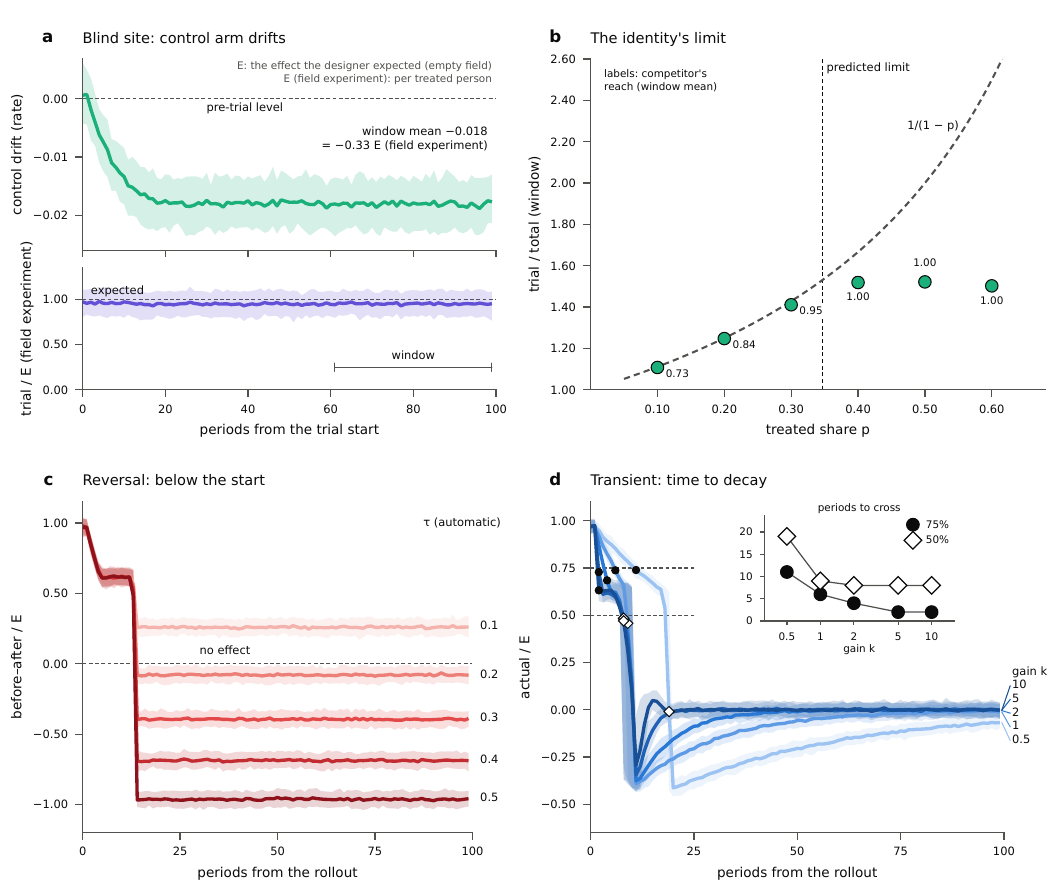}\caption{\textbf{The fates a trial cannot see from its treated $-$ control contrast leave marks in quantities a trial can record: the control arm's drift, the site rate after the rollout and the time course of the effect.}
\textbf{a}, Blind site, in a field experiment with 40\% of the site treated: top, the control arm's drift, as a change in the share of people acting (0 before the trial by construction); bottom, the trial's treated $-$ control contrast as a share of $E$ (field experiment).
\textbf{b}, The identity's limit: the trial's contrast over the actual effect (trial\,/\,total, window means) against the treated share $p$, with the curve $1/(1-p)$. Labels give the competitor's window-mean reach; from $p = 0.4$ it is at its limit. The vertical line marks the treated share, about 0.35, at which the derivation predicts the competitor reaches its limit (Methods).
\textbf{c}, Reversal: the before-and-after estimate over $E$ after the rollout against a maximiser competitor, one line per strength of its automatic nudge, $\tau$ (automatic) 0.1 to 0.5.
\textbf{d}, Transient: the actual effect over $E$ after the rollout, one line per response speed of the escalating competitor of Fig.~2b (its gain $k$, 0.5 to 10). Marks show where each line first falls below 75\% (filled circles) and 50\% (open diamonds) of $E$; the inset plots the periods to each crossing against $k$ (log scale).
$E$ is the effect the designer expected, the new nudge's effect on the site in the field with no other actors (\textbf{c}, \textbf{d}); $E$ (field experiment) is the same expectation per treated person (\textbf{a}). Actual (total): the site rate with the new nudge minus that without it, the other actors present in both; in \textbf{b}, per treated person, the treated people's rate with the new nudge minus their rate without it. A crossing is the first period in which a line, at or after its highest point following the rollout, lies below the level; periods are counted from the rollout. The window is periods 60--99, counted from the rollout or trial start at 0 (period 100 of the run). Lines are replicate means; bands span the 10th--90th percentiles of $n = 100$ replicate runs.}\end{figure}
\fi

A blind site shows in the untreated group (Fig.~4a). With 40\% of the site treated, the untreated people's rate falls by 0.018, or 1.8 percentage points, below its pre-trial level, a third of the effect expected for each treated person. Over the same window the trial's comparison stays at about 95\% of that expected effect (94\% in the runs of Fig.~2e), so the comparison alone shows nothing unusual. A trial that also compares its control group with that group's own pre-trial level could see the drift, provided the site would otherwise have stayed at that level.

How far the trial overstates the effect on the people it treats follows from an accounting identity, which holds whenever the other actors hold the site's rate where it was (Fig.~4b). The untreated group's fall then offsets the treated group's gain, and what the treated actually gain is a fraction $1-p$ of the trial's comparison~\cite{sobel2006,hudgens2008,crepon2013}:
\begin{equation}
\frac{\text{effect reported by the trial}}{\text{actual effect on the treated}} = \frac{1}{1-p}.
\end{equation}
In the calibrated setting the simulated trials follow this curve to within 1.3\% while up to 30\% of the site is treated. From 40\% the competitor is at its limit and can no longer hold the site's rate, and the points fall below the curve. The derivation puts the treated share at which the competitor reaches its limit at about 0.35 (Fig.~4b, vertical line). Across random settings the relation is not general. Whatever the treated share, blind sites occurred in 4--16\% of the field-experiment settings that received a fate, but the relation held in fewer than half of the settings even where few people were treated (\textit{SI Appendix}, section~\ref{si:sweep}). It appears to be a property of the calibrated setting, where the competitor holds the site's rate exactly.

A reversal shows in the site's rate after the rollout (Fig.~4c). In the reversed field a before-and-after comparison first falls, apparently as the competitor reaches more people. About 14 periods after the rollout the competitor's automatic nudge comes on. When that nudge is at least two-thirds as strong as the new nudge, the comparison drops below the pre-rollout level; at a third of its strength it stays above. A comparison taken early would report a positive effect.

A transient effect shows in its time course (Fig.~4d). The faster the competitor responds, the sooner the effect falls below three-quarters of $E$: 11, 6, 4, 2 and 2 periods after the rollout at the five speeds tried. It falls below half of $E$ within 8 to 19 periods, and the automatic nudge then removes the rest.

A larger pilot does not see more. Across pilots that treat 2\% to 60\% of the site, against the competitors of Fig.~2a, b and d, the pilot's own comparison stays within 0.91--1.01 of $E$ (\textit{SI Appendix}, section~\ref{si:pilotshare}). What changes is what the pilot does to the site. Until it reaches its limit, a steering competitor absorbs the same fraction of a pilot's effect on the site whatever the pilot's size (0.45 at the default speed of response). A 2\% pilot provokes this response as surely as a 40\% one, only smaller. What a small pilot cannot do is drive the competitor to its limit, as the rollout does. A pilot of 40\% or more does drive an escalating competitor to its limit, during the pilot itself. The competitor's automatic nudge then stays on, and the rollout that follows starts in an escalated field: its fate is null instead of transient. A larger pilot is itself a blind site, and a large enough one changes the rollout it was meant to forecast.

\section*{Discussion}

What becomes of a nudge, in this model, depends on who else is acting. The same nudge was attenuated, made transient, nulled or reversed, or it left a blind site, and the one thing that changed between those fates was the field: other actors who watch the rate of a behaviour and respond to it. Its fate was a property of the field it entered and not of the nudge alone. Of the five, only the null fate shows in the pilot; the other four appear after it. And what a trial reported could diverge from what the intervention delivered: a pilot's comparison could not see the response it provoked, a rollout met the full response, and a field experiment's comparison stayed intact while the site gained little.

The mechanism is familiar. An actor that pushes back against the gap between a rate and its goal, up to a limit, is negative feedback with a cap, a standard object in control engineering~\cite{astrom2008}. A system that defeats an intervention by responding to it is what system dynamics calls policy resistance~\cite{sterman2006}. The Lucas critique makes a parallel point in economics~\cite{lucas1976}: relations estimated under one policy shift when the policy changes. The subject here is not the feedback but the fate of a nudge inside it, and what a trial reports of that fate.

The model suggests a reading of four known anomalies. Effects shrink at scale~\cite{alubaydli2020}: a pilot provokes only a small response, which its comparison hides, and a rollout provokes the full one. Government trials report smaller effects than academic ones~\cite{dellavigna2022}: the same trial reports less in a crowded field than in a quiet one, and in the calibrated crowded field that gap comes from crowded attention alone. DellaVigna and Linos attribute most of the gap to selective publication and low power, so matching its size is not evidence for crowding; the two readings compete for the same gap. Effects vary between sites more than chance allows~\cite{bryan2021}: sites that differ only in their other actors give different effects, and often different fates. Meta-analyses find effects that vary widely between studies~\cite{mertens2022,maier2022,szaszi2025}: fields that differ from study to study could add a dispersion of their own. None of this rules out failures of implementation, selection or publication. A field of other actors can produce the same patterns without them.

What this model adds is an actor that regulates the rate of a behaviour: it offsets the site's gain while the trial still reports close to the expected effect. Earlier revivals of Lewin's idea take the other forces as given: the lift an advertising campaign shows depends on what competitors test at the same time~\cite{waisman2025}, and a policy's measured effect depends on the policies in place beside it~\cite{matthay2021}. Combined nudges can crowd each other out~\cite{brandon2019,alt2024}, and one intervention can take attention from another~\cite{rodemeier2025,altmann2022}, as sources that compete for scarce attention do~\cite{persson2018}. Where others respond, they mostly respond to the intervention itself~\cite{sunstein2017,spiegler2015}, or through prices as a programme grows~\cite{heckman1998,lise2004,munro2025}. In a companion agent-based model the same seeding intervention stalls, drifts or cascades with the rules by which people learn from each other~\cite{veltri2026learning}: there the field is the people's influence on one another, here it is the other actors. The nearest analogue to those actors is in online marketplaces. There the two arms of an A/B test compete in the same auctions~\cite{blake2014} and draw on the same advertising budgets, paid out at a set pace, so each arm biases the other's result~\cite{liu2021,liao2025}. The model's actors differ: they never see the nudge, respond only to the rate, and reach treated and untreated people alike.

The blind site sits closest to work on trials where treating some people changes outcomes for others. Its accounting is known~\cite{sobel2006,hudgens2008}, and untreated people in treated sites have been found to do worse while the net gain was small~\cite{crepon2013}. Munro, Kuang and Wager show that a trial randomised within one market estimates an intervention's direct effect, and that the spillover through prices needs price elasticities from an augmented design~\cite{munro2025}. The blind site is the same division, between a direct effect the trial reports and a total effect it does not. The channel differs: here it is an actor that steers a rate towards a goal and can reach its limit, not a price that clears a market, so the response has a cap and a course over time, and what recovers the total effect is a record of the actors' actions. The static part of the loss is what the earlier revivals describe. The reaction part is the response Lewin described~\cite{lewin1951}.

One property follows from the rules without being written into any of them: a field in stalemate does not react. When a competitor and an ally steer the same rate towards goals that cannot both be met, each pushes to its limit and neither can respond further. The new nudge's loss is then almost entirely static, a rollout keeps about what the pilot showed and, except where attention is ample, no blind site appears (\textit{SI Appendix}, Fig.~S4h,i, and section~\ref{si:record}, run 4). Only where attention is ample does the new nudge lift the rate past the ally's goal, and the ally then withdraws. Whether a field reacts thus depends mostly on the other actors' headroom, and crowding removes that headroom first. The crowded field calibrated on the academic--government gap is such a stalemate, so the model's match to that gap rests on crowding alone.

Do real fields react? Published trials give a qualified answer. Where an actor with a goal of its own was recorded, it responded, though to the intervention rather than to the rate of a behaviour, and not in a nudge setting. Households cut their own spending on schooling when they expected a school grant~\cite{das2013}, and civil-service teachers reduced their effort once their school was given an extra contract teacher~\cite{duflo2015}. Both behave like the model's ally, which withdraws as the rate rises. Party agents moved registration irregularities from observed centres to unobserved ones nearby~\cite{ichino2012}, and job seekers who were not counselled did worse where more of the others were~\cite{crepon2013}. Both show the pattern of the blind site, but through a fixed number of irregularities or of jobs to go round, not through an actor that regulates a rate. For nudges themselves the evidence runs the other way. Across more than 400 product categories, firms mostly did not respond to a rival's promotion or advertising~\cite{steenkamp2005}. And in every behavioural trial found that treated only part of a place, the people left untreated moved the same way as the treated, not against them. Women not visited by a voter-awareness campaign voted more where their neighbours were visited~\cite{gine2018}. Untreated taxpayers paid more where a high share of their block was sent a letter~\cite{cruces2025}. Precincts not shown information ads moved with those shown them where the ads reached most of the electorate~\cite{enriquez2024}. And households that received no enforcement mailing complied more when their neighbours received one~\cite{drago2020}. Positive spillovers point to influence between people, which this model leaves out and the companion model studies~\cite{veltri2026learning}. No open dataset was found in which a behavioural intervention, a varied treated share and a named actor come together, so whether and how often nudge fields react remains to be measured.

The model has limits. It is calibrated to published magnitudes and fitted to no data, so it shows what a reacting field can do, not how often real fields do it. Apart from the ratcheting competitor, whose gains in reach are permanent, each field settles into one state by design: there is no social norm that grows with the rate, no habit that forms with repetition and no influence between people, which the companion model shows can decide a fate on its own~\cite{veltri2026learning}. Actors follow fixed response rules without modelling one another. Three assumptions were held fixed in every run: an escalation, once made, is never withdrawn; the actors observe the rate without error; and within a pilot or a field experiment the same people stay treated. Attention costs and budgets have no published anchor, so they were varied. Under priority sharing a nudge's strength also decides how early a person attends to it, so the finding that strength shifts the fate most partly reflects that rule. Two results also proved less general across random settings than in the calibrated setting and on the maps. The relation between a trial's report and the treated share held in fewer than half of the settings. And reversal was not confined to ratcheting competitors. In rollouts it almost always came with the competitor's automatic nudge switched on (98.7\% of reversed rollouts, over all three response rules). A reading made after the fact is that the effect reverses when the automatic nudge pushes the rate down by more than the competitor can give back by withdrawing its other nudge. A ratcheting competitor's reversal also depends on how far back it looks. It holds across runs when the competitor compares the rate with the lowest level it has ever seen, or with the lowest over about 25 periods or more; with about half that look-back the effect is attenuated instead.

Trials could keep two records. One is the field: an inventory of the other nudges and actors at a site, and whether they are already at their limit. The other is the marks the fates leave: the control group's rate against its own level before the trial, the site's rate after a rollout, and the effect's course over time, which sequence methods for trials can read~\cite{veltri2026time}. Neither record identifies the mechanism on its own: drift, decay and a fall after the rollout can also come from trends, substitution or habituation. The test that would identify it is a trial that randomises the treated share across sites far enough apart not to share actors, keeps untreated sites, and measures both the behaviour and the named actors' actions over time, in the manner of randomised-saturation designs~\cite{baird2018}; the dataset screening found no open trial of that shape (\textit{SI Appendix}, section~\ref{si:methods}). The model also suggests a test not yet run. A ratcheting competitor keeps the reach it gains, so withdrawing the new nudge from such a field should not restore the earlier rate, while a field whose competitor steers to a fixed goal should return to it. A pilot's comparison of treated and untreated people cannot show any of this: the response the pilot provokes falls on both groups, and a small pilot does not exhaust the competitor as the rollout does. Making the pilot larger does not help. For pilots of up to 60\% of the site, the comparison stays near the full effect, since a steering competitor absorbs the same fraction of any pilot's effect on the site. And a pilot large enough to exhaust an escalating competitor hands the rollout a field already escalated. A pilot that also recorded the site's rate and the other actors' reach would show more.

The answer the model gives to the question it asked is this. What becomes of a nudge is decided by the headroom of the actors around it: where they have room to respond, the rollout is absorbed, made transient or reversed, and a well-run trial can report the full effect while the site gains little; where they have none, the field is a backdrop and the pilot forecasts the rollout. The record of the field is therefore as much part of an evaluation as the record of the effect.

\section*{Materials and Methods}

This section gives what is needed to judge the results. The rules in full and the numbers behind each check are in \textit{SI Appendix}, section~\ref{si:methods}; a description of the model in the ODD format (Overview, Design concepts, Details, the standard protocol for agent-based models~\cite{grimm2020}), every parameter with its source and the dated record of runs are in sections~\ref{si:odd}, \ref{si:parameters} and~\ref{si:record}.

\subsection*{The model}

The model simulates one site of $N = 20{,}000$ people, the new nudge and the other actors over 200 periods, each read as a week.

\paragraph{People and nudges.} Each person $i$ has a baseline propensity $a_i$, the standing inclination of the Results, towards the behaviour the new nudge targets. It lies on an unobserved scale with the threshold at 0 and is drawn from a mixture of a habitual group (30\% of people) and a non-habitual one. A nudge $k$ has a direction $d_k \in \{+1, -1\}$, a strength $\tau_k \ge 0$ and a reach, the share of the site it reaches in a period. It pushes person $i$ by $\tau_k v_{ik}$; the valence $v_{ik}$, how the person receives the nudge, differs between people and is negative (the nudge repels) for about 5\% of them. Person $i$ acts in period $t$ when
\begin{equation}
\begin{split}
Y_{it} = \mathbf{1}\Big\{\, & a_i + s_i^{\mathrm{auto}} \sum_{k \in \mathcal{A}_{it}} d_k \tau_k v_{ik} \\
& + s_i^{\mathrm{delib}} \sum_{k \in \mathcal{D}_{it}} w_{ikt}\, d_k \tau_k v_{ik} + \varepsilon_{it} > 0 \Big\}, \qquad \varepsilon_{it} \sim \mathcal{N}(0, 1),
\end{split}
\end{equation}
where $\mathcal{A}_{it}$ and $\mathcal{D}_{it}$ are the automatic and deliberative (attention-needing) nudges reaching $i$ in period $t$, $s_i^{\mathrm{auto}}$ and $s_i^{\mathrm{delib}}$ the person's susceptibility to each type and $w_{ikt} \in [0, 1]$ the attention the person gives the nudge. The noise is independent across people and periods. The new nudge is deliberative, raises the behaviour and has $\tau = 0.30$. With no other actor, control take-up (the rate without the nudge) is 0.309 and the new nudge's effect is 0.152 on the probit scale, which allows for how common the behaviour already is. Both lie inside the bounds set for the model from published trials (probit effects 0.05--0.4, control take-up 0.05--0.9)~\cite{dellavigna2022,milkman2021}.

\paragraph{Attention.} Each person has an attention budget per period, and each deliberative nudge an attention cost (1.0 for every nudge here). Under priority sharing, the default, a person attends to the deliberative nudges reaching them strongest first, by perceived push $\tau_k |v_{ik}|$ (the salience of Fig.~1b), until the budget is spent; the rest go unattended. Under proportional sharing every deliberative nudge reaching the person gets the same weight, the budget divided by their summed cost, at most 1. Automatic nudges bypass attention.

\paragraph{Actors.} Each other actor owns one deliberative nudge and sets its reach. A competitor pushes the targeted behaviour down and an ally pushes it up. A bystander promotes a different behaviour: it competes for attention, does not push the targeted behaviour and does not respond; its nudge reaches 60\% of the site from the start and never changes. Competitors and allies start at reach 0 and begin to respond when their first, delayed observation of the rate arrives; the maximiser (below) holds a reach of 0.60 through the burn-in, the opening periods before the nudge enters. An actor observes only the rate of its target behaviour at the site, two periods late, and never sees another actor's nudge or the new one. An actor that reaches the whole site is at its limit; one that reaches fewer people has headroom.

A competitor follows one of three response rules; the allies in the runs here are thermostats. A \emph{thermostat}, the steering competitor of the Results, steers the observed rate towards a goal: each period it changes its reach by a gain (2 by default) times the gap between the observed rate and the goal, within 0 and 1. A competitor's goal is the site's rate with no actors minus 0.03, and an ally's is that rate plus 0.03. A \emph{maximiser}, the ratcheting competitor of the Results, keeps a reference, the lowest rate it has seen; it raises its reach whenever the observed rate is more than 0.01 above the reference, and never lowers it: a ratchet. A \emph{threshold trigger}, the tolerance-band competitor of the Results, moves its reach by a fixed step of 0.05 per period, in the direction that closes the gap, while the rate is more than 0.01 from its goal, and holds it otherwise. Any of the three can also escalate, by switching on an automatic nudge against the behaviour, which reaches everyone and stays on for the rest of the run. A thermostat or a threshold trigger does so once the observed rate has been above its goal for ten consecutive periods and its reach is at the limit; a maximiser, after ten consecutive periods at its limit with the rate more than 0.01 above its reference.

\subsection*{Designs and effects}

Each run starts with a burn-in of 60 periods (0--59), in which the field settles before the nudge enters. A pilot treats a fixed random 2\% of the site over periods 60--79. The new nudge is then off until period 100, when the rollout reaches everyone to the end of the run. A field experiment has no pilot: from period 100 it treats a fixed random share $p$ of the site (0.4 by default), and the rest form the control group. Outcomes are read over the evaluation window, periods 160--199.

Each run is repeated in three worlds that share its people and its random draws: with no other actors, with the actors held at their reach before the rollout, and with the actors reacting. Each world is run with and without the new nudge, and the difference in rate is the effect in that world (in a field experiment, among the treated). $E$, the expected effect, is the effect with no other actors; the total effect, the actual effect of the Results and figures, is the effect with the actors reacting. The loss has a static part, already there with the actors held fixed, and a reaction part, added when they respond. What a trial reports is the treated $-$ control contrast (the trial's comparison, in the Results) in the reacting world. A rollout has no control group, and is judged by the before--after difference: the mean rate over the evaluation window minus that over periods 80--99.

\subsection*{Outcome classes (the fates) and robustness}

The classes, their thresholds and the robustness rule were fixed in a dated record before the first simulation. For each run, the ratio is the total effect divided by $E$. Two means of it are used: over the evaluation window (the window mean) and over the window's last ten periods (the end mean). Runs with $|E| < 0.02$ are dropped. Each run takes the first class whose rule it meets, in this order:
\begin{itemize}\itemsep0pt
\item reversed: window mean $\le -0.25$;
\item blind trial, the person-level counterpart of the blind site (field experiment only): the trial reports 0.75--1.25 of $E$ while the window mean is below 0.5;
\item transient: ratio $\ge 0.75$ in some period from the rollout on, and end mean $< 0.5$;
\item oscillating: a cycle in the window (an autocorrelation peak above 0.5 at a lag of 2--20 periods, with amplitude above $0.25|E|$);
\item delayed: ratio first $\ge 0.75$ 20 or more periods after the start, and end mean $\ge 0.75$;
\item null: $|$window mean$| < 0.25$;
\item attenuated: end mean in $[0.25, 0.75]$;
\item as expected: window mean in $[0.75, 1.25]$;
\item otherwise unclassified.
\end{itemize}
One amendment, made after the first runs, gave the classes \emph{as expected} and \emph{delayed} their present form; when the saved runs were classified again, 372 of 6,782 changed class. In a field experiment two site-level classes are tested first, against the site-level effect a designer expects from treating a share $p$, which is $p$ times the effect per treated person. A site is a reversed site if its change is at or below $-0.25$ of that, and a blind site if the trial reports 0.75--1.25 of the effect per treated person while the site's change is below half of the site-level effect expected.

Results come from sweeps, sets of runs over a grid of parameter values. A class is robust when it holds in at least 90\% of runs at two or more adjacent points of a sweep, and those points span at least 10\% of the swept range. The outline in Fig.~3 marks the cells where the rollout's total stays below 0.8 of the pilot contrast and the new nudge's effect with no other actor lies within the bounds above. Each pre-specified claim was rerun under the other attention rule, and the thermostat's map also under the threshold trigger. \emph{Erased} is not a class: the Results use it for a transient effect whose run-averaged end effect is 0.00\,$E$.

\subsection*{Checks against four known regularities}

Before predicting anything new, the model had to reproduce four documented regularities in direction and rough size, each a test whose threshold and run count (10 per setting) were fixed in advance: the loss at scale~\cite{alubaydli2020}, the gap between quiet and crowded fields~\cite{dellavigna2022}, the dispersion across sites~\cite{bryan2021} and the peak near one half (the inverted U)~\cite{saccardo2025}. All four tests were met, with three limits. The first is met by the ratio of run means: a 2\% pilot is noisy, and between 1 and 6 of the 10 single runs per setting do not show the loss. The second is a calibration, not a prediction: the number of bystanders in the crowded field was set to meet it. In the fourth, 10 runs per setting locate the peak only as a plateau. The crowded fields behind Figs.~2c and 3b,c are not the stalemate: they have a thermostat competitor of strength 0.35 plus bystanders, without an ally. At their calibrated point (two bystanders, the default budget) the competitor keeps headroom, and the second test is only just met: the quiet-to-crowded ratio of reported effects is 1.508, against an accepted range of 1.5--2.5. Heavier load, as in Fig.~2c, drives the competitor to its limit. \textit{SI Appendix}, section~\ref{si:methods} gives each test and its numbers.

\subsection*{The blind-site identity}

Write $D$ for the trial's treated~$-$~control contrast and $C$ for the control group's change against what it would have done without the trial. By accounting, the treated gain $G_{\text{treated}} = D + C$ and the site gains $G_{\text{site}} = pD + C$ when a share $p$ is treated. If a competitor with headroom holds the site's rate where it was, $G_{\text{site}} = 0$, so $C = -pD$ and $G_{\text{treated}} = (1-p)\,D$: the trial reports $1/(1-p)$ times the actual effect on the treated (Eq.~2). The relation is an identity, not a law: it holds only while the competitor has headroom and keeps the site's rate where it was. $C$ is measured against the counterfactual; the control group's pre-trial rate stands in for it only if the site would otherwise have stayed there (derivation in \textit{SI Appendix}, section~\ref{si:methods}).

\subsection*{Analytic reduction}

This is the derivation of the Results. An actor's nudge reaches each person independently each period, so the expected site rate is linear in each actor's reach. With one competitor the derivation needs four averages of the probability of acting: when neither nudge reaches a person, when the new nudge alone does, when the competitor's alone does and when both do. They give $E$ and the competitor's pull, on people the new nudge does not reach and on those it does; with the distance of the competitor's goal below the no-actor rate (0.03), these are the four quantities of the Results. For a thermostat, the reach at which it rests, the share of $E$ kept after a rollout and the treated share at which it reaches its limit in a field experiment follow in closed form (\textit{SI Appendix}, Table~\ref{tab:cases}); once its automatic nudge is on, that nudge's strength enters as a fifth quantity. For every rule, the prediction that is compared with the simulation applies the simulator's own actor rules to the expected rate, and so also gives the static and reaction parts. The averages are taken over 200,000 people drawn as the model's population is. Nothing is fitted.

A cell of a map agrees when the predicted class is the simulated majority class and the predicted size differs from the simulated one by at most 0.1 of $E$ (in a field experiment, of the expected site-level effect and of the expected effect per treated person); a map is predicted when at least 90\% of its cells agree. The rule was fixed before the ten distinct maps of Fig.~3 and \textit{SI Appendix}, Fig.~S4 were compared. The simulated maps were known before the reduction, which has no free parameter, was built; a feasibility check on a few settings already simulated and one unplanned simulator run are declared in \textit{SI Appendix}, section~\ref{si:reduction}. Eight maps pass. The maximiser under priority sharing (49 of 64 cells) and the threshold trigger (50 of 64) are predicted in part. Seven further sets of runs, behind Figs.~2a,b,d,e and 4 and the ally of \textit{SI Appendix}, Fig.~S1, agree at every point (33 of 33).

\subsection*{Sensitivity of the fates to the field's parameters}

For each of twelve combinations of response rule, attention rule and design, a sample of 4,096 settings spread evenly over the ranges (a Latin hypercube) was drawn: 49,152 settings, each with one competitor and no ally. Nine parameters of the nudge, the competitor, attention and take-up varied, and the treated share in a field experiment (ranges in \textit{SI Appendix}, section~\ref{si:sweep}). Each setting's class was read by the rules above from the reduction's prediction of the run, computed from expected rates without sampling noise. For each combination the share of each fate is reported and, for each parameter, how far those shares shift between the lowest and the highest third of the parameter's drawn values. Of each combination's settings, 24 drawn at random (288 in all) were simulated and compared with that prediction by the cell rule above. The ranges, the statistics and five expectations were fixed before the sample was drawn; \textit{SI Appendix}, section~\ref{si:sweep} reports each expectation as registered and as found, including those not met.

\subsection*{Pre-specification}

Every sweep behind a figure was specified before it ran, as a dated entry in a version-controlled record naming the scenario, design, swept values, runs, seeds, attention rule, the class expected at each point and the prior knowledge it drew on. The record is not a public preregistration. Deviations and unmet expectations are disclosed in the entry. One decision narrowed a map: Fig.~3a starts at a competitor strength of 0.20, because at 0.15 the thermostat reaches its limit in burn-in and escalates before the pilot; the sensitivity sweep includes such settings. Run 11, registered after all the other results were in, varies the pilot's share from 0.02 to 0.60 against the three competitors of Fig.~2 (\textit{SI Appendix}, section~\ref{si:pilotshare}). \textit{SI Appendix}, sections~\ref{si:methods} and~\ref{si:record} list the runs behind each figure.

\subsection*{Software and reproducibility}

The simulator is written in Python with PyTorch. Random draws differ between devices, so a run reproduces exactly only on the same kind of device; run 11 repeated the settings of Fig.~2a, b and d on a second kind of device, and the classes were the same and each ratio was within 0.1 of the first, the tolerance fixed before the run. Each run keeps a record (a manifest) of its configuration, random seeds, device, software versions and the exact version of the code, and the figure scripts refuse any run whose manifest shows that it did not finish or that the code had unrecorded changes. Every run behind a figure is pre-specified, complete and made with a recorded version of the code; the figures were drawn from such a version too, and drawing them a second time gave identical files.

\subsection*{Data, Materials, and Software Availability}

The simulator (Python with PyTorch), the configuration and dated record of every run, the figure scripts, the summary tables behind every figure and table, each run's manifest (configuration, random seeds, device, software versions and code version) and its per-run effect tables (1,635 Parquet files, about 1.2 GB) are deposited at Zenodo, doi:10.5281/zenodo.24150612. The model is calibrated to published effect sizes and uses no individual-level data; every number in the text is reported in the article or the \textit{SI Appendix}.

\section*{Acknowledgments}%
The author is grateful for the support of a start-up grant from the Yong Loo Lin School of Medicine, National University of Singapore.

\bibliographystyle{unsrtnat}
\bibliography{refs}

\ifdefined\arxivversion\else
\clearpage
\section*{Figure legends}
\begin{figure}[p]\centering\includegraphics[width=\textwidth]{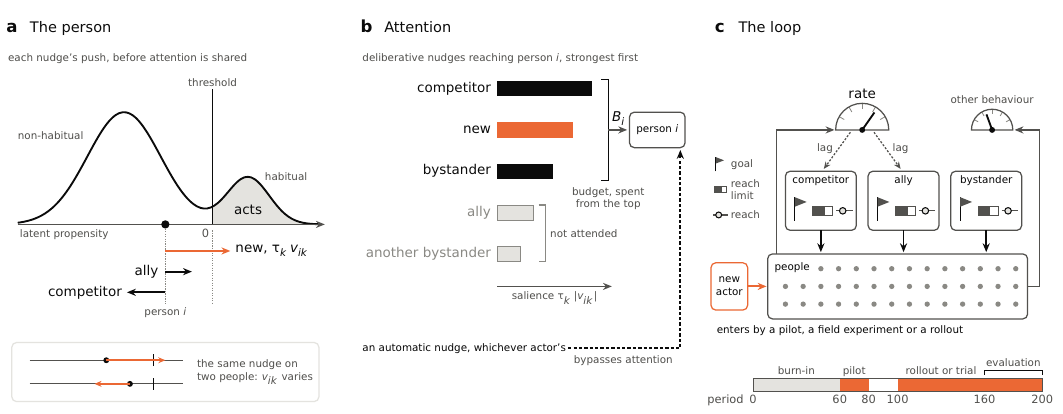}\caption{\textbf{A nudge is a force on a threshold; it shares a person's attention with other nudges; and the actors who own those nudges steer the rate of the behaviour with their own goals and reach limits, reacting to the rate, not to the new nudge.}
\textbf{a}, The person. The curve is the population's baseline propensity (non-habitual and habitual people); those beyond the threshold at 0 (shaded) act. Person $i$ is the dot. The arrows are three nudges on person $i$, the new nudge, an ally's and a competitor's, each its own push, $\tau_k v_{ik}$, before attention is shared; they are not summed. Inset: the same nudge on two people, longer for one and flipped for the other ($v_{ik}$ varies).
\textbf{b}, Attention. Person $i$'s deliberative nudges (those that need attention), ranked by salience $\tau_k|v_{ik}|$, spend the attention budget $B_i$ from the top (priority). Nudges beyond it (light) are not attended: here the bystander's nudge crowds out the ally's. An automatic nudge, whichever actor's (dashed), bypasses the budget.
\textbf{c}, The loop. The people of a site feed the rate (dial). A competitor, an ally and a bystander, each with a goal (flag), a limit on its reach (bar) and a reach (slider), observe a rate with a lag (dashed) and nudge the people (solid). The bystander promotes a different behaviour (second dial), keeps a fixed reach and competes for attention; nothing is fed back to it. No arrow runs from the new actor, the nudge's owner, to another actor. The timeline, in periods, marks burn-in, the pilot (60--79), the rollout or trial from 100 and the evaluation window (160--199); the new nudge is off over periods 80--99.
Response rules, not drawn: a thermostat steers the rate towards its goal; a maximiser pushes while the rate is above its reference. Lengths, the density and the dials are illustrative.}\end{figure}

\fi

\clearpage
\sisection{Methods in full}{si:methods}
\setcounter{equation}{0}
\renewcommand{\theequation}{S\arabic{equation}}
\renewcommand{\theHequation}{S\arabic{equation}}

This section gives the model, the designs, the outcome classes, the checks and the analyses in full; Materials and Methods in the main text summarises it. It keeps the model's own names: the thermostat, maximiser and threshold trigger are the Results' steering, ratcheting and tolerance-band competitors; the empty, frozen and reacting worlds are the worlds with no other actors, with the actors held at their reach before the rollout, and with the actors reacting; the entry ticket is the four checks against known regularities, T1--T4; burn-in is the 60 periods in which the field settles before the nudge enters; an actor's limit is the whole site.

\subsection*{The model}

The model simulates one site of $N = 20{,}000$ people, the new nudge and the other actors over 200 periods, each read as a week. Sections~\ref{si:odd} and~\ref{si:parameters} give an ODD description~\citeSI{SI-grimm2020} and every parameter with its source.

\paragraph{People.} Each person $i$ has a baseline propensity $a_i$ for the focal behaviour on a latent scale. Propensities follow a two-component normal mixture: with probability 0.3 a person is habitual, drawn from $\mathcal{N}(0.4, 0.8^2)$, and otherwise from $\mathcal{N}(-1.2, 0.8^2)$. Susceptibilities $s_i^{\mathrm{auto}}$ and $s_i^{\mathrm{delib}}$ scale the action of automatic and deliberative nudges; both are log-normal (mean 1, log-scale s.d.\ 0.3). The attention budget per period is $B_i = b\,u_i$, with $u_i$ log-normal (mean 1, log-scale s.d.\ 0.4) and the budget scale $b$ at 1.0 by default. A nudge $k$ pushes person $i$ by $\tau_k v_{ik}$. The valence $v_{ik}$ is drawn once per person and nudge: its magnitude is log-normal with mean 1 and s.d.\ 0.5, and its sign flips with probability 0.05, so the mean valence is 0.90. Person $i$ acts in period $t$ when
\begin{equation}
\begin{split}
Y_{it} = \mathbf{1}\Big\{\, & a_i + s_i^{\mathrm{auto}} \sum_{k \in \mathcal{A}_{it}} d_k \tau_k v_{ik} \\
& + s_i^{\mathrm{delib}} \sum_{k \in \mathcal{D}_{it}} w_{ikt}\, d_k \tau_k v_{ik} + \varepsilon_{it} > 0 \Big\}, \qquad \varepsilon_{it} \sim \mathcal{N}(0, 1),
\end{split}
\end{equation}
where $\mathcal{A}_{it}$ and $\mathcal{D}_{it}$ are the automatic and deliberative nudges reaching $i$ in period $t$, $d_k \in \{+1, -1\}$ is a nudge's direction and $w_{ikt} \in [0, 1]$ its attention weight. The noise is independent across people and periods (a probit), and habit updating is off.

\paragraph{Nudges.} A nudge has a direction $d_k$, a type (automatic or deliberative), a strength $\tau_k \ge 0$, an attention cost $c_k$ (deliberative only), a reach $r_{kt} \in [0, 1]$ (the share of the site it reaches in a period) and an owner. An actor's nudge reaches each person with probability $r_{kt}$, drawn afresh every period; the new nudge reaches a fixed treated set. The new nudge is deliberative, raises the behaviour, has $\tau = 0.30$ and costs 1.0, as do the other actors' nudges unless a run states otherwise. In the quiet field control take-up is 0.309 and the new nudge's probit effect 0.152, inside the design's bounds (probit effects 0.05--0.4, control take-up 0.05--0.9)~\citeSI{SI-dellavigna2022,SI-milkman2021}.

\paragraph{Attention.} Under priority sharing, the default, a person ranks the deliberative nudges reaching them by perceived push, $\tau_k |v_{ik}|$, strongest first. Each in turn takes $w_{ikt} = \min(1, B^{\mathrm{rem}}/c_k)$, where the remaining budget $B^{\mathrm{rem}}$ starts at $B_i$ and falls by $w_{ikt} c_k$; the rest go unattended. Under proportional sharing each deliberative nudge reaching the person gets $w_{ikt} = \min(1, B_i/D_{it})$, with $D_{it} = \sum_k c_k$ over those nudges. Automatic nudges bypass attention ($w = 1$).

\paragraph{Actors.} Each other actor owns one deliberative nudge and controls its reach. A competitor pushes the focal behaviour down, an ally pushes it up, and a bystander promotes a different behaviour. An actor's reach is at most the whole site; an actor with $r \ge 0.99$ is at its limit, and below it has headroom. An actor observes only the rate of its target behaviour at the site, with a lag of $L = 2$ periods, and never sees another actor's nudge or the new one.

The thermostat steers the observed rate towards a goal $G$ by bounded integral control. With the gap $g_t = r^{\mathrm{obs}}_{t-L} - G$, a competitor sets its reach to $r_t = \min\{1, \max\{0,\, r_{t-1} + \kappa g_t\}\}$, and an ally uses $-\kappa g_t$; the default gain is $\kappa = 2$. The goal is the run's no-actor baseline (the mean rate over periods 0--59 without other actors or the new nudge) minus 0.03 for a competitor and plus 0.03 for an ally, offsets at which a lone actor settles below its limit in burn-in. Competitors and allies start at reach 0. An ally whose goal is exceeded lowers its reach.

The maximiser pushes while the rate stays above the lowest level it has seen. Its reach is 0.60 until period 59. From period 60 it keeps a reference $q_t$, starting at the mean observed rate over periods 50--59. When $r^{\mathrm{obs}}_{t-L} > q_t + 0.01$ it raises its reach by $\kappa(r^{\mathrm{obs}}_{t-L} - q_t)$, limited to 1, and it never lowers it. The reference then becomes $\min(q_t, r^{\mathrm{obs}}_{t-L})$, a ratchet. When the ratchet is replaced by an exponential average, $q \leftarrow (1-\lambda) q + \lambda r^{\mathrm{obs}}_{t-L}$, the reversal stays robust at $\lambda = 0.04$, a memory of about $1/\lambda = 25$ periods, and gives way to attenuation by $\lambda = 0.08$, 12.5 periods (run 2c).

Any of the three competitors can escalate. A thermostat or a threshold trigger escalates when it is at its limit and its observed gap has been positive for ten consecutive periods; a maximiser, after ten consecutive periods each at its limit and above its reference plus 0.01. It then switches on an automatic nudge against the behaviour, with strength $\tau_{\mathrm{auto}}$ and reach 1, which stays on for the rest of the run. The threshold trigger, the alternative response rule, moves reach by a fixed 0.05 per period in the direction that closes the gap while $|g_t| > 0.01$, and holds it otherwise.

Bystanders do not respond: their gain is 0 and their reach is 0.6 throughout. A bystander's nudge pushes its own behaviour by the same threshold rule; on the focal behaviour it competes for attention at its own strength but pushes nothing.

\subsection*{Designs and counterfactual worlds}

Each run starts with burn-in, periods 0--59. A sweep stops unless, in every run, the fitted slope of the site rate over periods 40--59 (reacting world, without the new nudge) is below 0.002 per period; every run reported here passed. A pilot treats a fixed random 2\% of the site over periods 60--79. The new nudge is then off until period 100, when the rollout reaches everyone to the end of the run. A field experiment has no pilot: from period 100 it treats a fixed random share $p$ of the site (default 0.4), and the rest form the control group. Outcomes are read over the evaluation window, periods 160--199.

Three worlds (no other actors; actors held fixed; actors reacting), each run with and without the new nudge, share each run's people and per-period random draws. The empty world has no other actor; in the reacting world the actors respond; the frozen world holds each actor, from period 100, at its reach (and automatic-nudge reach) before the rollout, the reacting world's mean over periods 90--99. The difference in site rate between a world's two runs (in a field experiment, in the treated people's rate) is the empty-field, frozen-field or total effect. $E$ is the empty-field effect averaged over the evaluation window, per run; in a field experiment it is per treated person ($E_{\mathrm{fe}}$). The static part of the loss is (frozen $-$ empty)$/E$ and the reaction part (total $-$ frozen)$/E$, each a window mean per run, so that total$/E = 1 + {}$static${} + {}$reaction.

What a trial reports is the treated $-$ control contrast in the reacting world (pilot: periods 60--79; field experiment: from period 100), which carries the arms' difference in composition as a real trial's does. For a rollout, which has no control group, the recorded quantity is the before--after difference an evaluator would measure: the mean rate over the evaluation window minus that over periods 80--99. In a field experiment the pre-trial reference is the reacting world's mean site rate over periods 90--99.

\subsection*{Outcome classes and robustness}

The classes, their thresholds and the robustness rule were fixed in a dated record before the first simulation. Let $\mathrm{ratio}(t) = \mathrm{total}(t)/E$ for each run, with its window mean over periods 160--199 and its end-of-window mean over periods 190--199. Runs with $|E| < 0.02$ are dropped. Each run takes the first class whose rule it meets, in this order:
\begin{itemize}\itemsep0pt
\item reversed: window mean $\le -0.25$;
\item blind trial (field experiment only): window mean of trial$/E_{\mathrm{fe}}$ in $[0.75, 1.25]$ and window mean $< 0.5$;
\item transient: ratio $\ge 0.75$ in some period from the rollout on, and end-of-window mean $< 0.5$;
\item oscillating: in the detrended window series, an autocorrelation peak above 0.5 at a lag of 2--20 after a negative value, with amplitude above $0.25|E|$;
\item delayed: ratio first $\ge 0.75$ 20 or more periods after the start, and end-of-window mean $\ge 0.75$;
\item null: $|$window mean$| < 0.25$;
\item attenuated: end-of-window mean in $[0.25, 0.75]$;
\item as expected: window mean in $[0.75, 1.25]$;
\item otherwise unclassified.
\end{itemize}
An amendment made after the first runs gave as expected and delayed their present form, after noise near the band's edge had left runs unclassified or delayed. Re-classifying the saved runs (originals kept) changed 372 of 6,782, all from unclassified to as expected or from delayed to attenuated.

In a field experiment two site-level classes are tested first. The site-level effect a designer expects from treating a share $p$ is $pE_{\mathrm{fe}}$. A run is a reversed site if the window-mean site change is $\le -0.25\,pE_{\mathrm{fe}}$, and a blind site if the window mean of trial$/E_{\mathrm{fe}}$ lies in $[0.75, 1.25]$ while the window-mean site change is below $0.5\,pE_{\mathrm{fe}}$.

A class is robust when it holds in at least 90\% of runs at each point of a stretch of at least two adjacent sweep points spanning at least 10\% of the swept range. Robustness is evaluated per design, along escalation strength in Fig.~3a and the number of bystanders in Fig.~3b,c. The outline marks the cells where the rollout's total stays below 0.8 of the pilot contrast (as in T1 below) and the new nudge's quiet-field effect lies within the design's bounds. Each pre-specified claim was rerun under the other attention rule (the thermostat map also under the threshold trigger); a point is rule-dependent in class when the majority class or its robust mark differs, in size when only the mean ratios differ by more than 0.1, and rule-independent otherwise.

\emph{Erased} is not a class: the Results use it for a transient effect whose run-averaged end-of-window effect is 0.00\,$E$ (Fig.~2b; the thermostat half of Fig.~3a at zero escalation strength from a competitor strength of 0.45, against 0.39\,$E$ at 0.35). The time to decay (Fig.~4d) is the first period after the rollout, at or after the run-averaged total$/E$ peaks, at which it falls below 0.75 (then 0.5).

\subsection*{The entry ticket: four checks against known regularities}

Before predicting anything new, the model had to reproduce four documented regularities in direction and rough size, each a test whose threshold and run count (10 per setting) were fixed in advance.

\emph{T1, loss at scale}~\citeSI{SI-alubaydli2020}. Pilot then rollout, with one thermostat competitor or one thermostat ally at gain 2, under each attention rule. The statistic, the mean over runs of the rollout's window-mean total divided by the mean over runs of the pilot contrast, must be below 0.8. A ratio of means is used because a 2\% pilot (about 400 people) gives a noisy contrast. The ratios are 0.741 and 0.671 with the competitor (priority, proportional) and 0.621 and 0.738 with the ally; 4, 3, 1 and 6 of the 10 single runs reach 0.8 or more.

\emph{T2, quiet versus crowded field}~\citeSI{SI-dellavigna2022}. The statistic is the field experiment's probit trial effect, $\Phi^{-1}(\text{treated}) - \Phi^{-1}(\text{control})$ of the window-mean rates averaged over runs, in the quiet field over that in a crowded field. The anchor is the academic--government gap on the probit scale, about 2; the band is $[1.5, 2.5]$ under priority. The crowded field has a thermostat competitor and ally (gain 2, offsets $\mp 0.03$) plus $m$ bystanders, with $m$ the smallest value in the pre-specified set $\{2, 3, 4, 6, 8\}$ in the band: $m = 2$, ratio 2.25. T2 is therefore a calibration, not a prediction, and the band does not fix $m$ from below ($m = 0$ and 1 give 1.81 and 2.16). Both thermostats here reach their limits in burn-in, so no actor can respond (the stalemate fields). A second set of crowded fields, the headroom fields, keeps a reacting competitor: a thermostat of strength 0.35 plus $m$ bystanders, no ally. Its calibrated point (run 8), $m = 2$ at budget scale 1.0, keeps headroom and gives T2 1.508, just inside the band's lower edge, and T1 0.640. Figs.~2c and 3b,c come from this second set, not from the stalemate fields.

\emph{T3, dispersion across sites}~\citeSI{SI-bryan2021}. A field experiment at six sites (quiet; one competitor; one ally; 2 and 4 bystanders; the stalemate field). The s.d.\ across sites of the mean window-mean trial contrast, 0.0132, must exceed three times the mean within-site s.e., $3 \times 0.0012 = 0.0036$.

\emph{T4, the inverted U}~\citeSI{SI-saccardo2025}. No other actor; the new nudge automatic with $\tau = 0.5$; one propensity component, $a_i \sim \mathcal{N}(\mu, 0.8^2)$, with $\mu$ set so that control take-up spans 15 levels from 0.1 to 0.8 plus 9 levels from 0.30 to 0.50. The take-up at which the rollout effect peaks must be interior, below 0.5 and within 0.05 of the threshold mechanism's prediction $\Phi(-\beta/2) = 0.423$, with $\beta = \tau\,\mathbb{E}[s^{\mathrm{auto}}]/\sqrt{1 + 0.8^2}$. The peak lies at 0.376, but the top is flat within one s.e.\ over take-up 0.356--0.414, so at 10 runs it is resolved only to that plateau.

\subsection*{The blind-site identity}

In a field experiment treating a share $p$ of the site, write $T_t$ and $C_t$ for the treated and control rates in the reacting world, $S_t = pT_t + (1-p)C_t$ for the site rate and $\bar{S}$ for the pre-trial reference. The trial reports $T_t - C_t$, the control drift is $C_t - \bar{S}$ and the site change is $S_t - \bar{S}$. By definition,
\begin{equation}
\text{site change} = \text{control drift} + p \times \text{trial}.
\end{equation}
The total effect on the treated is $T_t - T^0_t$, where $T^0_t$ is the same people's rate in the paired world without the new nudge. Two conditions make it observable. First, without the intervention the site would have stayed at its pre-trial rate, $S^0_t \approx \bar{S}$. Second, the treated are a random sample of the site, so $T^0_t \approx S^0_t$. The total is then about $T_t - \bar{S}$, the trial plus the control drift, and substituting the definition gives
\begin{equation}
\text{total} \approx (1-p) \times \text{trial} + \text{site change}.
\end{equation}
When a competitor holds the site's rate at its goal, the site change is zero and trial$/$total $= 1/(1-p)$. In run 6 the ratio at $p = 0.4$ is 1.519, 8.9\% below 1.667: the competitor is at its limit (window-mean reach 0.9995), so the site change is no longer zero.

\subsection*{Analytic reduction}

An actor's nudge reaches each person independently each period, so the expected site rate is linear in each actor's reach. With one competitor it needs four averages of the probability of acting: $R_\emptyset$ when neither nudge reaches a person, $R_o$ when the new nudge alone does, $R_c$ when the competitor's alone does and $R_{oc}$ when both do. Then $E = R_o - R_\emptyset$; the competitor's pull is $D_0 = R_\emptyset - R_c$ on people the new nudge does not reach and $D_1 = R_o - R_{oc}$ on those it does; and with the new nudge on a share $p$ and the competitor at reach $r$ the site rate is $R_\emptyset + pE - rD(p)$, $D(p) = (1-p)D_0 + pD_1$. A thermostat, whose goal lies $\delta = 0.03$ below $R_\emptyset$, rests at $r^*(p) = \min\{1, \max\{0, (pE + \delta)/D(p)\}\}$. From its rest before and after a rollout, the share of $E$ kept without escalation is $1 - (D_1 - \delta)/E$ if $D_0 > \delta$ and $D_1 < E + \delta$; 0 if $D_0 > \delta$ and $D_1 \ge E + \delta$; $1 - (D_1 - D_0)/E$, all static, if $D_0 \le \delta$ and $D_1 < E + \delta$; and $(D_0 - \delta)/E \le 0$ otherwise. In a field experiment the thermostat reaches its limit at the treated share $p^* = (D_0 - \delta)/(E + D_0 - D_1)$; below it the site change is zero and trial$/$total $= 1/(1-p)$. Escalation applies in any phase of the run, so a thermostat too weak to reach its goal ($D_0 < \delta$) escalates in burn-in; once its automatic nudge is on, the four rates are recomputed with it, so its strength enters as a fifth quantity. Near its rest the thermostat's rule is stable at the two-period lag while gain times pull is below 1 (at most 0.51 in the runs, at gain 10).

The maximiser's reference, a running minimum of an observed rate that varies by about 0.003 per period, drifts below its expected value, so a prediction fed the expected rate can misplace where the maximiser stops or escalates. A second prediction for it lowers the reference to its starting value minus $\sigma\, m(n)$: $\sigma$ is the sampling s.d.\ of the rate it reads, $n$ the number of periods so far whose rate lay within $\sigma$ of the noise-free running minimum and $m(n)$ the expected maximum of $n$ standard normal draws. It is reported beside the first, outside the rule below; with it, 53 of 64 cells of the maximiser's map under priority sharing agree, against 49 without it (section~\ref{si:reduction}).

The rates (eight in the stalemate fields) are averages over 200,000 people drawn as the model's population is, under the run's attention rule. The predicted paths run the simulator's own actor rules on the expected rate. Nothing is fitted.

A cell agrees when the predicted class is the simulated majority class and the size the class rule reads differs by at most 0.1 (as a share of $E$; in a field experiment, both the site change as a share of $pE_{\mathrm{fe}}$ and the trial as a share of $E_{\mathrm{fe}}$); a map is predicted when at least 90\% of its cells agree. The rule and its application were fixed before the ten distinct maps of Fig.~3 and Fig.~S4 were compared. Seen beforehand, and declared in section~\ref{si:reduction}, were a feasibility check on a handful of settings already in the draft and one unplanned simulator run on one setting during the build. Eight maps pass; the maximiser under priority sharing (49 of 64 cells) and the threshold trigger (50 of 64) are predicted in part. Seven further runs, behind Figs.~2a,b,d,e and 4 and the ally of Fig.~S1, agree at every point (33 of 33).

\subsection*{Sensitivity of the fates to the field's parameters}

For each of twelve combinations of competitor (thermostat, maximiser or threshold trigger), attention rule and design, a Latin-hypercube sample of 4,096 settings was drawn, spread evenly over every parameter's range: 49,152 settings, each with one competitor and no ally. The new nudge's strength and the competitor's ranged over 0.10--0.60; the competitor's gain over 0.5--10 (log scale; for the threshold trigger, its step, 0.01--0.10 per period); its goal from 0 to 0.10 below the no-actor baseline; its lag over 1--5 periods; its automatic nudge's strength over 0--0.50; the budget scale over 0.35--4.0 (log scale); the bystanders over 0--16; control take-up over 0.10--0.80, set by one shift of both propensity components; and, in a field experiment, the treated share over 0.10--0.60. Everything else was as above. Each setting's class, and whether it lies inside the outline, was read by the rules above from the reduction's noise-free path (200,000 people, the same draws at every setting); a setting with $|E| < 0.02$ counts as excluded. Reported per combination are the share of each fate, over all settings and inside the outline, and, per parameter, how far the fate shares shift between the lowest and highest third of its drawn values (half the summed absolute differences, over classified settings). Of each combination's settings, 24 drawn at random (288 in all) were simulated, 10 runs of 20,000 people each on a cloud T4 GPU, and compared with the path by the cell rule above. The ranges, the statistics and five expectations were fixed before the sample was drawn; section~\ref{si:sweep} reports each expectation as registered and as found.

\subsection*{Pre-specified runs and parameters}

Every sweep behind a figure was specified before it ran, as a dated entry in a version-controlled record naming the scenario, design, swept values, runs, seeds, attention rule, the class expected at each point and the prior knowledge it drew on. The record is not a public preregistration. Deviations and unmet expectations are disclosed in the entry. Runs 1, 2 and 2b cover competitor gains 0.5--5, the ally, the bystanders, escalation strengths 0.1--0.5 and the escalating thermostat's gain 0.5--10 (Figs.~2a,b,d,e, 4c,d; Figs.~S1, S2); runs 1p and 2c add proportional sharing and the maximiser's memory. Run 3 maps competitor strength 0.20--0.60 against escalation strength 0--0.50 for the thermostat, maximiser and threshold trigger (Fig.~3a; Fig.~S4a--e). Run 4 maps the stalemate fields over budget scale 0.35--4.0 and 0--16 bystanders (Fig.~S4h,i), and run 6 the treated share 0.1--0.6 (Fig.~4a,b). Run 8 calibrates the headroom fields, and run 4h maps them over budget scale and bystanders in both designs and rules (Figs.~2c, 3b,c; Figs.~S1, S2, S4f,g). Runs 9 and 10 are the clean suite of the four checks and its finer T4 grid (Fig.~S3). The record, with the date of each entry, is section~\ref{si:record}. Run 11, registered after the final review of 12 November 2026, varies the pilot's share of the site against the three competitors of Fig.~2 (section~\ref{si:pilotshare}).

\subsection*{Software and reproducibility}

The simulator is written in Python with PyTorch, vectorised over runs and people, and was run on an Apple M4, an NVIDIA DGX Spark and, for run 8, cloud T4 GPUs. Seeds are set by scenario, design, actor structure and seed base; draws differ between devices, so runs reproduce per device. Each run directory has a manifest of its configuration, seeds, device, software versions, code commit and tree state. The figure scripts refuse any run whose manifest shows an incomplete run, a dirty tree, outdated class rules or the wrong attention rule, and a sidecar file lists each figure's runs and commits. Figs.~1--4 and S1--S4 were rendered at a clean commit, and a second render was byte-identical; Fig.~S5 was also rendered at a clean commit.

\subsection*{The dataset screenings}

No open dataset was found on which what is specific to a responding field could be tested, so every field-specific result in the paper rests on simulation. Three screenings were run, each with its eligibility gates fixed before the search began. The first looked for a published two-by-two field experiment with a binary outcome in which both interventions move the outcome and the arms are large enough to detect an interaction of 0.2 on the probit scale: nine candidates, none passing both gates; the precise ones have one factor without an effect, and the one with two active factors is too small. The second searched Zenodo, Figshare, OSF, Harvard Dataverse and three further archives for three designs: a factorial with two strong forces, the same intervention across backgrounds with a known propensity per person, and a trial with a logged field of other sources. About 50 candidates were scored; two are eligible for the location of the peak in take-up, none for a test that separates a responding field from a fixed one, because open trials field one weak force at a time and the crossed designs with strong forces have small cells. The third looked for the signatures of Results, ``Signatures a trial can record'': randomised treated shares across sites, a logged response by another actor or a pilot followed by a rollout, and multi-site trials with a site identifier. Of 66 designs, none was eligible. Three structural reasons emerged. Megastudies assign people individually, so the treated share of a site's population is about the same everywhere, and small. In behavioural trials that varied the treated share, the spillover onto the untreated is zero or positive; the blind site's opposite sign appears only where a fixed pool or a responding actor is named, and those trials are not of nudges. And open data record the other actor's outcome, not its action. Some candidates appear in more than one screening; about 120 were scored in all. The record of the screenings, with every candidate and gate, is in the deposit (doi:10.5281/zenodo.24150612).

\clearpage
\sisection{supplementary figures}{si:figures}
\setcounter{figure}{0}
\renewcommand{\figurename}{Fig.}
\renewcommand{\thefigure}{S\arabic{figure}}
\renewcommand{\theHfigure}{ED\arabic{figure}}
\begin{figure}[p]\centering\includegraphics[width=0.49\textwidth]{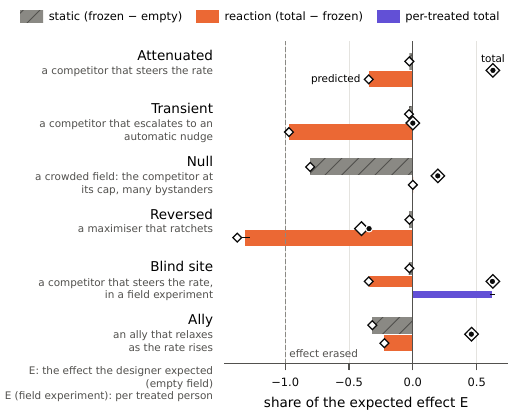}\caption{\textbf{The field's effect has a static part, which the earlier revivals of the field of forces describe, and a reaction part, the forces' response that Lewin described~\protect\citeSI{SI-lewin1951} and the revivals leave out; which part dominates is set by the field: crowding alone is static, a reacting actor with headroom is reaction.}
One row per field: the five fields of Fig.~2, in its order and with its settings, then a thermostat ally with response gain 2. The bars split the new nudge's shortfall from $E$ over the evaluation window into a static part (grey, hatched: frozen $-$ empty) and a reaction part (orange: total $-$ frozen), as shares of $E$. The frozen field holds the other actors' nudges at their reach before the rollout, so the static part is what the field costs by standing there and the reaction part is what the actors' responses to the rate add. The split fixes one order: the frozen field is the field as a pilot saw it, so the static part is what a pilot could have measured and the reaction part what it could not; the two parts add exactly to the whole shortfall. $E$ is the new nudge's effect on the site in the empty field. Since 1 + static + reaction = total\,/\,$E$, the ink dot in each row (``total'') is the window total over $E$. The dashed line at $-1$ (``effect erased'') marks where the two parts together cancel the expected effect; parts that pass it, as in the reversed row, turn the effect's sign. The null row has no reaction bar: its reaction is 0 in every replicate run, because the competitor is at its limit before the pilot and the bystanders do not respond to the rate. The blind-site row is per treated person, over $E$ (field experiment), and adds the per-treated total (violet). Open diamonds mark the analytic reduction's prediction for each part and, around the dot, for the total (section~\ref{si:methods}); they differ from the simulated values by about 0.06 at most in the reversed row (its reaction part) and by under 0.004 in the others. Bars are replicate means. Whiskers span the 10th--90th percentiles of $n = 100$ replicate runs per row; they are drawn only on bars long enough to carry them, so the static bars of about $-0.03\,E$ carry none. All runs are pre-specified, complete and from a clean commit.}\end{figure}
\begin{figure}[p]\centering\includegraphics[width=\textwidth]{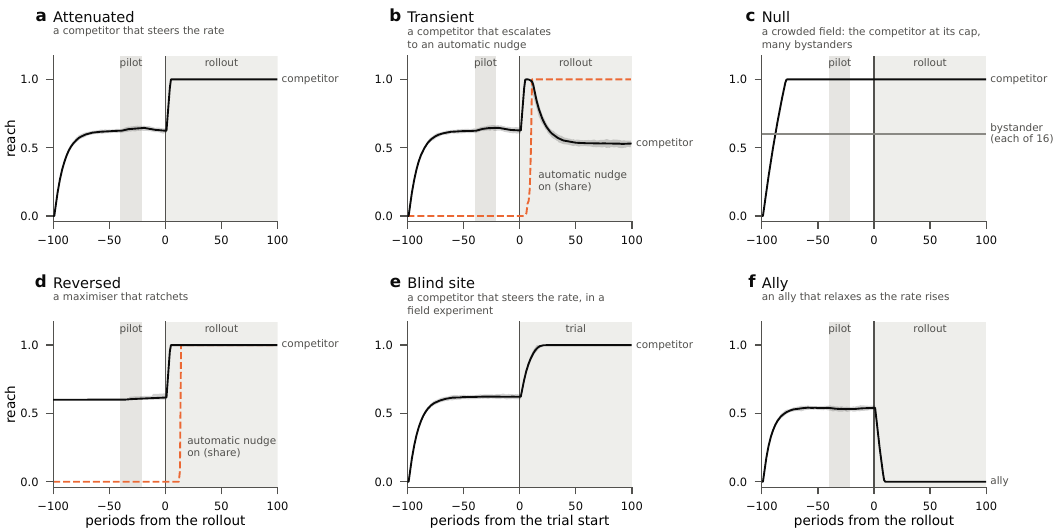}\caption{\textbf{Where the actors can respond, the loss is produced by their responses to the rate; where they cannot (c), it is static.}
What the other actors did in the six fields of Fig.~S1, at its settings: \textbf{a}, attenuated; \textbf{b}, transient; \textbf{c}, null; \textbf{d}, reversed; \textbf{e}, blind site; \textbf{f}, ally. Each panel draws every other actor's reach over the whole run, periods $-100$ to 99 from the rollout or trial start, with the pilot and rollout spans shaded (in \textbf{e}, a field experiment, there is no pilot, so only the trial span is shaded). Copies of one actor are averaged into one line (the 16 bystanders in \textbf{c}). The dashed line is the share of replicate runs in which the competitor's automatic nudge is on.
\textbf{a}, The competitor reaches its limit within a few periods of the rollout.
\textbf{b}, The automatic nudge comes on in every run about ten periods after the rollout, and the competitor relaxes.
\textbf{c}, The competitor reaches its limit early in burn-in, and the bystanders, which do not respond to the rate (gain 0), stay at their initial reach: no actor can respond, and the whole loss is static (Fig.~S1).
\textbf{d}, The competitor reaches its limit within a few periods of the rollout and stays there when its automatic nudge comes on, about fourteen periods after the rollout.
\textbf{e}, The competitor reaches its limit within about 20 periods of the trial start.
\textbf{f}, The ally relaxes to 0 within ten periods of the rollout.
Each reach line is the replicate mean (copies averaged within each run first); its band spans the 10th--90th percentiles of $n = 100$ replicate runs per panel. The share line has no band, being the mean of an on/off indicator. All runs are pre-specified, complete and from a clean commit.}\end{figure}
\begin{figure}[p]\centering\includegraphics[width=\textwidth]{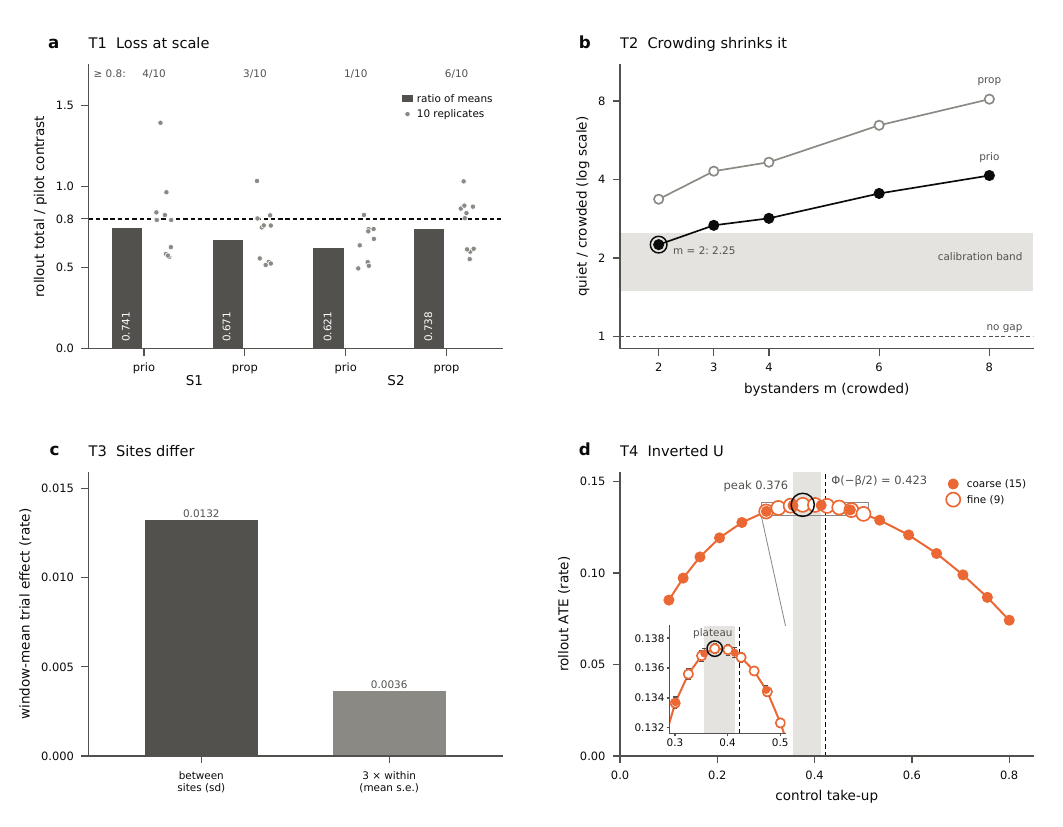}\caption{\textbf{The model reproduces the known regularities before it predicts new ones.}
The four tests of the entry-ticket suite.
\textbf{a}, T1, loss at scale. Per site and attention rule (S1, a thermostat competitor, and S2, a thermostat ally (scenario names, not SI sections), each with response gain 2; prio, priority; prop, proportional), the ratio of the rollout total to the pilot contrast for each replicate run (light grey dots), the ratio of the replicate means (dark grey bars; the test's statistic) and the bound of 0.8 (dashed). The top row counts the runs at or above 0.8; single runs cross it because the pilot contrast is noisy.
\textbf{b}, T2, crowding shrinks it. The trial effect on the probit scale in a quiet field over that in a crowded stalemate field (a thermostat competitor and ally whose goals cannot both be met, both at their limit), against the number of bystanders $m$ in the crowded field (log scale, base 2; priority filled, proportional open), with the calibration band [1.5, 2.5] for priority and the calibrated $m = 2$ ringed.
\textbf{c}, T3, sites differ. The between-site standard deviation of the window-mean trial effect over six sites (dark grey; the statistic) against three times the mean within-site standard error (light grey; the threshold).
\textbf{d}, T4, inverted U. The rollout average treatment effect against control take-up, on a coarse grid of 15 levels (filled) and a fine grid of 9 levels from 0.30 to 0.50 (open), with $\pm 1$ s.e. bars, the peak ringed and labelled, the plateau (the levels within one s.e. of the maximum) shaded and the oracle $\Phi(-\beta/2)$, the take-up at which the threshold alone puts the peak, dashed. The inset zooms on take-up 0.29--0.51, where the bars and the plateau are legible.
The suite runs at $n = 10$ replicate runs per setting, the number fixed in advance for it; every other figure uses 100. All runs are pre-specified, complete and from one clean commit.}\end{figure}
\begin{figure}[p]\centering\includegraphics[width=0.66\textwidth]{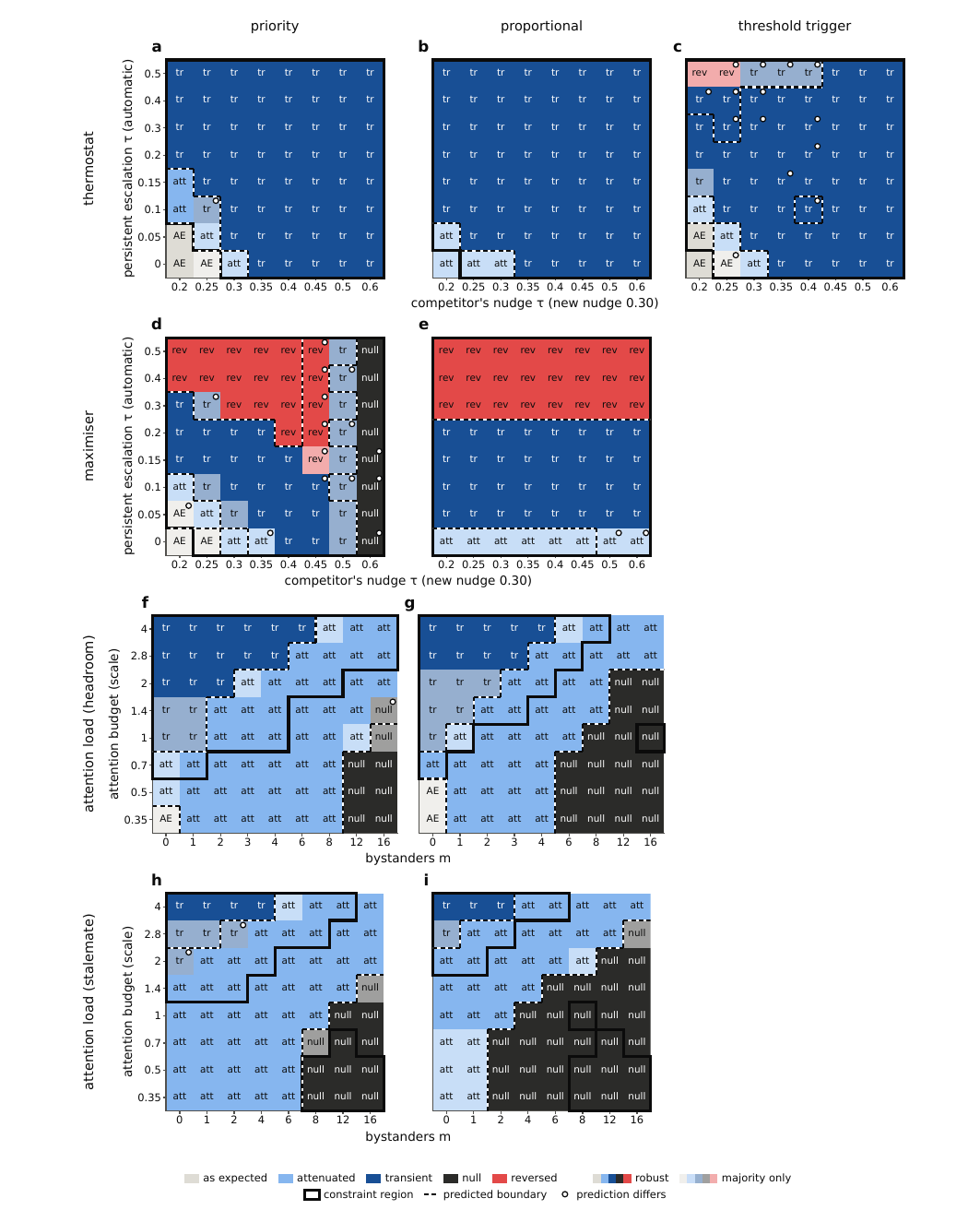}\caption{\textbf{The fates survive a change of attention or response rule; the boundaries between them move.}
Nine phase maps as in Fig.~3 (each cell the majority outcome over $n = 100$ replicate runs), in rows by kind of map and columns by rule, so that the maps in a row compare cell for cell. Columns: priority (attention goes to the strongest nudge first; the default), proportional (attention is shared in proportion) and threshold trigger (the competitor acts only once the rate leaves a tolerance band).
Row 1, the thermostat competitor: \textbf{a}, priority; \textbf{b}, proportional; \textbf{c}, threshold trigger.
Row 2, the maximiser: \textbf{d}, priority; \textbf{e}, proportional.
Row 3, attention load with a competitor that has headroom: \textbf{f}, priority; \textbf{g}, proportional.
Row 4, attention load in the stalemate fields (a competitor and an ally whose goals cannot both be met, both at their limit): \textbf{h}, priority; \textbf{i}, proportional.
The priority maps \textbf{a}, \textbf{d} and \textbf{f} are those of Fig.~3a and 3b, without the Fig.~2 markers. Rows 1 and 2 plot the strength $\tau$ of the competitor's nudge (the new nudge's is 0.30) against $\tau$ (automatic); rows 3 and 4 plot the number of bystanders $m$ against the attention budget (scale), as in Fig.~3. Row 3 has nine $m$ columns, row 4 eight. Solid (robust) and faded cells as in Fig.~3. The outline is as in Fig.~3; in \textbf{e} it covers the whole map. The maximiser has no threshold-trigger variant, and the threshold trigger is inert on the bystanders (gain 0), so rows 2--4 have no third map. In \textbf{h} and \textbf{i}, at budgets of 1 or less, the first of the four checks is set by noise (within two standard errors of 0.8 in 33 and 38 of 64 cells), so the outlined region there falls into pieces; it is drawn as the data give it. In \textbf{c}, two faded cells are reversed (58 and 52 of 100 runs), at the two weakest competitors under the strongest automatic nudge. The dashed lines and open dots are the analytic reduction's predicted boundaries and the cells where it differs from the simulation, as in Fig.~3. Cells that agree: \textbf{a}, 63 of 64; \textbf{b}, 64 of 64; \textbf{c}, 50 of 64 (predicted in part; in each of the 14 differing cells the predicted effect is lower than the simulated one); \textbf{d}, 49 of 64 (predicted in part); \textbf{e}, 62 of 64; \textbf{f}, 71 of 72; \textbf{g}, 72 of 72; \textbf{h}, 62 of 64; \textbf{i}, 64 of 64. In \textbf{c}, \textbf{d} and \textbf{e} a cell without a dot can still differ by up to 0.1: the simulated effect is the larger in every cell of \textbf{c} and \textbf{e} (median differences 0.069 and 0.065 of $E$) and in \textbf{d} at competitor strengths up to 0.40 (column medians 0.050 to 0.067), and mostly the smaller in \textbf{d} at 0.45 to 0.60. Oscillating, delayed, blind-trial and reversed-site outcomes were found robust nowhere in the explored runs and are not drawn. All runs are pre-specified, complete and from a clean commit.}\end{figure}

\clearpage
\sisection{the analytic reduction}{si:reduction}
\setcounter{table}{0}
\renewcommand{\tablename}{Table}
\renewcommand{\thetable}{S\arabic{table}}
\renewcommand{\theHtable}{S\arabic{table}}

This section derives the fates of the main text from the model's rules, with nothing fitted, and compares the result with the simulated maps. The derivation follows the simulator's rules as coded. Every quantity in it comes from a run's configuration; the simulation remains the source of the run-to-run spread, the noise in a trial's comparison and the robustness marks.

\subsection*{The site's rate is linear in each reach}

Person $i$ acts in period $t$ when the latent sum of the Methods' threshold equation plus standard normal noise is positive, so with probability $\Phi(\ell_{it})$, where $\ell_{it}$ is that sum without the noise. Each other actor's nudge reaches each person independently each period with probability equal to the actor's reach; the new nudge reaches a fixed treated set. The population is fixed, habit updating is off, bystanders keep a constant reach and an automatic nudge, once on, reaches everyone. The expected rate of any group is therefore an average of $\Phi$ over its members and over their exposure states, weighted by the probability of each state. With the new nudge and one competitor there are four states and four rates:
\[
R_\emptyset \ (\text{neither nudge}),\quad R_o \ (\text{the new nudge only}),\quad R_c \ (\text{the competitor's only}),\quad R_{oc} \ (\text{both}).
\]
Each is an average of $\Phi$ over the population's inclinations, susceptibilities, valences and attention budgets, under the attention rule that shares a budget between the nudges reaching a person. Write
\[
E = R_o - R_\emptyset, \qquad D_0 = R_\emptyset - R_c, \qquad D_1 = R_o - R_{oc}
\]
for the new nudge's effect in the empty field and the competitor's pull on people the new nudge does not reach and on those it does. With the new nudge on a share $p$ of the site and the competitor at reach $r$, the expected site rate is
\[
S(r, p) = R_\emptyset + pE - r\,D(p), \qquad D(p) = (1-p)\,D_0 + p\,D_1 .
\]
The expression is exact in expectation; at 20,000 people the period-to-period sampling s.d.\ of the rate is about 0.003. With $k$ reacting nudges the rate is multilinear in their reaches: the sum over subsets of the nudges of the rate when exactly that subset reaches everyone, times $\prod_i r_i$ over the subset and $\prod_i (1 - r_i)$ over the rest, so $2^k$ rates are needed. Bystanders' nudges enter every rate at their constant reach, through one draw per person and nudge shared by every subset. The new nudge enters as a reacting nudge whose reach is the treated share. Each rate is computed as an average over 200,000 people drawn with the simulator's own population functions, one draw per setting.

\subsection*{The resting reach of a steering competitor}

A thermostat competitor sets $r_t = \min\{1, \max\{0,\, r_{t-1} + \kappa\,(S^{\mathrm{obs}}_{t-L} - G)\}\}$ with gain $\kappa$, lag $L = 2$ and goal $G = R_\emptyset - \delta$, $\delta = 0.03$ (section~\ref{si:methods}). At rest the rate it observes equals its goal unless its reach is clipped, $S(r, p) = G$, so its resting reach is
\[
r^*(p) = \min\Big\{1, \max\Big\{0,\ \frac{pE + \delta}{D(p)}\Big\}\Big\}.
\]
Before the rollout ($p = 0$) it rests at $r_0 = \min\{1, \delta/D_0\}$; after a rollout to everyone ($p = 1$), at $r_1 = \min\{1, (E + \delta)/D_1\}$. Its headroom before the rollout is $1 - \delta/D_0$. As in the simulator, a reach of at least 0.99 counts as the limit.

\subsection*{Four cases}

Without the new nudge the competitor stays at $r_0$ and the rate at $R_\emptyset - r_0 D_0$; with it, the competitor moves to $r_1$ and the rate to $R_\emptyset + E - r_1 D_1$. The rollout's effect at rest is
\[
T = E - (D_1 r_1 - D_0 r_0),
\]
which, without escalation, takes one of four forms (Table~\ref{tab:cases}). The loss at scale needs room before the rollout, $D_0 > \delta$ (see the pilot, below). The effect is erased when the competitor's pull on the people the new nudge reaches is at least the new nudge's effect plus the distance of its goal, $D_1 \ge E + \delta$. The fourth case requires $D_1 - D_0 \ge E$, a competitor that pulls the people the new nudge reaches harder than the rest by at least the new nudge's whole effect.

\begin{table}[ht]\centering\small
\caption{\textbf{The four cases for a steering competitor without escalation.} $T/E$ is the share of the expected effect kept at rest.}\label{tab:cases}
\begin{tabular}{@{}lll@{}}
\toprule
Case & Condition & $T/E$ \\
\midrule
no room before or after & $D_0 \le \delta$ and $D_1 < E + \delta$ & $1 - (D_1 - D_0)/E$, all static \\
room that runs out & $D_0 > \delta$ and $D_1 < E + \delta$ & $1 - (D_1 - \delta)/E$ \\
room enough & $D_0 > \delta$ and $D_1 \ge E + \delta$ & $0$ \\
no room before, room after & $D_0 \le \delta$ and $D_1 \ge E + \delta$ & $(D_0 - \delta)/E \le 0$ \\
\bottomrule
\end{tabular}
\end{table}

In the field of Fig.~2a the reduction gives $E = 0.0555$, $R_\emptyset = 0.3090$, $D_0 = 0.0484$ and $D_1 = 0.0509$: the competitor rests at reach 0.620 before the rollout, the case is room that runs out, and the rollout keeps 0.623 of $E$ (simulated, 0.63).

\subsection*{The parts of the loss}

The frozen world holds the competitor at $r_0$, where the new nudge's effect is $E - r_0 (D_1 - D_0)$. The static and reaction parts are therefore
\[
\text{static} = -\frac{(D_1 - D_0)\, r_0}{E}, \qquad \text{reaction} = -\frac{D_1 (r_1 - r_0)}{E},
\]
and $1 + \text{static} + \text{reaction} = T/E$. The static part is small when the competitor pulls treated and untreated people about equally, $D_1 \approx D_0$: in the field of Fig.~2a it is $-0.028$, as simulated. Across the six rows of Fig.~S1 the predicted parts differ from the simulated ones by about 0.06 at most, in the reaction part of the reversed row, and by under 0.004 in the other rows.

\subsection*{The pilot}

A pilot on 2\% of the site does not move the competitor's reach. Its contrast is $E - r_0 (D_1 - D_0)$, close to $E$ because $D_1 \approx D_0$. Without room before the rollout ($r_0 = 1$) the pilot's contrast equals the rollout's effect in the first case of Table~\ref{tab:cases}, and is no larger than it in the fourth, so a loss at scale needs $D_0 > \delta$.

\subsection*{The blind site and its limit}

In a field experiment on a share $p$ the competitor rests at $r^*(p)$. The treated rate is $R_\emptyset + E - r^* D_1$ and the control rate $R_\emptyset - r^* D_0$, so
\begin{gather*}
\text{trial} = E - r^*(p)\,(D_1 - D_0), \qquad \text{control drift} = -\big(r^*(p) - r_0\big)\,D_0, \\
\text{site change} = pE - r^*(p)\,D(p) + r_0 D_0 .
\end{gather*}
While $r^*(p) < 1$ (and $r_0 < 1$) the site change is zero, and the identity of section~\ref{si:methods} gives trial$/$total $= 1/(1-p)$. The competitor reaches its limit where $pE + \delta = D(p)$, at
\[
p^* = \frac{D_0 - \delta}{E + D_0 - D_1},
\]
meaningful when $D_0 > \delta$ and $E + D_0 - D_1 > 0$. For the setting of Fig.~4b, $p^* = 0.347$ from one draw of 200,000 people. At $p = 0.4$ the competitor is at its limit, the trial's contrast is 0.954 of the expected effect per treated person, the control drift $-0.018$ and the site change $+0.003$. The simulated trials follow $1/(1-p)$ to within 1.3\% up to $p = 0.3$ and fall 8.9\% below it at $p = 0.4$.

\subsection*{Escalation}

A thermostat with an automatic nudge escalates when it is at its limit and the rate it observes has been above its goal for ten consecutive periods, in any phase of the run. After the rollout that is the case $D_1 < E + \delta$. Before the rollout it is the case $D_0 < \delta$: a competitor too weak to reach its goal escalates during burn-in, with or without the new nudge. The reduction applies the rule in both phases. Once the automatic nudge is on, the four rates are recomputed with it (written with a superscript $a$) and the competitor relaxes to
\[
r^{**} = \frac{R_o^a - G}{D_1^a}.
\]
If $0 \le r^{**} \le 1$ the effect ends at zero, the erased effect of Fig.~2b. If $r^{**} > 1$ the competitor stays at its limit and some effect remains. If $r^{**} < 0$ the automatic nudge alone holds the rate below the goal and the effect ends negative, at $(R_o^a - G)/E$.

\subsection*{The ratcheting competitor}

A maximiser has reach 0.60 until period 59. Fed the expected rate, its reference starts at the rate then, $q = R_\emptyset - 0.60\,D_0$. After the rollout it raises its reach while the rate it observes is more than 0.01 above the reference, and it never lowers it. If the rate at its limit, $R_{oc}$, is more than 0.01 above $q$, it escalates, and the effect ends at $(R_{oc}^a - q)/E$, negative once the automatic nudge is strong enough. If the rate comes within 0.01 of $q$ before the limit, the competitor stops there and a small positive effect remains, near $0.01/E$. The predictions use the path, which runs the simulator's own maximiser rule.

The reference is a running minimum of the observed rate from period 60. Sampling noise of about 0.003 per period carries a running minimum below its expected value, so the simulated reference falls below $q$. A path fed the expected rate (the noise-free prediction, used for every map) keeps the reference at $q$, and so can misplace both where the competitor stops and the line between stopping and escalating. For this competitor a second prediction was fixed before the comparison. At each period the competitor's reference is lowered to its starting value minus $\sigma\, m(n)$, with $\sigma$ the sampling s.d.\ of the rate it read (20,000 people), $n$ the number of periods so far whose read rate lay within $\sigma$ of the noise-free running minimum, and $m(n)$ the expected maximum of $n$ standard normal draws. It applies only where the reference is a true running minimum. It is reported beside the noise-free prediction and does not enter the comparison's rule.

With this term the maximiser's map under priority sharing has 53 of 64 cells agreeing, against 49 without it. At competitor strengths 0.45 and 0.60 every cell then agrees except one at 0.45, which matches in class and differs in size ($-1.27$ predicted, $-1.17$ simulated). At strength 0.50 all seven cells with an automatic nudge disagree (three predicted reversed and three null where the simulation is transient, one differing in size only), against three without the term. Under proportional sharing, and on the line behind Fig.~2d, the two predictions give the same cells.

\subsection*{The ally}

An ally steers towards $R_\emptyset + \delta$. Write $U_0$ and $U_1$ for its lift on people the new nudge does not and does reach ($U = -D$). It rests at $\min\{1, \delta/U_0\}$ before the rollout and at $\min\{1, \max\{0, (\delta - E)/U_1\}\}$ after it, so it withdraws fully when $E \ge \delta$, and the share of $E$ kept is then $1 - \delta/E$. The parts follow the formulas above with $D = -U$. The predicted total is 0.46 of $E$, as simulated (Fig.~S1).

\subsection*{Two reacting actors: the stalemate fields}

With a competitor and an ally the site rate is linear in each reach separately, so it needs eight exposure rates, and the resting state solves the two bounded integrators together. Both actors sit at their limits while the rate with both at full reach lies between their goals. After the rollout, where the new nudge lifts the rate past the ally's goal (as with ample attention), the ally withdraws. With bystanders and a varied attention budget, as on the attention-load maps, the rates become functions of the number of bystanders and the budget scale and the same algebra holds: the competitor loses its room to respond where $D_0 = \delta$, and the null fate begins where $T/E$ falls to 0.25.

\subsection*{Other rules}

Proportional sharing changes the attention weights inside the rates and nothing else. The threshold trigger moves its reach only while the rate is more than 0.01 from its goal, so it rests at the edge of that band, in practice $G + 0.01$ approached from above, and it escalates only when the rate at its limit stays more than 0.01 above the goal, $D_1 < E + \delta - 0.01$.

\subsection*{The time course and the stability bound}

The simulator sets the reach at period $t$ from the rate observed at $t - L$, before the rate of period $t$ is formed. Counting the rollout period as 0, periods $0, \dots, L-1$ keep $r_0$, and from $t = L$ on
\[
r_t = \min\{1, \max\{0,\ r_{t-1} + \kappa\,(E + \delta - D_1 r_{t-L})\}\}, \qquad \text{effect}_t = E - D_1 r_t + D_0 r_0 .
\]
Iterated on the expected rates, the effect falls below three-quarters of $E$ 11, 6, 4, 2 and 2 periods after the rollout at gains 0.5, 1, 2, 5 and 10, as in the simulation (Fig.~4d).

Away from the limit, a deviation $x_t$ of the reach from its rest follows $x_t = x_{t-1} - k\, x_{t-L}$ with $k = \kappa D_1$. Its characteristic equation is $z^L - z^{L-1} + k = 0$, so $k = z^{L-1}(1 - z)$. On the unit circle, $z = e^{i\theta}$ and $1 - z = 2\sin(\theta/2)\, e^{i(\theta - \pi)/2}$, so $k$ is real and positive when $(L-1)\theta + (\theta - \pi)/2 = 0$, that is $\theta = \pi/(2L-1)$ and
\[
k = 2 \sin\!\Big(\frac{\pi}{2(2L-1)}\Big),
\]
the first $k > 0$ at which a root leaves the unit disc; for small $k > 0$ all roots lie inside it. A lag of one period gives 2 and the model's lag of two periods gives 1: the response dies away while gain times pull is below 1. The largest value in the explored runs is 0.51, at gain 10.

\subsection*{What was known before the comparison}

\emph{A feasibility check.} Before the reduction was built, a script outside the project (not kept) computed the four rates by Monte Carlo over 2 million people drawn as section~\ref{si:methods} describes and applied the closed forms above to settings whose simulated values were already in the draft (Table~\ref{tab:prior}). These comparisons were made knowing the simulated numbers. The reduction has no free parameter, so there was nothing to tune, but the rows are prior knowledge. The ally and the field experiment's contrast and control drift at $p = 0.4$ were not computed, although they follow by arithmetic from these rows; no attention-load cell, no proportional-sharing cell and no phase map as a whole was computed. The check's lag timing gave crossings one period earlier than the simulation; the reduction used in the comparison takes its timing from the simulator's code.

\begin{table}[ht]\centering\small
\caption{\textbf{The feasibility check, made before the reduction was built and knowing the simulated values.} Competitor strengths are the strength of the competitor's nudge (the new nudge's is 0.30); ``automatic'' is the strength of its automatic nudge.}\label{tab:prior}
\begin{tabularx}{\textwidth}{@{}>{\raggedright\arraybackslash}X>{\raggedright\arraybackslash}p{0.25\textwidth}>{\raggedright\arraybackslash}p{0.3\textwidth}@{}}
\toprule
Quantity & Feasibility check & Simulation (draft) \\
\midrule
$R_\emptyset$; $E$ & 0.309; 0.0553 & 0.309; 0.055 \\
competitor's reach before the rollout (strength 0.30) & 0.623 & 0.626 \\
effect kept, competitor strength 0.20 / 0.30 / 0.35 / $\ge 0.45$ & 1.10 / 0.625 / 0.39 / 0.00 & 1.10 / 0.63 / 0.39 / 0.00 \\
static part, strength 0.30 & $-0.029$ & about $-0.03$ \\
reach after escalation (automatic 0.30) & 0.526 & 0.53 \\
reversal, automatic 0.30; 0.10 & $-0.41$; $+0.25$ & $-0.39$ (end); $+0.26$ \\
steering competitor, automatic 0.50 & $-0.21$ (short of the $-0.25$ that defines reversed) & transient under the default rule; reversed in 58 and 52 of 100 runs under the threshold trigger at two weaker strengths \\
blind site, $p^*$ & 0.35 (held at 0.3, at the limit at 0.4) & holds to 0.3, breaks from 0.4 \\
periods to fall below $0.75\,E$, gains 0.5 / 1 / 2 / 5 / 10 & 10 / 5 / 3 / 1 / 1 & 11 / 6 / 4 / 2 / 2 \\
\bottomrule
\end{tabularx}
\end{table}

\emph{An unplanned simulator run.} During the build, against the build's own rules, the simulator was run once on the setting of Fig.~2b (a steering competitor of strength 0.30, gain 2, automatic nudge 0.30), with 20,000 people, 8 runs and seed 3. The competitor's reach was compared with the path's at three periods: 99 (path 0.623; runs 0.608--0.648), 120 (0.731; 0.66--0.76) and 199 (0.526; 0.50--0.56); the automatic nudge was on by period 120 in both. By the implementer's statement the run came after the code and its tests were written and passing, and no code or test value changed afterwards. No table of a run behind the paper's figures was read. The setting is one cell of the steering competitor's map in Fig.~3a. That cell is counted, as every cell is; with it left out, the map has 62 of 63 cells agreeing (98.4\%, unchanged), and the cell itself agrees.

\emph{Who knew what, and when.} The simulated maps were drawn, and known to the author, before the reduction was built. The comparison rule, and how it would be applied (what is compared, the simulated side of a cell, agreement without rounding allowance, the second prediction for the maximiser, the number of people), were fixed and committed before any map was predicted, with the reduction's code frozen. The implementers of the reduction's code were kept from the run outputs, with the one exception stated above. Between the first computation of the predictions and the comparison, one correction was made, to how the class-rule version of the older runs is read: the manifests of runs 1, 2 and 2b, made before the class rules were amended, carry no version, and it is now read from the reclassification record beside each run, as the figure code already does. The prediction files of the ten maps, and of the two finite-size maps, are identical before and after it (their hashes are with the data). Nothing was changed after the comparison was computed.

\subsection*{The comparison}

A cell agrees when the predicted class equals the simulated majority class and every size the class rule reads differs by at most 0.1: the window-mean effect over $E$ in a pilot-then-rollout map, and in a field-experiment map both the site's change over its expected change and the trial's contrast over the expected effect. A map passes when at least 90\% of its cells agree, on the noise-free prediction. Table~\ref{tab:comparison} gives the result for the ten distinct maps and for the seven further runs compared point by point, for information and outside the rule. Across the ten maps 624 of 664 cells agree (94.0\%), a descriptive total that is not the rule's statistic.

\begin{table}[ht]\centering\small
\caption{\textbf{The comparison of the reduction with the simulation.} Differences are in the size each class rule reads (shares of $E$; in the field-experiment map, the larger of the site and trial differences). The lines are compared point by point, outside the rule.}\label{tab:comparison}
\begin{tabularx}{\textwidth}{@{}>{\raggedright\arraybackslash}Xlrrrrl@{}}
\toprule
Map or line & Drawn in & Cells & Agree & Median & Largest & Result \\
\midrule
Steering competitor, priority & Fig.~3a left; Fig.~S4a & 64 & 63 & 0.0002 & 0.017 & passes \\
Ratcheting competitor, priority & Fig.~3a right; Fig.~S4d & 64 & 49 & 0.060 & 1.336 & partial \\
Attention load, pilot then rollout, priority & Fig.~3b; Fig.~S4f & 72 & 71 & 0.002 & 0.008 & passes \\
Attention load, field experiment, priority & Fig.~3c & 72 & 67 & 0.008 & 0.038 & passes \\
Steering competitor, proportional & Fig.~S4b & 64 & 64 & 0.0004 & 0.008 & passes \\
Steering competitor, threshold trigger & Fig.~S4c & 64 & 50 & 0.069 & 0.203 & partial \\
Ratcheting competitor, proportional & Fig.~S4e & 64 & 62 & 0.065 & 0.090 & passes \\
Attention load, pilot then rollout, proportional & Fig.~S4g & 72 & 72 & 0.0003 & 0.004 & passes \\
Stalemate, priority & Fig.~S4h & 64 & 62 & 0.001 & 0.011 & passes \\
Stalemate, proportional & Fig.~S4i & 64 & 64 & 0.0004 & 0.009 & passes \\
\midrule
Steering competitor, gain & Fig.~2a & 3 & 3 & 0.001 & 0.005 & \\
Escalating steering competitor & Fig.~2b & 3 & 3 & 0.000 & 0.003 & \\
Ratcheting competitor, memory & Figs.~2d, 4c & 10 & 10 & 0.030 & 0.059 & \\
Steering competitor, field experiment & Fig.~2e & 3 & 3 & 0.012 & 0.013 & \\
Treated share & Fig.~4a,b & 6 & 6 & 0.009 & 0.020 & \\
Escalating competitor, gain & Fig.~4d & 5 & 5 & 0.000 & 0.000 & \\
Ally & Fig.~S1 (ally row) & 3 & 3 & 0.002 & 0.002 & \\
\bottomrule
\end{tabularx}
\end{table}

In the eight maps that pass, the 11 disagreeing cells all differ in class only. Their sizes differ by at most 0.011, except two cells of the ratcheting competitor's map under proportional sharing, attenuated in the simulation and transient in the prediction (0.572 against 0.493 and 0.541 against 0.462, a difference of 0.079). Several sit on a class boundary in the size itself: 0.499 simulated against 0.502 predicted (transient or attenuated), 0.248 against 0.250 (null or attenuated) and a trial contrast of 0.749 against 0.760 (blind site or transient). In the ratcheting competitor's map under priority sharing, 9 of the 15 disagreeing cells lie at competitor strengths 0.45 and 0.50. At 0.45 the simulated effect is negative ($-0.09$ at the weakest automatic nudge of the six, classed transient; from $-0.25$ to $-1.17$ at the other five, classed reversed) where the noise-free path stops short of escalating ($+0.16$ to $+0.17$). At 0.50 three cells differ in class (transient simulated, null predicted), with sizes of about 0.10 against 0.13. Three cells at 0.60 match in class (null) with sizes of about $-0.006$ against $+0.096$, and the other three are single cells at 0.20, 0.25 and 0.35. Under the threshold trigger, 13 of the 14 disagreeing cells have an automatic nudge, and in all 14 the predicted effect is lower than the simulated one, by 0.05 to 0.20; among these cells, where the prediction passes $-0.25$ it is classed reversed and, in all but one of them, the simulation transient.

On three maps the differences have one sign, in agreeing cells as well. On the threshold-trigger map the simulated effect is above the predicted one in all 64 cells (median $+0.069$; from $+0.008$ to $+0.203$), and on the ratcheting competitor's map under proportional sharing, which passes, in all 64 cells (median $+0.065$; from $+0.047$ to $+0.090$). On the ratcheting competitor's map under priority sharing it is above at competitor strengths 0.20 to 0.40 (column medians $+0.050$ to $+0.067$) and mostly below at 0.45 to 0.60 (21 of 24 cells). On the other seven maps the differences have no common sign and their median size is at most 0.008. On those three maps, then, a cell that agrees is within 0.1, not necessarily a close match. Table~\ref{tab:cells} lists every cell that disagrees under the noise-free prediction, with both classes and both sizes.

{\small
\setlength{\tabcolsep}{4pt}
\begin{longtable}{@{}>{\raggedright\arraybackslash}p{0.18\textwidth}llll>{\raggedright\arraybackslash}p{0.12\textwidth}>{\raggedright\arraybackslash}p{0.12\textwidth}r@{}}
\caption{\textbf{Every cell that disagrees under the noise-free prediction (40 cells).} Map: steering or ratcheting competitor, or attention load (``Load'') with a competitor that has headroom, or in the stalemate fields; the attention rule or the competitor's response rule. Axis 1 and axis 2: the competitor's strength $\tau$ and the strength of its automatic nudge, $\tau$ (automatic), on the competitor maps; the number of bystanders $m$ and the attention budget (scale) on the load maps. Sizes are what the class rule reads: the window-mean effect over $E$, or, in the field experiment, the site's change over its expected change (site) and the trial's contrast over the expected effect (trial); the difference is the larger of the two. Rows per map: 1, 15, 1, 5, 14, 2 and 2; the steering competitor under proportional sharing, attention load under proportional sharing and the stalemate under proportional sharing have none. The rows of the second, finite-size prediction for the ratcheting competitor are not typeset; they are in the comparison's generated report (\texttt{docs/analysis/2026-11-06-analytic-reduction-comparison/comparison.md}), to be deposited with the data.}\label{tab:cells}\\
\toprule
Map & Axis 1 & Axis 2 & Simulated & Predicted & Simulated size & Predicted size & Difference \\
\midrule
\endfirsthead
\multicolumn{8}{@{}l}{\emph{Table~\thetable, continued}}\\
\toprule
Map & Axis 1 & Axis 2 & Simulated & Predicted & Simulated size & Predicted size & Difference \\
\midrule
\endhead
\bottomrule
\endfoot
Steering, priority & 0.25 & 0.1 & transient & attenuated & 0.499 & 0.502 & 0.003 \\
Ratcheting, priority & 0.35 & 0 & attenuated & transient & 0.507 & 0.447 & 0.060 \\
Ratcheting, priority & 0.6 & 0 & null & null & $-$0.005 & 0.095 & 0.101 \\
Ratcheting, priority & 0.2 & 0.05 & as expected & attenuated & 0.770 & 0.720 & 0.050 \\
Ratcheting, priority & 0.45 & 0.1 & transient & transient & $-$0.087 & 0.166 & 0.252 \\
Ratcheting, priority & 0.5 & 0.1 & transient & null & 0.105 & 0.130 & 0.025 \\
Ratcheting, priority & 0.6 & 0.1 & null & null & $-$0.007 & 0.096 & 0.103 \\
Ratcheting, priority & 0.45 & 0.15 & reversed & transient & $-$0.250 & 0.166 & 0.415 \\
Ratcheting, priority & 0.6 & 0.15 & null & null & $-$0.006 & 0.096 & 0.102 \\
Ratcheting, priority & 0.45 & 0.2 & reversed & transient & $-$0.400 & 0.167 & 0.567 \\
Ratcheting, priority & 0.5 & 0.2 & transient & null & 0.107 & 0.131 & 0.024 \\
Ratcheting, priority & 0.25 & 0.3 & transient & reversed & $-$0.219 & $-$0.276 & 0.057 \\
Ratcheting, priority & 0.45 & 0.3 & reversed & transient & $-$0.692 & 0.161 & 0.852 \\
Ratcheting, priority & 0.45 & 0.4 & reversed & transient & $-$0.940 & 0.165 & 1.104 \\
Ratcheting, priority & 0.5 & 0.4 & transient & null & 0.103 & 0.130 & 0.027 \\
Ratcheting, priority & 0.45 & 0.5 & reversed & transient & $-$1.169 & 0.167 & 1.336 \\
Load, rollout, priority & 16 & 1.4 & null & attenuated & 0.248 & 0.250 & 0.002 \\
Load, field experiment & 16 & 1.4 & null & attenuated & site 0.252, trial 0.243 & site 0.250, trial 0.250 & 0.007 \\
Load, field experiment & 2 & 2 & transient & blind site & site $-$0.001, trial 0.749 & site 0.000, trial 0.760 & 0.011 \\
Load, field experiment & 3 & 2 & transient & attenuated & site $-$0.002, trial 0.694 & site 0.000, trial 0.693 & 0.002 \\
Load, field experiment & 4 & 2.8 & blind site & transient & site 0.001, trial 0.752 & site 0.000, trial 0.742 & 0.011 \\
Load, field experiment & 8 & 4 & transient & attenuated & site 0.003, trial 0.690 & site 0.000, trial 0.698 & 0.008 \\
Steering, threshold trigger & 0.25 & 0 & as expected & attenuated & 0.824 & 0.725 & 0.099 \\
Steering, threshold trigger & 0.4 & 0.1 & transient & reversed & $-$0.067 & $-$0.263 & 0.197 \\
Steering, threshold trigger & 0.35 & 0.15 & transient & transient & $-$0.086 & $-$0.215 & 0.129 \\
Steering, threshold trigger & 0.4 & 0.2 & transient & transient & $-$0.066 & $-$0.198 & 0.132 \\
Steering, threshold trigger & 0.25 & 0.3 & transient & reversed & $-$0.111 & $-$0.313 & 0.203 \\
Steering, threshold trigger & 0.3 & 0.3 & transient & transient & $-$0.098 & $-$0.207 & 0.109 \\
Steering, threshold trigger & 0.4 & 0.3 & transient & transient & $-$0.077 & $-$0.206 & 0.129 \\
Steering, threshold trigger & 0.2 & 0.4 & transient & reversed & $-$0.130 & $-$0.279 & 0.149 \\
Steering, threshold trigger & 0.25 & 0.4 & transient & reversed & $-$0.116 & $-$0.266 & 0.150 \\
Steering, threshold trigger & 0.3 & 0.4 & transient & transient & $-$0.101 & $-$0.244 & 0.143 \\
Steering, threshold trigger & 0.25 & 0.5 & reversed & reversed & $-$0.250 & $-$0.350 & 0.100 \\
Steering, threshold trigger & 0.3 & 0.5 & transient & reversed & $-$0.245 & $-$0.318 & 0.073 \\
Steering, threshold trigger & 0.35 & 0.5 & transient & reversed & $-$0.243 & $-$0.292 & 0.049 \\
Steering, threshold trigger & 0.4 & 0.5 & transient & reversed & $-$0.240 & $-$0.297 & 0.057 \\
Ratcheting, proportional & 0.5 & 0 & attenuated & transient & 0.572 & 0.493 & 0.079 \\
Ratcheting, proportional & 0.6 & 0 & attenuated & transient & 0.541 & 0.462 & 0.079 \\
Stalemate, priority & 0 & 2 & transient & attenuated & 0.394 & 0.395 & 0.001 \\
Stalemate, priority & 2 & 2.8 & transient & attenuated & 0.390 & 0.390 & 0.000 \\
\end{longtable}
}

\clearpage
\sisection{the sensitivity sweep}{si:sweep}
\setcounter{figure}{4}
\renewcommand{\figurename}{Fig.}
\renewcommand{\thefigure}{S\arabic{figure}}
\renewcommand{\theHfigure}{SF\arabic{figure}}

The maps of Fig.~3 and Fig.~S4 vary two parameters of the field at a time around one calibrated setting. This section asks how much of the parameter space each fate occupies, which parameters decide it, and whether the noise-free path of the analytic reduction, which reads the fates here, agrees with the simulator at random settings. The sample, the statistics and five expectations were fixed in a design signed off before the sample was drawn; the expectations are reported below as registered and as found, four of the five not met. The shares are over the ranges below, which the calibration bounds only in part; they are not frequencies of real fields.

\subsection*{The sample}

Twelve combinations: the competitor's response rule (a thermostat, which steers to a goal; a maximiser, which ratchets; a threshold trigger, which steers only while the rate is more than 0.01 from its goal), the attention rule (priority; proportional) and the design (pilot then rollout; field experiment). Each combination has one competitor, no ally and the bystanders of the headroom fields (gain 0, reach 0.6). For each, a Latin-hypercube sample of 4,096 settings (seed 20261108 plus the combination's index) spans nine parameters, ten in a field experiment: the new nudge's strength, 0.10--0.60; the competitor's strength, 0.10--0.60; its gain, 0.5--10, log-uniform (for the threshold trigger, a fixed step of 0.01--0.10 per period); its goal, 0 to 0.10 below the no-actor baseline; its lag, 1--5 periods; the strength of its automatic nudge, 0--0.50; the attention budget scale, 0.35--4.0, log-uniform; the number of bystanders, 0--16; control take-up, 0.10--0.80, set by one shift of both components of the propensity mixture; and, in a field experiment, the treated share, 0.10--0.60. The maximiser never reads its goal, so that parameter is inert in its four combinations and serves as a noise floor. Everything else is at the values of section~\ref{si:methods}. A competitor too weak to reach its goal is part of the sample. The maps left that case out: a ruling made before run 3 dropped the competitor strength 0.15 from them, because there the thermostat reaches its limit in burn-in and escalates before the pilot (section~\ref{si:record}, entry \S10). In all, 49,152 settings.

\subsection*{What was computed}

Each setting is one run of the noise-free path at 200,000 people, with the same people and draws at every setting, so that two settings differ only in their parameters. The path gives the class by the rules of section~\ref{si:methods}, whether the setting lies inside the outline (the loss-at-scale test and the magnitude bounds, on the path's own series; a field experiment takes its first test from a pilot-then-rollout path at the same setting) and whether the competitor's automatic nudge switched on. Four statistics, fixed in advance and named S1--S4 in the registered design (not SI sections): \emph{S1}, the share of each fate over all settings, with excluded ($|E| < 0.02$) and unclassified settings counted, and over the classified settings inside the outline (Table~\ref{tab:sens-shares}); \emph{S2}, for each parameter, the total variation distance between the fate distributions of its lowest and highest third of drawn values (thirds by rank, ties broken by setting), over classified settings, with the parameters ranked by it (Fig.~\ref{fig:sf1}); \emph{S3}, in field experiments, the blind-site share by third of the treated share, and the share of settings where trial$/$total lies within 10\% of $1/(1-p)$; \emph{S4}, the check. From each combination 24 settings were drawn at random with the sample (288 in all; seed 20261108) and simulated with 10 runs of 20,000 people each, on a cloud T4 GPU. A setting agrees when the simulated majority class equals the path's and the sizes differ by at most 0.1, the larger of the two sizes counting in a field experiment and a setting excluded on one side only disagreeing (the cell rule of the reduction; Table~\ref{tab:sens-check}). In 20 of the 288 simulated settings every run was excluded, and their stored class tables lacked the size columns, which the comparison code refused until the missing columns were added as empty values, changing no value or row (the simulator was not re-run; an excluded setting on both sides agrees).

\subsection*{The expectations, as registered and as found}

The five expectations were written and signed off before the sample was drawn. Five points of their wording were fixed after the sign-off, all before the sample was drawn. E1 was confined to settings at which the automatic nudge never switches on, the domain of the reduction's table, which the first wording had left implicit. An automatic nudge counts as switched on only where its strength is positive, since at zero strength it pushes nothing. A reversal in a field experiment includes a reversed site. ``Every field parameter'' in E3 means the competitor's five parameters, the attention budget and the bystanders, so take-up and the treated share do not enter it. E1's share is read pooled over the four steering combinations, with the share of each beside it. The last four were fixed during the build and ratified by the author after the run. None changes the verdict on an expectation, though two results for single combinations rest on them. Read without the first, E1 fails pooled and in all four steering combinations, because the escalated settings are then counted (below): in a rollout under priority sharing, where E1 as amended is met, 930 of 3,747 settings break the rule. The automatic nudge's strength is positive at every setting drawn, so the second moves none. Read without the third, E2 fails in the steering and threshold-trigger rollouts, where no site is read, and so in four of the eight tested combinations, not in all eight. Take-up and the treated share shift the fates by at most 0.20 and 0.28, less than the best field parameter in every combination (0.40 or more). And E1 fails in three of the four steering combinations taken singly. Nothing in the sample, the ranges, the statistics or the rule was changed after the results were seen. One of the five was met.

\emph{E1, the reduction's table made global} (steering competitor). Registered: among classified settings at which the automatic nudge never switches on, as expected occurs only where the competitor has no room before the rollout, and null only where it has room after it, with at most 2\% of such settings breaking the rule. Found: \textbf{not met}. Pooled over the four steering combinations, 15.3\% break it (349 of 2,277); per combination, 1.0\% under priority sharing in a rollout (5 of 489; met), 11.9\% under priority in a field experiment (112 of 945), 16.1\% under proportional sharing in a rollout (33 of 205) and 31.2\% under proportional in a field experiment (199 of 638). The breaches are of two kinds: as expected with room before the rollout, only in the field experiments (66 and 63 settings), and null without room after it, in every combination (5, 46, 33 and 136). A reading made after the tables, not registered: the table was derived for a rollout's total effect, and in a field experiment a trial that reports the expected effect while the competitor has room is the blind-site mechanism, so E1 should have been confined to rollouts; and the 38 rollout breaches, all null without room after, have 3--16 bystanders, a median ratio of the effect under the bystanders' load to the effect without them of 0.21, and 33 of them proportional sharing, so the class reads crowding there, not the competitor's room. E1 restricted to rollouts without bystanders was not registered and is not claimed. The escalated settings, outside the table's domain because the escalated case has its own closed form, are reported as registered and not tested: they break the table's rule in 925 of 3,258, 632 of 2,815, 1,936 of 3,562 and 1,562 of 3,115 classified settings, almost all as null without room after the rollout.

\emph{E2, where reversal occurs.} Registered: reversal occurs against a ratcheting competitor; against a steering competitor it does not occur among classified settings; against a threshold trigger it is confined to the top third of the automatic nudge's strength (the corner of Fig.~S4c). Found: \textbf{not met} in all eight tested combinations. A steering competitor reverses the effect in 3.2\%, 6.2\%, 5.4\% and 5.5\% of classified settings (119 of 3,747, 235 of 3,764, 204 of 3,767 and 207 of 3,756; priority rollout, priority field experiment, proportional rollout, proportional field experiment). Against a threshold trigger only 56\%, 41\%, 54\% and 42\% of reversed settings lie in the top third of the automatic nudge's strength (126 of 225, 148 of 357, 176 of 326, 133 of 313). The ratcheting competitor, reported without a test, reverses the effect in 24.3\%, 17.1\%, 26.9\% and 5.5\% of classified settings. In the steering rollouts, 322 of the 323 reversals have the automatic nudge on, at strengths from 0.14 to 0.50 (median 0.39); over the rollouts of all three competitors, 98.7\% do (2,768 of 2,805). In field experiments it is on less often: at 399 of the 442 reversed sites against a steering competitor, and at 1,697 of 1,963 (86.4\%) against all three. A reading, not registered: where the automatic nudge pushes harder than withdrawing the competitor's deliberate nudge can give back, a reach being unable to fall below zero, the rate settles below its level before the rollout. The maps did not show this because, at their competitor strengths and gain, the withdrawal offsets the automatic push; the trigger's corner and the ratcheting competitor of Fig.~2d are cases of the same condition. The check bears out the path's reversals: at all six steering settings where the path predicted a reversal (five reversed, one reversed site), the simulation reversed the effect, in class at all six and within 0.1 in size at four (the other two differ by 0.59 and 0.14); over all combinations, 24 of 25 predicted reversals and 10 of 11 predicted reversed sites were simulated so.

\emph{E3, what sets the fate.} Registered: the new nudge's strength ranks first in no combination, where first means a distance more than 0.03 above every field parameter's; against a steering competitor the top field parameter is its strength, its automatic nudge or its goal. Found: \textbf{not met} under priority sharing, \textbf{met} under proportional sharing. Under priority the new nudge's strength gives 0.54--0.63, against 0.40--0.52 for the best field parameter (margins 0.07--0.15) in all six combinations. Under proportional sharing it gives 0.07--0.21, against 0.58--0.74 for the number of bystanders, first in all six; against a steering competitor the new nudge's strength ranks seventh, of nine in a rollout and of ten in a field experiment. The second part was met in one of four steering combinations (priority, rollout: the competitor's strength); the bystanders come first in the other three. The inert goal of the ratcheting competitor gives 0.01--0.04, the noise floor. A reading, not registered: under priority sharing the stronger nudge takes a person's attention first, so the new nudge's strength relative to the competitor's decides which is heard; under proportional sharing it does not.

\emph{E4, the blind site across fields.} Registered: the blind-site share falls as the treated share rises, and the blind-site identity, trial$/$total within 10\% of $1/(1-p)$, holds for most settings in the lowest third of the treated share and for a minority in the highest. Found: \textbf{not met}. The blind-site share falls from the lowest to the highest third in one field-experiment combination of six (0.11 to 0.08, steering competitor, proportional sharing); in the others it is 0.15 and 0.15, 0.10 and 0.15, 0.04 and 0.07, 0.10 and 0.14, 0.06 and 0.07. The identity holds for 0.45, 0.34, 0.11, 0.05, 0.23 and 0.19 of the lowest third, never a majority, and less often in the highest third in all six. Across random fields the blind site occurs in 4--16\% of classified field-experiment settings in every third of the treated share. A reading, not registered: the identity of Fig.~4b is a property of the calibrated setting, where the competitor holds the site's rate exactly.

\emph{E5, the check.} Registered: at least 90\% of the steering settings agree; the ratcheting and threshold-trigger combinations are reported as they come, expected below the bar and expected to show the one-sided difference of the reduction's maps (simulated above predicted). Found: \textbf{met}. The steering settings agree at 24, 22, 22 and 22 of 24, pooled 90 of 96 (93.75\%). The ratcheting combinations agree at 37.5--79.2\% (55 of 96), below the bar as expected. The threshold-trigger combinations agree at 75.0--91.7\% (79 of 96, 82.3\%): below the bar pooled and in three combinations, but above it under proportional sharing in a rollout (22 of 24), where the registered sub-expectation ``below the bar'' failed. The one-sided difference did not reappear either, so that sub-expectation failed too (Table~\ref{tab:sens-check}); E5 is met by its registered bar. Of the six steering disagreements, two differ in class (a blind site simulated as attenuated; a null simulated as a reversed site on a five--five tie of runs) and four only in size, by 0.14 to 0.59.

\subsection*{What sets the fate, drawn}

\begin{figure}[ht]\centering\includegraphics[width=\textwidth]{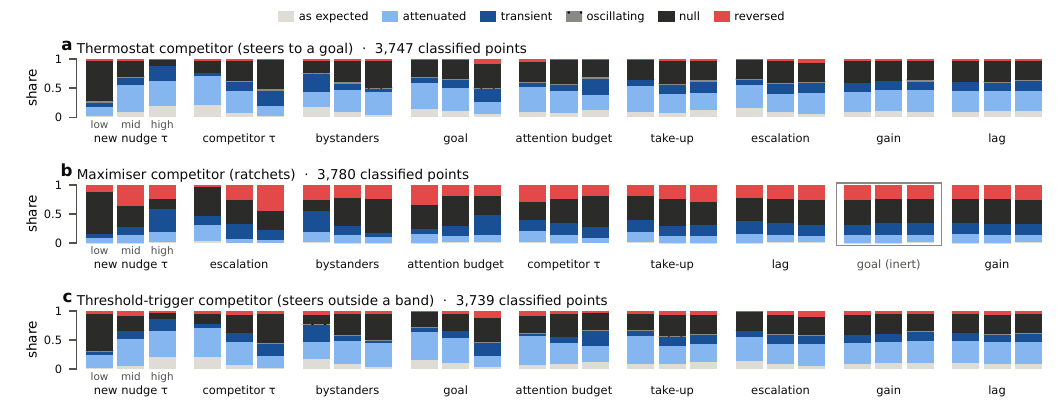}\caption{\textbf{In a rollout under the default attention rule the fate shifts most with the nudge's own strength, then with the competitor's strength (for a ratcheting competitor, its automatic nudge), the bystanders, the competitor's goal where it has one, and the attention budget, and little with take-up or the competitor's gain or lag.}
One panel per response rule of the competitor, pilot then rollout, priority sharing: \textbf{a}, a thermostat (steers to a goal), 3,747 classified settings; \textbf{b}, a maximiser (ratchets), 3,780; \textbf{c}, a threshold trigger (steers outside a tolerance band), 3,739; each of 4,096 settings drawn. For each parameter, three stacked bars give the fate shares among the classified settings in the lowest (low), middle (mid) and highest (high) third of its drawn values; excluded and unclassified settings are left out. The groups are ordered by how far the shares shift between the lowest and the highest third, largest first: for the new nudge's strength 0.63, 0.56 and 0.56 (\textbf{a}, \textbf{b}, \textbf{c}); for the attention budget 0.28, 0.26 and 0.27; for take-up 0.16, 0.09 and 0.17; for the competitor's gain 0.07, 0.02 and 0.08; for its lag 0.04, 0.08 and 0.03. Three shifts lie between the groups of the first sentence: the maximiser's strength (0.22, \textbf{b}) and the automatic nudge's strength against the thermostat (0.15, \textbf{a}) and the threshold trigger (0.11, \textbf{c}). The maximiser never reads its goal, so in \textbf{b} the goal is inert (framed) and its spread, 0.04, is the noise floor of the thirds; \textbf{b} has no oscillating settings. Colours as in Figs~2 and 3. Fates read from the analytic reduction's noise-free path at 200,000 people, the same draws at every setting.}\label{fig:sf1}\end{figure}

Fig.~\ref{fig:sf1} draws S2 for the competitor's three response rules under priority sharing, in a rollout: the fate shares in the lowest, middle and highest third of each parameter. The other nine combinations are in Table~\ref{tab:sens-shares} (shares) and in the generated report with the data (distances).

\subsection*{Tables}

{\scriptsize
\setlength{\tabcolsep}{2.5pt}
\begin{longtable}{@{}>{\raggedright\arraybackslash}p{0.2\textwidth}lrrrrrrrrrrrr@{}}
\caption{\textbf{Fate shares per combination (statistic S1), in percent.} ``All'': over the 4,096 settings of the combination, with excluded ($|E| < 0.02$) and unclassified settings as categories, the share inside the outline (region) and the share whose burn-in had not settled (unsettled) beside them. ``Region'': over the classified settings inside the outline. AE, as expected; att, attenuated; tr, transient; osc, oscillating; rev, reversed; RS, reversed site; BS, blind site (field experiments only; a field experiment's reversal is read at the site, so its rev column is 0). Blind trial (1 setting) and delayed (2 settings) are not shown. Rollout: pilot then rollout; field exp.: field experiment. Inside the outline, as expected is all but excluded in a rollout by the definitions (Results). Typeset from \texttt{shares.csv} of the run of record.}\label{tab:sens-shares}\\
\toprule
Combination & Scope & AE & att & tr & osc & null & rev & RS & BS & excl. & uncl. & region & unsettled \\
\midrule
\endfirsthead
\multicolumn{14}{@{}l}{\emph{Table~\thetable, continued}}\\
\toprule
Combination & Scope & AE & att & tr & osc & null & rev & RS & BS & excl. & uncl. & region & unsettled \\
\midrule
\endhead
\bottomrule
\endfoot
Steering, priority, rollout & all & 9.0 & 32.1 & 14.6 & 0.9 & 31.9 & 2.9 & -- & -- & 8.4 & 0.1 & 49.9 & 1.7 \\
 & region & 0.9 & 14.1 & 30.2 & 1.9 & 46.9 & 6.0 & -- & -- & -- & -- & -- & -- \\
Steering, priority, field exp. & all & 9.3 & 40.4 & 1.2 & 0.3 & 20.8 & 0.0 & 5.7 & 14.2 & 7.9 & 0.2 & 49.2 & 1.9 \\
 & region & 2.8 & 31.7 & 2.4 & 0.6 & 23.9 & 0.0 & 10.6 & 28.0 & -- & -- & -- & -- \\
Steering, proportional, rollout & all & 1.8 & 25.1 & 9.3 & 0.2 & 50.6 & 5.0 & -- & -- & 7.9 & 0.1 & 40.6 & 2.0 \\
 & region & 0.5 & 11.5 & 23.4 & 0.6 & 51.5 & 12.6 & -- & -- & -- & -- & -- & -- \\
Steering, proportional, field exp. & all & 2.9 & 34.3 & 0.4 & 0.2 & 40.4 & 0.0 & 5.1 & 8.5 & 8.1 & 0.2 & 40.5 & 1.9 \\
 & region & 3.0 & 34.9 & 1.0 & 0.5 & 28.7 & 0.0 & 10.5 & 21.4 & -- & -- & -- & -- \\
\midrule
Ratcheting, priority, rollout & all & 0.9 & 12.3 & 18.1 & 0.0 & 38.6 & 22.4 & -- & -- & 7.7 & 0.0 & 71.4 & 0.0 \\
 & region & 0.5 & 11.4 & 25.4 & 0.0 & 31.2 & 31.5 & -- & -- & -- & -- & -- & -- \\
Ratcheting, priority, field exp. & all & 6.5 & 38.8 & 0.7 & 0.0 & 18.3 & 0.0 & 15.8 & 12.1 & 7.9 & 0.0 & 70.1 & 0.0 \\
 & region & 8.3 & 49.1 & 1.0 & 0.0 & 2.1 & 0.0 & 22.6 & 16.8 & -- & -- & -- & -- \\
Ratcheting, proportional, rollout & all & 0.1 & 8.5 & 7.3 & 0.0 & 51.2 & 24.8 & -- & -- & 8.0 & 0.0 & 61.9 & 0.0 \\
 & region & 0.0 & 7.4 & 11.9 & 0.0 & 40.3 & 40.3 & -- & -- & -- & -- & -- & -- \\
Ratcheting, proportional, field exp. & all & 4.6 & 37.8 & 0.3 & 0.0 & 37.8 & 0.0 & 5.0 & 6.0 & 8.3 & 0.3 & 62.3 & 0.0 \\
 & region & 7.4 & 54.1 & 0.5 & 0.0 & 20.1 & 0.0 & 8.1 & 9.7 & -- & -- & -- & -- \\
\midrule
Threshold trigger, priority, rollout & all & 9.0 & 33.8 & 12.9 & 0.4 & 29.6 & 5.5 & -- & -- & 8.3 & 0.4 & 44.0 & 4.8 \\
 & region & 0.9 & 15.3 & 29.9 & 1.0 & 40.2 & 12.7 & -- & -- & -- & -- & -- & -- \\
Threshold trigger, priority, field exp. & all & 11.7 & 40.5 & 0.6 & 0.0 & 19.1 & 0.0 & 8.7 & 11.1 & 8.1 & 0.2 & 43.4 & 5.6 \\
 & region & 5.1 & 36.2 & 1.3 & 0.1 & 14.4 & 0.0 & 18.0 & 24.9 & -- & -- & -- & -- \\
Threshold trigger, proportional, rollout & all & 1.7 & 25.5 & 7.6 & 0.2 & 49.0 & 8.0 & -- & -- & 7.8 & 0.1 & 38.1 & 6.5 \\
 & region & 0.2 & 12.1 & 20.4 & 0.5 & 45.6 & 21.3 & -- & -- & -- & -- & -- & -- \\
Threshold trigger, proportional, field exp. & all & 3.6 & 36.0 & 0.3 & 0.0 & 38.3 & 0.0 & 7.6 & 6.0 & 8.0 & 0.1 & 38.5 & 6.7 \\
 & region & 5.6 & 37.1 & 0.9 & 0.1 & 24.0 & 0.0 & 16.4 & 15.9 & -- & -- & -- & -- \\
\end{longtable}
}

\begin{table}[ht]\centering\footnotesize
\caption{\textbf{The simulated check per combination (statistic S4).} 24 settings per combination drawn at random with the sample, each simulated with 10 runs of 20,000 people. A setting agrees when the simulated majority class equals the path's and the sizes differ by at most 0.1 (in a field experiment, the larger of the site's and the trial's differences). Differences (diff.) are simulated minus predicted, in the size the class rule reads (shares of $E$); ``largest'' is the largest $|$diff.$|$. The 90\% bar was registered for the steering competitor only; the other rows are reported as they came. Typeset from \texttt{check\_summary.csv} of the run of record.}\label{tab:sens-check}
\begin{tabular}{@{}lrrrrr@{}}
\toprule
Combination & Agree & \% & Median $|$diff.$|$ & Largest & Median diff. \\
\midrule
Steering, priority, rollout & 24 of 24 & 100.0 & 0.001 & 0.034 & $-0.0001$ \\
Steering, priority, field exp. & 22 of 24 & 91.7 & 0.022 & 0.238 & $-0.0063$ \\
Steering, proportional, rollout & 22 of 24 & 91.7 & 0.001 & 0.590 & $-0.0006$ \\
Steering, proportional, field exp. & 22 of 24 & 91.7 & 0.017 & 0.143 & $-0.0100$ \\
Ratcheting, priority, rollout & 19 of 24 & 79.2 & 0.033 & 0.993 & $+0.0041$ \\
Ratcheting, priority, field exp. & 12 of 24 & 50.0 & 0.180 & 1.320 & $-0.1796$ \\
Ratcheting, proportional, rollout & 15 of 24 & 62.5 & 0.051 & 2.268 & $+0.0234$ \\
Ratcheting, proportional, field exp. & 9 of 24 & 37.5 & 0.187 & 1.377 & $-0.1784$ \\
Threshold trigger, priority, rollout & 20 of 24 & 83.3 & 0.009 & 0.167 & $-0.0007$ \\
Threshold trigger, priority, field exp. & 18 of 24 & 75.0 & 0.038 & 0.777 & $-0.0308$ \\
Threshold trigger, proportional, rollout & 22 of 24 & 91.7 & 0.013 & 0.144 & $-0.0005$ \\
Threshold trigger, proportional, field exp. & 19 of 24 & 79.2 & 0.024 & 4.984 & $+0.0029$ \\
\midrule
Steering, all four (pooled) & 90 of 96 & 93.8 & 0.007 & 0.590 & $-0.0006$ \\
\bottomrule
\end{tabular}
\end{table}

\clearpage
\sisection{the model in the ODD format}{si:odd}

This section reproduces the project's description of the model in the ODD format (Overview, Design concepts, Details)~\citeSI{SI-grimm2020}. The documents reproduced in the SI Appendix were written during the project and use its working names. F1--F4 are Figs.~1--4, ED1--ED4 are Figs.~S1--S4 and SF1 is Fig.~S5; ``our nudge'' is the new nudge; T1--T4 are the four checks against known regularities; ``spec'', ``brief'' and ``spike note'' are design documents in the repository, cited by path and section. Figure and panel labels in the record are those planned when an entry was registered (``Extended Data'' is the earlier name of the SI figures); the runs behind each published figure are listed in section~\ref{si:methods}. R1--R9 are the rulings that fixed the model's rules, R1--R8 in the spike note and R9 the amended outcome-class rules; G1--G5 are the goals of the design documents. The Build phase tested the model and the Explore phase ran the registered sweeps. Machines: M4, an Apple laptop (device ``mps''); Spark, an NVIDIA DGX Spark (``cuda''); Modal, a cloud service with T4 GPUs. ``cls\_combined'' is a run's combined outcome-class column; ``levels'' are a saved run's rate levels, re-run from its manifest for the probit contrast; a ``twin'' is a run's counterpart under the other design at the same setting; the ``oracle'' of T4 is the analytic peak $\Phi(-\beta/2)$; the crowded, stalemate and headroom families are the crowded fields of section~\ref{si:methods}; ``S1\_competitor'', ``S2\_ally'' and the like are scenario files, not SI sections. Source: \texttt{docs/nudgefield/odd-protocol.md}, last changed at commit 5e32b61, 2026-10-29; reproduced unchanged.

\begingroup\footnotesize\sloppy\setlength{\emergencystretch}{3em}\setlength{\tabcolsep}{4pt}
This note describes the \texttt{nudgefield} model in the ODD format (Overview, Design concepts, Details; Grimm et al.~2020, \emph{JASSS} 23(2) 7). Functions are named \texttt{module.\allowbreak{}function} relative to \texttt{code/\allowbreak{}src/\allowbreak{}nudgefield/\allowbreak{}}. Parameter values and their sources are in \texttt{docs/\allowbreak{}nudgefield/\allowbreak{}parameters.\allowbreak{}md} and are not repeated here. ``Spec'' is the design document; ``spike note'' is \texttt{docs/\allowbreak{}analysis/\allowbreak{}2026-\allowbreak{}10-\allowbreak{}26-\allowbreak{}spike-\allowbreak{}note.\allowbreak{}md}, whose §1--§3 fixed the rules before any run.

\subsection*{1. Overview}

\subsubsection*{1.1 Purpose and patterns}

The model asks what happens to one behavioural intervention (``our nudge''), designed for a neutral field, when it enters a field of other actors' nudges whose owners react to the behaviour rates they observe (spec §1). It compares four effects per run: the effect the designer expected (empty field), the effect a randomised trial reports, the static effect of the field (frozen field) and the true total effect (field free to react). Outcomes are classified into the registered classes of spike note §1.3 (as expected, attenuated, null, reversed, delayed, transient, oscillating, blind trial) and the site-level classes of rule R8 (\texttt{reversed\_\allowbreak{}site}, \texttt{blind\_\allowbreak{}site}).

Patterns the model must reproduce (spec §6; \texttt{code/\allowbreak{}tests/\allowbreak{}nudgefield/\allowbreak{}test\_\allowbreak{}targets.\allowbreak{}py}): T1 loss of effect from pilot to rollout; T2 a quiet-versus-crowded probit ratio of about 2 (calibrated, so a constraint, not a prediction; spec §12); T3 dispersion across sites above sampling error; T4 an inverted U in control take-up peaking below one half.

\subsubsection*{1.2 Entities, state variables and scales}

\textbf{People} (\texttt{people.\allowbreak{}Population}), \texttt{N} per run. State, each a tensor of shape \texttt{(R,\ N)}: baseline propensity \texttt{a} (two-component normal mixture); attention budget \texttt{B} (log-normal); susceptibilities \texttt{s\_\allowbreak{}auto}, \texttt{s\_\allowbreak{}delib} (log-normal, mean 1); tercile labels of \texttt{a}, \texttt{B} and \texttt{s\_\allowbreak{}delib}, fixed at the draw. Per person and nudge, a valence \texttt{v\_\allowbreak{}ik} (log-normal magnitude with mean 1, sign flipped with probability \texttt{flip\_\allowbreak{}prob}), drawn on first use and cached by nudge name. Per person and bystander, a propensity for that bystander's own behaviour. Only \texttt{a} changes over time, and only when habit is on.

\textbf{Nudges} (\texttt{nudges.\allowbreak{}Nudge}). Name, direction \texttt{d\ ∈\ \{+1,\ −1\}}, kind (\texttt{auto} or \texttt{delib}), and per-run tensors \texttt{(R,)} of latent strength \texttt{τ}, attention cost \texttt{c} and salience (the strength the priority rule ranks by; equal to \texttt{τ} except for a bystander's nudge, §3.3.3). Reach is held by the owner, not by the nudge.

\textbf{Actors} (\texttt{actors.\allowbreak{}base.\allowbreak{}ActorSpec}, the scalar description; \texttt{actors.\allowbreak{}base.\allowbreak{}Actor}, the per-run state). Role (\texttt{competitor}, \texttt{ally}, \texttt{bystander}), response rule (from the registry \texttt{actors.\allowbreak{}RULES}), one deliberative nudge, lag, gain, goal offset and the other rule parameters. State, each \texttt{(R,)}: reach, goal, the thermostat's previous gap, the maximiser's reference, the escalation counter, the automatic-nudge switch and its trigger period. Our own actor has no rule: its nudge's reach is set by the study design.

\textbf{Worlds} (\texttt{world.\allowbreak{}World}). One population (shared, or a copy when habit is on), a list of actors, our nudge or none, the design and its treated sets, and the recorded histories. A world is \texttt{free} (actors react), \texttt{frozen} (actors held) or \texttt{empty} (no actors).

\textbf{Scales.} A period is one week (spec §4). A run lasts \texttt{T\ =\allowbreak{}\ 200} periods. A run is one site of \texttt{N} people (default 20,000; 50,000 at the Spark target); \texttt{R} runs form the leading tensor axis. No space and no network: people interact only through the aggregate rates the actors watch (spec §2). Sites are not a simulator dimension; a multi-site comparison runs several scenario configurations (plan decision; T3).

\subsubsection*{1.3 Process overview and scheduling}

Each world advances one period at a time in \texttt{world.\allowbreak{}World.\allowbreak{}step}, in this order (the spike's order):

\begin{enumerate}
\def\labelenumi{\arabic{enumi}.}
\tightlist
\item
  \textbf{Rule.} For each actor in turn (steps 1--3), its response rule reads its observed rate at \texttt{t\ −\ lag} and sets its reach.
\item
  \textbf{Escalation.} Rule R6 updates the actor's counter and may switch its automatic nudge on.
\item
  \textbf{Exposure masks.} Actor \texttt{k} reaches person \texttt{i} when a fresh uniform draw \texttt{u\_\allowbreak{}ik\ \textless{}\ reach\_\allowbreak{}k} (membership redrawn every period); an active automatic nudge reaches everyone; our nudge reaches the design's treated set (\texttt{designs.\allowbreak{}Design.\allowbreak{}our\_\allowbreak{}mask}).
\item
  \textbf{Attention.} The sharing rule computes one weight per person and deliberative nudge.
\item
  \textbf{Latent value.} \texttt{a\ +\ s\_\allowbreak{}auto\ ·\ Σ\_\allowbreak{}auto\ d\ τ\ v\ +\ s\_\allowbreak{}delib\ ·\ Σ\_\allowbreak{}delib\ w\ d\ τ\ v}, over the nudges reaching the person.
\item
  \textbf{Behaviour.} \texttt{y\ =\allowbreak{}\ 1\{latent\ +\ ε\ \textgreater{}\ 0\}}, \texttt{ε\ \textasciitilde{}\ N(0,\ 1)}.
\item
  \textbf{Records.} Site rate; treated and control rates in randomised periods; rates by tercile; reach per actor; the running sum of \texttt{y} over the evaluation window.
\item
  \textbf{Own behaviours.} Each bystander's own-behaviour rate (free worlds only).
\item
  \textbf{Observations.} Each actor's observed rate for this period: the site rate (or its own rate, for a bystander) plus \texttt{obs\_\allowbreak{}noise\_\allowbreak{}sd\ ·\ ε\_\allowbreak{}obs}.
\item
  \textbf{Habit.} \texttt{a\ ←\ a\ +\ habit\_\allowbreak{}rate\ ·\ (2y\ −\ 1)}; no-op at the default \texttt{habit\_\allowbreak{}rate\ =\allowbreak{}\ 0}.
\end{enumerate}

\textbf{Paired worlds and passes} (\texttt{simulate.\allowbreak{}run\_\allowbreak{}paired}). Six worlds share one population and one draw stream:

{\def\LTcaptype{none} 
\begin{longtable}[]{@{}lll@{}}
\toprule\noalign{}
World & Actors & Our nudge \\
\midrule\noalign{}
\endhead
\bottomrule\noalign{}
\endlastfoot
\texttt{empty\_\allowbreak{}none} & none & no \\
\texttt{empty} & none & yes \\
\texttt{none} & free & no \\
\texttt{free} & free & yes \\
\texttt{frozen\_\allowbreak{}none} & held & no \\
\texttt{frozen} & held & yes \\
\end{longtable}
}

Pass 1 steps \texttt{empty\_\allowbreak{}none} and \texttt{empty} over all periods. The mean \texttt{empty\_\allowbreak{}none} site rate over periods 0--59 per run is the no-actor baseline, and each focal-observing actor's goal is set to baseline + goal offset (\texttt{actors.\allowbreak{}base.\allowbreak{}Actor.\allowbreak{}set\_\allowbreak{}goal}; ruling R1). Pass 2 resets the stream, replays the identical draws and steps \texttt{none} and \texttt{free}. At \texttt{rollout\_\allowbreak{}start} (period 100) the two frozen worlds are built from clones of the free world's actors, each held at its mean reach (and automatic reach) over periods 90--99 (\texttt{simulate.\allowbreak{}\_\allowbreak{}held\_\allowbreak{}clones}, \texttt{actors.\allowbreak{}base.\allowbreak{}Actor.\allowbreak{}hold}), and stepped from then on.

\textbf{Designs} (\texttt{designs.\allowbreak{}Design}; spike note §3.3, spec §4.4). Burn-in 0--59 in all designs. \texttt{pilot\_\allowbreak{}rollout}: a fixed 2\% set treated in 60--79, nobody in 80--99, everyone from 100. \texttt{field\_\allowbreak{}experiment}: a fixed 40\% set treated from 100, no pilot. \texttt{pilot\_\allowbreak{}field\_\allowbreak{}experiment}: the pilot, then an independently drawn 40\% set from 100. Control is everyone outside the active treated set. The evaluation window is 160--199.

\subsection*{2. Design concepts}

\textbf{Basic principles.} Lewin's field of forces as inspiration: behaviour is an equilibrium of forces that respond to each other (spec §1). Forces add on a latent scale and behaviour is a threshold crossing (probit), so effects depend on where people sit relative to the threshold. Deliberative nudges compete for a limited attention budget; automatic nudges bypass it (spec §4.1). Actors are controllers on observed rates, not strategists (spec §2).

\textbf{Emergence.} The gaps between expected, trial-reported, frozen and total effects, hence the outcome class, emerge from the feedback between behaviour and the actors' rules; no rule encodes attenuation, reversal or a blind trial.

\textbf{Adaptation.} Actors adapt reach to the observed rate (§3.3.4) and may switch irreversibly to a persistent automatic nudge (§3.3.5). People do not adapt, except through habit when it is on.

\textbf{Objectives.} Thermostat and trigger actors target a rate: baseline − 0.03 for a competitor, baseline + 0.03 for an ally (ruling R1); bystanders observe their own behaviour and, at gain 0 in the registered scenarios, do not react. A reacting bystander (non-zero \texttt{gain} or \texttt{gain\_\allowbreak{}p}) is rejected in v1 (\texttt{actors.\allowbreak{}base.\allowbreak{}ActorSpec}): its goal would be the absolute \texttt{goal\_\allowbreak{}offset}, not a rate relative to its own behaviour's baseline. The maximiser has a direction (spec §4.3, rule R7): it wants the focal rate lower and never gives back reach.

\textbf{Learning.} None. The maximiser's reference is a memory of the best rate it achieved (a ratchet), not learning.

\textbf{Prediction.} None. Actors react to lagged observations only.

\textbf{Sensing.} An actor senses one quantity: the aggregate rate of its target behaviour at its site, with lag (default 2) and optional Gaussian observation noise (spec §4.3, §12). It does not sense other actors' nudges (author, spec §10). People sense only the nudges that reach them.

\textbf{Interaction.} Indirect only. Actors interact through the rate they all watch. Nudges interact inside each person through the latent sum and through competition for attention. There is no person-to-person interaction.

\textbf{Stochasticity.} Drawn once per run: the population (\texttt{people.\allowbreak{}Population.\allowbreak{}draw}), valences per nudge name (\texttt{people.\allowbreak{}Population.\allowbreak{}valence}), bystanders' own propensities (\texttt{people.\allowbreak{}Population.\allowbreak{}own\_\allowbreak{}propensity}) and our treated sets (\texttt{designs.\allowbreak{}draw\_\allowbreak{}treated}). Drawn every period (\texttt{draws.\allowbreak{}DrawStream.\allowbreak{}period}), in this order: behaviour noise \texttt{ε}, the uniform \texttt{u\_\allowbreak{}ours} for our nudge (drawn every period to keep the order fixed; the registered designs reach fixed treated sets and do not read it), actor exposure uniforms \texttt{u\_\allowbreak{}actor}, automatic-nudge uniforms \texttt{u\_\allowbreak{}auto}, bystanders' own-behaviour noise \texttt{eps\_\allowbreak{}own} and observation noise \texttt{obs}. There is one draw stream per paired run, shared by all six worlds, so the paired differences contain no noise from separate draws. Every name's draws are made each period whichever world uses them.

\textbf{Seeds.} A sweep groups its grid points by actor structure; each group is one batch of \texttt{points\ ×\ n\_\allowbreak{}rep} runs with one seed, \texttt{stable\_\allowbreak{}hash("scenario\textbar{}design\textbar{}structure\textbar{}seed\_\allowbreak{}base")} masked to 63 bits (\texttt{sweep.\allowbreak{}batch\_\allowbreak{}seed}; \texttt{stable\_\allowbreak{}hash} is an 8-byte blake2b). With \texttt{run.\allowbreak{}max\_\allowbreak{}batch\_\allowbreak{}runs\ \textgreater{}\ 0} a larger batch is split into chunks, chunk \texttt{i} seeded from the same string with the suffix \texttt{\textbar{}chunk\{i\}}; a batch that fits keeps the unsuffixed seed. The population seed and the treated-set generators derive from the batch seed (\texttt{draws.\allowbreak{}population\_\allowbreak{}seed}, \texttt{draws.\allowbreak{}treated\_\allowbreak{}generator}); the draw stream uses it directly. Runs within a batch are rows of the same tensors, hence independent. Draws are reproducible per device, not across devices. The manifest records every batch seed and its runs.

\textbf{Collectives.} None (arms and terciles are recording groups).

\textbf{Observation.} §3.1 (outputs) and §3.3.10 (metrics).

\subsection*{3. Details}

\subsubsection*{3.1 Implementation and outputs}

PyTorch, float32 for per-person tensors, all per-period operations vectorised over \texttt{(R,\ N)}; no tensor with a time axis over people is ever allocated. Device by flag (\texttt{cpu}, \texttt{mps}, \texttt{cuda}, \texttt{auto}; \texttt{device.\allowbreak{}resolve\_\allowbreak{}device}). Histories are kept on the device as \texttt{(T,\ R)} and converted to CPU float64 at the end (\texttt{world.\allowbreak{}to\_\allowbreak{}cpu64}). A sweep (\texttt{sweep.\allowbreak{}run\_\allowbreak{}sweep}) writes \texttt{effects.\allowbreak{}parquet} (long table \texttt{t,\ replicate,\ quantity,\ value}), \texttt{classes.\allowbreak{}parquet} (one row per run), \texttt{robust.\allowbreak{}parquet} and \texttt{manifest.\allowbreak{}json} (configs, seeds, batches, git commit, device, environment) to \texttt{code/\allowbreak{}outputs/\allowbreak{}\textless{}scenario\textgreater{}\_\allowbreak{}\textless{}design\textgreater{}\_\allowbreak{}\textless{}YYYYMMDD\_\allowbreak{}HHMMSS\textgreater{}/\allowbreak{}}; every row carries \texttt{scenario,\ variant,\ design,\ point,\ param\_\allowbreak{}name,\ param\_\allowbreak{}value,\ replicate,\ seed,\ run}. Scenario files live in \texttt{code/\allowbreak{}conf/\allowbreak{}nudgefield/\allowbreak{}scenario/\allowbreak{}} and load through \texttt{scenarios.\allowbreak{}load\_\allowbreak{}scenario}; the F2 class-to-scenario map is \texttt{code/\allowbreak{}conf/\allowbreak{}nudgefield/\allowbreak{}figures/\allowbreak{}F2.\allowbreak{}yaml}.

\subsubsection*{3.2 Initialisation and input data}

At period 0: people are drawn from the population parameters; competitors and allies start at reach 0 and respond from period \texttt{lag}; bystanders start at reach 0.6 and keep it (gain 0); the maximiser holds 0.60 until period 60 (R7). Goals are set from the pass-1 baseline before pass 2 starts. Burn-in is periods 0--59. \textbf{Burn-in settled test} (\texttt{world.\allowbreak{}burn\_\allowbreak{}in\_\allowbreak{}settled}; spike note §1.2): the absolute least-squares slope of the \texttt{none} world's site rate over 40--59 must be below 0.002 per period in every run; otherwise the sweep stops after that batch, writes the manifest with the failing runs and raises \texttt{sweep.\allowbreak{}BurnInNotSettledError}.

\textbf{Input data:} none. The model is calibrated to published magnitudes (spec §6), not fitted; every value and its source is in \texttt{docs/\allowbreak{}nudgefield/\allowbreak{}parameters.\allowbreak{}md}.

\subsubsection*{3.3 Submodels}

\paragraph{3.3.1 Population draw}

\texttt{people.\allowbreak{}Population.\allowbreak{}draw}. Draw order from one seeded generator: component indicator, \texttt{a}, \texttt{B}, \texttt{s\_\allowbreak{}auto}, \texttt{s\_\allowbreak{}delib}; tercile labels by within-run rank. Valences (\texttt{people.\allowbreak{}Population.\allowbreak{}valence}) come from a generator seeded by the population seed XOR the nudge name's hash, so they do not depend on call order. Fixed by spike note §3.1 and the valence reading of §1.6; spec §4.1.

\paragraph{3.3.2 Latent value and behaviour}

\texttt{attention.\allowbreak{}latent}, \texttt{attention.\allowbreak{}behave}. Probit threshold of spec §4.1 with the push \texttt{d\_\allowbreak{}k\ τ\_\allowbreak{}k\ v\_\allowbreak{}ik}, automatic nudges scaled by \texttt{s\_\allowbreak{}auto}, deliberative by \texttt{s\_\allowbreak{}delib\ ·\ w}. Fixed by spec §4.1 and spike note §3.1.

\paragraph{3.3.3 Attention sharing}

Registry \texttt{attention.\allowbreak{}ATTENTION\_\allowbreak{}RULES}; the rule is chosen by \texttt{run.\allowbreak{}attention}. - \texttt{attention.\allowbreak{}priority} (default; author decision of 2026-10-27, spec §12): each person ranks the deliberative nudges reaching them by perceived push \texttt{salience\_\allowbreak{}k\ ·\ \textbar{}v\_\allowbreak{}ik\textbar{}} (\texttt{salience\ =\allowbreak{}\ τ}), strongest first, ties broken by exposure order; each takes \texttt{w\ =\allowbreak{}\ min(1,\ remaining\ /\allowbreak{}\ c\_\allowbreak{}k)} until the budget \texttt{B} is spent. Ranking by perceived push means equal strengths do not tie. - \textbf{Bystander salience} (controller ruling of Task 4a, commit 6309d9a, from spec §4.3 ``affects ours only through attention''): a bystander's nudge enters the focal behaviour with \texttt{τ\ =\allowbreak{}\ 0} (it pushes nothing) and salience equal to its own \texttt{τ}, with its own name and hence its own valences, so under priority it competes for attention at the strength the person perceives (\texttt{actors.\allowbreak{}base.\allowbreak{}Actor.\allowbreak{}focal\_\allowbreak{}nudge}). - \texttt{attention.\allowbreak{}proportional} (the spike's rule, spike note §3.2; spec §4.1): \texttt{w\ =\allowbreak{}\ min(1,\ B\ /\allowbreak{}\ D)}, \texttt{D} the summed cost of the deliberative nudges reaching the person. The parity runs use it (the spike's results were registered under it); both rules stay in the sensitivity sweep.

\paragraph{3.3.4 Actor response rules}

Registry \texttt{actors.\allowbreak{}RULES}; each rule reads \texttt{observed{[}t\ −\ lag{]}}, updates reach in place and returns the per-run ``losing'' flag that R6 counts. - \texttt{actors.\allowbreak{}rules.\allowbreak{}thermostat}: \texttt{gap\ =\allowbreak{}\ r{[}t\ −\ lag{]}\ −\ goal}; a competitor moves reach by \texttt{+(gain\ ·\ gap\ +\ gain\_\allowbreak{}p\ ·\ Δgap)}, allies and bystanders by the negative, clipped to {[}0, 1{]}. With \texttt{gain\_\allowbreak{}p\ =\allowbreak{}\ 0} it is the spike's integral rule (spike note §2); the proportional term is spec §4.3's PI form, and it is zero at the first observation (period \texttt{lag}: \texttt{Δgap} starts from that period's gap, so there is no kick). - \texttt{actors.\allowbreak{}rules.\allowbreak{}maximiser}: rule R7 (spike note §1.6). From period 60, \texttt{ref} starts at the mean observed rate over 50--59; when \texttt{r{[}t\ −\ 2{]}\ \textgreater{}\ ref\ +\ tolerance} reach rises by \texttt{gain\ ·\ (r\ −\ ref)}, never falls; \texttt{ref\ ←\ min(ref,\ r)} (the registered ratchet), or an exponential average with \texttt{ref\_\allowbreak{}smoothing\ \textgreater{}\ 0} (spike note §8 (ii), Explore). Competitors only. - \texttt{actors.\allowbreak{}rules.\allowbreak{}threshold\_\allowbreak{}trigger}: spec §4.3's alternative rule; reach moves by \texttt{gain} per period while \texttt{\textbar{}gap\textbar{}\ \textgreater{}\ tolerance}. In the rule registry for the sensitivity sweep of rule forms (spec §5 SI); no registered scenario uses it.

\paragraph{3.3.5 Escalation to a persistent automatic nudge}

\texttt{actors.\allowbreak{}escalation.\allowbreak{}escalate}; rule R6 (spike note §1.6), and R7's escalation condition. After the rule, the counter of consecutive ``losing'' periods increments (for the maximiser, only while reach is also at the cap) or resets; at the first period with reach ≥ 0.99 and a count ≥ 10, an automatic nudge in the competitor's direction (−1 in every registered scenario), strength \texttt{escalation\_\allowbreak{}tau}, reach 1 switches on and stays on. Competitors only.

\paragraph{3.3.6 Observation and noise}

\texttt{world.\allowbreak{}World.\allowbreak{}step}. The observed rate is the true rate plus \texttt{obs\_\allowbreak{}noise\_\allowbreak{}sd\ ·\ ε\_\allowbreak{}obs}, \texttt{ε\_\allowbreak{}obs\ \textasciitilde{}\ N(0,\ 1)} per run and period from the shared stream. Author decision of 2026-10-27 (spec §12, implementing spec §4.3's ``lag and noise''); default 0 for parity, swept in Explore. Without it an actor's sensitivity is an artefact of \texttt{N}.

\paragraph{3.3.7 Bystanders' own behaviour}

\texttt{world.\allowbreak{}World.\allowbreak{}\_\allowbreak{}own\_\allowbreak{}rate}, \texttt{people.\allowbreak{}Population.\allowbreak{}own\_\allowbreak{}propensity}. Reduced form of spec §4.1: each bystander's behaviour has its own propensities (drawn like \texttt{a}) and the same threshold mechanism, pushed only by the bystander's own nudge with the attention weight that nudge received on the focal scale. Only its rate is kept; it is what the bystander observes. Simulated in free worlds only, as in the spike (spike note §2; \texttt{code/\allowbreak{}spike/\allowbreak{}worlds.\allowbreak{}py}).

\paragraph{3.3.8 Study designs and treated sets}

\texttt{designs.\allowbreak{}Design}, \texttt{designs.\allowbreak{}draw\_\allowbreak{}treated}. Exactly \texttt{round(share\ ·\ N)} people per run, drawn uniformly once per run from a generator tagged \texttt{pilot} or \texttt{fe}. Fixed by spike note §2 (``arm assignment'') and §3.3; spec §4.4. The timeline and reaches are the \texttt{design\_\allowbreak{}params} section of \texttt{conf/\allowbreak{}nudgefield/\allowbreak{}config.\allowbreak{}yaml} (defaults those of spike note §3.3; the CLI's \texttt{design=\allowbreak{}} names the kind).

\paragraph{3.3.9 Counterfactual worlds and habit}

\texttt{simulate.\allowbreak{}run\_\allowbreak{}paired}, \texttt{simulate.\allowbreak{}\_\allowbreak{}held\_\allowbreak{}clones}. The frozen field holds each actor at its free-world mean reach over 90--99 (spike note §1.1; spec §4.4). With habit off, branching at period 100 equals running from period 0 (people are memoryless). \textbf{Frozen-world counterfactual under habit} (author decision of 2026-10-27, spec §12): with \texttt{habit\_\allowbreak{}rate\ \textgreater{}\ 0} every world owns a population copy, and the frozen pair branches at period 100 from a copy of its source world's population state (\texttt{free} for \texttt{frozen}, \texttt{none} for \texttt{frozen\_\allowbreak{}none}), then holds the field: same history to the rollout, then the field stops reacting. Habit is \texttt{people.\allowbreak{}Population.\allowbreak{}update\_\allowbreak{}habit}; off in v1 (author, spec §11), a sensitivity switch.

\paragraph{3.3.10 Metrics, classes and robustness}

\begin{itemize}
\tightlist
\item
  \texttt{metrics.\allowbreak{}effects.\allowbreak{}effects}: per-period quantities of spike note §1.1. Pilot → rollout: \texttt{total} (free − none), \texttt{frozen}, \texttt{empty}, \texttt{trial} (pilot), \texttt{before\_\allowbreak{}after}. Field experiment, per treated person: \texttt{total}, \texttt{frozen}, \texttt{empty} (\texttt{E\_\allowbreak{}fe}, ruling R2), \texttt{trial}, \texttt{control\_\allowbreak{}drift} (ruling R3), \texttt{site\_\allowbreak{}effect} (rule R8). All designs: each actor's reach and subgroup contrasts by tercile of \texttt{a}, \texttt{B}, \texttt{s\_\allowbreak{}delib}.
\item
  \texttt{metrics.\allowbreak{}classes.\allowbreak{}classify}: the classes, thresholds, evaluation order and readings of spike note §1.3 and §1.5 (oscillation lags by ruling R4, \texttt{metrics.\allowbreak{}classes.\allowbreak{}oscillation}), the R8 site classes (\texttt{metrics.\allowbreak{}classes.\allowbreak{}site\_\allowbreak{}class}) and the recorded, unclassified \texttt{metrics.\allowbreak{}classes.\allowbreak{}subgroup\_\allowbreak{}flip}.
\item
  \texttt{metrics.\allowbreak{}robust.\allowbreak{}robust}: spike note §1.4 (≥ 90\% of replicates at each of ≥ 2 adjacent sweep points). \textbf{Robustness thresholds as parameters} (author decision of 2026-10-27, spec §12): \texttt{min\_\allowbreak{}points} and \texttt{min\_\allowbreak{}range\_\allowbreak{}share} are set by \texttt{run.\allowbreak{}robust\_\allowbreak{}min\_\allowbreak{}points} and \texttt{run.\allowbreak{}robust\_\allowbreak{}min\_\allowbreak{}range\_\allowbreak{}share} (defaults 2 and 0.10 since the author's restatement of 2026-10-29, spec §13: a run of ≥ 2 adjacent qualifying points spanning ≥ 10\% of the swept range; 2 and 0.0 was the spike's rule; \texttt{conf/\allowbreak{}nudgefield/\allowbreak{}config.\allowbreak{}yaml}), passed to both labels' calls and recorded in the manifest (\texttt{run} and \texttt{robust\_\allowbreak{}rule}). In a multi-parameter sweep the rule runs along \texttt{param\_\allowbreak{}value} within each slice of the other swept paths (one \texttt{robust.\allowbreak{}parquet} row per slice and class).
\end{itemize}

\paragraph{3.3.11 Sweeps}

\texttt{sweep.\allowbreak{}expand}, \texttt{sweep.\allowbreak{}run\_\allowbreak{}sweep}. A scenario's grid is the cartesian product of its sweep lists; points that share an actor structure form one batch whose swept floats vary across runs as \texttt{(R,)} overrides; a swept \texttt{count} changes the structure and so forms its own batch. Per-batch seeding (plan decision) avoids \texttt{R} kernel launches per tensor per period.

\endgroup

\clearpage
\sisection{every parameter and its source}{si:parameters}

This section reproduces the project's table of parameters: each value, its unit, its source and whether a sweep varies it. The working names are explained at the head of section~\ref{si:odd}. Source: \texttt{docs/nudgefield/parameters.md}, last changed at commit 63f16a0, 2026-11-01; reproduced unchanged.

\begingroup\footnotesize\sloppy\setlength{\emergencystretch}{3em}\setlength{\tabcolsep}{4pt}
One row per parameter of \texttt{PopulationConfig}, \texttt{NudgeSpec} (ours and the other actors'), \texttt{ActorSpec}, \texttt{DesignConfig} and \texttt{RunConfig} (\texttt{code/\allowbreak{}src/\allowbreak{}nudgefield/\allowbreak{}config.\allowbreak{}py}, \texttt{code/\allowbreak{}src/\allowbreak{}nudgefield/\allowbreak{}actors/\allowbreak{}base.\allowbreak{}py}), plus the calibrated crowded field. \textbf{Source} is one of: spike note §3 (or the ruling of spike note §1.6 that fixed it), spec §6 magnitude, author decision (spec §12), ``swept, no anchor'' (spec §6: no published anchor) or ``calibrated to T2 band''. \textbf{Explore} gives the sweep range foreseen for the Explore phase where one exists; ``range to be registered'' means the parameter is foreseen for a sweep but no range has been fixed yet. Latent units are the probit scale of \texttt{Y\ =\allowbreak{}\ 1\{a\ +\ F\ +\ ε\ \textgreater{}\ 0\}}, \texttt{ε\ \textasciitilde{}\ N(0,\ 1)}. A period is a week (spec §4).

\subsubsection*{\texorpdfstring{1. Population (\texttt{PopulationConfig})}{1. Population (PopulationConfig)}}

{\def\LTcaptype{none} 
\begin{longtable}[]{@{}
  >{\raggedright\arraybackslash}p{(\linewidth - 8\tabcolsep) * \real{0.1245}}
  >{\raggedright\arraybackslash}p{(\linewidth - 8\tabcolsep) * \real{0.1245}}
  >{\raggedright\arraybackslash}p{(\linewidth - 8\tabcolsep) * \real{0.1651}}
  >{\raggedright\arraybackslash}p{(\linewidth - 8\tabcolsep) * \real{0.2814}}
  >{\raggedright\arraybackslash}p{(\linewidth - 8\tabcolsep) * \real{0.2845}}@{}}
\toprule\noalign{}
\begin{minipage}[b]{\linewidth}\raggedright
Parameter
\end{minipage} & \begin{minipage}[b]{\linewidth}\raggedright
Value
\end{minipage} & \begin{minipage}[b]{\linewidth}\raggedright
Unit
\end{minipage} & \begin{minipage}[b]{\linewidth}\raggedright
Source
\end{minipage} & \begin{minipage}[b]{\linewidth}\raggedright
Explore
\end{minipage} \\
\midrule\noalign{}
\endhead
\bottomrule\noalign{}
\endlastfoot
\texttt{n} & 20\_\allowbreak{}000 & people per site & spike note §3.1 & 50\_\allowbreak{}000 on the Spark (spec §8); a scale, not a model parameter \\
\texttt{t\_\allowbreak{}total} & 200 & periods & spike note §3.3 & --- \\
\texttt{mix\_\allowbreak{}share} & 0.3 & share of the high-propensity component & spike note §3.1; spec §6 magnitude (control take-up 0.05--0.9: realised quiet control take-up 0.309, see §6) & --- (T4 uses 0: one component) \\
\texttt{mu\_\allowbreak{}low} & −1.2 & latent & spike note §3.1 & --- (T4 sweeps it, see §5) \\
\texttt{mu\_\allowbreak{}high} & 0.4 & latent & spike note §3.1 & --- \\
\texttt{sd\_\allowbreak{}a} & 0.8 & latent & spike note §3.1 & --- \\
\texttt{budget\_\allowbreak{}mean} & 1.0 & attention-cost units per period & spike note §3.1; swept, no anchor & attention budget: range to be registered (spec §5 F3 names bystanders × budget); sweepable per run since 1898717 (sweep path \texttt{population.\allowbreak{}budget\_\allowbreak{}mean}, a scenario's \texttt{population:\allowbreak{}\ \{budget\_\allowbreak{}mean:\allowbreak{}\ {[}.\allowbreak{}.\allowbreak{}{]}\}}) \\
\texttt{budget\_\allowbreak{}sd\_\allowbreak{}log} & 0.4 & log scale & spike note §3.1; swept, no anchor & range to be registered \\
\texttt{susc\_\allowbreak{}sd\_\allowbreak{}log} & 0.3 & log scale (mean 1) & spike note §3.1 & --- \\
\texttt{valence\_\allowbreak{}sd} & 0.5 & sd of the log-normal magnitude (mean 1) & spike note §3.1 and valence reading of §1.6 & --- \\
\texttt{flip\_\allowbreak{}prob} & 0.05 & probability per person-nudge & spike note §3.1 (mean valence 0.90) & --- \\
\texttt{habit\_\allowbreak{}rate} & 0.0 & per period & spec §4.1 (author: habit off in v1); global constraints & habit on as a switch; range to be registered \\
\end{longtable}
}

\subsubsection*{\texorpdfstring{2. Nudges (\texttt{NudgeSpec})}{2. Nudges (NudgeSpec)}}

{\def\LTcaptype{none} 
\begin{longtable}[]{@{}
  >{\raggedright\arraybackslash}p{(\linewidth - 8\tabcolsep) * \real{0.1396}}
  >{\raggedright\arraybackslash}p{(\linewidth - 8\tabcolsep) * \real{0.1245}}
  >{\raggedright\arraybackslash}p{(\linewidth - 8\tabcolsep) * \real{0.1525}}
  >{\raggedright\arraybackslash}p{(\linewidth - 8\tabcolsep) * \real{0.3446}}
  >{\raggedright\arraybackslash}p{(\linewidth - 8\tabcolsep) * \real{0.2188}}@{}}
\toprule\noalign{}
\begin{minipage}[b]{\linewidth}\raggedright
Parameter
\end{minipage} & \begin{minipage}[b]{\linewidth}\raggedright
Value
\end{minipage} & \begin{minipage}[b]{\linewidth}\raggedright
Unit
\end{minipage} & \begin{minipage}[b]{\linewidth}\raggedright
Source
\end{minipage} & \begin{minipage}[b]{\linewidth}\raggedright
Explore
\end{minipage} \\
\midrule\noalign{}
\endhead
\bottomrule\noalign{}
\endlastfoot
ours: \texttt{direction} & +1 & --- & spike note §3.2 & --- \\
ours: \texttt{kind} & \texttt{delib} & --- & spike note §3.2 & --- (T4 uses \texttt{auto}, \texttt{tau} 0.5) \\
ours: \texttt{tau} & 0.30 & latent & spike note §3.2; spec §6 magnitude (probit effects 0.05--0.4: realised quiet field-experiment probit effect 0.152, see §6) & --- \\
ours: \texttt{cost} & 1.0 & attention-cost units & spike note §3.2; swept, no anchor & attention costs: range to be registered \\
others: \texttt{tau} & 0.30 & latent & spike note §3.2 & \texttt{escalation\_\allowbreak{}tau} below \\
others: \texttt{cost} & 1.0 & attention-cost units & spike note §3.2; swept, no anchor & attention costs: range to be registered \\
\end{longtable}
}

\subsubsection*{\texorpdfstring{3. Actors (\texttt{ActorSpec} defaults; scenario files override per actor)}{3. Actors (ActorSpec defaults; scenario files override per actor)}}

{\def\LTcaptype{none} 
\begin{longtable}[]{@{}
  >{\raggedright\arraybackslash}p{(\linewidth - 8\tabcolsep) * \real{0.1245}}
  >{\raggedright\arraybackslash}p{(\linewidth - 8\tabcolsep) * \real{0.1245}}
  >{\raggedright\arraybackslash}p{(\linewidth - 8\tabcolsep) * \real{0.1462}}
  >{\raggedright\arraybackslash}p{(\linewidth - 8\tabcolsep) * \real{0.3616}}
  >{\raggedright\arraybackslash}p{(\linewidth - 8\tabcolsep) * \real{0.2232}}@{}}
\toprule\noalign{}
\begin{minipage}[b]{\linewidth}\raggedright
Parameter
\end{minipage} & \begin{minipage}[b]{\linewidth}\raggedright
Default
\end{minipage} & \begin{minipage}[b]{\linewidth}\raggedright
Unit
\end{minipage} & \begin{minipage}[b]{\linewidth}\raggedright
Source
\end{minipage} & \begin{minipage}[b]{\linewidth}\raggedright
Explore
\end{minipage} \\
\midrule\noalign{}
\endhead
\bottomrule\noalign{}
\endlastfoot
\texttt{rule} & \texttt{thermostat} & --- & spike note §2 (integral thermostat); \texttt{maximiser} R7; \texttt{threshold\_\allowbreak{}trigger} spec §4.3 & rule form in the sensitivity sweep (spec §5 SI) \\
\texttt{gain} & 2.0 & reach per unit rate gap per period & spike note §2 (S1/S2 sweep k ∈ \{0.5, 2, 5\}); swept, no anchor & range to be registered \\
\texttt{gain\_\allowbreak{}p} & 0.0 & reach per unit change of the gap & spec §4.3 (proportional--integral form); 0 = spike's integral rule & \texttt{gain\_\allowbreak{}p}: range to be registered \\
\texttt{goal\_\allowbreak{}offset} & 0.0 & rate & scenario files: competitor −0.03, ally +0.03 (ruling R1, spike note §1.6) & --- \\
\texttt{lag} & 2 & periods & spike note §3.2 & --- \\
\texttt{reach0} & 0.0 & share of people & spike note §3.2 (bystanders 0.6, maximiser 0.60 through burn-in, R7) & --- \\
\texttt{tolerance} & 0.01 & rate & R7 (three per-period sampling sds of the site rate at n = 20\_\allowbreak{}000) & tied to \texttt{obs\_\allowbreak{}noise\_\allowbreak{}sd} \\
\texttt{ref\_\allowbreak{}window} & (50, 60) & periods, half-open & R7 & --- \\
\texttt{start} & 60 & period & R7 & --- \\
\texttt{ref\_\allowbreak{}smoothing} & 0.0 & weight per period (0 = R7 ratchet) & R7; spike note §8 (ii) & \texttt{ref\_\allowbreak{}smoothing} λ \textgreater{} 0 (smoothed reference): range to be registered \\
\texttt{escalation\_\allowbreak{}tau} & None (no escalation) & latent & R6, R7 (registered sweeps 0.10, 0.20, 0.30) & \texttt{escalation\_\allowbreak{}tau} to 0.40 and beyond (spike note §8 (i)) \\
\texttt{patience} & 10 & periods & R6 & --- \\
\texttt{reach\_\allowbreak{}threshold} & 0.99 & share of people (reach cap) & R6 & --- \\
\texttt{count} & 1 & copies & scenario files (S3 \texttt{m} ∈ \{0, 2, 4, 8, 16\}) & bystander count with the budget (spec §5 F3) \\
\texttt{obs\_\allowbreak{}noise\_\allowbreak{}sd} & 0.0 & rate & author decision 2026-10-27 (spec §12; 0 for spike parity) & \texttt{obs\_\allowbreak{}noise\_\allowbreak{}sd}: range to be registered \\
\end{longtable}
}

\subsubsection*{\texorpdfstring{4. Timeline and run settings (\texttt{DesignConfig}, \texttt{RunConfig})}{4. Timeline and run settings (DesignConfig, RunConfig)}}

The timeline rows are the YAML section \texttt{design\_\allowbreak{}params} (\texttt{DesignParams}, the fields of \texttt{DesignConfig} but its kind); the others are \texttt{run}.

{\def\LTcaptype{none} 
\begin{longtable}[]{@{}
  >{\raggedright\arraybackslash}p{(\linewidth - 8\tabcolsep) * \real{0.1305}}
  >{\raggedright\arraybackslash}p{(\linewidth - 8\tabcolsep) * \real{0.1245}}
  >{\raggedright\arraybackslash}p{(\linewidth - 8\tabcolsep) * \real{0.1626}}
  >{\raggedright\arraybackslash}p{(\linewidth - 8\tabcolsep) * \real{0.2987}}
  >{\raggedright\arraybackslash}p{(\linewidth - 8\tabcolsep) * \real{0.2637}}@{}}
\toprule\noalign{}
\begin{minipage}[b]{\linewidth}\raggedright
Parameter
\end{minipage} & \begin{minipage}[b]{\linewidth}\raggedright
Value
\end{minipage} & \begin{minipage}[b]{\linewidth}\raggedright
Unit
\end{minipage} & \begin{minipage}[b]{\linewidth}\raggedright
Source
\end{minipage} & \begin{minipage}[b]{\linewidth}\raggedright
Explore
\end{minipage} \\
\midrule\noalign{}
\endhead
\bottomrule\noalign{}
\endlastfoot
\texttt{pilot\_\allowbreak{}start}, \texttt{pilot\_\allowbreak{}end} & 60, 80 & periods (half-open: 60--79) & spike note §3.3 & --- \\
\texttt{pilot\_\allowbreak{}reach} & 0.02 & share of people (fixed set drawn at 60) & spike note §3.3; spec §4.4 (1--5\%) & --- \\
\texttt{rollout\_\allowbreak{}start} & 100 & period & spike note §3.3 & --- \\
\texttt{fe\_\allowbreak{}reach} & 0.4 & share of people (fixed set drawn at 100) & spike note §3.3; spec §4.4 (20--50\%) & treated share (spike note disclosure: 0.2--0.5 probed); sweepable per run since b7dfa39 (sweep path \texttt{design.\allowbreak{}fe\_\allowbreak{}reach}, a scenario's \texttt{design:\allowbreak{}\ \{fe\_\allowbreak{}reach:\allowbreak{}\ {[}.\allowbreak{}.\allowbreak{}{]}\}}; field-experiment kinds only, the \texttt{fe} set of \texttt{pilot\_\allowbreak{}field\_\allowbreak{}experiment}; R8 site classes use each run's share; the manifest's \texttt{design.\allowbreak{}fe\_\allowbreak{}reach} stays this default) \\
\texttt{eval\_\allowbreak{}window} & 40 & periods (160--199) & spike note §3.3 & --- \\
\texttt{frozen\_\allowbreak{}window} & (90, 100) & periods, half-open & spike note §3.3, ruling R3 & --- \\
\texttt{settle\_\allowbreak{}window} & (40, 60) & periods, half-open & spike note §3.3 & --- \\
\texttt{settle\_\allowbreak{}tol} & 0.002 & rate per period (absolute slope) & spike note §3.3 & --- \\
\texttt{n\_\allowbreak{}rep} & 10 & replicates per point & spike note §2 & more on the Spark (spec §8) \\
\texttt{seed\_\allowbreak{}base} & 20261027 & --- & plan (global constraints, seeding) & --- \\
\texttt{attention} & \texttt{priority} & rule name & author decision 2026-10-27 (spec §12); \texttt{proportional} for parity & both rules in the sensitivity sweep \\
\texttt{max\_\allowbreak{}batch\_\allowbreak{}runs} & 0 (no split) & runs & plan (Task 6) & --- \\
\texttt{robust\_\allowbreak{}min\_\allowbreak{}points} & 2 & adjacent sweep points & spike note §1.4 (registered rule); restated in range terms 2026-10-29 (spec §13) & --- \\
\texttt{robust\_\allowbreak{}min\_\allowbreak{}range\_\allowbreak{}share} & 0.10 & share of the swept range & spec §7 (≥ 10\% of each swept parameter's range); author ruling 2026-10-29 (spec §13); 0.0 before that & binds only on grids finer than 10 points \\
class rules & r9 & --- & R9 (spike note §1.6) & \texttt{run.\allowbreak{}class\_\allowbreak{}rules} v1 for the ported rules \\
\texttt{resume} & false & --- & plan 2026-10-30 (Spark half, Task 1): continue a stopped call from its \texttt{parts/\allowbreak{}} & --- \\
\end{longtable}
}

\subsubsection*{\texorpdfstring{5. Calibrated crowded field (\texttt{code/\allowbreak{}conf/\allowbreak{}nudgefield/\allowbreak{}scenario/\allowbreak{}crowded.\allowbreak{}yaml})}{5. Calibrated crowded field (code/conf/nudgefield/scenario/crowded.yaml)}}

{\def\LTcaptype{none} 
\begin{longtable}[]{@{}
  >{\raggedright\arraybackslash}p{(\linewidth - 8\tabcolsep) * \real{0.1686}}
  >{\raggedright\arraybackslash}p{(\linewidth - 8\tabcolsep) * \real{0.2442}}
  >{\raggedright\arraybackslash}p{(\linewidth - 8\tabcolsep) * \real{0.1245}}
  >{\raggedright\arraybackslash}p{(\linewidth - 8\tabcolsep) * \real{0.3182}}
  >{\raggedright\arraybackslash}p{(\linewidth - 8\tabcolsep) * \real{0.1245}}@{}}
\toprule\noalign{}
\begin{minipage}[b]{\linewidth}\raggedright
Parameter
\end{minipage} & \begin{minipage}[b]{\linewidth}\raggedright
Value
\end{minipage} & \begin{minipage}[b]{\linewidth}\raggedright
Unit
\end{minipage} & \begin{minipage}[b]{\linewidth}\raggedright
Source
\end{minipage} & \begin{minipage}[b]{\linewidth}\raggedright
Explore
\end{minipage} \\
\midrule\noalign{}
\endhead
\bottomrule\noalign{}
\endlastfoot
crowded: competitor & thermostat, gain 2, \texttt{goal\_\allowbreak{}offset} −0.03 & --- & spike note §2 S1 at k = 2, ruling R1 & --- \\
crowded: ally & thermostat, gain 2, \texttt{goal\_\allowbreak{}offset} +0.03 & --- & spike note §2 S2 at k = 2, ruling R1 & --- \\
crowded: bystanders \texttt{m} & \textbf{2} (gain 0, \texttt{reach0} 0.6) & bystanders & calibrated to T2 band {[}1.5, 2.5{]} (spec §12), default rule \texttt{priority} & --- \\
quiet & no actor & --- & spec §6 T2 & --- \\
\end{longtable}
}

\textbf{Calibration run.} \texttt{test\_\allowbreak{}crowded\_\allowbreak{}calibration} in \texttt{code/\allowbreak{}tests/\allowbreak{}nudgefield/\allowbreak{}test\_\allowbreak{}targets.\allowbreak{}py}, outputs \texttt{code/\allowbreak{}outputs/\allowbreak{}targets\_\allowbreak{}20261027\_\allowbreak{}201230/\allowbreak{}calibration/\allowbreak{}} and the table \texttt{calibration.\allowbreak{}parquet} there. Field-experiment design, \texttt{n\ =\allowbreak{}\ 20\_\allowbreak{}000}, 10 replicates, device \texttt{mps}. The probit trial effect is \texttt{Φ⁻¹(treated)\ −\ Φ⁻¹(control)} of the free world's window means (periods 160--199) per replicate, averaged over replicates; the ratio is quiet / crowded. Quiet: treated 0.3645, control 0.3090, probit effect 0.1524 (identical under both rules: one deliberative nudge, so both rules give \texttt{w\ =\allowbreak{}\ min(1,\ B)}). The rule picks the smallest \texttt{m} with the \texttt{priority} ratio in {[}1.5, 2.5{]}. The levels behind the probit effects are each run directory's \texttt{levels.\allowbreak{}parquet}, written since 86b6317 by the package tool \texttt{nudgefield.\allowbreak{}levels} (before it, by a private helper of \texttt{test\_\allowbreak{}targets.\allowbreak{}py}); the T2 test pins the \texttt{m\ =\allowbreak{}\ 2} ratio under \texttt{priority} to 2.25 ± 0.01 (2.2506 at 86b6317).

{\def\LTcaptype{none} 
\begin{longtable}[]{@{}
  >{\raggedright\arraybackslash}p{(\linewidth - 10\tabcolsep) * \real{0.1245}}
  >{\raggedright\arraybackslash}p{(\linewidth - 10\tabcolsep) * \real{0.1998}}
  >{\raggedright\arraybackslash}p{(\linewidth - 10\tabcolsep) * \real{0.1566}}
  >{\raggedright\arraybackslash}p{(\linewidth - 10\tabcolsep) * \real{0.1245}}
  >{\raggedright\arraybackslash}p{(\linewidth - 10\tabcolsep) * \real{0.2089}}
  >{\raggedright\arraybackslash}p{(\linewidth - 10\tabcolsep) * \real{0.1657}}@{}}
\toprule\noalign{}
\begin{minipage}[b]{\linewidth}\raggedright
\texttt{m}
\end{minipage} & \begin{minipage}[b]{\linewidth}\raggedright
crowded probit (\texttt{priority})
\end{minipage} & \begin{minipage}[b]{\linewidth}\raggedright
ratio (\texttt{priority})
\end{minipage} & \begin{minipage}[b]{\linewidth}\raggedright
in band
\end{minipage} & \begin{minipage}[b]{\linewidth}\raggedright
crowded probit (\texttt{proportional})
\end{minipage} & \begin{minipage}[b]{\linewidth}\raggedright
ratio (\texttt{proportional})
\end{minipage} \\
\midrule\noalign{}
\endhead
\bottomrule\noalign{}
\endlastfoot
2 & 0.0677 & \textbf{2.25} & yes & 0.0453 & 3.36 \\
3 & 0.0571 & 2.67 & no & 0.0354 & 4.30 \\
4 & 0.0537 & 2.84 & no & 0.0327 & 4.66 \\
6 & 0.0431 & 3.54 & no & 0.0236 & 6.46 \\
8 & 0.0368 & 4.14 & no & 0.0187 & 8.13 \\
\end{longtable}
}

The ratio rises with \texttt{m} under both rules; \texttt{m\ =\allowbreak{}\ 2}, the smallest candidate, is the only one in the band. Under \texttt{proportional} every ratio is above the band (the band applies to the default rule only; the direction holds).

\textbf{T2 at \texttt{m} = 0 and 1 under \texttt{priority} (fix wave of 2026-10-27).} The same calibration code (\texttt{\_\allowbreak{}quiet\_\allowbreak{}crowded} with \texttt{\_\allowbreak{}crowded\_\allowbreak{}swept((0,\ 1))} from \texttt{test\_\allowbreak{}targets.\allowbreak{}py}), \texttt{priority} only, same settings and seed derivation (scenario \texttt{crowded}, \texttt{seed\_\allowbreak{}base} 20261027; \texttt{m} = 0 and 1 are new actor structures and so new batch seeds), device \texttt{mps}, code at 90a1370 (\texttt{git\_\allowbreak{}dirty:\allowbreak{}\ false}). Outputs in a new directory, \texttt{code/\allowbreak{}outputs/\allowbreak{}targets\_\allowbreak{}20261027\_\allowbreak{}213643/\allowbreak{}calibration/\allowbreak{}priority/\allowbreak{}\{quiet,crowded\}/\allowbreak{}}, table \texttt{calibration\_\allowbreak{}m01.\allowbreak{}parquet} there; the committed \texttt{targets\_\allowbreak{}20261027\_\allowbreak{}201230} runs are unchanged. The quiet run reproduces the committed quiet probit effect, 0.1524.

{\def\LTcaptype{none} 
\begin{longtable}[]{@{}llll@{}}
\toprule\noalign{}
\texttt{m} & crowded probit (\texttt{priority}) & ratio (\texttt{priority}) & in band \\
\midrule\noalign{}
\endhead
\bottomrule\noalign{}
\endlastfoot
0 & 0.0841 & 1.81 & yes \\
1 & 0.0704 & 2.16 & yes \\
\end{longtable}
}

The band does not constrain \texttt{m} from below: the competitor and the ally alone (\texttt{m\ =\allowbreak{}\ 0}) already put the ratio in {[}1.5, 2.5{]}, and every \texttt{m} in \{0, 1, 2\} is in the band; only \texttt{m\ ≥\ 3} is ruled out. \texttt{m\ =\allowbreak{}\ 2} is the smallest member of the registered candidate set \{2, 3, 4, 6, 8\} in the band, not the smallest \texttt{m} in the band. \textbf{Author ruling (2026-10-29, spec §13): m = 2 stands.} It is the outcome of the registered rule on the registered candidate set; the bystander count is swept in the F3 phase maps, where the sensitivity is shown. \texttt{crowded.\allowbreak{}yaml} is unchanged.

\textbf{T4 settings (test only, not a baseline).} No actor; our nudge automatic, \texttt{tau} 0.5; \texttt{mix\_\allowbreak{}share} 0, \texttt{sd\_\allowbreak{}a} 0.8; \texttt{mu\_\allowbreak{}low} on seven equally spaced values between take-up 0.1 and 0.8, \texttt{mu\ =\allowbreak{}\ sqrt(1\ +\ 0.\allowbreak{}8²)\ ·\ Φ⁻¹(q)}: −1.641, −1.188, −0.735, −0.282, 0.171, 0.625, 1.078.

\textbf{Headroom family (run 8, registry §14; 2026-10-31).} A second crowded family, \texttt{crowded\_\allowbreak{}headroom} (the S1 competitor at \texttt{nudge.\allowbreak{}tau} 0.35 plus m bystanders, no ally), is calibrated at m = 2, budget 1.0, by the joint-cell rule of §14: quiet / crowded probit ratio 1.508 (at the band's lower edge; the adjacent cell m = 3 gives 1.733), T1 0.640, competitor burn-in reach 0.735, attenuated 100 of 100, under priority (\texttt{docs/\allowbreak{}analysis/\allowbreak{}2026-\allowbreak{}10-\allowbreak{}31-\allowbreak{}run8-\allowbreak{}calibration-\allowbreak{}map.\allowbreak{}md}; Modal T4, cuda). It is the family in which T1 and T2 hold together with a reacting competitor; the \texttt{crowded.\allowbreak{}yaml} calibration above stays as the stalemate case (both of its thermostats at full reach from burn-in; \texttt{docs/\allowbreak{}analysis/\allowbreak{}2026-\allowbreak{}10-\allowbreak{}31-\allowbreak{}t1-\allowbreak{}probe-\allowbreak{}two-\allowbreak{}actor-\allowbreak{}field.\allowbreak{}md}). Run 4h (registry §15) maps the headroom family over m × budget.

\subsubsection*{6. Magnitude checks against spec §6 and the T1 spread}

From \texttt{levels.\allowbreak{}parquet} and \texttt{effects.\allowbreak{}parquet} of the runs above (window means, replicate means):

\begin{itemize}
\tightlist
\item
  Control take-up in the quiet field 0.309 (spec §6: 0.05--0.9).
\item
  Probit effect of our nudge in the quiet field experiment 0.152; in the crowded field 0.068 (\texttt{priority}) (spec §6: 0.05--0.4).
\item
  Quiet / crowded ratio 2.25 under \texttt{priority}: a constraint, not a prediction, since \texttt{m} was chosen to meet it (spec §12).
\end{itemize}

\textbf{T1, loss of effect at scale} (pilot → rollout; \texttt{total} mean over 160--199 divided by pilot \texttt{trial} mean over 60--79; from \texttt{T1/\allowbreak{}\textless{}rule\textgreater{}/\allowbreak{}\textless{}site\textgreater{}/\allowbreak{}effects.\allowbreak{}parquet}, per replicate in \texttt{T1\_\allowbreak{}\textless{}site\textgreater{}\_\allowbreak{}\textless{}rule\textgreater{}.\allowbreak{}parquet}):

{\def\LTcaptype{none} 
{\scriptsize
\begin{longtable}[]{@{}
  >{\raggedright\arraybackslash}p{(\linewidth - 12\tabcolsep) * \real{0.1791}}
  >{\raggedright\arraybackslash}p{(\linewidth - 12\tabcolsep) * \real{0.1251}}
  >{\raggedright\arraybackslash}p{(\linewidth - 12\tabcolsep) * \real{0.1250}}
  >{\raggedright\arraybackslash}p{(\linewidth - 12\tabcolsep) * \real{0.1091}}
  >{\raggedright\arraybackslash}p{(\linewidth - 12\tabcolsep) * \real{0.1436}}
  >{\raggedright\arraybackslash}p{(\linewidth - 12\tabcolsep) * \real{0.1891}}
  >{\raggedright\arraybackslash}p{(\linewidth - 12\tabcolsep) * \real{0.1091}}@{}}
\toprule\noalign{}
\begin{minipage}[b]{\linewidth}\raggedright
Site / rule
\end{minipage} & \begin{minipage}[b]{\linewidth}\raggedright
Ratio of replicate means (test, \textless{} 0.8)
\end{minipage} & \begin{minipage}[b]{\linewidth}\raggedright
Mean of per-replicate ratios
\end{minipage} & \begin{minipage}[b]{\linewidth}\raggedright
Replicates ≥ 0.8
\end{minipage} & \begin{minipage}[b]{\linewidth}\raggedright
Per-replicate min--max
\end{minipage} & \begin{minipage}[b]{\linewidth}\raggedright
Pilot contrast mean (sd across replicates)
\end{minipage} & \begin{minipage}[b]{\linewidth}\raggedright
Rollout \texttt{total}
\end{minipage} \\
\midrule\noalign{}
\endhead
\bottomrule\noalign{}
\endlastfoot
S1 k = 2 / \texttt{priority} & 0.741 & 0.795 & 4/10 & 0.565--1.393 & 0.0467 (0.0117) & 0.0346 \\
S1 k = 2 / \texttt{proportional} & 0.671 & 0.705 & 3/10 & 0.515--1.034 & 0.0462 (0.0106) & 0.0310 \\
S2 k = 2 / \texttt{priority} & 0.621 & 0.639 & 1/10 & 0.494--0.824 & 0.0405 (0.0070) & 0.0252 \\
S2 k = 2 / \texttt{proportional} & 0.738 & 0.767 & 6/10 & 0.552--1.031 & 0.0341 (0.0068) & 0.0252 \\
\end{longtable}
}
}

The test statistic is the ratio of replicate means because a 2\% pilot of 20,000 people (about 400 treated) gives a noisy contrast (sd 0.007--0.012 against means 0.034--0.047), while the rollout \texttt{total} varies little across replicates (sd ≤ 0.0007). The per-replicate reading is weaker: in three of the four cells at least one replicate has rollout ≥ pilot (ratio ≥ 1: one replicate each in S1/\texttt{priority}, S1/\texttt{proportional} and S2/\texttt{proportional}), and in S2/\texttt{proportional} 6 of 10 replicates are at or above 0.8.

\subsubsection*{7. Run provenance}

The 25 \texttt{manifest.\allowbreak{}json} files under \texttt{code/\allowbreak{}outputs/\allowbreak{}targets\_\allowbreak{}20261027\_\allowbreak{}201230/\allowbreak{}} record \texttt{git\_\allowbreak{}commit} 5882f76 with \texttt{git\_\allowbreak{}dirty:\allowbreak{}\ true}, because \texttt{quiet.\allowbreak{}yaml}, \texttt{crowded.\allowbreak{}yaml} and \texttt{test\_\allowbreak{}targets.\allowbreak{}py} were untracked when the runs were made. Commit c47b114 holds the code that generated them, and the Task 8 fix-round-1 commit 9bcf9d6 holds it with named constants and an index-alignment check; neither changes a run input (the T4 grid and every seed are unchanged), and the targets suite re-run after the fix passed 9/9.

\textbf{Clean re-run (run 9, registry §16, 2026-11-01).} \texttt{code/\allowbreak{}outputs/\allowbreak{}targets\_\allowbreak{}20261101\_\allowbreak{}200002/\allowbreak{}} holds the targets suite re-run at c220bcf with \texttt{git\_\allowbreak{}dirty:\allowbreak{}\ false} in all 33 manifests (M4, \texttt{mps}), with one registered input change: \texttt{T4\_\allowbreak{}POINTS} 15 (was 7). The T1 ratios of replicate means reproduce the table above to three decimals (0.741, 0.671, 0.621, 0.738) with the same per-replicate counts ≥ 0.8 (4, 3, 1, 6 of 10); T2 2.2506 under priority; T3 0.0132 against 0.0036; T4 peak at control take-up 0.414 against the oracle 0.423. ED3 is drawn from this directory; the dirty Build-phase runs above remain the record of the calibration's origin.

\textbf{Fine T4 grid (run 10, registry §17, 2026-11-01).} \texttt{code/\allowbreak{}outputs/\allowbreak{}targets\_\allowbreak{}20261101\_\allowbreak{}223633/\allowbreak{}} re-runs the suite at 4b1b94e (clean; 42 manifests) with a second T4 grid of 9 take-up levels 0.300--0.500 under the same seeds; the coarse tables equal run 9's exactly. The fine grid's peak is at take-up 0.376, 0.047 from the oracle 0.423 (inside the 0.05 tolerance, not within one grid step: the inverted U's top is flat --- within one replicate s.e. of the maximum over take-up 0.356--0.414, within 0.0006 of it over 0.351--0.426 --- so the peak is resolved only to that plateau at n\_\allowbreak{}rep 10; registry §17 records this and a second unmet expectation on \texttt{ate\_\allowbreak{}sd} as deviations). Both T4 tables now carry \texttt{ate\_\allowbreak{}sd} and \texttt{n\_\allowbreak{}rep}. ED3 panel (d) is drawn from this directory.

\endgroup

\clearpage
\sisection{the dated record of runs}{si:record}

This section reproduces the record in which each sweep was specified before it ran, with the date of each entry, its expectations, and the deviations and unmet expectations disclosed afterwards. It is a version-controlled file, not a public preregistration; the date of each entry is that of its commit. The working names are explained at the head of section~\ref{si:odd}. Source: \texttt{docs/nudgefield/explore-registry.md}, last changed at commit dae2903, 2026-11-12; reproduced unchanged.

\begingroup\footnotesize\sloppy\setlength{\emergencystretch}{3em}\setlength{\tabcolsep}{4pt}
Every Explore sweep is registered here as a dated entry \textbf{before it is run}: scenario file, design, swept values, replicates, population size, seed base, attention rule, output directory, the class each point is expected to show, and what counts as prior knowledge. The entry is committed, the author signs it off, and only then is the command run. A change to an entry after its run is a new dated entry that states the reason; the original stays. Results are not written here: they go to the run directory's \texttt{classes.\allowbreak{}parquet} / \texttt{robust.\allowbreak{}parquet} and to the analysis notes. The figures' selection of points (\texttt{code/\allowbreak{}conf/\allowbreak{}nudgefield/\allowbreak{}figures/\allowbreak{}F2.\allowbreak{}yaml} and the F3/F4 YAMLs) is a separate, ratified step after the runs.

Rules in force for every entry: spike note §1--§3 with rulings R1--R8; the restated robustness rule of spec §13 (≥ 90\% of replicates at each point of a run of ≥ 2 adjacent points spanning ≥ 10\% of the swept range; \texttt{run.\allowbreak{}robust\_\allowbreak{}min\_\allowbreak{}points} 2, \texttt{run.\allowbreak{}robust\_\allowbreak{}min\_\allowbreak{}range\_\allowbreak{}share} 0.10); \texttt{run.\allowbreak{}attention\ =\allowbreak{}\ priority} (the package default) unless the entry says otherwise; \texttt{run.\allowbreak{}max\_\allowbreak{}batch\_\allowbreak{}runs\ =\allowbreak{}\ 0} (no chunking: chunking changes the batch seeds and so the draws). Device is not a registered parameter (the fast suite pins \texttt{cpu}, \texttt{mps} and \texttt{cuda} to the same results within float32 tolerance). If a batch does not fit in memory, the fallback is \texttt{run.\allowbreak{}max\_\allowbreak{}batch\_\allowbreak{}runs} = one point's worth of runs (100), recorded in the entry's ledger line, and the run is reported as made under that setting.

\subsubsection*{0. Settings shared by runs 1, 2 and 2b (registered 2026-10-29)}

{\def\LTcaptype{none} 
\begin{longtable}[]{@{}
  >{\raggedright\arraybackslash}p{(\linewidth - 4\tabcolsep) * \real{0.1357}}
  >{\raggedright\arraybackslash}p{(\linewidth - 4\tabcolsep) * \real{0.3814}}
  >{\raggedright\arraybackslash}p{(\linewidth - 4\tabcolsep) * \real{0.4630}}@{}}
\toprule\noalign{}
\begin{minipage}[b]{\linewidth}\raggedright
Setting
\end{minipage} & \begin{minipage}[b]{\linewidth}\raggedright
Value
\end{minipage} & \begin{minipage}[b]{\linewidth}\raggedright
Reason
\end{minipage} \\
\midrule\noalign{}
\endhead
\bottomrule\noalign{}
\endlastfoot
\texttt{run.\allowbreak{}n\_\allowbreak{}rep} & 100 & brief §3: final figures at n\_\allowbreak{}rep = 100 \\
\texttt{population.\allowbreak{}n} & 20,000 & brief §9; the M4 is enough at this size \\
\texttt{run.\allowbreak{}seed\_\allowbreak{}base} & 20261029 & new base for the Explore phase so that its draws are distinct from the Build-phase runs (seed base 20261027); batches are seeded \texttt{scenario\textbar{}design\textbar{}structure\textbar{}seed\_\allowbreak{}base} \\
\texttt{run.\allowbreak{}attention} & priority & the default rule; the F2 points are being re-established under it (F2.yaml header) \\
\texttt{run.\allowbreak{}max\_\allowbreak{}batch\_\allowbreak{}runs} & 0 & one batch per actor structure \\
windows & \texttt{eval\_\allowbreak{}window} 40 (periods 160--199), \texttt{frozen\_\allowbreak{}window} {[}90, 100{]}, \texttt{settle\_\allowbreak{}window} {[}40, 60{]}, \texttt{settle\_\allowbreak{}tol} 0.002 & unchanged from \texttt{config.\allowbreak{}yaml} \\
\texttt{design\_\allowbreak{}params} & pilot 60--80 at reach 0.02; rollout / field experiment from 100; \texttt{fe\_\allowbreak{}reach} 0.4 & unchanged \\
output & \texttt{code/\allowbreak{}outputs/\allowbreak{}explore\_\allowbreak{}20261029/\allowbreak{}\textless{}run\textgreater{}/\allowbreak{}\textless{}scenario\textgreater{}\_\allowbreak{}\textless{}design\textgreater{}/\allowbreak{}} via \texttt{out\_\allowbreak{}dir=\allowbreak{}} & one directory per sweep call; the figure YAMLs point here \\
code & the commit that contains this entry (recorded in every \texttt{manifest.\allowbreak{}json} as \texttt{git\_\allowbreak{}commit}, \texttt{git\_\allowbreak{}dirty:\allowbreak{}\ false}) & provenance rule of the brief §3 \\
\end{longtable}
}

\subsubsection*{1. Run 1: the registered sweeps under priority (feeds F2, ED1, ED2)}

Purpose: re-establish, under the default attention rule and at n\_\allowbreak{}rep = 100, the outcome class at every point of the sweeps that were registered in the spike (under proportional sharing at n\_\allowbreak{}rep = 10). These are the F2 candidate points. Nothing in the scenario files changes.

{\def\LTcaptype{none} 
\begin{longtable}[]{@{}
  >{\raggedright\arraybackslash}p{(\linewidth - 8\tabcolsep) * \real{0.1245}}
  >{\raggedright\arraybackslash}p{(\linewidth - 8\tabcolsep) * \real{0.1884}}
  >{\raggedright\arraybackslash}p{(\linewidth - 8\tabcolsep) * \real{0.1861}}
  >{\raggedright\arraybackslash}p{(\linewidth - 8\tabcolsep) * \real{0.3565}}
  >{\raggedright\arraybackslash}p{(\linewidth - 8\tabcolsep) * \real{0.1245}}@{}}
\toprule\noalign{}
\begin{minipage}[b]{\linewidth}\raggedright
Call
\end{minipage} & \begin{minipage}[b]{\linewidth}\raggedright
Scenario file
\end{minipage} & \begin{minipage}[b]{\linewidth}\raggedright
Design
\end{minipage} & \begin{minipage}[b]{\linewidth}\raggedright
Swept values
\end{minipage} & \begin{minipage}[b]{\linewidth}\raggedright
Runs
\end{minipage} \\
\midrule\noalign{}
\endhead
\bottomrule\noalign{}
\endlastfoot
1a & \texttt{S1\_\allowbreak{}competitor} & pilot\_\allowbreak{}rollout & \texttt{competitor.\allowbreak{}gain} \{0.5, 2, 5\} & 300 \\
1b & \texttt{S1\_\allowbreak{}competitor} & field\_\allowbreak{}experiment & \texttt{competitor.\allowbreak{}gain} \{0.5, 2, 5\} & 300 \\
1c & \texttt{S2\_\allowbreak{}ally} & pilot\_\allowbreak{}rollout & \texttt{ally.\allowbreak{}gain} \{0.5, 2, 5\} & 300 \\
1d & \texttt{S3\_\allowbreak{}bystanders} & pilot\_\allowbreak{}rollout & \texttt{bystander.\allowbreak{}count} \{0, 2, 4, 8, 16\} & 5 batches × 100 (one structure per m) \\
1e & \texttt{S1\_\allowbreak{}persistent\_\allowbreak{}auto} & pilot\_\allowbreak{}rollout & \texttt{competitor.\allowbreak{}escalation\_\allowbreak{}tau} \{0.1, 0.2, 0.3\} & 300 \\
1f & \texttt{S1\_\allowbreak{}persistent\_\allowbreak{}auto} & field\_\allowbreak{}experiment & \texttt{competitor.\allowbreak{}escalation\_\allowbreak{}tau} \{0.1, 0.2, 0.3\} & 300 \\
1g & \texttt{S1\_\allowbreak{}maximiser} & pilot\_\allowbreak{}rollout & \texttt{competitor.\allowbreak{}escalation\_\allowbreak{}tau} \{0.1, 0.2, 0.3\} & 300 \\
1h & \texttt{S1\_\allowbreak{}maximiser} & field\_\allowbreak{}experiment & \texttt{competitor.\allowbreak{}escalation\_\allowbreak{}tau} \{0.1, 0.2, 0.3\} & 300 \\
\end{longtable}
}

Range check: three equally spaced points, two adjacent points span 50\% of the range (1a--1c, 1e--1h); for 1d the smallest adjacent gap (0 → 2) is 12.5\% of 0--16. The rule applies as written everywhere.

\textbf{Expected classes, stated before running} (prior knowledge: the spike's results under proportional sharing, spike note §4--§7; the Spark run of 1,002 S1 field-experiment runs under priority, blind\_\allowbreak{}site 1,002/1,002, attenuated 1,002/1,002, \texttt{ratio\_\allowbreak{}window} 0.63; the parity runs P1--P6; the targets runs' T1 spread, \texttt{parameters.\allowbreak{}md} §6):

\begin{itemize}
\tightlist
\item
  1a: attenuated at every k (the spike: 10/10 at each; under priority the Spark FE run gave 0.63 of E, so pilot → rollout should sit near the spike's 0.6--0.7). Robust expected.
\item
  1b: \texttt{cls\_\allowbreak{}combined} blind\_\allowbreak{}site at every k (the Spark: 1,002/1,002). Robust expected. \texttt{cls} attenuated.
\item
  1c: attenuated or as expected; the spike (proportional) found the ally relaxing to zero and removing about a tenth; under priority the ally's copy of our direction may share attention differently (spike note §4.6 item 3 is to be re-established here). Expectation: attenuated at k = 2, 5 with \texttt{ratio\_\allowbreak{}window} 0.85--0.95; as expected possible at k = 0.5. Not stated as robust.
\item
  1d: as expected at m = 0 (by construction; single point); attenuated at m = 2, 4; null at m = 8, 16 (the spike under proportional: 10/10; under priority the crowding is sharper, \texttt{parameters.\allowbreak{}md} §5, so null may begin at m = 4). The brief's question whether priority gives a tipping line is answered here first.
\item
  1e: transient at 0.1, 0.2, 0.3 (the spike: 10/10 at each, robust). Robust expected.
\item
  1f: \texttt{cls\_\allowbreak{}combined} blind\_\allowbreak{}site at 0.1 and 0.2; blind\_\allowbreak{}site or reversed\_\allowbreak{}site at 0.3 (R8's expectation; the spike's site window means 0.0002 / 0.0001 / −0.0040).
\item
  1g: 0.1 transient or null; 0.2 transient (the spike: −0.12·E, above the −0.25 threshold); 0.3 reversed (the spike: 10/10, −0.44·E). Reversal at one point only, hence not robust in this run; run 2 extends the axis.
\item
  1h: \texttt{cls\_\allowbreak{}combined} blind\_\allowbreak{}site at 0.1, 0.2; blind\_\allowbreak{}site or reversed\_\allowbreak{}site at 0.3.
\end{itemize}

What would count against the model's story: a class flipping between the rules at the registered F2 points (e.g.~no transient in 1e, or no blind\_\allowbreak{}site in 1b) would mean the F2 entries are rule-dependent, and the figure would have to say so (ED4).

\subsubsection*{2. Run 2: reversal robustness and the reference rule (feeds F2 reversed, F4b)}

Scenario file \texttt{S1\_\allowbreak{}maximiser\_\allowbreak{}smoothing} (new, this commit): the R7 maximiser of \texttt{S1\_\allowbreak{}maximiser} with \texttt{competitor.\allowbreak{}escalation\_\allowbreak{}tau} extended to \{0.1, 0.2, 0.3, 0.4, 0.5\} and \texttt{competitor.\allowbreak{}ref\_\allowbreak{}smoothing} ∈ \{0, 0.1\} as the slice axis; both designs; one batch of 1,000 runs per design.

{\def\LTcaptype{none} 
\begin{longtable}[]{@{}
  >{\raggedright\arraybackslash}p{(\linewidth - 8\tabcolsep) * \real{0.1245}}
  >{\raggedright\arraybackslash}p{(\linewidth - 8\tabcolsep) * \real{0.2051}}
  >{\raggedright\arraybackslash}p{(\linewidth - 8\tabcolsep) * \real{0.2243}}
  >{\raggedright\arraybackslash}p{(\linewidth - 8\tabcolsep) * \real{0.2948}}
  >{\raggedright\arraybackslash}p{(\linewidth - 8\tabcolsep) * \real{0.1314}}@{}}
\toprule\noalign{}
\begin{minipage}[b]{\linewidth}\raggedright
Call
\end{minipage} & \begin{minipage}[b]{\linewidth}\raggedright
Design
\end{minipage} & \begin{minipage}[b]{\linewidth}\raggedright
Robustness axis
\end{minipage} & \begin{minipage}[b]{\linewidth}\raggedright
Slice
\end{minipage} & \begin{minipage}[b]{\linewidth}\raggedright
Runs
\end{minipage} \\
\midrule\noalign{}
\endhead
\bottomrule\noalign{}
\endlastfoot
2p & pilot\_\allowbreak{}rollout & \texttt{escalation\_\allowbreak{}tau} 0.1 \ldots{} 0.5 & \texttt{ref\_\allowbreak{}smoothing} 0 (R7 ratchet), 0.1 (exponential average) & 1,000 \\
2f & field\_\allowbreak{}experiment & same & same & 1,000 \\
\end{longtable}
}

Choices and reasons. \emph{Upper bound 0.5:} the brief's proposal; above the calibrated single-arm probit range of spec §6 (0.05--0.4) by the same judgement R5 used (an escalating actor may exceed a single arm's effect by half; R5 took 0.60). \emph{Spacing:} 0.1, so two adjacent points span 25\% of the range. \emph{λ = 0.1:} an exponential average \texttt{ref\ ←\ 0.\allowbreak{}9\ ref\ +\ 0.\allowbreak{}1\ r{[}t\ −\ 2{]}} has a memory of about ten periods, the same horizon as the escalation's patience (10 periods); it removes the ratchet that spike note §7.4 and §7.7 identified as the reason the maximiser's reach creeps before the rollout. One value, not a sweep: the question is whether the reversal depends on the ratchet, not how it scales with λ.

\textbf{Expected classes, stated before running} (prior knowledge: the spike's S1max and S1maxfe, note §7; run 1g/1h is not yet run):

\begin{itemize}
\tightlist
\item
  2p, λ = 0: 0.1 null or transient; 0.2 transient (borderline); 0.3, 0.4, 0.5 reversed. If 0.3--0.5 are each ≥ 90\% reversed, the class is robust (three adjacent points, 50\% of the range) and F2's ``reversed'' panel loses its ``single point'' marker. The depth of the reversal should grow with τ\_\allowbreak{}auto (about −0.44·E at 0.3 in the spike; the automatic push is on everyone).
\item
  2p, λ = 0.1: the reach climb before the rollout should be smaller (no ratchet to noise minima), so the pre-rollout state is closer to the registered 0.60; the trigger (reach ≥ 0.99 and ten periods above \texttt{ref} + 0.01) should still fire after the rollout, because the rollout lifts the rate by more than 0.01 and a ten-period average lags it. Expectation: the same classes as λ = 0 at 0.3--0.5; at 0.1--0.2 the same or one step toward ``as expected''; a later trigger period in \texttt{classes.\allowbreak{}parquet} (\texttt{trigger\_\allowbreak{}free\_\allowbreak{}competitor@auto}). If the reversal vanishes under λ = 0.1, the result is that R7's reversal depends on the ratchet, and the paper says so.
\item
  2f: \texttt{cls\_\allowbreak{}combined} blind\_\allowbreak{}site at 0.1 and 0.2; reversed\_\allowbreak{}site at 0.3 or above once the automatic push exceeds \texttt{p\ ·\ E\_\allowbreak{}fe} on the site (expected from 0.3 up under both λ). The trial contrast stays near \texttt{E\_\allowbreak{}fe} throughout (the identity trial/total ≈ 1/(1 − p) while the competitor neutralises the site rate).
\end{itemize}

\subsubsection*{3. Run 2b: time to decay against the competitor's gain (feeds F4c)}

Scenario file \texttt{S1\_\allowbreak{}persistent\_\allowbreak{}auto\_\allowbreak{}gain} (new, this commit): the R6 competitor of \texttt{S1\_\allowbreak{}persistent\_\allowbreak{}auto} at \texttt{escalation\_\allowbreak{}tau} 0.3 with \texttt{competitor.\allowbreak{}gain} ∈ \{0.5, 1, 2, 5, 10\}; pilot\_\allowbreak{}rollout only; one batch of 500 runs.

Reasons. The brief §7c: the decay time is set by the lag and the gain, not by τ\_\allowbreak{}auto (spike note §6.4: at τ\_\allowbreak{}auto 0.1--0.3 the 75\% / 50\% crossings fall on the same periods). 0.5--5 covers the spike's registered gains; 10 extends toward the oscillation region seen only at k = 20--40 in the disclosed probes. Spacing is uneven; the smallest adjacent gap (0.5 → 1) is 5.3\% of the range, so the rule's 10\% share is met by any two adjacent points from 1 upward (1 → 2 is 10.5\%); the 0.5 → 1 pair alone does not qualify. \texttt{escalation\_\allowbreak{}tau} 0.3 because that is the registered transient point (F2.yaml) and the spike's clearest decay.

\textbf{Expected, stated before running:} transient at k = 1, 2, 5, 10 (the automatic push erases the rollout and the thermostat offsets its own push); at k = 0.5 transient or attenuated, since a slow thermostat may not reach its cap plus ten periods of gap before the evaluation window (the spike's first 75\% crossing at k = 0.5 was period 109, so transient is the better bet). Time to the 75\% and 50\% crossings decreasing in gain and flattening by k = 5 (the spike: 109, 103, 102 periods at k = 0.5, 2, 5). Oscillating is not expected below k = 20.

\subsubsection*{4. Registered rule: the subgroup flip on attention-budget terciles (author, 2026-10-29)}

Applies to every Explore run from this entry on; it is read first on run 4 (F4d) and may be read on run 1d.

\emph{Rule.} Within one replicate, for a grouping variable \texttt{v} ∈ \{\texttt{budget} (registered), \texttt{a} (SI variant)\} with terciles g ∈ \{0, 1, 2\} formed on the population at period 0 (as \texttt{metrics/\allowbreak{}effects.\allowbreak{}py} forms them), the flip indicator is 1 when the tercile with the largest window mean (periods 160--199) of \texttt{group\_\allowbreak{}total\_\allowbreak{}v\_\allowbreak{}g} differs from the tercile with the largest window mean of \texttt{group\_\allowbreak{}empty\_\allowbreak{}v\_\allowbreak{}g}; it is 0 otherwise, and 0 when any group series is missing or its window mean is not finite. The flip is counted like a class under the restated robustness rule: robust at a sweep point when ≥ 90\% of the replicates flip, over ≥ 2 adjacent points spanning ≥ 10\% of the range. \emph{Prediction (F4d):} under priority sharing, people with small attention budgets lose our nudge first, so the argmax tercile moves from low- or equal-budget in the empty field to high-budget in a crowded field; no flip is predicted on \texttt{a}.

\emph{Implementation.} \texttt{classes.\allowbreak{}parquet} carries \texttt{subgroup\_\allowbreak{}flip} on \texttt{a} (\texttt{metrics/\allowbreak{}classes.\allowbreak{}py}, \texttt{FLIP\_\allowbreak{}VAR\ =\allowbreak{}\ "a"}). The budget variant is computed from the \texttt{group\_\allowbreak{}total\_\allowbreak{}budget\_\allowbreak{}\textless{}g\textgreater{}} and \texttt{group\_\allowbreak{}empty\_\allowbreak{}budget\_\allowbreak{}\textless{}g\textgreater{}} series of \texttt{effects.\allowbreak{}parquet}, which every run already writes, so runs 1--2b need no code change; a \texttt{subgroup\_\allowbreak{}flip\_\allowbreak{}budget} column is added to \texttt{classes.\allowbreak{}parquet} (test-first) before run 4, computing the same rule on the same series. Both columns are reported wherever one is. Run directories classified before 63af731 (runs 1, 1p, 2, 2b, 2c) carry \texttt{subgroup\_\allowbreak{}flip} only unless re-classified with \texttt{-\allowbreak{}\/\allowbreak{}-\allowbreak{}force}. Implemented at 63af731: \texttt{subgroup\_\allowbreak{}flip(wide,\ win,\ var)} in \texttt{metrics/\allowbreak{}classes.\allowbreak{}py} gives \texttt{subgroup\_\allowbreak{}flip} (\texttt{a}) and \texttt{subgroup\_\allowbreak{}flip\_\allowbreak{}budget} (\texttt{budget}, right after it), and \texttt{robust.\allowbreak{}parquet} and \texttt{robust\_\allowbreak{}r9.\allowbreak{}parquet} carry one row per slice for each, \texttt{label} = the column and \texttt{cls\ =\allowbreak{}\ "flip"} (\texttt{metrics.\allowbreak{}robust.\allowbreak{}robust\_\allowbreak{}flag}).

\subsubsection*{5. Prior knowledge for runs 1, 2 and 2b (disclosed)}

\begin{itemize}
\tightlist
\item
  The spike's registered results under proportional sharing (spike note §4--§7) and the disclosed probes of §1.6 (R5-like strength escalation; gains 20 and 40 oscillating in 2/5 and 1/5; treated shares 0.2--0.5).
\item
  The Build-phase parity runs (\texttt{docs/\allowbreak{}analysis/\allowbreak{}2026-\allowbreak{}10-\allowbreak{}27-\allowbreak{}nudgefield-\allowbreak{}parity.\allowbreak{}md}, six pins under proportional) and the Spark run under priority (\texttt{docs/\allowbreak{}analysis/\allowbreak{}2026-\allowbreak{}10-\allowbreak{}27-\allowbreak{}spark-\allowbreak{}timing.\allowbreak{}md}: S1 field experiment, attenuated and blind\_\allowbreak{}site 1,002/1,002, \texttt{ratio\_\allowbreak{}window} 0.63).
\item
  The targets runs (\texttt{parameters.\allowbreak{}md} §5--§6): T1 holds as a ratio of means under priority with 4 of 10 S1 replicates ≥ 0.8; the quiet/crowded ratio 1.81 / 2.16 / 2.25 at m = 0 / 1 / 2.
\item
  No run of \texttt{S1\_\allowbreak{}maximiser\_\allowbreak{}smoothing} or \texttt{S1\_\allowbreak{}persistent\_\allowbreak{}auto\_\allowbreak{}gain} has been made before this entry; no probe of \texttt{ref\_\allowbreak{}smoothing} \textgreater{} 0 or of gain 10 exists in any run folder or in the assistant's session notes.
\end{itemize}

\subsubsection*{\texorpdfstring{6. Commands (from \texttt{code/\allowbreak{}}, M4, device \texttt{mps})}{6. Commands (from code/, M4, device mps)}}

\begin{Verbatim}[breaklines,breakanywhere,fontsize=\scriptsize]
R="run.n_rep=100 population.n=20000 run.seed_base=20261029 run.device=mps"
O=outputs/explore_20261029
uv run python -m nudgefield scenario=S1_competitor      design=pilot_rollout    $R out_dir=$O/run1/S1_competitor_pilot_rollout
uv run python -m nudgefield scenario=S1_competitor      design=field_experiment $R out_dir=$O/run1/S1_competitor_field_experiment
uv run python -m nudgefield scenario=S2_ally            design=pilot_rollout    $R out_dir=$O/run1/S2_ally_pilot_rollout
uv run python -m nudgefield scenario=S3_bystanders      design=pilot_rollout    $R out_dir=$O/run1/S3_bystanders_pilot_rollout
uv run python -m nudgefield scenario=S1_persistent_auto design=pilot_rollout    $R out_dir=$O/run1/S1_persistent_auto_pilot_rollout
uv run python -m nudgefield scenario=S1_persistent_auto design=field_experiment $R out_dir=$O/run1/S1_persistent_auto_field_experiment
uv run python -m nudgefield scenario=S1_maximiser       design=pilot_rollout    $R out_dir=$O/run1/S1_maximiser_pilot_rollout
uv run python -m nudgefield scenario=S1_maximiser       design=field_experiment $R out_dir=$O/run1/S1_maximiser_field_experiment
uv run python -m nudgefield scenario=S1_maximiser_smoothing  design=pilot_rollout    $R out_dir=$O/run2/S1_maximiser_smoothing_pilot_rollout
uv run python -m nudgefield scenario=S1_maximiser_smoothing  design=field_experiment $R out_dir=$O/run2/S1_maximiser_smoothing_field_experiment
uv run python -m nudgefield scenario=S1_persistent_auto_gain design=pilot_rollout    $R out_dir=$O/run2b/S1_persistent_auto_gain_pilot_rollout
\end{Verbatim}

Cost estimate (M4 mps, timing note §4: 2.5 s per 10 runs at N = 20,000, linear in runs): run 1 about 2,600 runs ≈ 11 min; run 2 2,000 runs ≈ 8 min; run 2b 500 runs ≈ 2 min; about 25 min in all, plus the S3 m = 16 batch's extra actors.

\subsubsection*{7. Ledger}

{\def\LTcaptype{none} 
\begin{longtable}[]{@{}
  >{\raggedright\arraybackslash}p{(\linewidth - 6\tabcolsep) * \real{0.1244}}
  >{\raggedright\arraybackslash}p{(\linewidth - 6\tabcolsep) * \real{0.5974}}
  >{\raggedright\arraybackslash}p{(\linewidth - 6\tabcolsep) * \real{0.1338}}
  >{\raggedright\arraybackslash}p{(\linewidth - 6\tabcolsep) * \real{0.1244}}@{}}
\toprule\noalign{}
\begin{minipage}[b]{\linewidth}\raggedright
Date
\end{minipage} & \begin{minipage}[b]{\linewidth}\raggedright
Entry
\end{minipage} & \begin{minipage}[b]{\linewidth}\raggedright
Commit
\end{minipage} & \begin{minipage}[b]{\linewidth}\raggedright
Status
\end{minipage} \\
\midrule\noalign{}
\endhead
\bottomrule\noalign{}
\endlastfoot
2026-10-29 & §0--§6: runs 1, 2, 2b registered; flip rule §4 & 316abf0 & registered \\
2026-10-29 & author signed off (``spark is busy now so we can run here''); runs 1, 2, 2b started on the M4 (mps) in the §6 order & d8b0118 & run \\
2026-10-29 & runs 1, 2, 2b complete (eleven calls, 23 min, every manifest at d8b0118 clean, no chunking); results in \texttt{docs/\allowbreak{}analysis/\allowbreak{}2026-\allowbreak{}10-\allowbreak{}29-\allowbreak{}explore-\allowbreak{}runs-\allowbreak{}1-\allowbreak{}2.\allowbreak{}md}; expectations met except: ally ratio 0.46 (not 0.85--0.95), null at m = 16 only, reversal absent under ref\_\allowbreak{}smoothing 0.1 (trigger never fires); F2.yaml updated, author to ratify & this commit & done \\
2026-10-29 & §8: run 1p registered (the ten calls of runs 1 and 2 under \texttt{run.\allowbreak{}attention=\allowbreak{}proportional}, output \texttt{explore\_\allowbreak{}20261029/\allowbreak{}run1p/\allowbreak{}}); awaits the author's sign-off & f3329d2 (amended 1565a75) & registered \\
2026-10-29 & §9: run 2c registered (\texttt{S1\_\allowbreak{}maximiser\_\allowbreak{}memory}, λ 0.02--0.16 × τ\_\allowbreak{}auto \{0.3, 0.5\}, both designs, output \texttt{explore\_\allowbreak{}20261029/\allowbreak{}run2c/\allowbreak{}}); awaits the author's sign-off & f3329d2 (amended 1565a75) & registered \\
2026-10-29 & author signed off §8 and §9 as written (the λ = 0.1 slice of 1pp/1pq reported descriptively, no expectation registered); runs 1p and 2c start on the M4 (mps) & fc1d622 & run \\
2026-10-29 & runs 1p and 2c complete (twelve calls, 18 min, every manifest at fc1d622 clean); results in \texttt{docs/\allowbreak{}analysis/\allowbreak{}2026-\allowbreak{}10-\allowbreak{}29-\allowbreak{}explore-\allowbreak{}runs-\allowbreak{}1p-\allowbreak{}2c.\allowbreak{}md}; table v1 in \texttt{docs/\allowbreak{}analysis/\allowbreak{}2026-\allowbreak{}10-\allowbreak{}29-\allowbreak{}rule-\allowbreak{}independence-\allowbreak{}v1.\allowbreak{}md}; the reversal's boundary lies between λ 0.04 (reversed robust) and 0.08 (attenuated robust) at both τ\_\allowbreak{}auto, one step below the predicted 0.06--0.10 (reversed not robust at 0.06: 68 / 69 of 100), and the field experiment's blind\_\allowbreak{}site is robust at no λ; 7 of 8 claims (19 of 21 points) rule-independent, null rule-dependent in class, the persistent-auto site class at τ\_\allowbreak{}auto 0.3 rule-independent against its registered expectation; F2.yaml unchanged & f476b2e & analysed \\
2026-10-30 & author ratified the R9 re-classification diff table \texttt{docs/\allowbreak{}analysis/\allowbreak{}2026-\allowbreak{}10-\allowbreak{}29-\allowbreak{}r9-\allowbreak{}reclassification.\allowbreak{}md} as is (spec §4.6 step 1; 372 of 6,782 replicates, the pre-computed set); \texttt{F2.\allowbreak{}yaml} as\_\allowbreak{}expected entry held until run 3's thermostat half, delayed note restated (``not exhibited so far; run 5 is the registered search''); reversal text = condition with measured boundary; Spark-half design §6 rulings 1--4 (\texttt{docs/\allowbreak{}superpowers/\allowbreak{}specs/\allowbreak{}2026-\allowbreak{}10-\allowbreak{}30-\allowbreak{}modelling-\allowbreak{}solidity-\allowbreak{}spark-\allowbreak{}design.\allowbreak{}md}) & this commit & ratified \\
2026-10-30 & §10: run 3 registered (\texttt{S1\_\allowbreak{}headroom\_\allowbreak{}thermostat}, \texttt{S1\_\allowbreak{}headroom\_\allowbreak{}maximiser}, \texttt{S1\_\allowbreak{}headroom\_\allowbreak{}trigger}; tau 0.20--0.60 × τ\_\allowbreak{}auto 0--0.50, pilot → rollout; five calls, \texttt{run.\allowbreak{}max\_\allowbreak{}batch\_\allowbreak{}runs} 2,000, Spark cuda; output \texttt{explore\_\allowbreak{}20261029/\allowbreak{}run3/\allowbreak{}}); awaits the author's sign-off & this commit & registered \\
2026-10-30 & §11: run 4 registered (\texttt{crowded\_\allowbreak{}load}; budget 0.35--4.0 × m 0--16, both designs, both rules; four calls, \texttt{run.\allowbreak{}max\_\allowbreak{}batch\_\allowbreak{}runs} 800 = one m-batch, so unchunked; Spark cuda; output \texttt{explore\_\allowbreak{}20261029/\allowbreak{}run4/\allowbreak{}}); awaits the author's sign-off & this commit & registered \\
2026-10-30 & §12: run 5 registered (\texttt{S2\_\allowbreak{}ally\_\allowbreak{}cycles}; ally gain 0.5--40 × lag \{1, 2, 5, 10\} by override, both rules; eight calls, M4 mps; output \texttt{explore\_\allowbreak{}20261029/\allowbreak{}run5/\allowbreak{}}); awaits the author's sign-off & this commit & registered \\
2026-10-30 & §13: run 6 registered (\texttt{S1\_\allowbreak{}treated\_\allowbreak{}share}; p 0.1--0.6, field experiment, both rules; two calls, M4 mps; output \texttt{explore\_\allowbreak{}20261029/\allowbreak{}run6/\allowbreak{}}); awaits the author's sign-off & this commit & registered \\
2026-10-30 & author signed off §10, §12 and §13 as written and §11 with three ``count against'' clauses removed (each entry put singly with its pros and cons); runs 3 and 4 start on the Spark (cuda, tmux), runs 5 and 6 on the M4 (mps), at this commit & this commit & run \\
2026-10-30 & runs 3--6 complete (nineteen calls; no resume, no burn-in stop; every manifest \texttt{status} complete, \texttt{burn\_\allowbreak{}in\_\allowbreak{}settled} true; manifests at 181c143 (docs-only commit on the sign-off commit 65914f2; no code change between them: HANDOFF.md and the daily note only), \texttt{git\_\allowbreak{}dirty} false). Spark (cuda, one tmux script, sequential, 3 h 40 min; \texttt{elapsed\_\allowbreak{}s} / peak memory \texttt{torch.\allowbreak{}cuda.\allowbreak{}max\_\allowbreak{}memory\_\allowbreak{}allocated}): 3a 1,759 s / 7.04 GiB, 3b 258 s / 4.28 GiB, 3c 1,764 s / 7.04 GiB, 3d 258 s / 4.28 GiB, 3e 1,765 s / 7.04 GiB (four chunks 2,000/2,000/2,000/400 each); 4a 3,110 s / 19.36 GiB, 4b 532 s / 7.36 GiB, 4c 3,122 s / 19.35 GiB, 4d 594 s / 7.36 GiB (eight unchunked m-batches of 800 each; the m = 16 priority batch at 20.79 GB against the predicted 20.75 GB, no fallback). M4 (mps, 35 min): 5a--5d 448 / 445 / 445 / 445 s, 5e--5h 36 s each, 6a 133 s, 6b 30 s. Manifests tracked (runs 5 and 6 in f0a1384, runs 3 and 4 here); parquet git-ignored, the Spark files copied to the same paths with SHA-256 equal on both machines & this commit & done \\
2026-10-30 & run 7 (\texttt{targets} on runs 3--6; run 6's T1 broadcast from run 1 / 1p S1 gain 2 by \texttt{-\allowbreak{}\/\allowbreak{}-\allowbreak{}twin-\allowbreak{}cell}, faf42c5), rule-independence table v2 (\texttt{docs/\allowbreak{}analysis/\allowbreak{}2026-\allowbreak{}10-\allowbreak{}30-\allowbreak{}rule-\allowbreak{}independence-\allowbreak{}v2.\allowbreak{}md}: 82 rows, 28 rule-independent, 48 rule-dependent in class, 6 in size; v1's rows unchanged), not-found list (\texttt{docs/\allowbreak{}analysis/\allowbreak{}2026-\allowbreak{}10-\allowbreak{}30-\allowbreak{}not-\allowbreak{}found.\allowbreak{}md}: oscillating, delayed, blind\_\allowbreak{}trial, reversed\_\allowbreak{}site robust nowhere) and the results note \texttt{docs/\allowbreak{}analysis/\allowbreak{}2026-\allowbreak{}10-\allowbreak{}30-\allowbreak{}explore-\allowbreak{}runs-\allowbreak{}3-\allowbreak{}6.\allowbreak{}md} produced (eight deviations, §1; F2.yaml unchanged: no robust as\_\allowbreak{}expected cell of 3a inside the constraint region) & this commit & analysed \\
2026-10-31 & T1 probe (\texttt{docs/\allowbreak{}analysis/\allowbreak{}2026-\allowbreak{}10-\allowbreak{}31-\allowbreak{}t1-\allowbreak{}probe-\allowbreak{}two-\allowbreak{}actor-\allowbreak{}field.\allowbreak{}md}): the crowded family's thermostats saturate in burn-in, reaction ≈ 0, T1 ≈ 1 by construction, T2's 2.25 static; ten-replicate feasibility probes (note §7, disclosed): competitor + bystanders keeps headroom only to m ≈ 2, a stronger competitor erases. Author chose the calibration map as a feasibility study; \textbf{run 8 registered (§14), scenario \texttt{crowded\_\allowbreak{}headroom}; stops for the author's sign-off}; engineering item \texttt{nudgefield.\allowbreak{}levels} before the run & this commit & signed off 2026-10-31 as written; levels tool next \\
2026-10-31 & run 8 complete (six calls, 9,800 runs, plus the levels re-runs of 8c, 8d, 8q, 8qp; Modal, one T4 container per call, \texttt{run.\allowbreak{}device} cuda, the §14 device deviation; every manifest at 59e8b74, \texttt{status} complete, \texttt{git\_\allowbreak{}dirty} false; \texttt{elapsed\_\allowbreak{}s} 8a 2,077 s, 8b 211 s, 8c 2,072 s, 8d 249 s, 8q 30 s, 8qp 8 s; launched 15:42, exit 16:53 local). Results in \texttt{docs/\allowbreak{}analysis/\allowbreak{}2026-\allowbreak{}10-\allowbreak{}31-\allowbreak{}run8-\allowbreak{}calibration-\allowbreak{}map.\allowbreak{}md} (\texttt{code/\allowbreak{}scripts/\allowbreak{}calibration\_\allowbreak{}map.\allowbreak{}py}): \textbf{a joint cell exists under priority}: (tau 0.35, m 2) and (0.35, 3) meet criterion 1--5; calibrated point (0.35, 2), T2 1.508 at the band's lower edge (s.e. 0.022), T1 0.640, headroom 0.735, attenuated 100; proportional: (0.40, 2) and (0.45, 2) meet 1--4, no joint cell (descriptive). Decision rule: first branch (T2 calibration of \texttt{crowded\_\allowbreak{}headroom} at (0.35, 2); run 4h its own entry; \texttt{crowded.\allowbreak{}yaml} stays as the stalemate case), not yet taken & 59e8b74 & analysed \\
2026-10-31 (evening) & run 8 result applied per §14's decision rule: the headroom family \texttt{crowded\_\allowbreak{}headroom} at competitor tau 0.35 is calibrated at m = 2 (T2 1.508, band edge; m 3: 1.733), the stalemate family's calibration stays; \textbf{run 4h registered (§15), scenario \texttt{crowded\_\allowbreak{}headroom\_\allowbreak{}load}, signed off by the author (rulings 1--4 in \texttt{docs/\allowbreak{}superpowers/\allowbreak{}specs/\allowbreak{}2026-\allowbreak{}10-\allowbreak{}31-\allowbreak{}run4h-\allowbreak{}draft-\allowbreak{}entry.\allowbreak{}md})}; runs on the Spark next & this commit & signed off; Spark run next \\
2026-10-31 (night) & \textbf{run 4h complete} (four calls, 28,800 runs, the Spark GB10 cuda, 17:36--19:43 local, 7,641 s; every manifest at ce1107d complete, clean, burn-in settled; GPU max 77 °C, guard never waited; \texttt{targets} on the four). Results in \texttt{docs/\allowbreak{}analysis/\allowbreak{}2026-\allowbreak{}10-\allowbreak{}31-\allowbreak{}run4h-\allowbreak{}headroom-\allowbreak{}load.\allowbreak{}md} (\texttt{code/\allowbreak{}scripts/\allowbreak{}headroom\_\allowbreak{}load\_\allowbreak{}map.\allowbreak{}py}): the two contours in the registered order under both rules (saturation at m 0, 1, 3, 6, 12, 16, --, -- and null at 12, 12, 12, 16, 16, --, --, -- per budget under priority; 0, 1, 2, 3, 6, 8, 12, 16 and 6, 6, 6, 8, 12, 12, --, -- under proportional); T1 region 37 / 72 (priority, every cell interior) and 26 / 72 (proportional); classes inside attenuated and transient under both rules → brief §6 rule met: \textbf{F3b from run 4h; run 4 to Extended Data}; blind\_\allowbreak{}site robust at budget 4.0 in the field experiment (both rules); four expectation deviations (T1 also holds at (1.0, 4); the T1 region empty at budgets 0.35--0.5; blind\_\allowbreak{}site at m 0 for budgets 0.7--1.0; count-against item 3 in one proportional cell, (1.0, 16), null at the cap with T1 0.732 s.e. 0.136); table v2 +32 rows (114; old rows unchanged) & this commit & analysed; F3b decision taken \\
2026-11-01 & §16: run 9 registered (entry ticket, clean; \texttt{T4\_\allowbreak{}POINTS} 15) and signed off by the author as written & 0639070 & registered \\
2026-11-01 & §16: command corrected (\texttt{NUDGEFIELD\_\allowbreak{}TARGETS\_\allowbreak{}OUT}, absolute path); the first execution's deviation disclosed & c220bcf & corrected \\
2026-11-01 & §16: run 9 of record (M4 mps, 20:00--20:03; 33 manifests clean; 9 passed; every expectation met; tables, levels, manifests, logs committed under \texttt{outputs/\allowbreak{}targets\_\allowbreak{}20261101\_\allowbreak{}200002/\allowbreak{}}) & 8eab24e & run; feeds ED3 \\
2026-11-01 & §17: run 10 drafted --- fine T4 grid near the peak (9 levels, take-up 0.30--0.50) for sign-off & a913951 & drafted \\
2026-11-01 & §17: run 10 signed off by the author as written (fine T4 grid, 9 levels 0.30--0.50) & 4b1b94e & registered \\
2026-11-01 & §17: run 10 of record (M4 mps, 22:36--22:53; 42 manifests clean; 10 passed; coarse tables = run 9's; fine peak at take-up 0.376, 0.047 from the oracle: inside the tolerance, not within one grid step --- deviations stated, see the amendment rows) & dddc1dd (artefacts), 6589748 (run line) & run; feeds ED3 \\
2026-11-01 & §17: reading paragraph amended (plateau → ED3's computed one-s.e. range 0.356--0.414, within 0.0006 over 0.351--0.426; dated marker) & d9b93e6, 0bcec5a & amended \\
2026-11-01 & §17: run line amended after the final whole-branch review --- the \texttt{ate\_\allowbreak{}sd} expectation is met only near the peak (the four outer neighbouring differences exceed it): a second deviation, stated; timing clause; parameters.md §7 aligned & 63f16a0 & amended \\
2026-11-02 & §17: first neighbouring difference printed 0.0020, the table gives 0.00195 → 0.0019 (scoped re-review, minor); classification unchanged & 66458d5 & corrected \\
\end{longtable}
}

\subsubsection*{8. Run 1p: the registered points under proportional sharing (registered 2026-10-29; feeds the rule-independence table v1)}

Purpose: the modelling-solidity spec §4.2 (G2, first half). The eight calls of run 1 and the two calls of run 2, each under \texttt{run.\allowbreak{}attention=\allowbreak{}proportional} and otherwise identical, so that every registered claim (the F2 entries, the ally point, the persistent-auto site class at τ\_\allowbreak{}auto 0.3, and the F2 reversed point \texttt{S1\_\allowbreak{}maximiser\_\allowbreak{}smoothing} λ = 0 at τ\_\allowbreak{}auto 0.3--0.5) has its twin under the other attention rule. No scenario file changes. The rule-independence table v1 (\texttt{python\ -\allowbreak{}m\ nudgefield.\allowbreak{}reports.\allowbreak{}rule\_\allowbreak{}independence}, claims in \texttt{code/\allowbreak{}conf/\allowbreak{}nudgefield/\allowbreak{}reports/\allowbreak{}rule\_\allowbreak{}independence.\allowbreak{}yaml}, whose rule-B paths are the output directories below) reads these directories as rule B and runs 1--2 as rule A; it is built after the run, not part of this entry's commands.

Settings: §0 throughout (\texttt{run.\allowbreak{}n\_\allowbreak{}rep} 100, \texttt{population.\allowbreak{}n} 20,000, \texttt{run.\allowbreak{}seed\_\allowbreak{}base} 20261029, \texttt{run.\allowbreak{}max\_\allowbreak{}batch\_\allowbreak{}runs} 0, the same windows and design parameters) except \texttt{run.\allowbreak{}attention} = proportional. The batch seed does not contain the attention rule (\texttt{scenario\textbar{}design\textbar{}structure\textbar{}seed\_\allowbreak{}base}), so each call's batch seeds are those of its run 1 / run 2 twin. Class rules: \texttt{run.\allowbreak{}class\_\allowbreak{}rules} r9 (the package default since 9466eed; spike note §1.6 R9), with \texttt{cls\_\allowbreak{}v1} recorded beside \texttt{cls}; the twins of runs 1--2 are compared through their R9 sidecars (\texttt{classes\_\allowbreak{}r9.\allowbreak{}parquet}, \texttt{robust\_\allowbreak{}r9.\allowbreak{}parquet}). Code: the clean commit at which the run starts, recorded in every \texttt{manifest.\allowbreak{}json}.

{\def\LTcaptype{none} 
\begin{longtable}[]{@{}
  >{\raggedright\arraybackslash}p{(\linewidth - 10\tabcolsep) * \real{0.1245}}
  >{\raggedright\arraybackslash}p{(\linewidth - 10\tabcolsep) * \real{0.1303}}
  >{\raggedright\arraybackslash}p{(\linewidth - 10\tabcolsep) * \real{0.1245}}
  >{\raggedright\arraybackslash}p{(\linewidth - 10\tabcolsep) * \real{0.2215}}
  >{\raggedright\arraybackslash}p{(\linewidth - 10\tabcolsep) * \real{0.1245}}
  >{\raggedright\arraybackslash}p{(\linewidth - 10\tabcolsep) * \real{0.2547}}@{}}
\toprule\noalign{}
\begin{minipage}[b]{\linewidth}\raggedright
Call
\end{minipage} & \begin{minipage}[b]{\linewidth}\raggedright
Scenario file
\end{minipage} & \begin{minipage}[b]{\linewidth}\raggedright
Design
\end{minipage} & \begin{minipage}[b]{\linewidth}\raggedright
Swept values
\end{minipage} & \begin{minipage}[b]{\linewidth}\raggedright
Runs
\end{minipage} & \begin{minipage}[b]{\linewidth}\raggedright
Output (\texttt{code/\allowbreak{}outputs/\allowbreak{}explore\_\allowbreak{}20261029/\allowbreak{}run1p/\allowbreak{}…})
\end{minipage} \\
\midrule\noalign{}
\endhead
\bottomrule\noalign{}
\endlastfoot
1pa & \texttt{S1\_\allowbreak{}competitor} & pilot\_\allowbreak{}rollout & \texttt{competitor.\allowbreak{}gain} \{0.5, 2, 5\} & 300 & \texttt{S1\_\allowbreak{}competitor\_\allowbreak{}pilot\_\allowbreak{}rollout} \\
1pb & \texttt{S1\_\allowbreak{}competitor} & field\_\allowbreak{}experiment & \texttt{competitor.\allowbreak{}gain} \{0.5, 2, 5\} & 300 & \texttt{S1\_\allowbreak{}competitor\_\allowbreak{}field\_\allowbreak{}experiment} \\
1pc & \texttt{S2\_\allowbreak{}ally} & pilot\_\allowbreak{}rollout & \texttt{ally.\allowbreak{}gain} \{0.5, 2, 5\} & 300 & \texttt{S2\_\allowbreak{}ally\_\allowbreak{}pilot\_\allowbreak{}rollout} \\
1pd & \texttt{S3\_\allowbreak{}bystanders} & pilot\_\allowbreak{}rollout & \texttt{bystander.\allowbreak{}count} \{0, 2, 4, 8, 16\} & 5 batches × 100 & \texttt{S3\_\allowbreak{}bystanders\_\allowbreak{}pilot\_\allowbreak{}rollout} \\
1pe & \texttt{S1\_\allowbreak{}persistent\_\allowbreak{}auto} & pilot\_\allowbreak{}rollout & \texttt{competitor.\allowbreak{}escalation\_\allowbreak{}tau} \{0.1, 0.2, 0.3\} & 300 & \texttt{S1\_\allowbreak{}persistent\_\allowbreak{}auto\_\allowbreak{}pilot\_\allowbreak{}rollout} \\
1pf & \texttt{S1\_\allowbreak{}persistent\_\allowbreak{}auto} & field\_\allowbreak{}experiment & \texttt{competitor.\allowbreak{}escalation\_\allowbreak{}tau} \{0.1, 0.2, 0.3\} & 300 & \texttt{S1\_\allowbreak{}persistent\_\allowbreak{}auto\_\allowbreak{}field\_\allowbreak{}experiment} \\
1pg & \texttt{S1\_\allowbreak{}maximiser} & pilot\_\allowbreak{}rollout & \texttt{competitor.\allowbreak{}escalation\_\allowbreak{}tau} \{0.1, 0.2, 0.3\} & 300 & \texttt{S1\_\allowbreak{}maximiser\_\allowbreak{}pilot\_\allowbreak{}rollout} \\
1ph & \texttt{S1\_\allowbreak{}maximiser} & field\_\allowbreak{}experiment & \texttt{competitor.\allowbreak{}escalation\_\allowbreak{}tau} \{0.1, 0.2, 0.3\} & 300 & \texttt{S1\_\allowbreak{}maximiser\_\allowbreak{}field\_\allowbreak{}experiment} \\
1pp & \texttt{S1\_\allowbreak{}maximiser\_\allowbreak{}smoothing} & pilot\_\allowbreak{}rollout & \texttt{escalation\_\allowbreak{}tau} 0.1 \ldots{} 0.5 × \texttt{ref\_\allowbreak{}smoothing} \{0, 0.1\} & 1,000 & \texttt{S1\_\allowbreak{}maximiser\_\allowbreak{}smoothing\_\allowbreak{}pilot\_\allowbreak{}rollout} \\
1pq & \texttt{S1\_\allowbreak{}maximiser\_\allowbreak{}smoothing} & field\_\allowbreak{}experiment & same & 1,000 & \texttt{S1\_\allowbreak{}maximiser\_\allowbreak{}smoothing\_\allowbreak{}field\_\allowbreak{}experiment} \\
\end{longtable}
}

4,600 runs; about 18 minutes on the M4 (spec §4.2). Range check: as in §1 and §2 (the grids are unchanged); the restated robustness rule applies as written.

\textbf{Expected classes, stated before running.} Spec §4.2, verbatim:

\begin{quote}
Registered with the expectation from the spike's results under proportional (n\_\allowbreak{}rep 10) and the parity pins: S1 attenuated at 0.56 (priority 0.63: a size difference under the 0.1 threshold); ally 0.45--0.47 (same as priority); null at m = 8 and 16 (priority: 16 only: a robustness difference); transient at every τ\_\allowbreak{}auto; reversed at 0.3 and, by extension, 0.4--0.5 under the ratchet; blind site at every S1 and S1max field-experiment point; the persistent-auto field experiment at τ 0.3 split (spike: blind\_\allowbreak{}site 5, reversed\_\allowbreak{}site 4, attenuated 1 of 10; priority: 25 / 75), so the site class at that point is expected to be rule-dependent.
\end{quote}

Per call. Each point's expectation has one of three sources, named in its cell: \emph{spec} = the quoted §4.2 sentence (every point it names); \emph{§1 / §2} = the expectation registered in this file for runs 1--2 (registered for priority, stated from the spike's proportional results at n\_\allowbreak{}rep 10), used where the spec sentence names no class (1pd m = 0, 2, 4; 1pf τ\_\allowbreak{}auto 0.1, 0.2; 1pg τ\_\allowbreak{}auto 0.1, 0.2; 1pp λ = 0 at τ\_\allowbreak{}auto 0.1, 0.2); \emph{none} = no expectation registered under proportional: the λ = 0.1 slice of 1pp and 1pq (10 of the 46 points). §2's λ = 0.1 expectation (the same classes as λ = 0) was falsified under priority by run 2, so it is not carried over; these 10 points are reported descriptively, outside the rule-independence table. The spec's ``blind site at every S1max field-experiment point'' does not cover 1pq at λ = 0.1 under this reading. The author may replace any \emph{§1 / §2} or \emph{none} cell with a stated expectation at sign-off. One cell carries a derived inference, marked as such (1pc). Rule A is the run 1 / run 2 result under priority, R9 classes (\texttt{docs/\allowbreak{}analysis/\allowbreak{}2026-\allowbreak{}10-\allowbreak{}29-\allowbreak{}explore-\allowbreak{}runs-\allowbreak{}1-\allowbreak{}2.\allowbreak{}md}, \texttt{docs/\allowbreak{}analysis/\allowbreak{}2026-\allowbreak{}10-\allowbreak{}29-\allowbreak{}r9-\allowbreak{}reclassification.\allowbreak{}md}).

{\def\LTcaptype{none} 
\begin{longtable}[]{@{}
  >{\raggedright\arraybackslash}p{(\linewidth - 4\tabcolsep) * \real{0.1245}}
  >{\raggedright\arraybackslash}p{(\linewidth - 4\tabcolsep) * \real{0.5706}}
  >{\raggedright\arraybackslash}p{(\linewidth - 4\tabcolsep) * \real{0.2849}}@{}}
\toprule\noalign{}
\begin{minipage}[b]{\linewidth}\raggedright
Call
\end{minipage} & \begin{minipage}[b]{\linewidth}\raggedright
Expected under proportional
\end{minipage} & \begin{minipage}[b]{\linewidth}\raggedright
Rule A (priority) result
\end{minipage} \\
\midrule\noalign{}
\endhead
\bottomrule\noalign{}
\endlastfoot
1pa & attenuated at every k, ratio about 0.56 (spec) & attenuated 100 / 100 / 100, ratio 0.62--0.63 \\
1pb & \texttt{cls\_\allowbreak{}combined} blind\_\allowbreak{}site at every k (spec) & blind\_\allowbreak{}site 100 / 98 / 100 \\
1pc & ratio 0.45--0.47 at every k (spec); class attenuated derived under R9, not stated by the spec (window mean 0.46 outside {[}0.75, 1.25{]}, end ratio ≤ 0.75) & attenuated 96 / 99 / 97, ratio 0.46 \\
1pd & null at m = 8 and 16 (spec); m = 0, 2, 4 not named by the spec: as expected at m = 0 by construction, attenuated at m = 2, 4 (§1, 1d, the spike under proportional) & as\_\allowbreak{}expected 100; attenuated 100; attenuated 100; attenuated 100; null 100 \\
1pe & transient at every τ\_\allowbreak{}auto (spec) & transient 100 / 100 / 100 \\
1pf & τ\_\allowbreak{}auto 0.3: split between blind\_\allowbreak{}site and reversed\_\allowbreak{}site, site class expected rule-dependent (spec); 0.1, 0.2 not named by the spec: blind\_\allowbreak{}site (§1, 1f) & \texttt{cls\_\allowbreak{}combined} blind\_\allowbreak{}site 99 / 93 / 25, reversed\_\allowbreak{}site 0 / 6 / 75 \\
1pg & reversed at 0.3 (spec); 0.1, 0.2 not named by the spec: 0.1 transient or null, 0.2 transient (§1, 1g) & transient 100 / 100 / 1, reversed 0 / 0 / 99 \\
1ph & \texttt{cls\_\allowbreak{}combined} blind\_\allowbreak{}site at every τ\_\allowbreak{}auto (spec) & blind\_\allowbreak{}site 98 / 97 / 99 \\
1pp & λ = 0: reversed at 0.3 and, by extension, 0.4--0.5 (spec); 0.1 null or transient, 0.2 transient (§2). λ = 0.1: none (no expectation registered under proportional; §2's λ = 0.1 expectation was falsified under priority by run 2; reported descriptively, outside the rule-independence table) & λ = 0: transient 100 / 100, reversed 99 / 100 / 100; λ = 0.1: attenuated 100 at every τ\_\allowbreak{}auto, ratio 0.61, automatic nudge never triggers \\
1pq & λ = 0: \texttt{cls\_\allowbreak{}combined} blind\_\allowbreak{}site at every τ\_\allowbreak{}auto (spec: ``every S1max field-experiment point''). λ = 0.1: none (as 1pp λ = 0.1; the spec's ``every S1max field-experiment point'' does not cover it) & \texttt{cls\_\allowbreak{}combined}: λ = 0: blind\_\allowbreak{}site 97 / 99 / 98 / 97 / 96; λ = 0.1: as\_\allowbreak{}expected 100 at every τ\_\allowbreak{}auto (R9) \\
\end{longtable}
}

The rule-independence tags (spec §4.2: rule-dependent in class, rule-dependent in size at Δratio \textgreater{} 0.1, rule-independent otherwise; first match wins) are computed by the table script from the directories, not stated here. Tags implied by the quoted sentence: S1 attenuated rule-independent (a size difference under the 0.1 threshold); ally rule-independent; null rule-dependent in class (a robustness difference); persistent-auto site class at τ\_\allowbreak{}auto 0.3 rule-dependent.

What would count against the model's story: a registered F2 class (attenuated, blind site, transient, reversed) that changes majority class or robustness between the rules; the claim is then tagged rule-dependent and the text says so.

\textbf{Prior knowledge (disclosed).} Everything in §5; the results of runs 1, 2 and 2b under priority (\texttt{docs/\allowbreak{}analysis/\allowbreak{}2026-\allowbreak{}10-\allowbreak{}29-\allowbreak{}explore-\allowbreak{}runs-\allowbreak{}1-\allowbreak{}2.\allowbreak{}md}); their R9 re-classification (\texttt{docs/\allowbreak{}analysis/\allowbreak{}2026-\allowbreak{}10-\allowbreak{}29-\allowbreak{}r9-\allowbreak{}reclassification.\allowbreak{}md}); the spike's proportional results at n\_\allowbreak{}rep 10 and the parity pins P1--P6 (proportional). No run of \texttt{S1\_\allowbreak{}maximiser\_\allowbreak{}smoothing} under proportional has been made; no proportional run at n\_\allowbreak{}rep 100 exists for any of the ten calls.

Commands (from \texttt{code/\allowbreak{}}, M4, device \texttt{mps}):

Run the block with \texttt{bash} (paste it into a \texttt{bash} shell, or wrap it as \texttt{bash\ \textless{}\textless{}\textquotesingle{}EOF\textquotesingle{}} \ldots{} \texttt{EOF}): \texttt{\$R} unquoted is word-split by bash but not by zsh, the macOS Terminal default.

\begin{Verbatim}[breaklines,breakanywhere,fontsize=\scriptsize]
R="run.n_rep=100 population.n=20000 run.seed_base=20261029 run.device=mps run.attention=proportional"
O=outputs/explore_20261029/run1p
uv run python -m nudgefield scenario=S1_competitor      design=pilot_rollout    $R out_dir=$O/S1_competitor_pilot_rollout
uv run python -m nudgefield scenario=S1_competitor      design=field_experiment $R out_dir=$O/S1_competitor_field_experiment
uv run python -m nudgefield scenario=S2_ally            design=pilot_rollout    $R out_dir=$O/S2_ally_pilot_rollout
uv run python -m nudgefield scenario=S3_bystanders      design=pilot_rollout    $R out_dir=$O/S3_bystanders_pilot_rollout
uv run python -m nudgefield scenario=S1_persistent_auto design=pilot_rollout    $R out_dir=$O/S1_persistent_auto_pilot_rollout
uv run python -m nudgefield scenario=S1_persistent_auto design=field_experiment $R out_dir=$O/S1_persistent_auto_field_experiment
uv run python -m nudgefield scenario=S1_maximiser       design=pilot_rollout    $R out_dir=$O/S1_maximiser_pilot_rollout
uv run python -m nudgefield scenario=S1_maximiser       design=field_experiment $R out_dir=$O/S1_maximiser_field_experiment
uv run python -m nudgefield scenario=S1_maximiser_smoothing design=pilot_rollout    $R out_dir=$O/S1_maximiser_smoothing_pilot_rollout
uv run python -m nudgefield scenario=S1_maximiser_smoothing design=field_experiment $R out_dir=$O/S1_maximiser_smoothing_field_experiment
\end{Verbatim}

\subsubsection*{9. Run 2c: the reversal's boundary in the competitor's memory (registered 2026-10-29; feeds G1 and the Extended Data panel on the reference rule)}

Purpose: the modelling-solidity spec §4.3 (G1). Measure the memory length of the maximiser's exponential reference at which the reversal stops: the reversal is then stated as a condition with a boundary. This run replaces the λ = 0.1 half of run 2 as the Extended Data panel on the reference rule; run 2 stays as the registered τ\_\allowbreak{}auto robustness run.

Scenario file \texttt{S1\_\allowbreak{}maximiser\_\allowbreak{}memory} (new, this commit): the R7 maximiser of \texttt{S1\_\allowbreak{}maximiser} (maximiser, gain 2, lag 2, reach0 0.60, R6's persistent automatic escalation) with \texttt{competitor.\allowbreak{}escalation\_\allowbreak{}tau} ∈ \{0.3, 0.5\} as the slice and \texttt{competitor.\allowbreak{}ref\_\allowbreak{}smoothing} (λ) ∈ \{0.02, 0.04, 0.06, 0.08, 0.10, 0.12, 0.14, 0.16\} as the robustness axis (the last sweep path); 16 points; both designs; one batch of 1,600 runs per design. Settings: §0 (priority attention, n\_\allowbreak{}rep 100, N 20,000, seed base 20261029, no chunking); class rules R9 (\texttt{run.\allowbreak{}class\_\allowbreak{}rules} r9, \texttt{cls\_\allowbreak{}v1} recorded beside).

{\def\LTcaptype{none} 
\begin{longtable}[]{@{}
  >{\raggedright\arraybackslash}p{(\linewidth - 10\tabcolsep) * \real{0.1244}}
  >{\raggedright\arraybackslash}p{(\linewidth - 10\tabcolsep) * \real{0.1244}}
  >{\raggedright\arraybackslash}p{(\linewidth - 10\tabcolsep) * \real{0.2125}}
  >{\raggedright\arraybackslash}p{(\linewidth - 10\tabcolsep) * \real{0.1244}}
  >{\raggedright\arraybackslash}p{(\linewidth - 10\tabcolsep) * \real{0.1244}}
  >{\raggedright\arraybackslash}p{(\linewidth - 10\tabcolsep) * \real{0.2700}}@{}}
\toprule\noalign{}
\begin{minipage}[b]{\linewidth}\raggedright
Call
\end{minipage} & \begin{minipage}[b]{\linewidth}\raggedright
Design
\end{minipage} & \begin{minipage}[b]{\linewidth}\raggedright
Robustness axis
\end{minipage} & \begin{minipage}[b]{\linewidth}\raggedright
Slice
\end{minipage} & \begin{minipage}[b]{\linewidth}\raggedright
Runs
\end{minipage} & \begin{minipage}[b]{\linewidth}\raggedright
Output (\texttt{code/\allowbreak{}outputs/\allowbreak{}explore\_\allowbreak{}20261029/\allowbreak{}run2c/\allowbreak{}…})
\end{minipage} \\
\midrule\noalign{}
\endhead
\bottomrule\noalign{}
\endlastfoot
2cp & pilot\_\allowbreak{}rollout & \texttt{ref\_\allowbreak{}smoothing} 0.02 \ldots{} 0.16, step 0.02 (memory 1/λ from 50 to about 6 periods) & \texttt{escalation\_\allowbreak{}tau} 0.3, 0.5 & 1,600 & \texttt{S1\_\allowbreak{}maximiser\_\allowbreak{}memory\_\allowbreak{}pilot\_\allowbreak{}rollout} \\
2cf & field\_\allowbreak{}experiment & same & same & 1,600 & \texttt{S1\_\allowbreak{}maximiser\_\allowbreak{}memory\_\allowbreak{}field\_\allowbreak{}experiment} \\
\end{longtable}
}

About 7 minutes per design on the M4 (spec §4.3). Range check: eight evenly spaced points; two adjacent points span 0.02 / 0.14 = 14\% of the range, so the registered robustness rule (\texttt{run.\allowbreak{}robust\_\allowbreak{}min\_\allowbreak{}points} 2, \texttt{run.\allowbreak{}robust\_\allowbreak{}min\_\allowbreak{}range\_\allowbreak{}share} 0.10) applies as written: a class is robust over any two or more adjacent λ at ≥ 90\% of replicates each. The grid is the author's ruling 2 of 2026-10-29 (spec §6); the earlier grid \{0, 0.01, 0.02, 0.05, 0.1, 0.2, 0.5\} could not make any boundary robust under the 10\% range share (spec §4.3).

The reference case, spec §4.3 verbatim:

\begin{quote}
The ratchet (λ = 0) is a different update rule (\texttt{ref\ ←\ min(ref,\ r)}, a running minimum, not the λ → 0 limit of the average, which would be a frozen reference) and is not a point on this axis: its results are run 2's λ = 0 slice, reported beside the sweep as the reference case.
\end{quote}

Run 2's λ = 0 slice at the two τ\_\allowbreak{}auto of this run: pilot → rollout reversed 99 of 100 at 0.3 (ratio −0.34) and 100 of 100 at 0.5 (ratio −0.91), automatic nudge on at period 114 in every replicate; field experiment \texttt{cls\_\allowbreak{}combined} blind\_\allowbreak{}site 98 at 0.3 and 96 at 0.5.

\textbf{Expected classes, stated before running.} Spec §4.3, verbatim (expectation and mechanism):

\begin{quote}
Expected, stated before running. The automatic nudge triggers only after ten consecutive periods with the observed rate above \texttt{ref} + 0.01 while reach is at its cap. Under run 2's λ = 0.1 the reach reached its cap at a median period 107 (105 under the ratchet), so the ten-period count ends 15--17 periods after the rollout, and the exponential reference has closed the rollout's gap of about 0.05 to 0.05 (1 − λ)\^{}16, while the competitor's own reach climb narrows it further. Solving 0.05 (1 − λ)\^{}16 = 0.01 gives λ ≈ 0.096, and the climb pushes the boundary lower. Prediction: reversed (robust) at λ ≤ 0.06, attenuated from λ ≥ 0.10, with 0.08 the uncertain point, in pilot → rollout at both τ; the field experiment moves from blind site (small λ) to as expected (large λ, under R9) across the same boundary. The boundary is reported as the largest λ at which reversal is robust and the smallest at which attenuated is robust; the memory length 1/λ is the number the text uses. The actor is then described in the text as one whose reference is a high-water mark (the ratchet) or a long memory, with the boundary as the condition; naming real actors of this kind (sales targets anchored on the best past period, budgets that ratchet) is a writing-phase item for the author.
\end{quote}

Per point, the same prediction:

{\def\LTcaptype{none} 
{\scriptsize
\begin{longtable}[]{@{}
  >{\raggedright\arraybackslash}p{(\linewidth - 16\tabcolsep) * \real{0.1287}}
  >{\raggedright\arraybackslash}p{(\linewidth - 16\tabcolsep) * \real{0.1064}}
  >{\raggedright\arraybackslash}p{(\linewidth - 16\tabcolsep) * \real{0.1064}}
  >{\raggedright\arraybackslash}p{(\linewidth - 16\tabcolsep) * \real{0.1064}}
  >{\raggedright\arraybackslash}p{(\linewidth - 16\tabcolsep) * \real{0.1064}}
  >{\raggedright\arraybackslash}p{(\linewidth - 16\tabcolsep) * \real{0.1064}}
  >{\raggedright\arraybackslash}p{(\linewidth - 16\tabcolsep) * \real{0.1064}}
  >{\raggedright\arraybackslash}p{(\linewidth - 16\tabcolsep) * \real{0.1064}}
  >{\raggedright\arraybackslash}p{(\linewidth - 16\tabcolsep) * \real{0.1064}}@{}}
\toprule\noalign{}
\begin{minipage}[b]{\linewidth}\raggedright
Call
\end{minipage} & \begin{minipage}[b]{\linewidth}\raggedright
λ 0.02
\end{minipage} & \begin{minipage}[b]{\linewidth}\raggedright
0.04
\end{minipage} & \begin{minipage}[b]{\linewidth}\raggedright
0.06
\end{minipage} & \begin{minipage}[b]{\linewidth}\raggedright
0.08
\end{minipage} & \begin{minipage}[b]{\linewidth}\raggedright
0.10
\end{minipage} & \begin{minipage}[b]{\linewidth}\raggedright
0.12
\end{minipage} & \begin{minipage}[b]{\linewidth}\raggedright
0.14
\end{minipage} & \begin{minipage}[b]{\linewidth}\raggedright
0.16
\end{minipage} \\
\midrule\noalign{}
\endhead
\bottomrule\noalign{}
\endlastfoot
2cp, τ\_\allowbreak{}auto 0.3 and 0.5 (\texttt{cls}) & reversed (robust) & reversed (robust) & reversed (robust) & uncertain & attenuated (robust) & attenuated (robust) & attenuated (robust) & attenuated (robust) \\
2cf, τ\_\allowbreak{}auto 0.3 and 0.5 (\texttt{cls\_\allowbreak{}combined}) & blind\_\allowbreak{}site & blind\_\allowbreak{}site & blind\_\allowbreak{}site & uncertain & as\_\allowbreak{}expected & as\_\allowbreak{}expected & as\_\allowbreak{}expected & as\_\allowbreak{}expected \\
\end{longtable}
}
}

Predicted boundary: between λ 0.06 and 0.10 (memory 1/λ between about 17 and 10 periods). Reported: the largest λ at which reversed is robust and the smallest at which attenuated is robust, with the trigger period (\texttt{trigger\_\allowbreak{}free\_\allowbreak{}competitor@auto}, \texttt{classes.\allowbreak{}parquet}) at each point.

What would count against the prediction: reversed robust at λ ≥ 0.10, or not robust at λ ≤ 0.06, at either τ\_\allowbreak{}auto; no single boundary (reversed and attenuated interleaved along λ); in the field experiment (\texttt{cls\_\allowbreak{}combined}), blind\_\allowbreak{}site not robust at λ ≤ 0.06, or no shift to as\_\allowbreak{}expected by λ ≥ 0.10. Any of these is reported as found, and the condition in the text is stated from the run, not from the prediction.

\textbf{Prior knowledge (disclosed).} Run 2 under priority (\texttt{docs/\allowbreak{}analysis/\allowbreak{}2026-\allowbreak{}10-\allowbreak{}29-\allowbreak{}explore-\allowbreak{}runs-\allowbreak{}1-\allowbreak{}2.\allowbreak{}md} §2): with the ratchet, reversed robust over τ\_\allowbreak{}auto 0.3--0.5; with λ = 0.1 the automatic nudge never triggers and every point is attenuated at ratio 0.61 (pilot → rollout) and as expected under R9 in the field experiment. Run 2's λ = 0.1 at τ\_\allowbreak{}auto 0.3 and 0.5 is the same parameter point as this run's λ = 0.10 under a different scenario name, hence different batch seeds. The median periods at which the reach reached its cap (107 under λ = 0.1, 105 under the ratchet) come from run 2 and enter the derivation above. No run with λ ∈ \{0.02, 0.04, 0.06, 0.08, 0.12, 0.14, 0.16\} exists in any run folder; no run of \texttt{S1\_\allowbreak{}maximiser\_\allowbreak{}memory} has been made before this entry.

Commands (from \texttt{code/\allowbreak{}}, M4, device \texttt{mps}):

Run the block with \texttt{bash} (paste it into a \texttt{bash} shell, or wrap it as \texttt{bash\ \textless{}\textless{}\textquotesingle{}EOF\textquotesingle{}} \ldots{} \texttt{EOF}): \texttt{\$R} unquoted is word-split by bash but not by zsh, the macOS Terminal default.

\begin{Verbatim}[breaklines,breakanywhere,fontsize=\scriptsize]
R="run.n_rep=100 population.n=20000 run.seed_base=20261029 run.device=mps"
O=outputs/explore_20261029/run2c
uv run python -m nudgefield scenario=S1_maximiser_memory design=pilot_rollout    $R out_dir=$O/S1_maximiser_memory_pilot_rollout
uv run python -m nudgefield scenario=S1_maximiser_memory design=field_experiment $R out_dir=$O/S1_maximiser_memory_field_experiment
\end{Verbatim}

\subsubsection*{10. Run 3: headroom × persistence, three halves (registered 2026-10-30; feeds F3a, F2's as-expected point, ED4 and the response-rule rows of table v2)}

Purpose: the Spark-half design \texttt{docs/\allowbreak{}superpowers/\allowbreak{}specs/\allowbreak{}2026-\allowbreak{}10-\allowbreak{}30-\allowbreak{}modelling-\allowbreak{}solidity-\allowbreak{}spark-\allowbreak{}design.\allowbreak{}md} §4.4 (G4). Map the competitor's headroom against the persistence of its automatic push, for a rate-targeting competitor (thermostat), a reach-maximising competitor (the R7 maximiser) and the thermostat under a second response rule (threshold trigger), so that the classes along persistence are known in each headroom slice and under both attention rules. Author's rulings of 2026-10-30 (design §6): 1, \texttt{competitor.\allowbreak{}nudge.\allowbreak{}tau} 0.20--0.60 (0.15 dropped: the thermostat saturates in burn-in and its escalation fires before the pilot, design §7 finding 3); 4, threshold-trigger gain 0.05, one value. Ruling (c) of the same sitting: F2's as-expected entry stays unchanged until this run's thermostat half is run.

Scenario files (new, this commit), one grid, pilot → rollout only:

\begin{itemize}
\tightlist
\item
  \texttt{S1\_\allowbreak{}headroom\_\allowbreak{}thermostat}: the S1 competitor (thermostat, gain 2, goal\_\allowbreak{}offset −0.03, lag 2, reach0 0) with the escalation.
\item
  \texttt{S1\_\allowbreak{}headroom\_\allowbreak{}maximiser}: the R7 maximiser (gain 2, lag 2, reach0 0.60, ratchet reference \texttt{ref\_\allowbreak{}smoothing} 0) with the escalation.
\item
  \texttt{S1\_\allowbreak{}headroom\_\allowbreak{}trigger}: the thermostat half with \texttt{rule:\allowbreak{}\ threshold\_\allowbreak{}trigger}, \texttt{gain:\allowbreak{}\ 0.\allowbreak{}05}, \texttt{tolerance} 0.01 (ActorSpec default).
\item
  Grid in each: slice \texttt{competitor.\allowbreak{}nudge.\allowbreak{}tau} ∈ \{0.20, 0.25, 0.30, 0.35, 0.40, 0.45, 0.50, 0.60\}; robustness axis (the last sweep path) \texttt{competitor.\allowbreak{}escalation\_\allowbreak{}tau} ∈ \{0, 0.05, 0.10, 0.15, 0.20, 0.30, 0.40, 0.50\}; \texttt{sweep:\allowbreak{}\ {[}competitor.\allowbreak{}nudge.\allowbreak{}tau,\ competitor.\allowbreak{}escalation\_\allowbreak{}tau{]}}; 64 points, 6,400 runs per call. \texttt{escalation\_\allowbreak{}tau} 0 switches the escalation on at zero strength, so the trigger is still recorded (design §7, confirmed by the builder check).
\end{itemize}

Settings: §0 (\texttt{run.\allowbreak{}n\_\allowbreak{}rep} 100, \texttt{population.\allowbreak{}n} 20,000, \texttt{run.\allowbreak{}seed\_\allowbreak{}base} 20261029, the same windows and design parameters) except: \texttt{run.\allowbreak{}max\_\allowbreak{}batch\_\allowbreak{}runs} 2,000 and \texttt{run.\allowbreak{}device} cuda (the Spark), \texttt{run.\allowbreak{}attention} per call. Chunking: each call is one actor structure, hence one batch of 6,400 runs; at about 4 MB per single-actor run (the M4 manifests, design §3) one batch would be about 25 GB, above the 20 GB chunk ceiling of design §3, so it is split into four chunks of 2,000, 2,000, 2,000 and 400 runs, seeded \texttt{scenario\textbar{}design\textbar{}structure\textbar{}seed\_\allowbreak{}base\textbar{}chunk\textless{}c\textgreater{}} (c = 0\ldots3). This departs from the preamble's \texttt{run.\allowbreak{}max\_\allowbreak{}batch\_\allowbreak{}runs\ =\allowbreak{}\ 0}: the chunk seeds differ from the unchunked batch seed, so the draws are not those of an unchunked call; no earlier run uses these scenario files, so no comparison depends on it. The 4 MB figure is an mps number; the cuda slope was measured for the crowded family only (Task 7, \texttt{docs/\allowbreak{}analysis/\allowbreak{}2026-\allowbreak{}10-\allowbreak{}27-\allowbreak{}spark-\allowbreak{}timing.\allowbreak{}md} §7), so the design's alternative (one batch if the cuda measurement comes in under the mps figure) is not taken. The batch seed does not contain the attention rule, so each proportional call's chunk seeds are those of its priority twin. Class rules R9 (\texttt{run.\allowbreak{}class\_\allowbreak{}rules} r9, \texttt{cls\_\allowbreak{}v1} recorded beside). Code: the clean commit at which the run starts, recorded in every \texttt{manifest.\allowbreak{}json}.

{\def\LTcaptype{none} 
\begin{longtable}[]{@{}
  >{\raggedright\arraybackslash}p{(\linewidth - 10\tabcolsep) * \real{0.1244}}
  >{\raggedright\arraybackslash}p{(\linewidth - 10\tabcolsep) * \real{0.1631}}
  >{\raggedright\arraybackslash}p{(\linewidth - 10\tabcolsep) * \real{0.1244}}
  >{\raggedright\arraybackslash}p{(\linewidth - 10\tabcolsep) * \real{0.1337}}
  >{\raggedright\arraybackslash}p{(\linewidth - 10\tabcolsep) * \real{0.1244}}
  >{\raggedright\arraybackslash}p{(\linewidth - 10\tabcolsep) * \real{0.3100}}@{}}
\toprule\noalign{}
\begin{minipage}[b]{\linewidth}\raggedright
Call
\end{minipage} & \begin{minipage}[b]{\linewidth}\raggedright
Scenario file
\end{minipage} & \begin{minipage}[b]{\linewidth}\raggedright
Design
\end{minipage} & \begin{minipage}[b]{\linewidth}\raggedright
Attention
\end{minipage} & \begin{minipage}[b]{\linewidth}\raggedright
Runs
\end{minipage} & \begin{minipage}[b]{\linewidth}\raggedright
Output (\texttt{code/\allowbreak{}outputs/\allowbreak{}explore\_\allowbreak{}20261029/\allowbreak{}run3/\allowbreak{}…})
\end{minipage} \\
\midrule\noalign{}
\endhead
\bottomrule\noalign{}
\endlastfoot
3a & \texttt{S1\_\allowbreak{}headroom\_\allowbreak{}thermostat} & pilot\_\allowbreak{}rollout & priority & 6,400 (4 chunks) & \texttt{S1\_\allowbreak{}headroom\_\allowbreak{}thermostat\_\allowbreak{}pilot\_\allowbreak{}rollout} \\
3b & \texttt{S1\_\allowbreak{}headroom\_\allowbreak{}thermostat} & pilot\_\allowbreak{}rollout & proportional & 6,400 (4 chunks) & \texttt{S1\_\allowbreak{}headroom\_\allowbreak{}thermostat\_\allowbreak{}pilot\_\allowbreak{}rollout\_\allowbreak{}proportional} \\
3c & \texttt{S1\_\allowbreak{}headroom\_\allowbreak{}maximiser} & pilot\_\allowbreak{}rollout & priority & 6,400 (4 chunks) & \texttt{S1\_\allowbreak{}headroom\_\allowbreak{}maximiser\_\allowbreak{}pilot\_\allowbreak{}rollout} \\
3d & \texttt{S1\_\allowbreak{}headroom\_\allowbreak{}maximiser} & pilot\_\allowbreak{}rollout & proportional & 6,400 (4 chunks) & \texttt{S1\_\allowbreak{}headroom\_\allowbreak{}maximiser\_\allowbreak{}pilot\_\allowbreak{}rollout\_\allowbreak{}proportional} \\
3e & \texttt{S1\_\allowbreak{}headroom\_\allowbreak{}trigger} & pilot\_\allowbreak{}rollout & priority & 6,400 (4 chunks) & \texttt{S1\_\allowbreak{}headroom\_\allowbreak{}trigger\_\allowbreak{}pilot\_\allowbreak{}rollout} \\
\end{longtable}
}

The trigger half runs under priority only: the response-rule leg of table v2 is read against the thermostat half under priority (3a against 3e); its proportional twin would be a fourth half and is not registered (design §4.4). 32,000 runs. Time (design §4.4, from 0.29 s per single-actor run under priority on the Spark and 0.05--0.06 s under proportional): the two priority halves (3a, 3c) about 1.0 h, the trigger half (3e) 0.5 h, the two proportional halves (3b, 3d) 0.2 h; about 1.7 h in all.

Range check: on the robustness axis \texttt{escalation\_\allowbreak{}tau} 0--0.50 the smallest adjacent gap, 0.05, is 10\% of the range, exactly the registered share; \texttt{metrics/\allowbreak{}robust.\allowbreak{}py} compares the span with a tolerance of 10⁻⁹ (\texttt{\_\allowbreak{}EPS}), so any two adjacent points qualify and the restated robustness rule (\texttt{run.\allowbreak{}robust\_\allowbreak{}min\_\allowbreak{}points} 2, \texttt{run.\allowbreak{}robust\_\allowbreak{}min\_\allowbreak{}range\_\allowbreak{}share} 0.10) applies as written in every headroom slice.

\textbf{Expected classes, stated before running.} Design §4.4, verbatim:

\begin{quote}
Thermostat half: at τ\_\allowbreak{}auto 0, as expected or attenuated at tau 0.20 (the competitor at reach 0.9--1 pushes two thirds of what it pushes at 0.30, which puts the ratio near the band's edge of 0.75; which side is the map's answer) and attenuated from tau 0.25 upward (the registered 0.63 at tau 0.30); transient at every tau once τ\_\allowbreak{}auto ≥ 0.10 (registered at tau 0.30; spike §6.6: a rate-targeting actor offsets its own persistent push, so no cell reverses). Maximiser half: attenuated at τ\_\allowbreak{}auto 0 (run 2 under smoothing: 0.61 with no trigger), transient at 0.10--0.20 (run 1), reversed from 0.30 at tau 0.30 (run 1, run 2), reversed from a lower τ\_\allowbreak{}auto as tau grows and from a higher one as tau falls (the automatic push is τ\_\allowbreak{}auto · reach · susceptibility; the reversal needs it to exceed what our nudge adds), with tau 0.20 the column where it may fail to reverse at any τ\_\allowbreak{}auto. Trigger half: the thermostat half's classes cell for cell (table v2's response-rule rows); oscillating is not expected at a step of 0.05 (the bang-bang overshoot is one step, below the oscillation detector's amplitude). The F2 as-expected point is the thermostat half's robust as-expected cell nearest tau 0.30 at τ\_\allowbreak{}auto 0, if one exists (brief §5; author item (c) of the handoff).
\end{quote}

Per call. Sources: \emph{spec} = the quoted §4.4 passage; \emph{none} = the passage names no class for the cell (no expectation registered; reported descriptively; the author may state one at sign-off). The passage names no attention rule; this entry reads it as applying to both rules (3a and 3b alike, 3c and 3d alike). Class \texttt{cls} (R9), pilot → rollout.

{\def\LTcaptype{none} 
\begin{longtable}[]{@{}
  >{\raggedright\arraybackslash}p{(\linewidth - 8\tabcolsep) * \real{0.1338}}
  >{\raggedright\arraybackslash}p{(\linewidth - 8\tabcolsep) * \real{0.4568}}
  >{\raggedright\arraybackslash}p{(\linewidth - 8\tabcolsep) * \real{0.1245}}
  >{\raggedright\arraybackslash}p{(\linewidth - 8\tabcolsep) * \real{0.1245}}
  >{\raggedright\arraybackslash}p{(\linewidth - 8\tabcolsep) * \real{0.1403}}@{}}
\toprule\noalign{}
\begin{minipage}[b]{\linewidth}\raggedright
Call
\end{minipage} & \begin{minipage}[b]{\linewidth}\raggedright
τ\_\allowbreak{}auto 0
\end{minipage} & \begin{minipage}[b]{\linewidth}\raggedright
τ\_\allowbreak{}auto 0.05
\end{minipage} & \begin{minipage}[b]{\linewidth}\raggedright
τ\_\allowbreak{}auto 0.10, 0.15, 0.20
\end{minipage} & \begin{minipage}[b]{\linewidth}\raggedright
τ\_\allowbreak{}auto 0.30, 0.40, 0.50
\end{minipage} \\
\midrule\noalign{}
\endhead
\bottomrule\noalign{}
\endlastfoot
3a, 3b (thermostat) & tau 0.20: as\_\allowbreak{}expected or attenuated; tau 0.25--0.60: attenuated (spec) & none & transient at every tau (spec) & transient at every tau; no cell reversed (spec) \\
3c, 3d (maximiser) & attenuated (spec; the source, run 2's λ = 0.1 slice, is at tau 0.30; read here for every tau, since at strength 0 nothing reverses) & none & tau 0.30: transient (spec, run 1). Other tau: see the onset row & tau 0.30: reversed (spec, runs 1 and 2). Other tau: see the onset row \\
3c, 3d, onset & the smallest τ\_\allowbreak{}auto at which the column is reversed is lower for tau above 0.30 and higher for tau below 0.30; at tau 0.20 the column may not reverse at any τ\_\allowbreak{}auto (spec). No per-cell class is stated off the tau 0.30 column: the passage gives the ordering, not the cells & & & \\
3e (trigger) & the class of 3a in the same cell, cell for cell (spec) & the class of 3a in the same cell (spec) & same (spec) & same (spec) \\
3e, everywhere & oscillating in no cell (spec) & & & \\
\end{longtable}
}

F2's as-expected point (ruling (c)): the robust as\_\allowbreak{}expected cell of 3a nearest tau 0.30 at τ\_\allowbreak{}auto 0, if one exists; if none exists, the fallback is decided by the author after the run (design §6, ruling (c)).

What would count against the expectations: a cell whose robust majority class differs from the one stated, a reversed cell in the thermostat half, a reversal onset in the maximiser half that does not move monotonically with tau, oscillating in 3e, or a 3e cell whose class differs from 3a's. Each is reported as found.

\textbf{Prior knowledge (disclosed).} Everything in §5 and §8; runs 1, 2, 2b, 1p and 2c (\texttt{docs/\allowbreak{}analysis/\allowbreak{}2026-\allowbreak{}10-\allowbreak{}29-\allowbreak{}explore-\allowbreak{}runs-\allowbreak{}1-\allowbreak{}2.\allowbreak{}md}, \texttt{docs/\allowbreak{}analysis/\allowbreak{}2026-\allowbreak{}10-\allowbreak{}29-\allowbreak{}explore-\allowbreak{}runs-\allowbreak{}1p-\allowbreak{}2c.\allowbreak{}md}, \texttt{docs/\allowbreak{}analysis/\allowbreak{}2026-\allowbreak{}10-\allowbreak{}29-\allowbreak{}r9-\allowbreak{}reclassification.\allowbreak{}md}), whose tau 0.30 points lie on this grid under other scenario names (1e / 1pe: the thermostat at τ\_\allowbreak{}auto 0.1--0.3; 1g / 1pg and 2p / 1pp λ = 0: the maximiser at τ\_\allowbreak{}auto 0.1--0.5); the builder check's cpu probes of design §7 (the thermostat at tau 0.15 and 0.20; the escalation at τ\_\allowbreak{}auto 0). No run of the three scenario files, of a competitor tau other than 0.30, of \texttt{escalation\_\allowbreak{}tau} 0 or 0.05, or of \texttt{threshold\_\allowbreak{}trigger} exists in any run folder (every \texttt{manifest.\allowbreak{}json} under \texttt{code/\allowbreak{}outputs/\allowbreak{}} checked on 2026-10-30).

Commands (from \texttt{\textasciitilde{}/\allowbreak{}MultiActorBPP/\allowbreak{}code} on the Spark, in tmux, the clone at the commit that records the author's sign-off with \texttt{git\ status} clean; device \texttt{cuda}). Run the block with \texttt{bash} (paste it into a \texttt{bash} shell, or wrap it as \texttt{bash\ \textless{}\textless{}\textquotesingle{}EOF\textquotesingle{}} \ldots{} \texttt{EOF}): \texttt{\$R} unquoted is word-split by bash but not by zsh. A call that stops is continued by the same command with \texttt{run.\allowbreak{}resume=\allowbreak{}true} added (design §4.1); the manifest records it. Runs 3 and 4 run in sequence on the Spark, run 3 first.

\begin{Verbatim}[breaklines,breakanywhere,fontsize=\scriptsize]
R="run.n_rep=100 population.n=20000 run.seed_base=20261029 run.device=cuda run.max_batch_runs=2000"
O=outputs/explore_20261029/run3
~/.local/bin/uv run python -m nudgefield scenario=S1_headroom_thermostat design=pilot_rollout $R out_dir=$O/S1_headroom_thermostat_pilot_rollout
~/.local/bin/uv run python -m nudgefield scenario=S1_headroom_thermostat design=pilot_rollout $R run.attention=proportional out_dir=$O/S1_headroom_thermostat_pilot_rollout_proportional
~/.local/bin/uv run python -m nudgefield scenario=S1_headroom_maximiser  design=pilot_rollout $R out_dir=$O/S1_headroom_maximiser_pilot_rollout
~/.local/bin/uv run python -m nudgefield scenario=S1_headroom_maximiser  design=pilot_rollout $R run.attention=proportional out_dir=$O/S1_headroom_maximiser_pilot_rollout_proportional
~/.local/bin/uv run python -m nudgefield scenario=S1_headroom_trigger    design=pilot_rollout $R out_dir=$O/S1_headroom_trigger_pilot_rollout
\end{Verbatim}

\subsubsection*{11. Run 4: attention load in the crowded field (registered 2026-10-30; feeds F3b, F4d, ED4 and the tipping rows of table v2)}

Purpose: design §4.5 (G4). Map the crowded field's classes over the number of bystanders and the population's attention budget, under both attention rules and in both designs, and answer whether priority sharing gives a tipping line that proportional sharing does not. Author's ruling 2 of 2026-10-30 (design §6): budget grid 0.35--4.0, the 0.35 boundary slice kept with its excluded share reported.

Scenario file \texttt{crowded\_\allowbreak{}load} (new, this commit): the calibrated crowded family of \texttt{crowded.\allowbreak{}yaml} (thermostat competitor and ally at gain 2, goal offsets ∓0.03, lag 2, reach0 0; bystanders at gain 0, reach0 0.6) with \texttt{population:\allowbreak{}\ \{budget\_\allowbreak{}mean:\allowbreak{}\ {[}0.\allowbreak{}35,\ 0.\allowbreak{}5,\ 0.\allowbreak{}7,\ 1.\allowbreak{}0,\ 1.\allowbreak{}4,\ 2.\allowbreak{}0,\ 2.\allowbreak{}8,\ 4.\allowbreak{}0{]}\}} (factor √2; the T2 point m = 2, budget 1.0 is on the grid) as the slice and \texttt{bystander.\allowbreak{}count} ∈ \{0, 1, 2, 4, 6, 8, 12, 16\} as the robustness axis (the last sweep path; the registered null claim runs along m); \texttt{sweep:\allowbreak{}\ {[}population.\allowbreak{}budget\_\allowbreak{}mean,\ bystander.\allowbreak{}count{]}}; 64 points; both designs.

Settings: §0 (\texttt{run.\allowbreak{}n\_\allowbreak{}rep} 100, \texttt{population.\allowbreak{}n} 20,000, \texttt{run.\allowbreak{}seed\_\allowbreak{}base} 20261029, the same windows and design parameters, \texttt{fe\_\allowbreak{}reach} 0.4) except: \texttt{run.\allowbreak{}max\_\allowbreak{}batch\_\allowbreak{}runs} 800 and \texttt{run.\allowbreak{}device} cuda (the Spark), \texttt{run.\allowbreak{}attention} per call. \texttt{bystander.\allowbreak{}count} enters the batch structure, so each call has eight batches, one per m, of 8 budgets × 100 replicates = 800 runs. The cap of 800 comes from the Task 7 measurement (\texttt{docs/\allowbreak{}analysis/\allowbreak{}2026-\allowbreak{}10-\allowbreak{}27-\allowbreak{}spark-\allowbreak{}timing.\allowbreak{}md} §7: crowded m = 16, field experiment, N 20,000, cuda: 25.9 MB and 0.806 s per run under priority, 9.8 MB and 0.160 s under proportional; 20 GiB / 25.9 MB = 828.6 runs, rounded down to 800). It equals one m-batch, so run 4 is in fact unchunked: every batch runs as one chunk with the unchunked batch seed \texttt{scenario\textbar{}design\textbar{}structure\textbar{}seed\_\allowbreak{}base}, as under the preamble's \texttt{run.\allowbreak{}max\_\allowbreak{}batch\_\allowbreak{}runs\ =\allowbreak{}\ 0} (checked on the scenario file: \texttt{expand(…,\ max\_\allowbreak{}batch\_\allowbreak{}runs=\allowbreak{}800)} gives eight batches of 800 with \texttt{chunk} None). Predicted peak of the m = 16 priority batch: 16 MB + 800 × 25.9 MB = 20.75 GB (19.32 GiB), within the Spark's 124 GB unified memory with room for the allocator's overhead, about 4\% above a decimal 20 GB (addendum §7, caveats). The measurement used the field-experiment design; the pilot → rollout calls were not measured (the tensor shapes depend on R, N and the actor count, not the design, so the slope is expected to carry over, unchecked; addendum §7). If a batch does not fit, the preamble's fallback applies (\texttt{run.\allowbreak{}max\_\allowbreak{}batch\_\allowbreak{}runs} 100, recorded in the ledger). The batch seed does not contain the attention rule, so each proportional call's batch seeds are those of its priority twin. The per-run budget override keeps every batch's draws (\texttt{a}, \texttt{s\_\allowbreak{}auto}, \texttt{s\_\allowbreak{}delib}, the lognormal budget factors) and scales the budget only (design §3, §4.3); no claim pairs cells. Class rules R9; the field-experiment calls read \texttt{cls\_\allowbreak{}combined}. \texttt{classes.\allowbreak{}parquet} carries \texttt{subgroup\_\allowbreak{}flip} (\texttt{a}) and \texttt{subgroup\_\allowbreak{}flip\_\allowbreak{}budget} (§4).

{\def\LTcaptype{none} 
\begin{longtable}[]{@{}
  >{\raggedright\arraybackslash}p{(\linewidth - 8\tabcolsep) * \real{0.1245}}
  >{\raggedright\arraybackslash}p{(\linewidth - 8\tabcolsep) * \real{0.1563}}
  >{\raggedright\arraybackslash}p{(\linewidth - 8\tabcolsep) * \real{0.1339}}
  >{\raggedright\arraybackslash}p{(\linewidth - 8\tabcolsep) * \real{0.2005}}
  >{\raggedright\arraybackslash}p{(\linewidth - 8\tabcolsep) * \real{0.3648}}@{}}
\toprule\noalign{}
\begin{minipage}[b]{\linewidth}\raggedright
Call
\end{minipage} & \begin{minipage}[b]{\linewidth}\raggedright
Design
\end{minipage} & \begin{minipage}[b]{\linewidth}\raggedright
Attention
\end{minipage} & \begin{minipage}[b]{\linewidth}\raggedright
Runs
\end{minipage} & \begin{minipage}[b]{\linewidth}\raggedright
Output (\texttt{code/\allowbreak{}outputs/\allowbreak{}explore\_\allowbreak{}20261029/\allowbreak{}run4/\allowbreak{}…})
\end{minipage} \\
\midrule\noalign{}
\endhead
\bottomrule\noalign{}
\endlastfoot
4a & pilot\_\allowbreak{}rollout & priority & 6,400 (8 batches × 800) & \texttt{crowded\_\allowbreak{}load\_\allowbreak{}pilot\_\allowbreak{}rollout} \\
4b & pilot\_\allowbreak{}rollout & proportional & 6,400 (8 batches × 800) & \texttt{crowded\_\allowbreak{}load\_\allowbreak{}pilot\_\allowbreak{}rollout\_\allowbreak{}proportional} \\
4c & field\_\allowbreak{}experiment & priority & 6,400 (8 batches × 800) & \texttt{crowded\_\allowbreak{}load\_\allowbreak{}field\_\allowbreak{}experiment} \\
4d & field\_\allowbreak{}experiment & proportional & 6,400 (8 batches × 800) & \texttt{crowded\_\allowbreak{}load\_\allowbreak{}field\_\allowbreak{}experiment\_\allowbreak{}proportional} \\
\end{longtable}
}

25,600 runs. Time, from the Task 7 per-run costs at m = 16 applied to every m (an upper bound: smaller m cost less): 6,400 × 0.806 s = 5,158 s, about 1.4 h per priority call; 6,400 × 0.160 s = 1,024 s, about 17 min per proportional call; about 3.4 h for the four calls. No call is budgeted above the two-hour rule of design §3 (read on measured time).

Range check: on the robustness axis m = 0--16 the adjacent gaps are 1, 1, 2, 2, 2, 4, 4, that is 6.25\%, 6.25\%, 12.5\%, 12.5\%, 12.5\%, 25\%, 25\% of the range. The pairs (0, 1) and (1, 2) alone do not meet the 10\% share; the three points 0, 1, 2 span 12.5\% and do; any two adjacent points from m = 2 upward do. The restated robustness rule applies as written within each budget slice.

The boundary slice (budget 0.35), a stated deviation. The grid's lower end is set by the exclusion rule, not by choice. Design §4.5, verbatim:

\begin{quote}
The lower end is set by the exclusion rule, not by choice: the builder check's probe at budget 0.25 gave \texttt{E} ≈ 0.016 in every replicate at m = 0 and m = 16 under both rules, below \texttt{SMALL\_\allowbreak{}E} 0.02, so every cell of that slice would be \texttt{excluded} (our own nudge is dropped by budget-starved people even in the quiet field) and its probit 0.044 is outside the magnitude bound anyway; at 0.35 \texttt{E} ≈ 0.022 and the probit 0.062, both just inside. The 0.35 slice is kept as the boundary slice with its excluded share reported per cell, and the \texttt{tipping} claim (§4.8) is undefined in any cell whose majority is \texttt{excluded}.
\end{quote}

So the 0.35 slice may hold \texttt{excluded} replicates in any cell, unlike the rest of the grid; its excluded share is reported per cell, and its cells enter the tipping claim only where the majority is not \texttt{excluded}.

\textbf{Expected classes, stated before running.} Design §4.5, verbatim:

\begin{quote}
Pilot → rollout: attenuated at m = 0 at every budget (the two reacting actors alone give attenuated: run 1's S1 and S2 are attenuated, and the ally's identity removes 0.03); attenuated across the middle; null where our weight falls below the null threshold, which under priority is a line in (m, budget) (our nudge, at cost 1 and tau 0.30, is dropped by every person whose budget is exhausted by higher-salience nudges: a tipping line), and under proportional a smooth decline reaching null at m ≥ 8 at budget 1.0 (run 1p) and at smaller m for smaller budgets; as expected nowhere on this map (the F2 as-expected point comes from run 3, not from here). Field experiment: blind site wherever the competitor neutralises the site rate and our nudge still reaches its treated (small m, any budget); as expected or attenuated at the T2 point (m = 2, budget 1.0: the calibrated ratio 2.25 against the quiet field); null at large m and small budgets in both labels. Whether priority gives a tipping line that proportional does not is the question the map answers either way (brief §6b); the claim's text follows the answer, and the table v2 row is ``tipping line present / absent'' per rule.
\end{quote}

The subgroup flip, registered rule §4 (author, 2026-10-29), read first on this run (F4d), verbatim from §4:

\begin{quote}
\emph{Prediction (F4d):} under priority sharing, people with small attention budgets lose our nudge first, so the argmax tercile moves from low- or equal-budget in the empty field to high-budget in a crowded field; no flip is predicted on \texttt{a}.
\end{quote}

Per call. Sources: \emph{spec} = the quoted §4.5 passage; \emph{§4} = the registered flip rule; \emph{none} = no class named for the cell. The passage states regions, not cell boundaries: where it says ``small m'', ``large m'' or ``the middle'', no m or budget is fixed by this entry, and the run reports where the boundaries fall.

{\def\LTcaptype{none} 
\begin{longtable}[]{@{}
  >{\raggedright\arraybackslash}p{(\linewidth - 2\tabcolsep) * \real{0.1549}}
  >{\raggedright\arraybackslash}p{(\linewidth - 2\tabcolsep) * \real{0.8251}}@{}}
\toprule\noalign{}
\begin{minipage}[b]{\linewidth}\raggedright
Call
\end{minipage} & \begin{minipage}[b]{\linewidth}\raggedright
Expected
\end{minipage} \\
\midrule\noalign{}
\endhead
\bottomrule\noalign{}
\endlastfoot
4a (pilot → rollout, priority) & m = 0: attenuated at every budget (spec); attenuated across the middle (spec); null in a region of (m, budget) bounded by a line, the tipping line (spec); as\_\allowbreak{}expected in no cell (spec). flip on budget (\texttt{subgroup\_\allowbreak{}flip\_\allowbreak{}budget}) in a crowded field, no flip on \texttt{a} (§4; the rule names no m or budget) \\
4b (pilot → rollout, proportional) & m = 0: attenuated at every budget (spec); attenuated across the middle (spec); a smooth decline reaching null at m ≥ 8 at budget 1.0 and at smaller m for smaller budgets (spec); as\_\allowbreak{}expected in no cell (spec). Flip: no prediction (§4 states one under priority only) \\
4c (field experiment, priority) & \texttt{cls\_\allowbreak{}combined} blind\_\allowbreak{}site at small m, any budget (spec); as\_\allowbreak{}expected or attenuated at the T2 point m = 2, budget 1.0 (spec); null at large m and small budgets in both labels (spec). Flip as 4a (§4; the rule names no design) \\
4d (field experiment, proportional) & as 4c (spec; the passage gives the field-experiment expectation without a rule). Flip: no prediction (§4) \\
Tipping (4a against 4b; 4c against 4d) & a tipping line under priority, a smooth decline under proportional (spec); table v2 row ``tipping line present / absent'' per rule, whichever the map shows (spec) \\
\end{longtable}
}

What would count against the expectations: as\_\allowbreak{}expected robust in any pilot → rollout cell; no null cell under priority. Each is reported as found, and the tipping claim's text follows the map either way (the author removed three clauses at sign-off, 2026-10-30, that added robustness conditions and a rule difference the design did not state: attenuated robust at m = 0, blind\_\allowbreak{}site robust at small m, a rule difference in the tipping claim).

\textbf{Prior knowledge (disclosed).} Everything in §5 and §8; runs 1d / 1pd (\texttt{S3\_\allowbreak{}bystanders}, bystanders without the two reacting actors: null at m = 16 only under priority, §8 table; at m = 8 and 16 under proportional, design §4.5 and §4.2); the targets runs (\texttt{parameters.\allowbreak{}md} §5--§6: the calibrated quiet/crowded ratio 2.25 at m = 2, budget 1.0); the builder check's cpu probes of the crowded field at budgets 0.25--0.35 (design §7 finding 2); the Task 7 measurement (\texttt{docs/\allowbreak{}analysis/\allowbreak{}2026-\allowbreak{}10-\allowbreak{}27-\allowbreak{}spark-\allowbreak{}timing.\allowbreak{}md} §7: \texttt{crowded}, field experiment, m = 16, budget 1.0, N 20,000, every one of the 100 and 400 runs null under both rules; not a registered run). The targets runs (\texttt{code/\allowbreak{}outputs/\allowbreak{}targets\_\allowbreak{}20261027\_\allowbreak{}*}, \texttt{parameters.\allowbreak{}md} §5) ran the crowded family at budget 1.0 with m ∈ \{0, 1\} and \{2, 3, 4, 6, 8\}. No run of \texttt{crowded\_\allowbreak{}load}, of a budget other than 1.0, or of the crowded family at m = 12 exists in any run folder (every \texttt{manifest.\allowbreak{}json} under \texttt{code/\allowbreak{}outputs/\allowbreak{}} checked on 2026-10-30: \texttt{budget\_\allowbreak{}mean} 1.0 throughout).

Commands (from \texttt{\textasciitilde{}/\allowbreak{}MultiActorBPP/\allowbreak{}code} on the Spark, in tmux, after run 3, the clone at the commit that records the author's sign-off with \texttt{git\ status} clean; device \texttt{cuda}; run the block with \texttt{bash}; a call that stops is continued with \texttt{run.\allowbreak{}resume=\allowbreak{}true} added):

\begin{Verbatim}[breaklines,breakanywhere,fontsize=\scriptsize]
R="run.n_rep=100 population.n=20000 run.seed_base=20261029 run.device=cuda run.max_batch_runs=800"
O=outputs/explore_20261029/run4
~/.local/bin/uv run python -m nudgefield scenario=crowded_load design=pilot_rollout    $R out_dir=$O/crowded_load_pilot_rollout
~/.local/bin/uv run python -m nudgefield scenario=crowded_load design=pilot_rollout    $R run.attention=proportional out_dir=$O/crowded_load_pilot_rollout_proportional
~/.local/bin/uv run python -m nudgefield scenario=crowded_load design=field_experiment $R out_dir=$O/crowded_load_field_experiment
~/.local/bin/uv run python -m nudgefield scenario=crowded_load design=field_experiment $R run.attention=proportional out_dir=$O/crowded_load_field_experiment_proportional
\end{Verbatim}

\subsubsection*{12. Run 5: ally gain × lag (registered 2026-10-30; feeds F3c, F4e if cycles, and the not-found list)}

Purpose: design §4.6 (G4, G5). Search the ally's gain and lag for the two classes not yet exhibited, oscillating and delayed (ruling (c) of 2026-10-30: the delayed note reads ``not exhibited so far; run 5 is the registered search''), with the competitor as a fixed background.

Scenario file \texttt{S2\_\allowbreak{}ally\_\allowbreak{}cycles} (new, this commit): the S1 thermostat competitor at gain 2 (goal\_\allowbreak{}offset −0.03, lag 2, reach0 0, no escalation) as actor 0 and the S2 ally (thermostat, goal\_\allowbreak{}offset +0.03, reach0 0) as actor 1, with \texttt{ally.\allowbreak{}gain} ∈ \{0.5, 1, 2, 5, 10, 20, 30, 40\} as the robustness axis; \texttt{sweep:\allowbreak{}\ {[}ally.\allowbreak{}gain{]}}; 8 points, 800 runs per call; pilot → rollout. The ally's \texttt{lag} is not sweepable (an integer of the observed-history indexing, design §4.3): one call per lag ∈ \{1, 2, 5, 10\} with the Hydra override \texttt{scenario.\allowbreak{}actors.\allowbreak{}1.\allowbreak{}lag=\allowbreak{}\textless{}l\textgreater{}} (the ally is actor index 1; checked by composing the four overrides and validating the result), one output directory per lag. The competitor's lag stays 2.

Settings: §0 throughout (\texttt{run.\allowbreak{}n\_\allowbreak{}rep} 100, \texttt{population.\allowbreak{}n} 20,000, \texttt{run.\allowbreak{}seed\_\allowbreak{}base} 20261029, \texttt{run.\allowbreak{}max\_\allowbreak{}batch\_\allowbreak{}runs} 0: one unchunked batch per call) on the M4 (\texttt{run.\allowbreak{}device} mps), \texttt{run.\allowbreak{}attention} per call. The batch seed \texttt{scenario\textbar{}design\textbar{}structure\textbar{}seed\_\allowbreak{}base} contains neither the lag nor the attention rule, so all eight calls share one batch seed: the four lags and the two rules are run on the same draws. No claim pairs cells. Class rules R9.

{\def\LTcaptype{none} 
\begin{longtable}[]{@{}
  >{\raggedright\arraybackslash}p{(\linewidth - 8\tabcolsep) * \real{0.1245}}
  >{\raggedright\arraybackslash}p{(\linewidth - 8\tabcolsep) * \real{0.1245}}
  >{\raggedright\arraybackslash}p{(\linewidth - 8\tabcolsep) * \real{0.1341}}
  >{\raggedright\arraybackslash}p{(\linewidth - 8\tabcolsep) * \real{0.1245}}
  >{\raggedright\arraybackslash}p{(\linewidth - 8\tabcolsep) * \real{0.4725}}@{}}
\toprule\noalign{}
\begin{minipage}[b]{\linewidth}\raggedright
Call
\end{minipage} & \begin{minipage}[b]{\linewidth}\raggedright
Ally lag
\end{minipage} & \begin{minipage}[b]{\linewidth}\raggedright
Attention
\end{minipage} & \begin{minipage}[b]{\linewidth}\raggedright
Runs
\end{minipage} & \begin{minipage}[b]{\linewidth}\raggedright
Output (\texttt{code/\allowbreak{}outputs/\allowbreak{}explore\_\allowbreak{}20261029/\allowbreak{}run5/\allowbreak{}…})
\end{minipage} \\
\midrule\noalign{}
\endhead
\bottomrule\noalign{}
\endlastfoot
5a & 1 & priority & 800 & \texttt{S2\_\allowbreak{}ally\_\allowbreak{}cycles\_\allowbreak{}pilot\_\allowbreak{}rollout\_\allowbreak{}lag1} \\
5b & 2 & priority & 800 & \texttt{S2\_\allowbreak{}ally\_\allowbreak{}cycles\_\allowbreak{}pilot\_\allowbreak{}rollout\_\allowbreak{}lag2} \\
5c & 5 & priority & 800 & \texttt{S2\_\allowbreak{}ally\_\allowbreak{}cycles\_\allowbreak{}pilot\_\allowbreak{}rollout\_\allowbreak{}lag5} \\
5d & 10 & priority & 800 & \texttt{S2\_\allowbreak{}ally\_\allowbreak{}cycles\_\allowbreak{}pilot\_\allowbreak{}rollout\_\allowbreak{}lag10} \\
5e & 1 & proportional & 800 & \texttt{S2\_\allowbreak{}ally\_\allowbreak{}cycles\_\allowbreak{}pilot\_\allowbreak{}rollout\_\allowbreak{}lag1\_\allowbreak{}proportional} \\
5f & 2 & proportional & 800 & \texttt{S2\_\allowbreak{}ally\_\allowbreak{}cycles\_\allowbreak{}pilot\_\allowbreak{}rollout\_\allowbreak{}lag2\_\allowbreak{}proportional} \\
5g & 5 & proportional & 800 & \texttt{S2\_\allowbreak{}ally\_\allowbreak{}cycles\_\allowbreak{}pilot\_\allowbreak{}rollout\_\allowbreak{}lag5\_\allowbreak{}proportional} \\
5h & 10 & proportional & 800 & \texttt{S2\_\allowbreak{}ally\_\allowbreak{}cycles\_\allowbreak{}pilot\_\allowbreak{}rollout\_\allowbreak{}lag10\_\allowbreak{}proportional} \\
\end{longtable}
}

6,400 runs; about 20 min on the M4 (design §4.6), in parallel with runs 3 and 4 on the Spark.

Range check: on \texttt{ally.\allowbreak{}gain} 0.5--40 (range 39.5) the adjacent gaps are 0.5, 1, 3, 5, 10, 10, 10, that is 1.3\%, 2.5\%, 7.6\%, 12.7\%, 25.3\%, 25.3\%, 25.3\% of the range. No pair below gain 5 meets the 10\% share alone; the runs \{1, 2, 5\} (10.1\%) and \{0.5, 1, 2, 5\} (11.4\%) do; any two adjacent points from gain 5 upward do. A class confined to gains 0.5--2 cannot be robust on this grid; the restated robustness rule applies as written within each lag directory.

\textbf{Expected classes, stated before running.} Design §4.6, verbatim:

\begin{quote}
Attenuated at gain ≤ 10 at every lag (the ally relaxes to zero and the competitor holds: run 1's 0.46 identity); oscillating, if anywhere, at gain ≥ 20 with lag ≥ 5 (the disclosure probes of spike §1.6 saw oscillation in 2 of 5 and 1 of 5 replicates at k = 20--40 with lag 2: the integral rule with a long lag overshoots); delayed nowhere (R9: a late arrival that is there at the end; the ally's relaxation is immediate). If neither oscillating nor delayed is robust inside the constraint region, F3c is replaced by run 6's treated-share map (brief §6c) and the classes go to the not-found list with this region.
\end{quote}

Per call. Sources: \emph{spec} = the quoted §4.6 passage; \emph{none} = no class named for the cell. The passage names no attention rule; this entry reads it as applying to both (5a--5d and 5e--5h alike).

{\def\LTcaptype{none} 
\begin{longtable}[]{@{}
  >{\raggedright\arraybackslash}p{(\linewidth - 4\tabcolsep) * \real{0.2503}}
  >{\raggedright\arraybackslash}p{(\linewidth - 4\tabcolsep) * \real{0.2473}}
  >{\raggedright\arraybackslash}p{(\linewidth - 4\tabcolsep) * \real{0.4825}}@{}}
\toprule\noalign{}
\begin{minipage}[b]{\linewidth}\raggedright
Calls
\end{minipage} & \begin{minipage}[b]{\linewidth}\raggedright
gain 0.5, 1, 2, 5, 10
\end{minipage} & \begin{minipage}[b]{\linewidth}\raggedright
gain 20, 30, 40
\end{minipage} \\
\midrule\noalign{}
\endhead
\bottomrule\noalign{}
\endlastfoot
5a, 5b, 5e, 5f (lag 1, 2) & attenuated (spec) & none; delayed nowhere (spec) \\
5c, 5d, 5g, 5h (lag 5, 10) & attenuated (spec) & oscillating possible, ``if anywhere'' (spec; not stated as robust); delayed nowhere (spec) \\
\end{longtable}
}

Delayed is expected in no cell of any call (spec). Fallback (spec): if neither oscillating nor delayed is robust inside the constraint region (run 7's \texttt{targets.\allowbreak{}parquet}, design §4.8), F3c is replaced by run 6's treated-share map and both classes enter the not-found list with this region.

What would count against the expectations: attenuated not robust at gains ≤ 10 at some lag; oscillating at a lag below 5 or a gain below 20; delayed robust anywhere. Each is reported as found.

\textbf{Prior knowledge (disclosed).} Everything in §5 and §8; run 1c / 1pc (\texttt{S2\_\allowbreak{}ally} at gains 0.5, 2, 5, lag 2, without the competitor: attenuated, ratio 0.46 under priority; §8 table); the disclosed probes of spike §1.6 (oscillation in 2 of 5 and 1 of 5 replicates at k = 20 and 40, lag 2); the builder check's cpu probe over ally gains 0.5--40 × lags 1--10 (design §7: every run passed the burn-in test; no classes are quoted from it). The crowded family at m = 0 (the targets runs, \texttt{code/\allowbreak{}outputs/\allowbreak{}targets\_\allowbreak{}20261027\_\allowbreak{}*}, budget 1.0) is the competitor and the ally at gain 2, lag 2: this grid's point gain 2 at lag 2, under another scenario name (quiet/crowded ratio 1.81, §5). No run of \texttt{S2\_\allowbreak{}ally\_\allowbreak{}cycles}, of an ally lag other than 2, or of an ally gain above 5 exists in any run folder (every \texttt{manifest.\allowbreak{}json} under \texttt{code/\allowbreak{}outputs/\allowbreak{}} checked on 2026-10-30).

Commands (from \texttt{code/\allowbreak{}}, M4, device \texttt{mps}; run the block with \texttt{bash}):

\begin{Verbatim}[breaklines,breakanywhere,fontsize=\scriptsize]
R="run.n_rep=100 population.n=20000 run.seed_base=20261029 run.device=mps"
O=outputs/explore_20261029/run5
uv run python -m nudgefield scenario=S2_ally_cycles design=pilot_rollout scenario.actors.1.lag=1  $R out_dir=$O/S2_ally_cycles_pilot_rollout_lag1
uv run python -m nudgefield scenario=S2_ally_cycles design=pilot_rollout scenario.actors.1.lag=2  $R out_dir=$O/S2_ally_cycles_pilot_rollout_lag2
uv run python -m nudgefield scenario=S2_ally_cycles design=pilot_rollout scenario.actors.1.lag=5  $R out_dir=$O/S2_ally_cycles_pilot_rollout_lag5
uv run python -m nudgefield scenario=S2_ally_cycles design=pilot_rollout scenario.actors.1.lag=10 $R out_dir=$O/S2_ally_cycles_pilot_rollout_lag10
uv run python -m nudgefield scenario=S2_ally_cycles design=pilot_rollout scenario.actors.1.lag=1  $R run.attention=proportional out_dir=$O/S2_ally_cycles_pilot_rollout_lag1_proportional
uv run python -m nudgefield scenario=S2_ally_cycles design=pilot_rollout scenario.actors.1.lag=2  $R run.attention=proportional out_dir=$O/S2_ally_cycles_pilot_rollout_lag2_proportional
uv run python -m nudgefield scenario=S2_ally_cycles design=pilot_rollout scenario.actors.1.lag=5  $R run.attention=proportional out_dir=$O/S2_ally_cycles_pilot_rollout_lag5_proportional
uv run python -m nudgefield scenario=S2_ally_cycles design=pilot_rollout scenario.actors.1.lag=10 $R run.attention=proportional out_dir=$O/S2_ally_cycles_pilot_rollout_lag10_proportional
\end{Verbatim}

\subsubsection*{13. Run 6: treated share (registered 2026-10-30; feeds F4a and the F3c fallback)}

Purpose: design §4.7. Measure the field experiment's site class and its trial / total ratio against the treated share p, under both attention rules. Author's ruling 3 of 2026-10-30 (design §6): run 6 in this phase.

Scenario file \texttt{S1\_\allowbreak{}treated\_\allowbreak{}share} (new, this commit): the S1 competitor of \texttt{S1\_\allowbreak{}competitor} at gain 2 (thermostat, goal\_\allowbreak{}offset −0.03, lag 2, reach0 0, no escalation) in the field experiment, with \texttt{design:\allowbreak{}\ \{fe\_\allowbreak{}reach:\allowbreak{}\ {[}0.\allowbreak{}1,\ 0.\allowbreak{}2,\ 0.\allowbreak{}3,\ 0.\allowbreak{}4,\ 0.\allowbreak{}5,\ 0.\allowbreak{}6{]}\}} as the one sweep path (\texttt{sweep:\allowbreak{}\ {[}design.\allowbreak{}fe\_\allowbreak{}reach{]}}, robustness along p; the brief's four values plus 0.3 and 0.5); 6 points, 600 runs per call; field experiment only (\texttt{design.\allowbreak{}fe\_\allowbreak{}reach} is defined for field-experiment kinds only). The swept share replaces §0's \texttt{fe\_\allowbreak{}reach} 0.4 per run; each row's treated set is the lowest \texttt{round(p\ ·\ N)} ranks of one uniform draw per row (design §4.3). The scenario name differs from \texttt{S1\_\allowbreak{}competitor}, so the batch seed, hence the draws, differ from run 1b's even at p = 0.4.

Settings: §0 throughout (\texttt{run.\allowbreak{}n\_\allowbreak{}rep} 100, \texttt{population.\allowbreak{}n} 20,000, \texttt{run.\allowbreak{}seed\_\allowbreak{}base} 20261029, \texttt{run.\allowbreak{}max\_\allowbreak{}batch\_\allowbreak{}runs} 0: one unchunked batch per call) on the M4 (\texttt{run.\allowbreak{}device} mps), \texttt{run.\allowbreak{}attention} per call; the proportional call shares the priority call's batch seed. Class rules R9; the site class is \texttt{cls\_\allowbreak{}combined}, computed with each row's share (\texttt{site\_\allowbreak{}ratio\ =\allowbreak{}\ site\_\allowbreak{}w\ /\allowbreak{}\ (share\_\allowbreak{}r\ ·\ e)}, design §4.3).

{\def\LTcaptype{none} 
\begin{longtable}[]{@{}
  >{\raggedright\arraybackslash}p{(\linewidth - 8\tabcolsep) * \real{0.1245}}
  >{\raggedright\arraybackslash}p{(\linewidth - 8\tabcolsep) * \real{0.1701}}
  >{\raggedright\arraybackslash}p{(\linewidth - 8\tabcolsep) * \real{0.1421}}
  >{\raggedright\arraybackslash}p{(\linewidth - 8\tabcolsep) * \real{0.1245}}
  >{\raggedright\arraybackslash}p{(\linewidth - 8\tabcolsep) * \real{0.4188}}@{}}
\toprule\noalign{}
\begin{minipage}[b]{\linewidth}\raggedright
Call
\end{minipage} & \begin{minipage}[b]{\linewidth}\raggedright
Design
\end{minipage} & \begin{minipage}[b]{\linewidth}\raggedright
Attention
\end{minipage} & \begin{minipage}[b]{\linewidth}\raggedright
Runs
\end{minipage} & \begin{minipage}[b]{\linewidth}\raggedright
Output (\texttt{code/\allowbreak{}outputs/\allowbreak{}explore\_\allowbreak{}20261029/\allowbreak{}run6/\allowbreak{}…})
\end{minipage} \\
\midrule\noalign{}
\endhead
\bottomrule\noalign{}
\endlastfoot
6a & field\_\allowbreak{}experiment & priority & 600 & \texttt{S1\_\allowbreak{}treated\_\allowbreak{}share\_\allowbreak{}field\_\allowbreak{}experiment} \\
6b & field\_\allowbreak{}experiment & proportional & 600 & \texttt{S1\_\allowbreak{}treated\_\allowbreak{}share\_\allowbreak{}field\_\allowbreak{}experiment\_\allowbreak{}proportional} \\
\end{longtable}
}

1,200 runs; about 5 min on the M4 (design §4.7).

Range check: six evenly spaced points on 0.1--0.6; two adjacent points span 0.1 / 0.5 = 20\% of the range, so the restated robustness rule applies as written.

\textbf{Expected classes, stated before running.} Design §4.7, verbatim:

\begin{quote}
Expected: blind site robust over p ≤ 0.4 (run 1: 100/98/100 at p = 0.4 across k), \texttt{trial\ /\allowbreak{}\ total} on the 1/(1 − p) curve there and departing at p = 0.6 (brief §7a; the competitor is at reach 0.998 at p = 0.4 and has no headroom left).
\end{quote}

Per call. Sources: \emph{spec} = the quoted passage; \emph{none} = no class named for the cell. The passage names no attention rule; this entry reads it as applying to both (6a and 6b alike).

{\def\LTcaptype{none} 
\begin{longtable}[]{@{}
  >{\raggedright\arraybackslash}p{(\linewidth - 6\tabcolsep) * \real{0.1245}}
  >{\raggedright\arraybackslash}p{(\linewidth - 6\tabcolsep) * \real{0.4501}}
  >{\raggedright\arraybackslash}p{(\linewidth - 6\tabcolsep) * \real{0.1245}}
  >{\raggedright\arraybackslash}p{(\linewidth - 6\tabcolsep) * \real{0.2809}}@{}}
\toprule\noalign{}
\begin{minipage}[b]{\linewidth}\raggedright
Call
\end{minipage} & \begin{minipage}[b]{\linewidth}\raggedright
p 0.1, 0.2, 0.3, 0.4
\end{minipage} & \begin{minipage}[b]{\linewidth}\raggedright
p 0.5
\end{minipage} & \begin{minipage}[b]{\linewidth}\raggedright
p 0.6
\end{minipage} \\
\midrule\noalign{}
\endhead
\bottomrule\noalign{}
\endlastfoot
6a, 6b & \texttt{cls\_\allowbreak{}combined} blind\_\allowbreak{}site, robust over the four points (spec); \texttt{trial\ /\allowbreak{}\ total} ≈ 1/(1 − p) (spec) & none & class none; \texttt{trial\ /\allowbreak{}\ total} departing from 1/(1 − p) (spec) \\
\end{longtable}
}

What would count against the expectations: blind\_\allowbreak{}site not robust over p ≤ 0.4, or \texttt{trial\ /\allowbreak{}\ total} off the 1/(1 − p) curve at p ≤ 0.4, or on it at p = 0.6. Each is reported as found.

\textbf{Prior knowledge (disclosed).} Everything in §5 and §8; runs 1b / 1pb (\texttt{S1\_\allowbreak{}competitor}, field experiment, p = 0.4, gains 0.5, 2, 5: \texttt{cls\_\allowbreak{}combined} blind\_\allowbreak{}site 100 / 98 / 100 under priority; §8 table); the disclosed probes of spike §1.6 (treated shares 0.2--0.5). No run with p ≠ 0.4, and no run of \texttt{S1\_\allowbreak{}treated\_\allowbreak{}share}, exists in any run folder (every \texttt{manifest.\allowbreak{}json} under \texttt{code/\allowbreak{}outputs/\allowbreak{}} checked on 2026-10-30: \texttt{fe\_\allowbreak{}reach} 0.4 throughout).

Commands (from \texttt{code/\allowbreak{}}, M4, device \texttt{mps}; run the block with \texttt{bash}):

\begin{Verbatim}[breaklines,breakanywhere,fontsize=\scriptsize]
R="run.n_rep=100 population.n=20000 run.seed_base=20261029 run.device=mps"
O=outputs/explore_20261029/run6
uv run python -m nudgefield scenario=S1_treated_share design=field_experiment $R out_dir=$O/S1_treated_share_field_experiment
uv run python -m nudgefield scenario=S1_treated_share design=field_experiment $R run.attention=proportional out_dir=$O/S1_treated_share_field_experiment_proportional
\end{Verbatim}

\subsubsection*{14. Run 8: the headroom calibration map, competitor strength × bystanders (registered 2026-10-31; a feasibility study; feeds the T2 recalibration if a joint cell exists, ED3 either way)}

Purpose: the T1 probe (\texttt{docs/\allowbreak{}analysis/\allowbreak{}2026-\allowbreak{}10-\allowbreak{}31-\allowbreak{}t1-\allowbreak{}probe-\allowbreak{}two-\allowbreak{}actor-\allowbreak{}field.\allowbreak{}md}, §3--§9; author's choice 2026-10-31: the most defensible option, the calibration map as a feasibility study). The calibrated crowded family (\texttt{crowded.\allowbreak{}yaml}) has no reacting actor at its T2 point: its two thermostats saturate at full reach in burn-in (an arms race between point goals that cannot both be met), the reaction part of the total is zero and T1 fails there by construction. This run asks whether any crowded field with a reacting competitor reproduces T1 and T2 together, under a criterion fixed here. It answers either way: a robust joint cell defines a new crowded family and the T2 point is recalibrated on it; no joint cell is reported as the finding that the two regularities belong to two field states, with the mechanism (attention crowding removes the actors' headroom before it removes the nudge) as the reason.

Scenario file \texttt{crowded\_\allowbreak{}headroom} (new, this commit): the S1 competitor (thermostat, gain 2, goal\_\allowbreak{}offset −0.03, lag 2, reach0 0) with its nudge strength \texttt{competitor.\allowbreak{}nudge.\allowbreak{}tau} ∈ \{0.30, 0.35, 0.40, 0.45\} as the slice, plus \texttt{bystander.\allowbreak{}count} m ∈ \{1, 2, 3, 4, 6, 8\} (gain 0, reach0 0.6, attention only) as the robustness axis; \texttt{sweep:\allowbreak{}\ {[}competitor.\allowbreak{}nudge.\allowbreak{}tau,\ bystander.\allowbreak{}count{]}}; 24 points, 2,400 runs per call; both designs; no ally. The budget scale stays at the default 1.0 (the T2 calibration's setting; \texttt{parameters.\allowbreak{}md} §5). m = 0 is not on the grid: it is run 1's S1 (tau 0.30) and run 3's τ\_\allowbreak{}auto = 0 column (tau 0.35--0.45), both registered. The grid's tau values are a subset of run 3's. Checked on this commit: the file loads and expands to six batches of 400 runs in each design.

Settings: §0 (\texttt{run.\allowbreak{}n\_\allowbreak{}rep} 100, \texttt{population.\allowbreak{}n} 20,000, \texttt{run.\allowbreak{}seed\_\allowbreak{}base} 20261029, \texttt{run.\allowbreak{}max\_\allowbreak{}batch\_\allowbreak{}runs} 0, the same windows and design parameters, \texttt{fe\_\allowbreak{}reach} 0.4) on the M4 (\texttt{run.\allowbreak{}device} mps), \texttt{run.\allowbreak{}attention} per call. \texttt{bystander.\allowbreak{}count} enters the batch structure, so each call has six batches (one per m) of 4 tau × 100 = 400 runs; the four tau values of a batch share its draws (paired along tau, as budgets were in run 4). Class rules R9.

{\def\LTcaptype{none} 
\begin{longtable}[]{@{}
  >{\raggedright\arraybackslash}p{(\linewidth - 10\tabcolsep) * \real{0.1245}}
  >{\raggedright\arraybackslash}p{(\linewidth - 10\tabcolsep) * \real{0.1306}}
  >{\raggedright\arraybackslash}p{(\linewidth - 10\tabcolsep) * \real{0.1421}}
  >{\raggedright\arraybackslash}p{(\linewidth - 10\tabcolsep) * \real{0.1338}}
  >{\raggedright\arraybackslash}p{(\linewidth - 10\tabcolsep) * \real{0.1245}}
  >{\raggedright\arraybackslash}p{(\linewidth - 10\tabcolsep) * \real{0.3245}}@{}}
\toprule\noalign{}
\begin{minipage}[b]{\linewidth}\raggedright
Call
\end{minipage} & \begin{minipage}[b]{\linewidth}\raggedright
Scenario file
\end{minipage} & \begin{minipage}[b]{\linewidth}\raggedright
Design
\end{minipage} & \begin{minipage}[b]{\linewidth}\raggedright
Attention
\end{minipage} & \begin{minipage}[b]{\linewidth}\raggedright
Runs
\end{minipage} & \begin{minipage}[b]{\linewidth}\raggedright
Output (\texttt{code/\allowbreak{}outputs/\allowbreak{}explore\_\allowbreak{}20261029/\allowbreak{}run8/\allowbreak{}…})
\end{minipage} \\
\midrule\noalign{}
\endhead
\bottomrule\noalign{}
\endlastfoot
8a & \texttt{crowded\_\allowbreak{}headroom} & pilot\_\allowbreak{}rollout & priority & 2,400 (6 batches × 400) & \texttt{crowded\_\allowbreak{}headroom\_\allowbreak{}pilot\_\allowbreak{}rollout} \\
8b & \texttt{crowded\_\allowbreak{}headroom} & pilot\_\allowbreak{}rollout & proportional & 2,400 & \texttt{crowded\_\allowbreak{}headroom\_\allowbreak{}pilot\_\allowbreak{}rollout\_\allowbreak{}proportional} \\
8c & \texttt{crowded\_\allowbreak{}headroom} & field\_\allowbreak{}experiment & priority & 2,400 & \texttt{crowded\_\allowbreak{}headroom\_\allowbreak{}field\_\allowbreak{}experiment} \\
8d & \texttt{crowded\_\allowbreak{}headroom} & field\_\allowbreak{}experiment & proportional & 2,400 & \texttt{crowded\_\allowbreak{}headroom\_\allowbreak{}field\_\allowbreak{}experiment\_\allowbreak{}proportional} \\
8q & \texttt{quiet} & field\_\allowbreak{}experiment & priority & 100 & \texttt{quiet\_\allowbreak{}field\_\allowbreak{}experiment} \\
8qp & \texttt{quiet} & field\_\allowbreak{}experiment & proportional & 100 & \texttt{quiet\_\allowbreak{}field\_\allowbreak{}experiment\_\allowbreak{}proportional} \\
\end{longtable}
}

9,800 runs, plus the levels re-run of 8c, 8d, 8q and 8qp (engineering item below: 5,000 runs). Time on the M4, from the Build-phase targets runs (\texttt{code/\allowbreak{}outputs/\allowbreak{}targets\_\allowbreak{}20261027\_\allowbreak{}201230}: the crowded family at m = 2--4 about 0.5--0.7 s per run under priority in the field experiment, about a fifth under proportional): about 25 min per priority call, 6 min per proportional call, about 1.5 h in all with the levels re-runs. Memory: at m = 8 and 400 runs per batch, well under the M4's limit (run 1's S3 m = 16 batch of 100 peaked at 4.1 GiB; per-batch parts are written as of 623f481).

Range check: on the robustness axis m = 1--8 (range 7) the adjacent gaps are 1, 1, 1, 2, 2, that is 14.3\%, 14.3\%, 14.3\%, 28.6\%, 28.6\% of the range; any two adjacent points meet the 10\% share and the restated robustness rule (\texttt{run.\allowbreak{}robust\_\allowbreak{}min\_\allowbreak{}points} 2, \texttt{run.\allowbreak{}robust\_\allowbreak{}min\_\allowbreak{}range\_\allowbreak{}share} 0.10) applies as written within each tau slice.

\textbf{Engineering before the run (one item, test-first).} The T2 ratio is a probit contrast of the treated and control window rates, which the sweep outputs do not record (\texttt{effects.\allowbreak{}parquet} carries differences). The Build-phase calibration recomputed them with a private test helper (\texttt{tests/\allowbreak{}nudgefield/\allowbreak{}test\_\allowbreak{}targets.\allowbreak{}py:\allowbreak{}:\allowbreak{}\_\allowbreak{}levels}: re-run the directory's batches with the same scenario, design, settings, seeds and device, keep the window means of the none, free, treated and control rates, check that the re-run reproduces \texttt{effects.\allowbreak{}parquet} to 10⁻⁹). That helper becomes a package tool, \texttt{nudgefield.\allowbreak{}levels}, run as \texttt{uv\ run\ python\ -\allowbreak{}m\ nudgefield.\allowbreak{}levels\ \textless{}run\_\allowbreak{}dir\textgreater{}}: it reads the directory's \texttt{manifest.\allowbreak{}json}, re-runs its batches and writes \texttt{levels.\allowbreak{}parquet} beside the originals (columns point, param\_\allowbreak{}value, the slice column, replicate, seed, run, none\_\allowbreak{}rate, free\_\allowbreak{}rate, free\_\allowbreak{}treated, free\_\allowbreak{}control; metadata: tool commit, source commit) after the reproduction check; it refuses a directory whose code commit differs from the tool's in any file under \texttt{src/\allowbreak{}nudgefield} (the parts rule of \texttt{run.\allowbreak{}resume}). The test suite's helper imports from the package; the T2 test's numbers must not change (a pinned assertion on the calibration table under priority). No class, threshold or output column of the sweep changes.

\textbf{The joint-cell criterion, fixed before the run.} Per cell (tau, m), under the default rule priority:

\begin{enumerate}
\def\labelenumi{\arabic{enumi}.}
\tightlist
\item
  \emph{Headroom:} the competitor's replicate-mean reach over the burn-in periods 90--99 in 8a is below 0.99 (\texttt{reach\_\allowbreak{}competitor} in \texttt{effects.\allowbreak{}parquet}; the share of replicates at or above 0.99 is reported beside it).
\item
  \emph{T1:} \texttt{t1\_\allowbreak{}ratio} \textless{} 0.8 in 8a (\texttt{nudgefield.\allowbreak{}targets.\allowbreak{}t1\_\allowbreak{}cells}, as in run 7).
\item
  \emph{T2:} the quiet / crowded probit trial effect, \texttt{probit(8q)\ /\allowbreak{}\ probit(8c)} at the cell, lies in {[}1.5, 2.5{]}, where \texttt{probit} is the replicate mean of Φ⁻¹(treated) − Φ⁻¹(control) of the window means (\texttt{levels.\allowbreak{}parquet}; the Build-phase definition, \texttt{parameters.\allowbreak{}md} §5; the band is the author's, spec §12, unchanged).
\item
  \emph{Class:} the cell's majority class in 8a (R9 \texttt{cls}) is attenuated or transient (not null, not excluded, not as\_\allowbreak{}expected: the voltage drop T1 names is partial).
\item
  \emph{Robustness:} conditions 1--4 also hold in an adjacent cell along m within the same tau slice (the restated rule: two adjacent points).
\end{enumerate}

A cell meeting 1--5 is a \textbf{joint cell}. The calibrated point, if any, is the smallest m with a joint cell, then the smallest tau. Conditions 1--4 are also computed and reported under proportional (8b, 8d, 8qp), descriptively; the calibration uses the default rule as the Build-phase calibration did. A script \texttt{code/\allowbreak{}scripts/\allowbreak{}calibration\_\allowbreak{}map.\allowbreak{}py} prints the 24 × 2 cell table (every condition's value and verdict) from the directories, committed with the results note.

\textbf{Decision rule after the run, fixed here.} If a joint cell exists: the point (tau\emph{, m}) is registered as the T2 calibration of a new family \texttt{crowded\_\allowbreak{}headroom} at that tau, the T2 band statement in \texttt{parameters.\allowbreak{}md} §5 is extended (not replaced: the \texttt{crowded.\allowbreak{}yaml} calibration stays as the stalemate case), and the m × budget map on the new family (run 4h) is registered as its own entry with its own expectations; F3b's candidates are then run 4 and run 4h. If no joint cell exists: no further crowded-family run; the results note states that in this model T1 and T2 are reproduced by two field states and why; F3b stays run 4 with its actors' saturation stated on the figure; ED3 says T1 comes from the single-competitor field and T2 from the crowded one.

\textbf{Expected, stated before running.} Sources: the ten-replicate probes of the T1 probe note §7 (disclosed, scenario \texttt{crowded} with the ally made inert, not registered, files not in the repository), run 1's S1, run 3's τ\_\allowbreak{}auto = 0 column, run 4.

{\def\LTcaptype{none} 
\begin{longtable}[]{@{}
  >{\raggedright\arraybackslash}p{(\linewidth - 10\tabcolsep) * \real{0.1339}}
  >{\raggedright\arraybackslash}p{(\linewidth - 10\tabcolsep) * \real{0.1851}}
  >{\raggedright\arraybackslash}p{(\linewidth - 10\tabcolsep) * \real{0.1786}}
  >{\raggedright\arraybackslash}p{(\linewidth - 10\tabcolsep) * \real{0.1302}}
  >{\raggedright\arraybackslash}p{(\linewidth - 10\tabcolsep) * \real{0.2277}}
  >{\raggedright\arraybackslash}p{(\linewidth - 10\tabcolsep) * \real{0.1245}}@{}}
\toprule\noalign{}
\begin{minipage}[b]{\linewidth}\raggedright
Slice
\end{minipage} & \begin{minipage}[b]{\linewidth}\raggedright
Headroom (cond. 1)
\end{minipage} & \begin{minipage}[b]{\linewidth}\raggedright
T1 (cond. 2)
\end{minipage} & \begin{minipage}[b]{\linewidth}\raggedright
T2 band (cond. 3)
\end{minipage} & \begin{minipage}[b]{\linewidth}\raggedright
Class (cond. 4)
\end{minipage} & \begin{minipage}[b]{\linewidth}\raggedright
Joint cell
\end{minipage} \\
\midrule\noalign{}
\endhead
\bottomrule\noalign{}
\endlastfoot
tau 0.30 & interior at m ≤ 2 (probe: 0.76, 0.93), at the cap from m = 3 or 4 & \textless{} 0.8 at m = 1 only (0.73); 0.88 at m = 2; ≥ 1 from m = 4 & by the proxy E / pilot (1.44 at m = 2, 2.14 at m = 4): m = 3--4, possibly 6 & attenuated at m ≤ 4--6; null by m = 8 or beyond (m = 16 null in the probe) & \textbf{none expected} \\
tau 0.45 & interior at every m (0.52--0.72 at m 2--8) & ≈ 0.01--0.35 everywhere & m = 2--4 (1.67, 2.39) & null at m 2--8 (window total ≤ 0.09 E); condition 4 fails & \textbf{none expected} \\
tau 0.35, 0.40 & no probe; between the two: interior to a larger m than at 0.30 & lower than at 0.30; the competitor erases at m = 0 (run 3: transient, ratio 0.39 and 0.17) & m = 2--4 & transient or null at small m; attenuated possible where crowding caps the competitor partially & \textbf{possible at m 2--4; not stated as expected} \\
proportional (8b, 8d) & no probe & unknown & unknown & unknown & descriptive only \\
\end{longtable}
}

What would count against the probes and this entry's reading: a robust joint cell at tau 0.30 (the probes put the competitor at the cap by m = 3--4); the competitor at the cap at tau 0.45 for some m ≤ 8; attenuated robust at tau 0.45; a T2 band under priority at m = 1. Each is reported as found. A joint cell at tau 0.35--0.40 does not contradict anything; its existence is the question.

\textbf{Prior knowledge (disclosed).} Everything in §5 and §8; run 1 (S1 at tau 0.30, m = 0: T1 0.64, reach 0.63 → 1.00); run 3 (tau 0.35--0.60 at m = 0: transient from 0.35, null at 0.60 under priority); run 4 (the stalemate family: competitor and ally at reach 1.00 from period 50 in every cell); the Build-phase calibration (\texttt{parameters.\allowbreak{}md} §5: quiet probit 0.152; crowded m = 2 ratio 2.25 under priority; m = 0: 1.81, m = 1: 2.16, all with the ally present); the T1 probe note §2 (static / reaction split of the single- and two-actor fields) and §7 (the ten-replicate probes at tau 0.30, 0.45, 0.60 × m 1--16, listed above). No run of \texttt{crowded\_\allowbreak{}headroom}, of the competitor-plus-bystanders field without the ally at n\_\allowbreak{}rep 100, or of \texttt{levels.\allowbreak{}parquet} outside the Build-phase targets runs exists in any run folder (every \texttt{manifest.\allowbreak{}json} under \texttt{code/\allowbreak{}outputs/\allowbreak{}} checked on 2026-10-31).

Commands (from \texttt{code/\allowbreak{}}, M4, device \texttt{mps}; run the block with \texttt{bash}; after the \texttt{nudgefield.\allowbreak{}levels} tool is committed and the author has signed off this entry; the levels re-run follows each field-experiment call):

\begin{Verbatim}[breaklines,breakanywhere,fontsize=\scriptsize]
R="run.n_rep=100 population.n=20000 run.seed_base=20261029 run.device=mps"
O=outputs/explore_20261029/run8
uv run python -m nudgefield scenario=crowded_headroom design=pilot_rollout     $R out_dir=$O/crowded_headroom_pilot_rollout
uv run python -m nudgefield scenario=crowded_headroom design=pilot_rollout     $R run.attention=proportional out_dir=$O/crowded_headroom_pilot_rollout_proportional
uv run python -m nudgefield scenario=crowded_headroom design=field_experiment  $R out_dir=$O/crowded_headroom_field_experiment
uv run python -m nudgefield scenario=crowded_headroom design=field_experiment  $R run.attention=proportional out_dir=$O/crowded_headroom_field_experiment_proportional
uv run python -m nudgefield scenario=quiet design=field_experiment             $R out_dir=$O/quiet_field_experiment
uv run python -m nudgefield scenario=quiet design=field_experiment             $R run.attention=proportional out_dir=$O/quiet_field_experiment_proportional
for d in crowded_headroom_field_experiment crowded_headroom_field_experiment_proportional quiet_field_experiment quiet_field_experiment_proportional; do
  uv run python -m nudgefield.levels $O/$d
done
\end{Verbatim}

Sign-off: this entry stops here for the author's sign-off of the grid, the criterion (conditions 1--5), the decision rule and the expectations, before the levels tool is built and the calls are run.

\textbf{Signed off by the author 2026-10-31 as written} (``ok proceed'').

\textbf{Stated deviation (2026-10-31, before the run): device and machine.} The entry registers the M4 (\texttt{run.\allowbreak{}device} mps). The first attempt on the M4 overheated the laptop after one call's four batches (author's instruction: stop and move to Modal); every partial M4 directory was discarded (none under the repository, none used). All six calls and the four levels re-runs run on Modal instead (\texttt{code/\allowbreak{}scripts/\allowbreak{}modal\_\allowbreak{}run8.\allowbreak{}py}: a shallow clone of the repository at the commit that holds this paragraph, \texttt{uv\ sync\ -\allowbreak{}\/\allowbreak{}-\allowbreak{}frozen}, one T4 GPU container per call, \texttt{run.\allowbreak{}device=\allowbreak{}cuda}, torch 2.14.1+cu130, driver 610.57; outputs on the volume \texttt{multiactorbpp-\allowbreak{}run8}, fetched with \texttt{modal\ volume\ get}). Every other setting is unchanged. Consequences: the batch seeds do not depend on the device, but the draws do (a cuda stream is not an mps stream), so the cells are not pairable with the M4 probes of the T1 probe note §7 (ten replicates, prior knowledge only); all six calls share one device, so the pairing along tau within a batch and the levels re-runs (same device as the manifest) hold as registered. The T2 pin of the slow test (2.25) is an mps value and is not touched by this run. Order: the \texttt{nudgefield.\allowbreak{}levels} tool (test-first, reviewed), then the six calls and four levels re-runs at the tool's commit, then \texttt{scripts/\allowbreak{}calibration\_\allowbreak{}map.\allowbreak{}py} and the results note \texttt{docs/\allowbreak{}analysis/\allowbreak{}\textless{}date\textgreater{}-\allowbreak{}run8-\allowbreak{}calibration-\allowbreak{}map.\allowbreak{}md}, then the decision rule.

Run 2026-10-31 at 59e8b74 (Modal T4, cuda; the levels tool's merge c631cbc plus \texttt{scripts/\allowbreak{}modal\_\allowbreak{}run8.\allowbreak{}py} and the device paragraph, no change under \texttt{code/\allowbreak{}src/\allowbreak{}nudgefield}); results in \texttt{docs/\allowbreak{}analysis/\allowbreak{}2026-\allowbreak{}10-\allowbreak{}31-\allowbreak{}run8-\allowbreak{}calibration-\allowbreak{}map.\allowbreak{}md}.

\subsubsection*{15. Run 4h: attention load in the headroom family (registered 2026-10-31 evening; feeds F3b from the headroom family, ED4, the T2 recalibration; run 4 becomes the stalemate comparison)}

Purpose: the decision rule of §14, branch ``joint cell exists'' (run 8: \texttt{docs/\allowbreak{}analysis/\allowbreak{}2026-\allowbreak{}10-\allowbreak{}31-\allowbreak{}run8-\allowbreak{}calibration-\allowbreak{}map.\allowbreak{}md}). The calibrated point of the headroom family is competitor \texttt{nudge.\allowbreak{}tau} 0.35 with m = 2 bystanders at budget 1.0 (T2 ratio 1.508, at the band's lower edge; the adjacent cell m = 3 gives 1.733; T1 0.640 and 0.698; competitor burn-in reach 0.735 and 0.828; attenuated 100 of 100). Run 4h maps this family over bystander count × attention budget, as run 4 mapped the stalemate family, so that F3b can show crowding and the competitor's reaction on one map, and the two regimes of crowding (the competitor's saturation boundary and the null boundary) appear as two contours.

Scenario file \texttt{crowded\_\allowbreak{}headroom\_\allowbreak{}load} (new, this commit; checked: loads and expands to nine batches of 800 in each design): the S1 competitor at \texttt{nudge.\allowbreak{}tau} 0.35 (thermostat, gain 2, goal\_\allowbreak{}offset −0.03, lag 2, reach0 0) plus bystanders (gain 0, reach0 0.6) with \texttt{population:\allowbreak{}\ \{budget\_\allowbreak{}mean:\allowbreak{}\ {[}0.\allowbreak{}35,\ 0.\allowbreak{}5,\ 0.\allowbreak{}7,\ 1.\allowbreak{}0,\ 1.\allowbreak{}4,\ 2.\allowbreak{}0,\ 2.\allowbreak{}8,\ 4.\allowbreak{}0{]}\}} as the slice and \texttt{bystander.\allowbreak{}count} ∈ \{0, 1, 2, 3, 4, 6, 8, 12, 16\} as the robustness axis (run 4's grid plus m = 3, ruling 2); \texttt{sweep:\allowbreak{}\ {[}population.\allowbreak{}budget\_\allowbreak{}mean,\ bystander.\allowbreak{}count{]}}; 72 points; both designs; no ally. The T2 point (budget 1.0, m = 2) and the run 8 column (budget 1.0, m 1--8 minus 3) are on the grid; m = 0 at budget 1.0 is run 3's tau 0.35 column at τ\_\allowbreak{}auto 0 (transient, ratio 0.39).

Settings: §0 (\texttt{run.\allowbreak{}n\_\allowbreak{}rep} 100, \texttt{population.\allowbreak{}n} 20,000, \texttt{run.\allowbreak{}seed\_\allowbreak{}base} 20261029, \texttt{fe\_\allowbreak{}reach} 0.4), \texttt{run.\allowbreak{}max\_\allowbreak{}batch\_\allowbreak{}runs} 0 (one batch per m: 8 budgets × 100 = 800 runs), class rules R9, both attention rules. Machine: the DGX Spark (ruling 3, below); \texttt{run.\allowbreak{}max\_\allowbreak{}batch\_\allowbreak{}runs} 0 (nine unchunked batches of 800 per call). Cost from run 4's Spark rates (0.806 s per run under priority at m = 16, 0.160 s under proportional): about 1.6 h per priority call, 20 min per proportional call, about 4 h in sequence. Memory: run 4's m = 16 priority batch peaked at 19.4 GiB of 124 GB.

Calls: 4ha pilot → rollout priority; 4hb pilot → rollout proportional; 4hc field experiment priority; 4hd field experiment proportional. Outputs \texttt{code/\allowbreak{}outputs/\allowbreak{}explore\_\allowbreak{}20261029/\allowbreak{}run4h/\allowbreak{}crowded\_\allowbreak{}headroom\_\allowbreak{}load\_\allowbreak{}\textless{}design\textgreater{}{[}\_\allowbreak{}proportional{]}/\allowbreak{}}. Run 7's \texttt{targets} producer on all four (the field-experiment calls with their pilot → rollout twins) for the constraint region; no levels re-run (T2 is not a per-cell criterion of this map).

Range check: see ruling 2 (nine counts; the pairs below m = 4 need a triple of adjacent points to span 10\% of the range; from m = 4 upward any two adjacent points do).

\textbf{Machine (ruling 3): the DGX Spark}, the machine run 4 was made on (GB10, cuda, 124 GB unified memory; run 4 peaked at 19.4 GiB and took 3 h 40 min for four calls). Unchunked batches with the standard seed, \texttt{run.\allowbreak{}device} cuda, one tmux job running the four calls in sequence (about four hours), then \texttt{targets} on the M4 after scp. Before the run: ship a bundle of \texttt{main} at the signed-off commit to the Spark (\texttt{git\ bundle\ create}, scp over \texttt{\textasciitilde{}/\allowbreak{}nudgefield.\allowbreak{}bundle}, \texttt{git\ fetch\ origin}, \texttt{git\ checkout\ -\allowbreak{}B\ main\ origin/\allowbreak{}main}, \texttt{uv\ sync\ -\allowbreak{}\/\allowbreak{}-\allowbreak{}frozen}), and check \texttt{git\ status} clean under \texttt{src/\allowbreak{}nudgefield}, \texttt{conf/\allowbreak{}nudgefield} and \texttt{uv.\allowbreak{}lock} (the parked housekeeping files are outside that scope). Thermal guard: a logger writes \texttt{nvidia-\allowbreak{}smi} temperature, power and utilisation every 60 s to \texttt{code/\allowbreak{}outputs/\allowbreak{}explore\_\allowbreak{}20261029/\allowbreak{}run4h/\allowbreak{}gpu.\allowbreak{}log} for the whole job, and the job waits before each call while the GPU reads above 85 °C; the log's maximum is quoted in the results note's provenance.

\textbf{Expected classes, stated before running} (sources: run 8's tau 0.35 row at budget 1.0; run 4 at other budgets; run 3's τ\_\allowbreak{}auto 0 column; the two-regimes reading of the T1 probe note §8). - Budget 1.0 (the run 8 column, other draws): m = 0 transient (run 3: ratio 0.39, robust); m 1 transient or attenuated (run 8: transient 75); m 2--4 attenuated, competitor interior to m 4 and at the cap from m 6; T1 \textless{} 0.8 at m ≤ 3; null not before m 12 under priority (run 8: attenuated at 8, ratio 0.33); under proportional null from m 8 (run 8: null 100 at m 8). - Smaller budgets (0.35--0.7): the static loss grows and the competitor saturates at smaller m (its nudge loses attention with ours); null from smaller m than at 1.0 under both rules (run 4's pattern); the T1 region shrinks to m ≤ 1--2. - Larger budgets (1.4--4.0): the competitor keeps headroom to larger m, our effect is less crowded, and the competitor with tau 0.35 erases more of it where it keeps headroom: transient at small m (run 3's m = 0 at budget 1.0 already transient), attenuated where crowding caps the competitor partially, no null at m ≤ 16 at budget ≥ 2.8. - Field experiment: as the pilot → rollout twin in the static regime; blind\_\allowbreak{}site possible at small m and large budgets where the competitor keeps headroom and the trial contrast stays near E\_\allowbreak{}fe (not expected at budget ≤ 1.0: the trial contrast is already attenuated there). - Two contours: the competitor's saturation boundary (burn-in reach ≥ 0.99) runs from small m at small budgets to large m at large budgets; the null boundary lies beyond it at every budget. T1 holds only inside the saturation boundary.

What would count against: a cell with the competitor interior where null is robust (reaction present but effect gone would need the competitor to erase, i.e.~transient, not null); the saturation boundary at larger m under proportional than under priority; T1 holding where the competitor is at the cap.

Table v2 rows: \texttt{F3b\_\allowbreak{}headroom\_\allowbreak{}pilot\_\allowbreak{}rollout} and \texttt{F3b\_\allowbreak{}headroom\_\allowbreak{}field\_\allowbreak{}experiment} walks per budget; tipping rows for the null boundary; a new \texttt{saturation} claim kind is not registered (the boundary is reported from the headroom column of the cell table, not as a class).

Decision for F3b after the run: the map is kept if it shows at least two classes inside the constraint region under both rules (brief §6 rule); run 4 (stalemate family) moves to Extended Data as the comparison. \texttt{parameters.\allowbreak{}md} §5 gains the headroom family's T2 calibration beside the stalemate family's (not replacing it).

Sign-off items for the author: (1) the family at tau 0.35, m = 2 as calibrated (or m = 3, the more central cell of the robust pair: a ruling, stated); (2) the grid (same as run 4); (3) Modal with chunking or a larger GPU; (4) the expectations as written.

\textbf{Rulings (author, 2026-10-31 evening, one at a time):} (1) \textbf{A}: the calibrated cell is (tau 0.35, m = 2), the pre-stated rule's answer; the T2 edge (1.508 against 1.5) is stated wherever the calibration is cited, and the m = 3 cell is reported beside it; run 4h's grid covers both. Reason: changing a pre-stated rule after seeing the result, in the direction that strengthens it, is the pattern the registry exists to prevent. (2) \textbf{B}: run 4's grid plus m = 3 (nine counts: 0, 1, 2, 3, 4, 6, 8, 12, 16; 72 cells), so the calibrated pair appears on the map and the saturation boundary gains a point where it lies; every run 4 cell is still present for the side-by-side comparison. Range check: gaps 1, 1, 1, 1, 2, 2, 4, 4 of a range of 16 → 6.25\% ×4, 12.5\% ×2, 25\% ×2; the pairs below m = 4 need a triple to reach 10\%, as in run 4; batches: nine per call (one per m) of 800 runs. (3) \textbf{C}: the Spark, with the thermal guard above; reason: run 4h is run 4's twin and the Spark removes every difference in how the two were made except the field. (4) \textbf{A}: the expectations as written.

Commands (on the Spark, from \texttt{\textasciitilde{}/\allowbreak{}MultiActorBPP/\allowbreak{}code} at the commit that records this entry with \texttt{git\ status} clean under \texttt{src/\allowbreak{}nudgefield}, \texttt{conf/\allowbreak{}nudgefield}, \texttt{uv.\allowbreak{}lock}; in tmux; run the block with \texttt{bash}; the GPU logger and the thermal guard are part of the job):

\begin{Verbatim}[breaklines,breakanywhere,fontsize=\scriptsize]
R="run.n_rep=100 population.n=20000 run.seed_base=20261029 run.device=cuda"
O=outputs/explore_20261029/run4h
mkdir -p $O
nvidia-smi --query-gpu=timestamp,temperature.gpu,power.draw,utilization.gpu,memory.used --format=csv -l 60 > $O/gpu.log 2>&1 &
LOGGER=$!
cool() { while [ "$(nvidia-smi --query-gpu=temperature.gpu --format=csv,noheader,nounits)" -gt 85 ]; do echo "$(date +%T) GPU above 85 C, waiting"; sleep 60; done; }
cool; uv run python -m nudgefield scenario=crowded_headroom_load design=pilot_rollout    $R out_dir=$O/crowded_headroom_load_pilot_rollout
cool; uv run python -m nudgefield scenario=crowded_headroom_load design=pilot_rollout    $R run.attention=proportional out_dir=$O/crowded_headroom_load_pilot_rollout_proportional
cool; uv run python -m nudgefield scenario=crowded_headroom_load design=field_experiment $R out_dir=$O/crowded_headroom_load_field_experiment
cool; uv run python -m nudgefield scenario=crowded_headroom_load design=field_experiment $R run.attention=proportional out_dir=$O/crowded_headroom_load_field_experiment_proportional
kill $LOGGER
\end{Verbatim}

Then on the M4, after \texttt{scp} of the four directories and \texttt{gpu.\allowbreak{}log}: \texttt{targets} on each pilot → rollout directory (\texttt{-\allowbreak{}\/\allowbreak{}-\allowbreak{}x\ population.\allowbreak{}budget\_\allowbreak{}mean\ -\allowbreak{}\/\allowbreak{}-\allowbreak{}y\ bystander.\allowbreak{}count}) and on each field-experiment directory with its pilot → rollout twin (\texttt{-\allowbreak{}\/\allowbreak{}-\allowbreak{}twin}), as in run 7; table v2 rows and the results note \texttt{docs/\allowbreak{}analysis/\allowbreak{}\textless{}date\textgreater{}-\allowbreak{}run4h-\allowbreak{}headroom-\allowbreak{}load.\allowbreak{}md} with deviations first.

\textbf{Signed off by the author 2026-10-31 evening}, rulings 1--4 recorded in \texttt{docs/\allowbreak{}superpowers/\allowbreak{}specs/\allowbreak{}2026-\allowbreak{}10-\allowbreak{}31-\allowbreak{}run4h-\allowbreak{}draft-\allowbreak{}entry.\allowbreak{}md}: (1) calibrated cell (0.35, 2) as the rule gives, edge stated; (2) run 4's grid plus m = 3; (3) the Spark, unchunked, thermal guard and GPU log; (4) the expectations as written.

\textbf{Run (2026-10-31 evening).} The four calls ran in the registered order in one tmux job on the Spark at ce1107d (\texttt{git\ status} clean in scope; \texttt{\textasciitilde{}/\allowbreak{}run4h.\allowbreak{}sh}, log \texttt{\textasciitilde{}/\allowbreak{}run4h.\allowbreak{}log}), 17:36--19:43 local time (the Spark's clock reads 2026-11-01 00:35--02:43: it runs seven hours ahead of local time, measured after the run), 7,641 s in all (2 h 7 min against the 4 h estimate, which used run 4's m = 16 rate for every batch); every manifest complete, \texttt{git\_\allowbreak{}dirty} false, cuda, burn-in settled; \texttt{gpu.\allowbreak{}log} maximum 77 °C (mean 69.7 °C, 128 samples), the guard never waited; priority peak memory 18.5 GiB. \texttt{targets} on the four as above. Results: \texttt{docs/\allowbreak{}analysis/\allowbreak{}2026-\allowbreak{}10-\allowbreak{}31-\allowbreak{}run4h-\allowbreak{}headroom-\allowbreak{}load.\allowbreak{}md} (\texttt{code/\allowbreak{}scripts/\allowbreak{}headroom\_\allowbreak{}load\_\allowbreak{}map.\allowbreak{}py}). \textbf{Decision taken per the rule above:} two classes (attenuated, transient) inside the constraint region under both rules, so F3b is drawn from run 4h and run 4 moves to Extended Data as the stalemate comparison.

\subsubsection*{16. Run 9: the entry ticket, clean (registered and signed off 2026-11-01 evening; run of record 20:00--20:03 at c220bcf; feeds ED3)}

Purpose: the brief's §8 ED3 and §9 run 9. The Build-phase targets runs (\texttt{code/\allowbreak{}outputs/\allowbreak{}targets\_\allowbreak{}20261027\_\allowbreak{}201230/\allowbreak{}}) carry \texttt{git\_\allowbreak{}dirty:\allowbreak{}\ true} (\texttt{parameters.\allowbreak{}md} §7: three files were untracked when they were made); the figure rules (brief §3) admit only clean run directories. Run 9 re-runs the targets suite (\texttt{code/\allowbreak{}tests/\allowbreak{}nudgefield/\allowbreak{}test\_\allowbreak{}targets.\allowbreak{}py}, the \texttt{slow} tests: T1 per site and rule, T2 under both rules, T3, T4, plus the crowded calibration check) at a clean commit, writing every table the suite produces to Parquet in one registered directory, so that ED3's four panels (T1--T4) and the T3 table come from clean provenance.

\textbf{The one input change:} \texttt{T4\_\allowbreak{}POINTS} 7 → 15 (the inverted U against control take-up on a 15-level grid, brief §8); \texttt{T4\_\allowbreak{}TAKE\_\allowbreak{}UP} (0.1, 0.8), \texttt{T4\_\allowbreak{}TAU} 0.5, \texttt{T4\_\allowbreak{}SD\_\allowbreak{}A} 0.8 and the oracle Φ(−β/2) unchanged, so the 15 \texttt{mu} values are equally spaced between the same ends. No threshold moves: \texttt{T1\_\allowbreak{}MAX} 0.8, \texttt{T2\_\allowbreak{}BAND} {[}1.5, 2.5{]}, \texttt{T2\_\allowbreak{}PIN} (m = 2, 2.25, ±0.01), \texttt{T3\_\allowbreak{}FACTOR} 3, \texttt{T4\_\allowbreak{}PEAK\_\allowbreak{}MAX} 0.5, \texttt{T4\_\allowbreak{}ORACLE\_\allowbreak{}TOL} 0.05.

Settings: as the suite is written (\texttt{N\_\allowbreak{}POP} 20,000, n\_\allowbreak{}rep 10, the suite's seeds; \texttt{priority} default, T1 and T2 also under \texttt{proportional}; \texttt{crowded.\allowbreak{}yaml} at its calibrated m). Machine: the M4 (\texttt{mps}); the earlier slow suite took about 3 min for 15 tests, T4 at 15 levels about doubles its own part. Output: \texttt{code/\allowbreak{}outputs/\allowbreak{}targets\_\allowbreak{}\textless{}YYYYMMDD\_\allowbreak{}HHMMSS\textgreater{}/\allowbreak{}} through the environment variable \texttt{NUDGEFIELD\_\allowbreak{}TARGETS\_\allowbreak{}OUT} (the suite's \texttt{TARGETS\_\allowbreak{}OUT} constant names it; the suite routes every run directory and \texttt{\_\allowbreak{}save} there; an absolute path, since the tests may change directory). The tables (\texttt{T1\_\allowbreak{}\textless{}site\textgreater{}\_\allowbreak{}\textless{}rule\textgreater{}.\allowbreak{}parquet}, \texttt{T2\_\allowbreak{}\textless{}rule\textgreater{}.\allowbreak{}parquet}, \texttt{T3.\allowbreak{}parquet}, \texttt{T4.\allowbreak{}parquet}, the per-run \texttt{levels.\allowbreak{}parquet}) and the run manifests are committed with \texttt{git\ add\ -\allowbreak{}f} (small; registered provenance); the effects Parquet stays ignored.

\textbf{Expected, stated before running.} Every test passes. T1 ratios of replicate means within 0.05 of \texttt{parameters.\allowbreak{}md} §6 at the same seeds (S1/priority 0.741, S1/proportional 0.671, S2/priority 0.621, S2/proportional 0.738), with the per-replicate spread as disclosed there (4 of 10 S1 replicates ≥ 0.8 under priority). T2 in the band under priority with the m = 2 ratio within 0.01 of 2.25, and \textgreater{} 1 under proportional. T3 between-site sd \textgreater{} 3 × the mean within-site s.e. T4: an interior peak of the rollout ATE on the 15-level grid, at a control take-up below 0.5 and within 0.05 of Φ(−β/2) (β = 0.5 / √(1 + 0.8²)); the 7 old levels are a subset in value only up to rounding, so the peak location is re-read on the finer grid.

What would count against: any test failing; a T1 ratio moving by more than 0.05 at the same seeds (that would mean the code path changed, not the field); a T4 peak that moves across the 0.5 line or away from the oracle on the finer grid.

Command (from \texttt{code/\allowbreak{}}, at the commit that records this entry with \texttt{git\ status} clean under \texttt{src/\allowbreak{}nudgefield}, \texttt{conf/\allowbreak{}nudgefield}, \texttt{tests/\allowbreak{}nudgefield/\allowbreak{}test\_\allowbreak{}targets.\allowbreak{}py}, \texttt{uv.\allowbreak{}lock}; \texttt{bash}):

\begin{Verbatim}[breaklines,breakanywhere,fontsize=\scriptsize]
export NUDGEFIELD_TARGETS_OUT=$PWD/outputs/targets_$(date +%Y%m%d_%H%M%S)
mkdir -p $NUDGEFIELD_TARGETS_OUT
uv run pytest -m slow tests/nudgefield/test_targets.py -q -p no:cacheprovider 2>&1 | tee $NUDGEFIELD_TARGETS_OUT/pytest.log
\end{Verbatim}

Then: check every \texttt{manifest.\allowbreak{}json} under \texttt{\$NUDGEFIELD\_\allowbreak{}TARGETS\_\allowbreak{}OUT} (status complete, \texttt{git\_\allowbreak{}dirty} false, one commit); \texttt{git\ add\ -\allowbreak{}f} the tables, manifests and \texttt{pytest.\allowbreak{}log}; \texttt{parameters.\allowbreak{}md} §7 gains the clean run's paragraph; the ledger row; ED3 (\texttt{figures/\allowbreak{}ed3.\allowbreak{}py}) reads only this directory.

Sign-off items for the author: (1) \texttt{T4\_\allowbreak{}POINTS} 15 as the brief says (or keep 7 and draw ED3's T4 on the Build grid); (2) the expectations as written; (3) committing the tables and manifests with \texttt{-\allowbreak{}f}.

\textbf{Signed off by the author 2026-11-01 evening, as written} (one question, three items): \texttt{T4\_\allowbreak{}POINTS} 15; the expectations as drafted; tables and manifests force-added. The run follows the Task 0 refactor commit of plan part 3 (so that the tree is clean under \texttt{src/\allowbreak{}nudgefield}) and precedes the F4 module work.

\textbf{Deviation (2026-11-01, 19:52--19:55 local, before the run of record).} The command block as signed off named the output variable \texttt{TARGETS\_\allowbreak{}OUT}; the suite reads \texttt{NUDGEFIELD\_\allowbreak{}TARGETS\_\allowbreak{}OUT} (the constant \texttt{TARGETS\_\allowbreak{}OUT} in \texttt{test\_\allowbreak{}targets.\allowbreak{}py} holds that name). The first execution therefore ran at 0639070 (clean in scope; 9 passed in 179 s) but wrote its tables and 33 run manifests to pytest's temporary directory (\texttt{pytest-\allowbreak{}81}), not to the registered directory. Nothing about the inputs, the seeds or the suite changed. The command block in this section is corrected (the variable's real name, and an absolute \texttt{\$PWD/\allowbreak{}.\allowbreak{}.\allowbreak{}.\allowbreak{}} path because the tests may change directory); the corrected block was not re-signed: it changes no input, and this note discloses it. The run of record is the re-run by the corrected command at the commit holding this note (c220bcf); the first execution's tables, recovered from the temporary directory, serve only as a cross-check and are not committed.

\textbf{Run of record (2026-11-01, 20:00--20:03 local, M4 \texttt{mps}).} \texttt{uv\ run\ pytest\ -\allowbreak{}m\ slow\ tests/\allowbreak{}nudgefield/\allowbreak{}test\_\allowbreak{}targets.\allowbreak{}py} at c220bcf (\texttt{git\ status} clean in scope), \texttt{NUDGEFIELD\_\allowbreak{}TARGETS\_\allowbreak{}OUT=\allowbreak{}code/\allowbreak{}outputs/\allowbreak{}targets\_\allowbreak{}20261101\_\allowbreak{}200002/\allowbreak{}}: 9 passed in 179 s; 33 run manifests, every one \texttt{status} complete, \texttt{git\_\allowbreak{}dirty} false, commit c220bcf. Against the expectations: T1 ratios of replicate means 0.741 / 0.671 / 0.621 / 0.738 (S1/priority, S1/proportional, S2/priority, S2/proportional), equal to \texttt{parameters.\allowbreak{}md} §6 to three decimals, with 4 / 3 / 1 / 6 of 10 replicates ≥ 0.8 as disclosed there; T2 quiet / crowded probit ratio 2.2506 under priority (band {[}1.5, 2.5{]}; pin 2.25 ± 0.01 met) and 3.3614 under proportional (\textgreater{} 1); T3 between-site sd 0.0132 against 3 × within-site s.e. 0.0036; T4 on 15 levels: interior peak at j = 7, control take-up 0.414 (\textless{} 0.5), oracle Φ(−β/2) = 0.423 (\textbar difference\textbar{} 0.009 ≤ 0.05). Every expectation met; nothing on the count-against list. The first execution's recovered tables (the deviation above) are identical to the run of record's T1--T4 and calibration tables (max \textbar difference\textbar{} 0). Committed with \texttt{-\allowbreak{}f}: the T1--T4 and calibration tables, the per-run \texttt{levels.\allowbreak{}parquet}, the 33 manifests, \texttt{pytest.\allowbreak{}log}, \texttt{run9.\allowbreak{}log}; the per-run \texttt{effects.\allowbreak{}parquet}, \texttt{classes.\allowbreak{}parquet}, \texttt{robust.\allowbreak{}parquet} stay ignored (5.8 MB in all on disk). ED3 reads this directory only.

\subsubsection*{17. Run 10: the fine T4 grid near the peak (drafted 2026-11-01 night for sign-off; feeds ED3 panel d)}

Purpose: ED3's T4 verdict (registry §16, run 9) holds at grid resolution only: the peak sits at take-up level 7 (0.414) with level 6 (0.356) only 0.00005 lower in ATE, and level 6 would lie 0.067 from the oracle Φ(−β/2) = 0.423, outside the 0.05 tolerance; the coarse grid's step near the peak (0.058) is wider than the tolerance and the table carries no ATE spread. The author ruled (2026-11-01 night) for a registered finer grid near the peak rather than a caption caveat.

\textbf{The addition (no change to any existing input):} a second T4 test, \texttt{test\_\allowbreak{}T4\_\allowbreak{}peak\_\allowbreak{}location\_\allowbreak{}fine}, on \texttt{T4\_\allowbreak{}FINE\_\allowbreak{}POINTS} = 9 levels equally spaced in control take-up between \texttt{T4\_\allowbreak{}FINE\_\allowbreak{}TAKE\_\allowbreak{}UP} = (0.30, 0.50) (step 0.025, half the tolerance; the oracle 0.4226 lies between levels 0.400 and 0.425), under the same scenario name \texttt{T4\_\allowbreak{}auto\_\allowbreak{}quiet} and therefore the same batch seeds as the coarse grid; both T4 tables gain \texttt{ate\_\allowbreak{}sd} (sd of the per-replicate window-mean rollout effect) and \texttt{n\_\allowbreak{}rep}. Thresholds unchanged: \texttt{T4\_\allowbreak{}PEAK\_\allowbreak{}MAX} 0.5, \texttt{T4\_\allowbreak{}ORACLE\_\allowbreak{}TOL} 0.05; the coarse grid (\texttt{T4\_\allowbreak{}TAKE\_\allowbreak{}UP} (0.1, 0.8), \texttt{T4\_\allowbreak{}POINTS} 15) untouched (a fast test pins it).

Settings: as §16 (the suite as written; n\_\allowbreak{}rep 10, N 20,000, the suite's seeds; M4 \texttt{mps}). The whole suite is re-run into a new directory \texttt{code/\allowbreak{}outputs/\allowbreak{}targets\_\allowbreak{}\textless{}YYYYMMDD\_\allowbreak{}HHMMSS\textgreater{}/\allowbreak{}} so that ED3 reads one directory at one clean commit; the coarse tables (T1--T4, calibration) are expected identical to run 9's. Committed with \texttt{-\allowbreak{}f}: every table, the per-run \texttt{levels.\allowbreak{}parquet}, the 33 + 9 manifests, \texttt{pytest.\allowbreak{}log}, \texttt{run10.\allowbreak{}log}.

\textbf{Expected, stated before running.} Every test passes, the coarse ones with tables identical to run 9's to machine precision. The fine grid's peak is interior and within 0.025 of the oracle in take-up (one grid step), below 0.5. The union of the two grids has its peak on the fine grid within the tolerance. \texttt{ate\_\allowbreak{}sd} on the fine grid is larger than the ATE differences between neighbouring fine levels (which is why the verdict is read on the grid, not between levels).

What would count against: the fine grid's peak outside the tolerance, or not interior; a coarse table differing from run 9's.

Command (from \texttt{code/\allowbreak{}}, at the commit recording this entry with \texttt{git\ status} clean under \texttt{src/\allowbreak{}nudgefield}, \texttt{conf/\allowbreak{}nudgefield}, \texttt{tests/\allowbreak{}nudgefield/\allowbreak{}test\_\allowbreak{}targets.\allowbreak{}py}, \texttt{uv.\allowbreak{}lock}; \texttt{bash}):

\begin{Verbatim}[breaklines,breakanywhere,fontsize=\scriptsize]
export NUDGEFIELD_TARGETS_OUT=$PWD/outputs/targets_$(date +%Y%m%d_%H%M%S)
mkdir -p $NUDGEFIELD_TARGETS_OUT
uv run pytest -m slow tests/nudgefield/test_targets.py -q -p no:cacheprovider 2>&1 | tee $NUDGEFIELD_TARGETS_OUT/pytest.log
\end{Verbatim}

Then: check every \texttt{manifest.\allowbreak{}json} under \texttt{\$NUDGEFIELD\_\allowbreak{}TARGETS\_\allowbreak{}OUT} (complete, clean, one commit); compare the coarse tables with run 9's; \texttt{git\ add\ -\allowbreak{}f} the artefacts; \texttt{parameters.\allowbreak{}md} §7; the ledger row; ED3 (\texttt{figures/\allowbreak{}ed3.\allowbreak{}py}) reads only this directory.

Sign-off items for the author: (1) the fine grid as stated (9 levels, 0.30--0.50); (2) the expectations as written; (3) the whole suite re-run into one new directory.

\textbf{Signed off by the author 2026-11-01 night, as written} (one question, three items). Implementation note recorded before the run: the coarse grid \texttt{T4\_\allowbreak{}MU} is equally spaced in \emph{mu}, so the fine grid is built as stated here, equally spaced in \emph{take-up} (mu\_\allowbreak{}j = s·Φ⁻¹(0.30 + 0.025 j), rounded to 3 dp; take-ups 0.300--0.500; the oracle between fine levels 4 and 5); both T4 tables gain \texttt{ate\_\allowbreak{}sd} and \texttt{n\_\allowbreak{}rep} after \texttt{ate}, so run 10's coarse table is compared with run 9's by column name. The slow suite now has 10 tests.

\textbf{Run of record (2026-11-01, 22:36--22:53 local wall clock, M4 \texttt{mps}; the suite's own timer 190 s --- the manifests show the machine idle between tests, 22:39--22:47 and 22:47--22:53).} The registered command at 4b1b94e (\texttt{git\ status} clean in scope), \texttt{NUDGEFIELD\_\allowbreak{}TARGETS\_\allowbreak{}OUT=\allowbreak{}code/\allowbreak{}outputs/\allowbreak{}targets\_\allowbreak{}20261101\_\allowbreak{}223633/\allowbreak{}}: 10 passed in 190 s; 42 run manifests, every one \texttt{status} complete, \texttt{git\_\allowbreak{}dirty} false, commit 4b1b94e. The coarse tables (T1 × 4, T2 × 2, T3, calibration, T4) equal run 9's on every shared column to machine precision (max \textbar difference\textbar{} 0). \textbf{Fine grid (9 levels, take-up 0.300--0.500):} ATE 0.1336, 0.1356, 0.1368, 0.1373, 0.1372, 0.1367, 0.1358, 0.1344, 0.1323; \texttt{ate\_\allowbreak{}sd} 0.0009--0.0012 (s.e. of the mean about 0.0003 at n\_\allowbreak{}rep 10). Peak at level 3, take-up 0.376 (interior; below 0.5); \textbar peak − oracle\textbar{} = 0.047, inside the tolerance 0.05, so the test passes; the union's peak is the same fine level. \textbf{Deviations from the expectations:} the fine peak is not within one grid step (0.025) of the oracle 0.423 --- it is 0.047 away, two steps. Reading, not a rule (amended 2026-11-01 night, after ED3 was drawn: the first version gave the plateau as 0.35--0.43 with differences ≤ 0.0005; corrected to ED3's computed one-s.e. range; the run's numbers are unchanged): the top of the inverted U is flat --- within one replicate s.e. (about 0.0003) of the maximum over take-up 0.356--0.414 (ED3's computed plateau), and within 0.0006 of it over 0.351--0.426 --- so the peak's location is resolved only to that plateau at n\_\allowbreak{}rep 10; the verdict ``holds'' stands on the registered tolerance, at its edge (the plateau's lower end, 0.356, would lie 0.067 from the oracle). The expectation that \texttt{ate\_\allowbreak{}sd} exceeds the neighbouring differences is met only near the peak, not at the grid's ends --- a second deviation (amended 2026-11-01 night, found in the final whole-branch review; the first version read ``is met (0.0009 against ≤ 0.0021 at the grid's ends, ≤ 0.0006 on the plateau)'', which contradicts itself): the eight neighbouring differences are 0.0019, 0.0012, 0.0005, 0.0001, 0.0005, 0.0009, 0.0014, 0.0021 against \texttt{ate\_\allowbreak{}sd} 0.0009--0.0012, so the four inner ones (take-up 0.351--0.451) lie below both levels' \texttt{ate\_\allowbreak{}sd} and the four outer ones (both pairs at each end) exceed it. No bearing on the verdict (the test reads the argmax; the count-against list did not fire), and the reading above (the plateau) is where the expectation holds. Nothing on the count-against list. Committed with \texttt{-\allowbreak{}f}: every table (both T4 tables with \texttt{ate\_\allowbreak{}sd}, \texttt{n\_\allowbreak{}rep}), the per-run \texttt{levels.\allowbreak{}parquet}, the 42 manifests, \texttt{pytest.\allowbreak{}log}, \texttt{run10.\allowbreak{}log}; effects/classes/robust Parquet ignored (6.9 MB on disk). ED3 reads this directory only; run 9's directory stays as the record of §16.

\subsubsection*{18. Run 11: the pilot's share of the site (drafted 2026-11-12 for sign-off; answers the final review's point M4)}

Purpose: the referee-style review of 2026-11-12 (point M4) notes that ``a pilot cannot forecast the rollout'' rests on the pilot being a small slice (2\%) of the population the competitor watches, and that the share was never varied. This run varies it. The question is not only whether a larger pilot provokes the competitor (it must, in proportion to the rate it moves) but what the pilot then \emph{reports}: the competitor's response reaches treated and untreated people alike, so the pilot's treated − control contrast should not fall as the share rises; the pilot should instead turn into a blind site, with its site-level gain falling while its contrast holds. If so, a larger pilot does not forecast the rollout through its own comparison, and the paper's claim stands with its condition made explicit. For a ratcheting competitor a large pilot should also damage the rollout that follows, because the reach the competitor gains during the pilot is never given back.

Settings: scenarios \texttt{S1\_\allowbreak{}competitor} (thermostat; its sweep \texttt{competitor.\allowbreak{}gain} \{0.5, 2, 5\}), \texttt{S1\_\allowbreak{}persistent\_\allowbreak{}auto} (escalating thermostat; \texttt{competitor.\allowbreak{}escalation\_\allowbreak{}tau} \{0.1, 0.2, 0.3\}) and \texttt{S1\_\allowbreak{}maximiser} (ratcheting; \texttt{competitor.\allowbreak{}escalation\_\allowbreak{}tau} \{0.1, 0.2, 0.3\}); design \texttt{pilot\_\allowbreak{}rollout} with \texttt{design\_\allowbreak{}params.\allowbreak{}pilot\_\allowbreak{}reach} ∈ \{0.02, 0.05, 0.10, 0.20, 0.40, 0.60\} (0.02 is the paper's pilot); \texttt{run.\allowbreak{}attention} priority; \texttt{run.\allowbreak{}n\_\allowbreak{}rep} 100; \texttt{population.\allowbreak{}n} 20,000; \texttt{run.\allowbreak{}seed\_\allowbreak{}base} 20261112; everything else at the scenario's registered values (§1--§2). 3 × 3 × 6 × 100 = 5,400 runs.

Machine: the DGX Spark (GB10, \texttt{cuda}), as runs 3, 4 and 4h, launched over SSH from the Linux machine; the Spark's clone takes the registering commit from a git bundle, as before. Timing from run 4h (28,800 runs in 7,641 s): about 25 min. \textbf{Device, declared in advance:} Figs. 2a, b, d come from runs 1 and 2 on the M4 (\texttt{mps}); draws differ between devices, so the 0.02 column of this run is also the run's own control: its classes and ratios must match Fig. 2a, b, d within the robustness rule (attenuated / transient / reversed, ratios within 0.1 E of the F2 values 0.63, 0.00 end, −0.34), or the device difference is material and the run is reported with that caveat. (A first draft of this entry named the Linux machine's CPU, benchmarked at 0.7 s per run; the author chose the Spark.)

Statistics, fixed in advance, per (scenario, point, share), from \texttt{effects.\allowbreak{}parquet} over the pilot window t ∈ {[}60, 80) and \texttt{classes.\allowbreak{}parquet}: - P1, the pilot's contrast over E: mean of \texttt{trial} over the window and replicates, divided by E (the per-person effect from \texttt{classes.\allowbreak{}E}). - P2, the pilot's site effect over the expected site effect: mean \texttt{total} over the window, divided by \texttt{pilot\_\allowbreak{}reach} × E. - P3, the competitor's response during the pilot: mean \texttt{reach\_\allowbreak{}competitor} over t ∈ {[}60, 80) minus its mean over t ∈ {[}40, 60). - P4, the rollout: \texttt{ratio\_\allowbreak{}window}, \texttt{cls}, and the robust class at each share. - P5, the forecast gap: P1 minus P4's \texttt{ratio\_\allowbreak{}window}.

Expected, stated before running: - E1 (the contrast holds): at every share P1 stays within 0.10 of its value at 0.02, for every scenario and point. A fall of more than 0.10 by share 0.4 would mean a larger pilot does forecast the rollout through its own comparison, against the paper's claim. - E2 (the pilot becomes a blind site): for the thermostat at gain 2 and 5, P2 falls with the share and is below 0.75 by share 0.4; at gain 0.5 the fall is smaller (the slow competitor responds less within twenty periods). - E3 (the response is provoked in proportion): P3 rises monotonically with the share in every thermostat scenario. - E4a (thermostats: the rollout is unchanged): for \texttt{S1\_\allowbreak{}competitor} and \texttt{S1\_\allowbreak{}persistent\_\allowbreak{}auto}, P4's class is the same at every share and \texttt{ratio\_\allowbreak{}window} moves by less than 0.10 (the nudge is off over 80--99 and the thermostat returns towards rest). - E4b (the ratchet remembers): for \texttt{S1\_\allowbreak{}maximiser}, \texttt{ratio\_\allowbreak{}window} falls with the share, by more than 0.10 between 0.02 and 0.6 at \texttt{escalation\_\allowbreak{}tau} 0.1, and the automatic nudge comes on earlier or already during the pilot at the largest shares.

What would count against: E1 failing anywhere (the claim's wording changes to ``a small pilot''); E4a failing (the pilot would then alter the rollout it is meant to forecast, a second mechanism worth reporting).

Prior knowledge, declared: one exploratory run on this machine on 2026-11-12 before this entry was written, \texttt{S1\_\allowbreak{}competitor} at N = 2,000, one replicate, share 0.4, \texttt{seed\_\allowbreak{}base} 1: at gain 2 the pilot's contrast was 0.073 against E 0.053 (above E: the control arm fell), the site effect 0.011 against 0.021 expected (P2 ≈ 0.5), and the competitor's reach rose from 0.68 to 0.89 during the pilot. The expectations above were written after seeing it; they are therefore confirmations at full size, not predictions, for that one cell.

Command (from \texttt{code/\allowbreak{}}, at the commit that records this entry, \texttt{git\ status} clean under \texttt{src/\allowbreak{}nudgefield} and \texttt{conf/\allowbreak{}nudgefield}; \texttt{bash}):

\begin{Verbatim}[breaklines,breakanywhere,fontsize=\scriptsize]
O=outputs/explore_20261112/run11; R="run.n_rep=100 population.n=20000 run.seed_base=20261112 run.device=cuda"
for s in S1_competitor S1_persistent_auto S1_maximiser; do for p in 0.02 0.05 0.10 0.20 0.40 0.60; do
  uv run python -m nudgefield scenario=$s design=pilot_rollout $R design_params.pilot_reach=$p out_dir=$O/${s}_pilot$p
done; done
\end{Verbatim}

Then: check every \texttt{manifest.\allowbreak{}json} (status complete, \texttt{git\_\allowbreak{}dirty} false, one commit, device cuda); a results note \texttt{docs/\allowbreak{}analysis/\allowbreak{}2026-\allowbreak{}11-\allowbreak{}12-\allowbreak{}run11-\allowbreak{}pilot-\allowbreak{}share.\allowbreak{}md} with P1--P5 per cell and E1--E4 as registered and as found; the tracked summary csv beside the runs; the paper's text changes only after the author reads the note.

Sign-off items for the author: (1) the six shares and the three scenarios; (2) n\_\allowbreak{}rep 100 at N = 20,000 on the Spark (\texttt{cuda}, about 25 min), with the 0.02 column as the device control; (3) the expectations E1--E4b as written; (4) that the one exploratory run is declared as prior knowledge.

\textbf{Signed off by the author 2026-11-12 afternoon, as written} (``signed off''; the four items). Launched on the Spark at the signing commit's successor (this note), detached in \texttt{tmux} session \texttt{run11}, log \texttt{code/\allowbreak{}outputs/\allowbreak{}explore\_\allowbreak{}20261112/\allowbreak{}run11/\allowbreak{}launch.\allowbreak{}log}.

\textbf{Run of record (2026-11-12, 15:30--15:56 local, DGX Spark \texttt{cuda}, at 6fd2e39).} 18 calls, 5,400 runs, 25 min 42 s; every manifest complete, \texttt{git\_\allowbreak{}dirty} false, one commit, device cuda. Summaries, manifests and the launch log tracked under \texttt{code/\allowbreak{}outputs/\allowbreak{}explore\_\allowbreak{}20261112/\allowbreak{}run11/\allowbreak{}}; the script \texttt{code/\allowbreak{}scripts/\allowbreak{}analyse\_\allowbreak{}run11.\allowbreak{}py}; the results note \texttt{docs/\allowbreak{}analysis/\allowbreak{}2026-\allowbreak{}11-\allowbreak{}12-\allowbreak{}run11-\allowbreak{}pilot-\allowbreak{}share.\allowbreak{}md}. \textbf{As registered and as found:} E1 met (the pilot's contrast stays within 0.08 of its 2\% value everywhere); E2 met (P2 0.45 and 0.28 at share 0.4); E3 not met for the escalating thermostat at τ auto 0.2 and 0.3 (the competitor escalates during a pilot of 0.4 or more and lowers its reach); E4a not met for the escalating thermostat at τ auto 0.3 (same window ratio, class null instead of transient at shares 0.4--0.6: the automatic nudge switched on during the pilot persists into the rollout); E4b not met (the maximiser's window ratio is identical across shares; its rollout outcome is set by its automatic nudge; the statistic was the wrong one). Not named in advance: P2 is flat in the share until the competitor's limit binds, so a steering competitor absorbs the same fraction of any pilot's site effect. Device control: the 2\% column matches Fig. 2a, b, d within the rule. The hub Mac's copy of the Parquet files is pending (offline at the run's end).

\endgroup

\clearpage
\sisection{the pilot's share of the site}{si:pilotshare}

The article's pilot treats 2\% of the site. Run 11, registered after the final review (section~\ref{si:record}, \S18) and run on 12 November 2026, varies that share over 0.02, 0.05, 0.10, 0.20, 0.40 and 0.60 against the three competitors of Fig.~2: the steering thermostat at gains 0.5, 2 and 5, the escalating thermostat at automatic-nudge strengths 0.1, 0.2 and 0.3, and the ratcheting maximiser at the same three strengths; pilot then rollout, priority sharing, 100 replicates of 20,000 people, seed base 20261112, on the DGX Spark (\texttt{cuda}). Five statistics were fixed before the run, per cell, over the pilot window (periods 60--79) and the replicates: P1, the pilot's treated~$-$~control contrast over $E$; P2, the pilot's site-level effect over the site-level effect expected from treating that share ($\text{share} \times E$); P3, the competitor's mean reach over the pilot window minus its mean over periods 40--59; P4, the rollout's evaluation-window ratio and class; P5, P1 minus P4's ratio (in the tracked summary). Five expectations were registered; two were met.

\begin{itemize}\itemsep0pt
\item E1, met: P1 stays within 0.10 of its value at share 0.02 in every cell (largest move 0.08). A larger pilot does not forecast the rollout through its own comparison.
\item E2, met: for the steering thermostat at gains 2 and 5, P2 falls below 0.75 by share 0.4 (0.45 and 0.28).
\item E3, not met for the escalating thermostat at strengths 0.2 and 0.3: P3 rises with the share up to 0.4 and falls at 0.6. Reading: a pilot of 0.4 or more holds the thermostat at its limit for ten periods within the pilot, so it escalates during the pilot and lowers its deliberative reach; at strength 0.3 and share 0.6 the pilot's site effect is negative (P2 $-0.09$).
\item E4a, not met for the escalating thermostat at strength 0.3: the rollout's window ratio is unchanged across shares (range 0.001) but its class is null instead of transient at shares 0.4 and 0.6 (87 and 100 of 100 runs). Reading: the automatic nudge switched on during the pilot is persistent, so the rollout starts in an escalated field and never reaches 0.75 of $E$.
\item E4b, not met: the maximiser's window ratio is identical across shares; its rollout outcome is set by its automatic nudge, which comes on in every case, so the registered statistic could not move. Its class changes with the share for the same reason as E4a.
\end{itemize}

Not named in advance: P2 is flat in the share until the competitor's limit binds (0.79 at every share for gain 0.5; 0.45 from 0.02 to 0.40 for gain 2; 0.20 up to 0.20 for gain 5). A steering competitor absorbs the same fraction of a pilot's site effect whatever the pilot's size, because it responds in proportion to the gap. The device control: the 0.02 column repeats the settings of Fig.~2a, b and d on \texttt{cuda} against the M4's \texttt{mps}: 0.625 against 0.63, $-0.001$ against 0.00 (end), $-0.346$ against $-0.34$, within the robustness rule.

{\footnotesize
\begin{longtable}{@{}llrrrrl@{}}
\caption{\textbf{Run 11: the pilot's share of the site.} Per cell, means over 100 replicates: P1, the pilot's contrast over $E$; P2, the pilot's site effect over the expected site effect; P3, the competitor's reach gain during the pilot; P4, the rollout's evaluation-window ratio and its modal class, with the share of runs in that class.}\label{tab:pilotshare}\\
\toprule
Competitor, setting & Pilot share & P1 & P2 & P3 & P4 ratio & P4 class (share of runs) \\
\midrule
\endfirsthead
\multicolumn{7}{@{}l}{\emph{Table~\thetable, continued}}\\
\toprule
Competitor, setting & Pilot share & P1 & P2 & P3 & P4 ratio & P4 class (share of runs) \\
\midrule
\endhead
\bottomrule
\endlastfoot
steering, gain 0.5 & 0.02 & 1.00 & 0.79 & 0.08 & 0.62 & attenuated (1.00) \\
 & 0.05 & 0.97 & 0.79 & 0.08 & 0.62 & attenuated (1.00) \\
 & 0.10 & 0.98 & 0.80 & 0.09 & 0.62 & attenuated (1.00) \\
 & 0.20 & 0.98 & 0.80 & 0.11 & 0.62 & attenuated (1.00) \\
 & 0.40 & 0.97 & 0.80 & 0.16 & 0.62 & attenuated (1.00) \\
 & 0.60 & 0.97 & 0.79 & 0.20 & 0.62 & attenuated (1.00) \\
steering, gain 2 & 0.02 & 0.98 & 0.45 & 0.02 & 0.62 & attenuated (1.00) \\
 & 0.05 & 0.98 & 0.45 & 0.03 & 0.62 & attenuated (1.00) \\
 & 0.10 & 0.97 & 0.45 & 0.06 & 0.62 & attenuated (1.00) \\
 & 0.20 & 0.96 & 0.45 & 0.12 & 0.62 & attenuated (1.00) \\
 & 0.40 & 0.96 & 0.45 & 0.24 & 0.62 & attenuated (1.00) \\
 & 0.60 & 0.96 & 0.54 & 0.29 & 0.62 & attenuated (1.00) \\
steering, gain 5 & 0.02 & 1.01 & 0.20 & 0.02 & 0.62 & attenuated (1.00) \\
 & 0.05 & 0.98 & 0.20 & 0.05 & 0.62 & attenuated (1.00) \\
 & 0.10 & 0.98 & 0.20 & 0.09 & 0.62 & attenuated (1.00) \\
 & 0.20 & 0.97 & 0.20 & 0.18 & 0.62 & attenuated (1.00) \\
 & 0.40 & 0.95 & 0.28 & 0.31 & 0.62 & attenuated (1.00) \\
 & 0.60 & 0.96 & 0.48 & 0.33 & 0.62 & attenuated (1.00) \\
ratcheting, $\tau$ (automatic) 0.1 & 0.02 & 1.01 & 0.94 & 0.00 & 0.31 & transient (1.00) \\
 & 0.05 & 0.98 & 0.89 & 0.01 & 0.31 & transient (1.00) \\
 & 0.10 & 0.96 & 0.81 & 0.02 & 0.31 & transient (1.00) \\
 & 0.20 & 0.97 & 0.60 & 0.09 & 0.31 & transient (1.00) \\
 & 0.40 & 0.96 & 0.47 & 0.23 & 0.31 & attenuated (0.87) \\
 & 0.60 & 0.96 & 0.48 & 0.30 & 0.31 & attenuated (1.00) \\
ratcheting, $\tau$ (automatic) 0.2 & 0.02 & 0.95 & 0.91 & 0.00 & -0.02 & transient (1.00) \\
 & 0.05 & 0.98 & 0.88 & 0.01 & -0.02 & transient (1.00) \\
 & 0.10 & 0.97 & 0.78 & 0.03 & -0.02 & transient (1.00) \\
 & 0.20 & 0.96 & 0.58 & 0.09 & -0.02 & transient (1.00) \\
 & 0.40 & 0.96 & 0.46 & 0.23 & -0.02 & null (0.95) \\
 & 0.60 & 0.96 & 0.44 & 0.30 & -0.02 & null (0.99) \\
ratcheting, $\tau$ (automatic) 0.3 & 0.02 & 0.97 & 0.92 & 0.00 & -0.35 & reversed (1.00) \\
 & 0.05 & 0.96 & 0.89 & 0.01 & -0.35 & reversed (1.00) \\
 & 0.10 & 0.95 & 0.81 & 0.02 & -0.35 & reversed (1.00) \\
 & 0.20 & 0.96 & 0.59 & 0.09 & -0.35 & reversed (1.00) \\
 & 0.40 & 0.96 & 0.46 & 0.23 & -0.35 & reversed (1.00) \\
 & 0.60 & 0.94 & 0.40 & 0.30 & -0.35 & reversed (1.00) \\
escalating, $\tau$ (automatic) 0.1 & 0.02 & 0.96 & 0.45 & 0.01 & 0.27 & transient (1.00) \\
 & 0.05 & 0.95 & 0.45 & 0.03 & 0.27 & transient (1.00) \\
 & 0.10 & 0.95 & 0.45 & 0.06 & 0.27 & transient (1.00) \\
 & 0.20 & 0.95 & 0.45 & 0.12 & 0.27 & transient (1.00) \\
 & 0.40 & 0.94 & 0.35 & 0.24 & 0.27 & transient (1.00) \\
 & 0.60 & 0.93 & 0.28 & 0.27 & 0.27 & transient (1.00) \\
escalating, $\tau$ (automatic) 0.2 & 0.02 & 0.97 & 0.45 & 0.01 & -0.00 & transient (1.00) \\
 & 0.05 & 0.96 & 0.45 & 0.03 & -0.00 & transient (1.00) \\
 & 0.10 & 0.96 & 0.45 & 0.06 & -0.00 & transient (1.00) \\
 & 0.20 & 0.95 & 0.45 & 0.12 & -0.00 & transient (1.00) \\
 & 0.40 & 0.95 & 0.27 & 0.23 & 0.00 & transient (0.98) \\
 & 0.60 & 0.94 & 0.09 & 0.22 & 0.00 & transient (0.98) \\
escalating, $\tau$ (automatic) 0.3 & 0.02 & 0.98 & 0.45 & 0.02 & -0.00 & transient (1.00) \\
 & 0.05 & 0.98 & 0.45 & 0.03 & -0.00 & transient (1.00) \\
 & 0.10 & 0.97 & 0.45 & 0.06 & -0.00 & transient (1.00) \\
 & 0.20 & 0.98 & 0.45 & 0.12 & -0.00 & transient (1.00) \\
 & 0.40 & 0.94 & 0.14 & 0.23 & 0.00 & null (0.87) \\
 & 0.60 & 0.91 & -0.09 & 0.17 & 0.00 & null (1.00) \\
\end{longtable}
}

\clearpage
\bibliographystyleSI{unsrtnat}
\bibliographySI{refs-si}

\end{document}